\documentclass[a4paper,11pt]{article}
\usepackage{jheppub} 
\usepackage{lineno}
\usepackage{color, amssymb}
\usepackage{enumitem}
\usepackage[dvipsnames]{xcolor}
\usepackage{hyperref}
\definecolor{amaranth}{rgb}{0.9, 0.17, 0.31}
\definecolor{forestgreen(web)}{rgb}{0.13, 0.55, 0.13}
\definecolor{blue(munsell)}{rgb}{0.0, 0.5, 0.69}
\definecolor{lav}{rgb}{0.45, 0.31, 0.59}
\definecolor{bblue}{rgb}{0.0, 0.58, 0.71}
\hypersetup{
pdfstartview={FitH}, 
pdftitle={}, 
colorlinks=true, 
linkcolor=blue(munsell), 
citecolor=blue, 
filecolor=magenta, 
urlcolor=blue 
}
\usepackage{pgfplots}
\usepackage{amssymb}
\usepackage{amsfonts}
\usepackage{graphicx}
\usepackage{epstopdf}
\usepackage{dcolumn}
\usepackage{amsmath}
\usepackage{environ}
\usepackage{latexsym,bm}
\usepackage{amsthm}
\usepackage{slashed}
\usepackage{float}
\usepackage{color}
\usepackage{url}
\usepackage{longtable}
\usepackage{rotating}
\usepackage[normalem]{ulem}
\usepackage{faktor}
\usepackage{tikz-cd}
\usepackage{tensor}
\usepackage{array}
\usepackage{tabularx}
\usepackage{makecell}
\usepackage{changepage}

\usepackage{tikz}
\usetikzlibrary{arrows.meta}
\usepackage{diagbox}

\usepackage[english]{babel} 
\usepackage[autostyle]{csquotes}
\usepackage{pifont}
\usepackage{xstring}
\usetikzlibrary{decorations.pathmorphing} 
\usetikzlibrary{decorations.markings} 
\usetikzlibrary{arrows} 
\usetikzlibrary{shapes} 
\usetikzlibrary{matrix} 
\usetikzlibrary{positioning} 
\usetikzlibrary{positioning}
\usetikzlibrary{calc}
\usetikzlibrary{decorations.pathreplacing,calligraphy}

\usetikzlibrary{shapes.multipart}

\tikzset{->-/.style={decoration={
  markings,
  mark=at position .5 with {\arrow{>}}},postaction={decorate}}}
  \usepackage{tikz,tikz-3dplot,tikz-cd,circuitikz}
\usetikzlibrary{patterns,decorations.markings, decorations.pathmorphing, arrows,calc,knots,hobby,
	decorations.pathreplacing,
	shapes.geometric,
	calc, arrows, arrows.meta, decorations.text}
\usepackage{tcolorbox}
\tcbuselibrary{breakable}
\usepackage{xcolor}
\definecolor{brickred}{rgb}{0.8, 0.25, 0.33}
\definecolor{byzantium}{rgb}{0.44, 0.16, 0.39}
\definecolor{darkcerulean}{rgb}{0.03, 0.27, 0.49}
\definecolor{darkslateblue}{rgb}{0.28, 0.24, 0.55}

\tcbuselibrary{breakable,xparse}
\NewTColorBox[auto counter, number within=section]{examplebox}{ O{} O{} }{
  breakable,
  colback=gray!5,
  colframe=darkslateblue,
  boxrule=0.6pt,
  arc=3mm,
  left=6pt,
  right=6pt,
  top=6pt,
  bottom=6pt,
  fonttitle=\bfseries,
  title={Example~\thetcbcounter\IfBlankF{#1}{: #1}},
  #2
}

\allowdisplaybreaks

\usepackage{tabularx}
\usepackage{siunitx,booktabs,array}

\newcommand{\be}{\begin{equation}}
\newcommand{\ee}{\end{equation}}
\newcommand{\ba}{\begin{aligned}}
\newcommand{\ea}{\end{aligned}}

\providecommand{\Qc}[2]{\mathcal{Q}_{[#1],#2}}

\definecolor{DarkGreen}{rgb}{0.1, 0.7, 0.3}

\newcommand{\Hom}{\operatorname{Hom}}
\newcommand{\End}{\operatorname{End}}
\newcommand{\Fun}{\operatorname{Fun}}
\renewcommand{\Vec}{\mathsf{Vec}}
\newcommand{\Rep}{\mathsf{Rep}}

\renewcommand{\dim}{\text{dim}}

\def\fT{\mathfrak{T}}
\def\fZ{\mathfrak{Z}}

\newcommand{\Vect}{\mathsf{Vec}}
\newcommand{\TwoVec}{2\mathsf{Vec}}

\newcommand{\Ising}{\mathsf{Ising}}

\newcommand{\bit}{\begin{itemize}}
\newcommand{\eit}{\end{itemize}}
\newcommand{\ben}{\begin{enumerate}}
\newcommand{\een}{\end{enumerate}}

\renewcommand{\ni}{\noindent}
\newcommand{\id}{\mathrm{id}}
\newcommand{\Id}{\mathrm{Id}}

\NewEnviron{eqs}{%
\begin{equation}\begin{split}
    \BODY
\end{split}\end{equation}
}

\newcommand{\wt}{\widetilde}

\newcommand{\Z}{{\mathbb Z}}

\newcommand{\bC}{{\mathbb C}}

\newcommand{\cB}{\mathcal{B}}
\newcommand{\cC}{\mathcal{C}}
\newcommand{\cD}{\mathcal{D}}
\newcommand{\cE}{\mathcal{E}}
\newcommand{\cF}{\mathcal{F}}

\newcommand{\cL}{\mathcal{L}}
\newcommand{\cM}{\mathcal{M}}
\newcommand{\cN}{\mathcal{N}}
\newcommand{\cO}{\mathcal{O}}

\newcommand{\cQ}{\mathcal{Q}}

\newcommand{\cT}{\mathcal{T}}

\newcommand{\cZ}{\mathcal{Z}}
\newcommand{\fB}{\mathfrak{B}}

\newcommand{\B}{\mathrm{B}}

\newcommand{\bone}{\mathbf{1}}

\def\Tr{\mathop{\mathrm{Tr}}\nolimits}

\def\unit{{1\kern-.65ex {\rm l}}}
\def\1{{1\kern-.65ex {\rm l}}}

\def\bbC{{\mathbb{C}}}

\def\bbG{{\mathbb{G}}}

\def\bbN{{\mathbb{N}}}

\def\bbZ{{\mathbb{Z}}}

\newcommand{\beq}{\begin{equation}}
\newcommand{\eeq}{\end{equation}}

\usepackage{verbatim}
\usepackage{tikz}
\usetikzlibrary{arrows,snakes,shapes.arrows,decorations.markings}
     \tikzset{>=triangle 90}
     \tikzstyle{gr}=[draw,circle,green!50!black,fill=green!50!black,scale=.6]
     \tikzstyle{Bl}=[draw,circle,blue,scale=.7]
     \tikzstyle{R}=[draw,circle,fill=red,scale=.7]
     \tikzstyle{bl}=[draw,circle,fill=black,scale=.2]
     \tikzstyle{bbc}=[draw,circle,fill=black,scale=.75]
     \tikzstyle{bbcs}=[draw,circle,fill=black,scale=.5]
     \tikzstyle{rc}=[circle,fill=red,scale=.6]
     \tikzstyle{wc}=[draw,circle,scale=.75]
\usepackage{array} 
\usepackage{multirow} 
\usepackage{colortbl} 
\usepackage{stfloats}  

\usepackage{xcolor}   

\newcommand{\graymidrule}{\arrayrulecolor{black!25}\midrule[0.3pt]\arrayrulecolor{black}}
\newcommand{\xdasharrow}[2][->]{
\tikz[baseline=-\the\dimexpr\fontdimen22\textfont2\relax]{
\node[anchor=south,font=\scriptsize, inner ysep=1.5pt,outer xsep=2.2pt](x){#2};
\draw[shorten <=3.4pt,shorten >=3.4pt,dashed,#1](x.south west)--(x.south east);
}
}
\theoremstyle{definition}
\newtheorem{definition}{Definition}[section]

\theoremstyle{plain}
\newtheorem{theorem}[definition]{Theorem}
\theoremstyle{remark}
\newtheorem{corollary}[definition]{Corollary}
\newtheorem{warning}[definition]{Warning}
\newtheorem{remark}[definition]{Remark}

\def\bar{\overline}

\def\hat{\widehat}

\def\^{\wedge}

\def\Tr{\mathop{\rm Tr}}

\def\Aut{\operatorname{Aut}}
\newcommand{\Inv}{\operatorname{Inv}}

\def\Z{\mathbb{Z}}

\def\cB{{\mathcal B}}

\def\cC{{\mathcal C}}

\def\cD{{\mathcal D}}
\def\cE{{\mathcal E}}

\def\cM{{\mathcal M}}
\def\cN{{\mathcal N}}
\def\cO{{\mathcal O}}

\newcommand{\newl}{\medskip\noindent}
\newcommand{\TUV}{\fT_{\rm UV}}
\newcommand{\TIR}{\fT_{\rm IR}}
\newcommand{\CUV}{{\cC_{\rm UV}}}
\newcommand{\CIR}{{\cC_{\rm IR}}}

\newcount\hour \newcount\minute
\hour=\time \divide \hour by 60
\minute=\time
\def\now{%
\ifnum \hour<13
  \ifnum \hour=0 \advance \hour by 12 \number\hour:\else \number\hour:\fi%
     \ifnum \minute<10 0\fi%
     \number\minute%
\ A.M.%
\else \advance \hour by -12 \number\hour:%
  \ifnum \minute<10 0\fi%
  \number\minute%
  \ P.M.%
\fi%
}

\usepackage{tikz}
\usetikzlibrary{arrows}
\usetikzlibrary{shapes.geometric,calc,arrows, positioning,shapes.misc,decorations.markings}
\tikzset{
  big arrow/.style={
    decoration={markings,mark=at position 1 with {\arrow[scale=2,#1]{>}}},
    postaction={decorate},
    shorten >=0.4pt},
  big arrow/.default=black}
\usetikzlibrary{external}
\usetikzlibrary{positioning}
\usetikzlibrary{calc}
\tikzset{gauge-node/.style={shape=circle, draw, minimum width=.6cm}}

\pgfdeclarelayer{edgelayer}
\pgfdeclarelayer{nodelayer}
\pgfsetlayers{edgelayer,nodelayer,main} 
\tikzstyle{none}=[inner sep=0pt] 

\makeatother

\title{\boldmath $(2+1)$-Dimensional Phases With a Unique Vacuum and Generalized Fiber 2-Functors}

\author{Jamie~J.~Pearson$^{1}$}
\emailAdd{jjpearson@imada.sdu.dk}
\author{Apoorv Tiwari$^1$}
\author{Hao Xu$^1$}

\affiliation{${}^1$
Center for Quantum Mathematics at IMADA, University of Southern Denmark, Campusvej 55, 5230 Odense, Denmark}

\abstract{
We leverage the fact that all finite bosonic symmetries in $(2+1)$ dimensions are group-theoretical to show that every such symmetry admits a gapped phase with a unique vacuum, in which the 0-form symmetry is preserved. We introduce the notion of a generalized fiber 2-functor to describe the RG flow of topological defects in these phases, and call such a 2-functor minimal if it gives rise to no emergent 1-form symmetry, i.e.\ if every topological line of the IR theory descends from a line of the UV theory. We provide a general construction of these phases and the corresponding 2-functors using the SymTFT, and give conditions under which they are minimal.}

\begin{document}

\toccontinuoustrue
\maketitle
\flushbottom

\section{Introduction}
\label{sec:intro}
\ni Given an ultraviolet (UV) quantum many body system $\TUV$ with a (higher) categorical symmetry $\CUV$, it is natural to ask how that symmetry constrains the spectrum of infrared (IR) phases that are realizable by the system.
Let us say that the theory $\TUV$ flows to a theory $\TIR$ in the IR, such that the category of topological defects in $\TIR$ is $\CIR$.
It is clear that a topological defect in the UV should flow to some other topological insertion in the IR.
Moreover, this UV to IR map of topological defects must be compatible with the (higher) fusion structure of topological defects.
Consequently, we see that the UV symmetry $\CUV$ is represented on the IR theory via a (higher) tensor functor
\begin{equation}
\label{eq:RG functor}
    \cF \, :\,  \CUV \longrightarrow \CIR\, ,
\end{equation}
which physically describes the symmetry related properties of the renormalization group (RG) flow from $\TUV$ to $\TIR$. We refer to functors that arise this way as \emph{RG functors}.
In the case when $\TUV$ realizes a gapped phase, the renormalization group flow ends on a fixed point realizing a TQFT \footnote{For simplicity, we ignore the possibility of gapped phases such as Fracton topological orders \cite{Vijay:2016phm, Nandkishore:2018sel, Pretko:2020cko} that are not described by TQFTs.}. 
Picking $\TIR$ to be that fixed point imposes strong constraints on the form of $\CIR$.

\newl{\bf Gapped Phases and Module Categories.} A large and important class of examples for such gapped flows, comes from asking that $\TIR$ is fully extended in $d\Vect$~\footnote{For completeness: we define $d\Vect$ iteratively, as the Morita subcategory of (separable) algebra objects in $(d-1)\Vec$, where $\Vect$ is simply the category of finite-dimensional vector spaces over $\bbC$.}, with $d$ the spacetime dimension.
Such gapped phases are labeled by module $(d\!-\!1)$-categories over $\CUV$~\cite{Ostrik2003, Thorngren:2019iar, Bhardwaj:2024qiv}, with the TQFT $\TIR$ characterized by a choice of (fully dualizable) object $\cM\in d\Vect$, and the $\CUV$ symmetry on $\TIR$ given by the $\CUV$-module action on $\cM$.

\newl More precisely, the (framed) cobordism hypothesis \cite{Baez:1995xq, Lurie:2009keu} formulates $\TIR$ functorially, as a symmetric monoidal $d$-functor on the $d$-category of fully extended framed $d$-bordisms \begin{equation}
    \cZ_\cM \,:\, \mathsf{Bord}^{\rm fr}_{d} \to d\Vect\, ,
\end{equation}
that assigns the positively oriented point to the object $\cM \in d\Vect$.
The corresponding $(d\!-\!1)$-category of IR topological defects is then given by
\begin{equation}
\label{eq:EndM}
    \CIR \simeq \End_{d\Vect}(\cM)\, ,
\end{equation}
and the RG functor
\begin{equation}
    \cF \,:\, \CUV \to \End_{d\Vect}(\cM)
\end{equation}
is obtained directly from a choice of (left) $\CUV$-module action on $\cM$.

\newl {\bf Single-vacuum gapped phases in 2d and fiber functors.} Let us momentarily specialize to the case where $\TUV$ is a gapped $(1\!+\!1)$-dimensional theory, so that $\CUV$ is a fusion 1-category. 
We first make the discussion of the previous paragraph explicit within this context.
A fully extended 2d TQFT valued in $2\Vec$ is characterized by a finite semisimple linear category,
\begin{equation}
    \cM \simeq \Vec^{\oplus n}\,,\qquad n \in \bbN\, .
\end{equation}
This is related to the familiar statement that fully extended two-dimensional TQFTs are classified by separable Frobenius algebras~\cite{schommer2009classification}; their values on the circle are centers of these algebras, which are more generally classified for non-extended two-dimensional TQFTs by commutative Frobenius algebras~\cite{Abrams:1996ty, Kock2004}. Here that algebra is $\mathbb{C}^{n}$, where $n$ is the number of vacua, and the corresponding TQFT is simply the direct sum of $n$ copies of the trivial theory.
Its topological line defects are the endomorphisms of $\cM$ in $2\Vec$,
\begin{equation}
    \CIR \simeq \End_{2\Vec}\!\big[\Vec^{\oplus n}\big] \simeq \mathrm{Mat}_{n}(\Vec) ,
\end{equation}
given by the multi-fusion category of $n\times n$ matrices valued in vector spaces. 
Its simple objects $1_{ij}$ are the domain walls between the $i$-th and $j$-th vacuum, with fusion $1_{ij}\otimes 1_{kl} \simeq \delta_{jk}1_{il}$.
In particular the unit $\bigoplus_i 1_{ii}$ is not simple, reflecting the presence of $n$ vacua. 
A monoidal functor $\cF:\CUV\to \mathrm{Mat}_n(\Vec)$ is precisely a $\CUV$-module structure on $\Vec^{\oplus n}$, so that gapped phases with $\CUV$ symmetry and $n$ vacua are labeled by $\CUV$-module categories of rank $n$ \cite{Ostrik:2001xnt, Thorngren:2019iar, Bhardwaj:2023fca, Bhardwaj:2023idu}.

\newl One question we may wish to answer is whether or not it is possible for $\TUV$ to realize a gapped phase with a single vacuum/ground state.
The IR physics of a unique-vacuum gapped phase is captured by the trivial two-dimensional TQFT with $\CIR=\Vec$.
In the above language, this translates to the existence of a \emph{fiber functor}
\begin{equation}
\cF \, :\, \CUV \longrightarrow \Vec \,.
\end{equation}
Physically, such a functor describes an RG flow to a gapped phase which is a symmetry protected topological phase in which all symmetry defects become trivial in the IR, i.e., the symmetry is fully preserved.

\newl A well-known class of fusion categories that do not admit fiber functors are invertible and of the form
\begin{equation}
    \CUV \simeq \Vec_{G}^{\pi}\, ,
\end{equation}
where $G$ is a finite group and the 3-cocycle $\pi \in \operatorname{Z}^{3}(G;U(1))$ captures the associator data~\footnote{Actually, the associator is valued in $\bC^\times$ but since this is homotopic to $U(1)$, which is more common in the unitary setting of QFT, we will use them interchangeably.}. Categories of this form admit a fiber functor if and only if $\pi$ is trivial in cohomology (i.e. $\pi$ is a coboundary).
Physically we say that $\CUV$ describes a $G$ global symmetry with 't Hooft anomaly $\pi$, and that it is possible to preserve the $G$ symmetry in the IR if and only if we can cancel out $\pi$ by adding local counter-terms.
This agrees with the physical expectation that in any gapped phase the symmetry $G$ must be spontaneously broken to a non-anomalous subgroup.

\newl Another example of a fusion category that does not admit a fiber functor is the Ising category.
It is known that any gapped phase realizable in a theory with Ising symmetry necessarily has $3n$ vacua with $n\in\mathbb Z_{>0}$ \cite{Ostrik:2001xnt, Thorngren:2019iar, Bhardwaj:2023idu}.

\newl The above two examples demonstrate an important fact in two dimensions: the only single-vacuum TQFT being the trivial one implies that a 0-form symmetry can only be preserved in the IR if the UV symmetry category it belongs to admits a fiber functor.

\newl {\bf Single-vacuum gapped phases in 3d and generalized fiber 2-functors.} In this work, we turn to the case where $\TUV$ is a 
$(2\!+\!1)$-dimensional quantum field theory and $\CUV$ is a fusion 2-category. Here, the question of whether $\CUV$ can be realized on a TQFT with a single vacuum becomes more nuanced since a TQFT having a single vacuum does not necessarily imply that the theory is the trivial $3$d TQFT, nor does it necessitate the existence of a fiber 2-functor. 
Rather, a single vacuum simply means that the theory contains a single local operator, and by modularity \cite{Lan:2018vjb, Johnson-Freyd:2020usu}, this is equivalent to the statement that every surface defect in the theory is a condensation defect.
The most general TQFT with this property has a symmetry of the form
\begin{equation}
    \CIR \simeq \Sigma B\, ,
\end{equation}
where $B$ is a modular tensor category and $\Sigma B$ denotes the $2$-category obtained by delooping $B$ and ``adding" all condensation defects (i.e. performing the idempotent completion) \cite{Gaiotto:2019xmp}.
%
%

\newl It is worth emphasizing that this is strictly more general than the fully extended setting of \eqref{eq:EndM}. Objects of $3\Vect=\Sigma\,2\Vect$ are Morita classes of (multi)fusion categories~\cite{douglas:2020dualizable,Gaiotto:2019xmp,Decoppet:2022dnz}, and the fully extended 3d TQFT determined by a fusion category $A$ is the Turaev--Viro theory $\mathrm{TV}[A]$ \cite{Turaev:1992hq, Barrett:1993ab, 2010arXiv1004.1533K}, with symmetry
\begin{equation}
    \CIR \simeq \End_{3\Vect}(\Sigma A)\simeq \Sigma\cZ(A)\, .
\end{equation}
In other words, these fully extended TQFTs correspond to Reshetikhin--Turaev theories ${\mathrm{RT}}[B]$ whose underlying modular tensor categories are of the form $B\simeq\cZ(A)$.

\newl A Reshetikhin--Turaev theory ${\mathrm{RT}}[B]$ \cite{Reshetikhin:1991tc} whose modular tensor category is not a Drinfeld center, that is, whose Witt class is non-trivial \cite{Davydov:2010wpa} such as $U(1)_2$, Fibonacci or Ising, is not fully extended in $3\Vect$. 
It is instead a relative TQFT in the sense of \cite{Freed:2012bs}, describing a boundary condition for the fully-extended invertible $4$-dimensional Crane--Yetter TQFT determined by $B$ \cite{Crane:1993if, Walker:2011mda, Haioun:2023oyb, Haioun:2025hen}, and correspondingly it carries a gravitational anomaly. 
Its surface defects nonetheless form $\Sigma B$, and this is the reason we take $\Sigma B$ for an arbitrary modular tensor category $B$, rather than $\End_{3\Vect}(\cM)$ for some arbitrary finite semisimple 2-category $\cM$, as the target of the RG 2-functor.

\newl Inspired by the RG functor we would attribute to these phases, we introduce the notion of a \emph{generalized} fiber 2-functor~\footnote{We choose this language to illustrate that $\cF$ is still a forgetful tensor functor, a fact we will cover in more detail in section~\ref{sec:gen-fib-fun-symtft}.}
\begin{equation}
    \cF \, : \, \CUV \to \Sigma B\, .
\end{equation}
The trivial TQFT corresponds of course to the special case $B=\Vec$, in which $\cF$ as above describes an ordinary fiber 2-functor.

\subsection{Main Results}
In this work we address the following question:
\begin{center}
\it For which fusion 2-categorical symmetries $\CUV$ in $(2\!+\!1)$ dimensions does there exist a gapped phase in which $\CUV$ acts on an IR TQFT with a unique vacuum?
\end{center}
In $(1\!+\!1)$ dimensions the analogous question has the answer: a fusion category symmetry is realizable on a single vacuum IR theory, which is then necessarily trivial, if and only if it admits a fiber functor.\footnote{Such a fusion category need not be group-theoretic: there are semisimple Hopf algebras whose representation categories are not group-theoretic \cite{Nikshych:2008}. In contrast, every fusion 2-category admitting a fiber 2-functor is group-theoretical \cite{Decoppet:2023bay}.} For a group-theoretical symmetry $\mathsf{C}(G,\pi|H,\psi)$, this is the case if and only if there is a subgroup $K\subseteq G$ with $G=HK$ and $\pi|_K$ trivializable, such that the induced 2-cocycle on $H\cap K$ is non-degenerate \cite{Ostrikmodule,Natale:2003}. In particular, it suffices that $H$ has a complement on which $\pi$ trivializes, but a non-trivial intersection $H\cap K$ is also allowed (see section~\ref{sec:warm-up}).
In this work, we focus on $(2\!+\!1)$ dimensions, where the situation is significantly different.
Our main results are the following.
\begin{enumerate}[label=(\arabic*)]
\item \emph{\bf Unique-vacuum phases always exist.} Every bosonic fusion 2-categorical symmetry, in the sense of having no emergent fermions among its transparent lines, admits a gapped phase with a unique vacuum in which the 0-form symmetry is preserved. Equivalently, every such symmetry admits a generalized fiber 2-functor. This is in sharp contrast with genuine fiber 2-functors, i.e.\ with phases that are completely trivial in the IR, which exist only for a restricted class of group-theoretical symmetries.
\item \emph{\bf Explicit constructions.} We give two constructions of such phases and of the corresponding RG functors. For invertible symmetries, i.e.\ finite groups with an 't Hooft anomaly, a generalized fiber 2-functor can be built from any symmetry extension that trivializes the anomaly. For general bosonic symmetries, we construct the phases from the SymTFT, by pairing the boundary that supports the symmetry with a suitable topological boundary on the other side, called the \emph{physical boundary}. We determine the UV symmetry, the IR TQFT and the RG functor explicitly, both in terms of module 2-categories and in terms of Lagrangian algebras.
\item \emph{\bf When is the IR as small as possible?} The IR TQFT of such a phase generally contains topological lines that are not inherited from the UV, i.e.\ emergent 1-form symmetries. We call a generalized fiber 2-functor \emph{minimal} if it produces none, so that every IR line comes from a UV line. We show that a minimal phase exists under conditions that closely parallel those for a genuine fiber 2-functor, in which case its IR topological order is canonically determined by the UV 1-form symmetry. When these conditions fail, there is in general no canonical smallest IR phase.
\end{enumerate}
We construct the generalized fiber 2-functors of invertible symmetries in section~\ref{sec:sym-ext}, establish existence and the general construction in section~\ref{sec:gen-fib-fun-symtft}, and discuss minimality in section~\ref{subsec:minimality}. Section~\ref{sec:warm-up} illustrates all constructions in the simpler setting of $(1\!+\!1)$-dimensional gapped phases.

\subsection{Organization of the Paper}
In section~\ref{sec:sym-ext} we study invertible symmetries $\CUV=2\Vect_G^\pi$ and  construct generalized fiber 2-functors for them from 3-group homomorphisms and symmetry extensions, based on \cite{Wang:2017loc}.
In section~\ref{sec:symtft} we review the SymTFT of a fusion 2-categorical symmetry, describe its bulk defects and topological boundaries and explain how gapped phases and their RG functors arise from the SymTFT sandwich construction.
As a first application we rederive the criterion for the existence of a genuine fiber 2-functor \cite{Decoppet:2023bay} using the SymTFT.
Section~\ref{sec:warm-up} is a warm-up one dimension lower: for $(1\!+\!1)$-dimensional theories with finite group symmetry we detail how to compute the UV symmetry, the IR symmetry and the RG functor of every gapped phase, first with module categories and then with Lagrangian algebras, with $G=S_3$ as a running example. Every construction there has a direct 2-categorical counterpart in the following section. Readers familiar with the 1-categorical story may skip section~\ref{sec:warm-up} and refer back to it as needed.
Section~\ref{sec:gen-fib-fun-symtft} contains our main results. We show that every bosonic fusion 2-categorical symmetry admits a phase with a unique vacuum, and hence a generalized fiber 2-functor. We then construct $\CUV$, $\CIR$ and the RG functor using module 2-categories, give conditions for the generalized fiber 2-functor to be minimal, and present the equivalent description in terms of Lagrangian algebras.
The appendices collect background material. 
Appendix~\ref{App:loops and suspensions} reviews loop spaces and suspensions of fusion $n$-categories, appendix~\ref{App:3-groups and homomorphisms} reviews 3-groups and their homomorphisms, and appendix~\ref{app:spec-seq} reviews the Lyndon--Hochschild--Serre spectral sequence. 
Finally, the 2-categorical notions used throughout, namely algebras and their modules, condensation and full centers, module 2-categories, Morita equivalence and the bulk-boundary correspondence, are collected in appendix~\ref{app:2cat}.

\subsection{Notation and Conventions}\label{subsec:notation}
We collect here the notation used throughout the paper.
\begin{itemize}
    \item \emph{Categories and their objects.} 
    \begin{itemize}[label={--}]
        \item 2-categories, such as fusion 2-categories and their module 2-categories, are denoted by calligraphic letters $\cC, \cM, \ldots$.
        \item Fusion, braided fusion and module 1-categories, as well as algebras and modules internal to 2-categories, are denoted by capital italic letters $A, B, M, \ldots$.
        \item Objects of 1-categories are denoted by lowercase letters.
        \item The Morita equivalence class of the fusion category $A$ is determined by its module 2-category $\mathsf{RMod}(A)$, regarded as an object of $3\Vect$; it is also equivalent to the suspension of $A$ and is written $\Sigma A$.
    \end{itemize}
    Two exceptions are the general $(d\!-\!1)$-categorical discussion at the beginning of this introduction and the $(1\!+\!1)$-dimensional warm-up in section~\ref{sec:warm-up}, where we follow the standard 1-categorical notation, with calligraphic fusion categories and module categories.
    \item \emph{Standard categories.} $\Vect^\pi_G$ is the fusion category of $G$-graded vector spaces with associator $\pi\in\operatorname{Z}^3(G;U(1))$, and $2\Vect^\pi_G$ is the fusion 2-category of $G$-graded 2-vector spaces with pentagonator $\pi\in\operatorname{Z}^4(G;U(1))$. We write $\Rep(G)$ and $2\Rep(G)$ for the symmetric fusion 1- and 2-categories of representations, and $\mathsf{C}(G,\pi|H,\psi)$ for the group-theoretical fusion $n$-category obtained from $n\Vect^\pi_G$ by gauging $H\subseteq G$ with the trivialization $\psi$ of $\pi|_H$. For an algebra $A$ in a (2-)category $\cC$, $\mathsf{LMod}_\cC(A)$, $\mathsf{RMod}_\cC(A)$ and $\mathsf{Bimod}_\cC(A)$ denote its (2-)categories of left modules, right modules and bimodules, and the subscript is dropped when $\cC$ is $\Vect$ or $2\Vect$.
    \item \emph{Group-theoretical symmetries.} When we say, as in the abstract, that a bosonic symmetry $\cC$ is \emph{group-theoretical}, we mean that its SymTFT is a Dijkgraaf--Witten theory up to an invertible factor, i.e.\ that $\cZ_1(\cC)\simeq\cZ_1(2\Vect^\pi_G)$ for some finite group $G$ and $\pi\in\operatorname{Z}^4(G;U(1))$ (theorem~\ref{thm:AB-center}). This is weaker than the notion of a group-theoretical fusion 2-category of \cite{Decoppet:2023bay}, which requires $\cC\simeq\mathsf{C}(G,\pi|H,\psi)$, i.e.\ $\cC\simeq\mathsf{Bimod}_{2\Vect^\pi_G}(A)$ for a pointed algebra $A=\Vect^\psi_H$ with identity component $A_1\simeq\Vect$, or equivalently $\Omega\cC$ to be Tannakian. Our notion relaxes this in two ways: the identity component $A_1$ of the graded fusion category $A$ may be any fusion category, so that $\Omega\cC\simeq\cZ(A_1)^H$, and more generally $\cC$ may differ from such a $\mathsf{Bimod}_{2\Vect^\pi_G}(A)$ by a non-degenerate braided fusion category $M$ of non-trivial Witt class (see warning~\ref{warn:center-does-not-determine-morita}). In this sense it is closer to a 2-categorical analogue of \emph{weakly} group-theoretical fusion categories \cite{Etingof:2011weakly}.
    \item \emph{Operations.} $\Sigma$ denotes the suspension (condensation completion) and $\Omega$ the loop space, so that $\Omega\cC$ is the braided fusion category of genuine lines of $\cC$ (see appendix~\ref{App:loops and suspensions}). $\cZ$ or $\cZ_1$ is the Drinfeld center, $\cZ_2$ the M\"uger center, $\boxtimes$ the Deligne tensor product, $(-)^H$ the $H$-equivariantization and $-\boxtimes_{\Rep(H)}\Vect$ the corresponding de-equivariantization. Equivalences of categories and isomorphisms internal to 2-categories are written as $\simeq$; $\cong$ is used only for isomorphisms internal to 1-categories.
    \item \emph{Theories.} TQFTs built from categorical data are written in roman capitals: the Turaev--Viro theory $\mathrm{TV}[A]$, the Reshetikhin--Turaev theory $\mathrm{RT}[B]$, the Crane--Yetter theory $\mathrm{CY}[B]$, the Douglas--Reutter theory $\mathrm{DR}[\cC]$ and the Dijkgraaf--Witten theory $\mathrm{DW}(G,\pi)$. We use TFT and TQFT interchangeably. The UV and IR theories are $\TUV$ and $\TIR$, their symmetries $\CUV$ and $\CIR$, and the RG functor is $\cF:\CUV\to\CIR$. Topological boundaries of the SymTFT are denoted by $\fB$, e.g.\ the Dirichlet boundary $\fB_{\rm Dir}$ and the boundaries $\fB_{(H,A)}$ of section~\ref{sec:symtft}.
    \item \emph{Cohomology.} We write $\operatorname{C}^n(G;U(1))$, $\operatorname{Z}^n(G;U(1))$ and $\operatorname{H}^n(G;U(1))$ for cochains, cocycles and cohomology classes, and $\delta$ for the coboundary. We use $U(1)$ and $\bC^\times$ interchangeably as coefficients.
    \item \emph{Conjugation.} For $g\in G$ and a subgroup $K\subseteq G$, we write ${}^{g}K=gKg^{-1}$, $K^g=g^{-1}Kg$, and ${}^{g}h=ghg^{-1}$.
    \item \emph{Tensor products.} We write $\otimes$ for the monoidal product of objects in a monoidal (2-)category, such as $\fZ$, $\CUV$ or a fusion category $A$. We write $\boxtimes$ for the Deligne product of linear (2-)categories and for the stacking of theories, and $\boxtimes_{A}$ for the relative Deligne product of module categories over $A$, i.e.\ the composition in $\mathsf{Bimod}(A)$.
\end{itemize}

\section{Invertible Symmetries and Symmetry Extensions}\label{sec:sym-ext}
In this section we restrict ourselves to invertible symmetries, and will begin by recalling some results pertaining to (higher) fiber functors and trivial gapped phases.

\newl Consider a QFT $\TUV$ in $d$ spacetime dimensions, with a global 0-form symmetry described by some finite group $G$ with an 't Hooft anomaly $\pi \in \operatorname{Z}^{d+1}(G; U(1))$.
Such a symmetry is captured by the fusion $(d\!-\!1)$-category
\begin{equation}
    \CUV = (d\!-\!1)\Vect_G^\pi\, .
\end{equation}
Suppose that the 't Hooft anomaly characterized by $\pi$ cannot be cancelled by the addition of local counter-terms---i.e. it belongs to a non-trivial group cohomology class---then there exists no fiber $(d\!-\!1)$-functor
\begin{equation}
\label{eq:higher FF}
    \cF \, : \, \CUV \to (d\!-\!1)\Vect\, . 
\end{equation}
Physically this means it is not possible for $\TUV$ to flow to a trivially gapped $G$-preserving phase. 
In $d=2$, the only TFT that has a unique vacuum state is (up to invertible theories, e.g., the Euler TFT) the trivial one, so the questions of whether or not a $G$ symmetry admits a fiber functor and whether or not we can preserve that symmetry in the IR are one and the same.
The non-existence of \eqref{eq:higher FF} implies that $G$ must always be spontaneously broken to a non-anomalous subgroup.

\newl For $d \geq 3$, there are many more single-vacuum TFTs than just the trivial one, therefore the above question becomes more nuanced.
In particular we can ask whether it is possible to find a gapped phase where $G$ is completely preserved, even when that symmetry does not admit a higher fiber functor like in~\eqref{eq:higher FF}.

\newl For $d=3$ this question can be formulated precisely. Three-dimensional single-vacuum TFTs are known to be labeled by modular tensor categories, so functionally this question is the same as asking whether or not there exists a \emph{generalized} fiber 2-functor
\begin{equation}\label{eq:gen-fib-fun}
    \cF \, : \, \CUV \to \Sigma B\, ,
\end{equation}
for some modular tensor category $B$. We claimed in the introduction that the answer to this question is always yes, not just for invertible symmetries, but for all (bosonic) fusion 2-categorical symmetries. In this section however, we restrict ourselves to $d=3$ and aim to demonstrate the existence of generalized fiber 2-functors like~\eqref{eq:gen-fib-fun} for invertible symmetries of the form
\begin{equation}
    \CUV = 2\Vect_G^\pi\, .
\end{equation}

\newl For $d=3$, we will review how to describe such functors explicitly, before presenting a method using symmetry extensions~\cite{Perez-Lona:2025ncg, Wang:2017loc} that allows us to construct examples for any choice of $(G,\pi)$, with $B$ of the form
\begin{equation}
    B \simeq \cZ(\Vect_K^\psi)\, ,
\end{equation}
for some particular pair of finite group $K$ and 3-cocycle $\psi \in \operatorname{Z}^3(K; U(1))$. 

\subsection{Generalized Fiber 2-Functors from 3-Group Homomorphisms}
In this subsection we provide a brief ``how-to" for extracting the data that labels generalized fiber 2-functors of the form
\begin{equation}
\label{eq:Gpi to SigmaB}
    \cF \, :\, 2\Vect^\pi_G \to \Sigma B\, ,
\end{equation}
where $B$ is a non-degenerate braided fusion category, following~\cite{etingof2009, Bullimore:2024khm, Pearson:2024snm}. 
For any monoidal 2-category $\cC$, the maximal sub-3-groupoid $\cC^\times$ of invertible objects, 1-morphisms and 2-morphisms is a 3-group (see appendix~\ref{App:3-groups and homomorphisms} for a definition).
Given a 2-functor $\cF:\cC\to \cD$, one can always consider the restriction to a 3-group homomorphism
\begin{equation}
    \cF^{\times}:\cC^{\times} \longrightarrow \cD^{\times}\,.
\end{equation}
In the case where the source category for the 2-functor $\cF$ is generated by its invertible defects, which in the present context is $\TwoVec_{G}^{\pi}$, the restriction from a 2-functor to a 3-group homomorphism is actually an equivalence.
In section~\ref{sec:gen-fib-fun-symtft}, we will consider the general case where the symmetry category is not necessarily invertible, wherein one needs the inherently SymTFT picture developed in section~\ref{sec:symtft}.

\newl Let us now make sense of \eqref{eq:Gpi to SigmaB} as a 3-group homomorphism
\begin{equation}
    \cF^{\times}:\bbG_{\pi} \longrightarrow \mathrm{Pic}(B)\, .
\end{equation}
The source 3-group $\bbG_{\pi}$ comprises the following groups in various degrees
\begin{equation}
    G_0=G\,, \quad G_1=1\,, \quad G_2=\mathbb C^{\times}\,.
\end{equation}
$\bbG_{\pi}$ has a single non-trivial $k$-invariant: the Postnikov class presented by $\pi\in \operatorname{Z}^{4}(G_0;G_2) = \operatorname{Z}^{4}(G;\mathbb C^{\times})$, corresponding physically to an 't Hooft anomaly.
$\TwoVec_{G}^{\pi}$ is the linearization (and condensation completion) of the 3-group $\bbG_{\pi}$.

Meanwhile the target 3-group $\mathrm{Pic}(B) = (\Sigma B)^\times$ is the automorphism 3-group of $B$, comprising the following data:
\begin{enumerate}
    \item {\bf Homotopy groups of $\mathrm{Pic}(B)$:} The 0-form component is the group $\Pi_0 = \Aut^{br}(B)$ of braided automorphisms of $B$.
    These correspond physically to invertible surface defects within $\Sigma B$, and their composition coincides with the fusion of such defects.
    The 1-form component is the Abelian group $\Pi_1 = \Inv(B)$ of invertible objects in $B$, corresponding to invertible topological lines in $\Sigma B$.
    The remaining 2-form component is just $\Pi_2=\bC^\times$, corresponding to (trivial) invertible topological local operators within $\Sigma B$.

    \item {\bf Actions:} There is a natural action of $\Pi_0$ on $\Pi_1$ which corresponds to the permutation of invertible lines via braided automorphisms. Physically this corresponds to the action induced on line defects by wrapping them with surfaces.
    $\Pi_0$ acts trivially on $\Pi_2$.

    \item {\bf $k$-Invariants:} There are two $k$-invariants. First is a 2-group extension $\Pi_1[1] \to \Pi_{\leq 1} \to \Pi_0$ determined by a 3-cocycle $\widetilde{\cO_3} \in \operatorname{Z}^3(\Pi_0; \Pi_1)$, twisted by the action of $\Pi_0$ on $\Pi_1$.
    Second is a central 3-group extension $\bC^\times[2] \to \Pi_{\leq 2} \to \Pi_{\leq 1}$ determined by a 4-cocycle~\footnote{In the context of this 2-group, a (cohomology class of) 4-cocycle simply means a (homotopy class of) map between the classifying spaces $\B\Pi_{\leq 1} \to \B^4\bC^\times$.}
    $\widetilde{\cO_4} \in \operatorname{Z}^4(\Pi_{\leq 1}; \bC^\times)$.
\end{enumerate}

\newl The 3-group homomorphism between $\bbG_{\pi}$ and $\mathrm{Pic}(B)$ then consists of the following data:
\begin{enumerate}
    \item An ordinary group homomorphism
    \begin{equation}
        \rho \, : \, G \to \Pi_0\, .
    \end{equation}
    Physically this captures the action of $G$ on the line defects of $B$.
    Since the 3-group $\bbG_{\pi}$ is trivial in degree 1, we do not need to provide any homomorphism into $\Pi_{1}$.
    Finally, the relevant homomorphism from $G_{2}$ to $\Pi_2$ is just the identity.
    \item A choice of (twisted) 2-cochain $\varphi \in \operatorname{C}^2(G; \Pi_1)$ satisfying
    \begin{equation}
        \delta\varphi = \rho^*\widetilde{\cO_3}^{-1}\, .
    \end{equation}
    Physically this corresponds to a symmetry fractionalization of $G$ by the (invertible) line defects of $B$.
    \item A choice of 3-cochain $\phi \in \operatorname{C}^3(G; U(1))$ satisfying
    \begin{equation}
        \delta\phi = (\rho, \varphi)^*\widetilde{\cO_4}^{-1}\pi\, .
    \end{equation}
    Physically this corresponds to a $G$-symmetric SPT phase.
\end{enumerate}
A key observation is that the final condition placed on the 3-cochain $\phi$ above \emph{does not require $\pi$ to be trivializable}; with the correct choice of $B$, $\rho$ and $\varphi$, it is possible to find a 4-cocycle $(\rho, \varphi)^*\widetilde{\cO_4}$ that cancels $\pi$ up to a coboundary. This phenomenon is vital to our construction, as it provides a mechanism to evade spontaneous breaking and construct $(2\!+\!1)$-dimensional gapped phases in which an anomalous 0-form $G$ symmetry is fully preserved.

\subsection{Symmetry Extensions}

The 3-group homomorphisms of the previous subsection tell us what data a generalized fiber 2-functor $2\Vect^\pi_G\to\Sigma B$ must consist of, but they do not tell us how to find one. 
In this subsection we describe a concrete recipe based on~\cite{Perez-Lona:2025ncg, Wang:2017loc}.
The central idea is to trade the anomalous symmetry $G$ for a larger, non-anomalous one. 
It is always possible to find a finite group $\Gamma$ surjecting onto $G$ such that the anomaly $\pi$ becomes trivial when pulled back to $\Gamma$.
A theory with a non-anomalous $\Gamma$ symmetry can flow to a trivially gapped phase, i.e.\ a $\Gamma$-SPT, in which the symmetry is fully preserved.
Gauging the kernel $K$ of $\Gamma\to G$ then produces a theory with the original $G$ symmetry and anomaly $\pi$, at the cost of introducing the topological degrees of freedom of a Dijkgraaf--Witten theory for $K$.
The result is a gapped phase with a single vacuum in which the anomalous $G$ symmetry is unbroken, realized on the surface defects of the Dijkgraaf--Witten theory, which is exactly a generalized fiber 2-functor with target $\Sigma\cZ(\Vect^{\psi|_K}_{K})$. 
In the remainder of the subsection we carry this out explicitly and extract the underlying 3-group homomorphism $(\rho,\varphi, \phi)$, showing in particular how the anomaly $\pi$ is cancelled by the fractionalization of $G$ on the lines of the Dijkgraaf--Witten theory rather than by a coboundary on $G$ alone.

\newl As mentioned, the core principle at work here is that for any choice of $(G, \pi)$, it is possible to find a finite extension
\begin{equation}
\label{eq:Gamma_SES}
    1 \to K \to \Gamma \xrightarrow{p} G \to 1\, ,
\end{equation}
such that the pull-back $[p^*\pi] \in \operatorname{H}^4(\Gamma;U(1))$ trivializes as a cohomology class.
It then follows from our earlier discussion that the associated fusion 2-category
\begin{equation}
    \widetilde{\CUV} = 2\Vect_{\Gamma}^{p^*\pi}
\end{equation}
admits a fiber 2-functor, and that there exists a trivial $(\Gamma, p^*\pi)$-symmetric gapped phase.
The construction of a generalized fiber 2-functor from the category $\TwoVec_{G}^{\pi}$ is carried out in steps as follows:
\begin{enumerate}
    \item We start with a $(2\!+\!1)$d SPT phase for $\Gamma$, specified by a choice of trivializing 3-cochain $\psi \in \operatorname{C}^3(\Gamma;U(1))$ satisfying
    \begin{equation}
        \delta\psi = p^*\pi\, .
    \end{equation}
    This is equivalent (up to coboundary) to specifying an equivalence class of fiber 2-functor $\widetilde{\CUV} \to 2\Vect$. 

    \item Next, we gauge the subgroup $K \,\subseteq\,\Gamma$ together with a trivial $K$-symmetric SPT phase (discrete torsion). 
    The effect of this gauging operation on the fiber 2-functor $\widetilde{\CUV} \to 2\Vect$ is two-fold.

    \newl Firstly, in the UV one obtains a symmetry described by the group-theoretical fusion 2-category $\mathsf{C}(\Gamma, p^*\pi| K, 1)$.
    This contains the original $\pi$-anomalous $G$ symmetry as a 2-subcategory
    \begin{equation}
        \CUV \hookrightarrow \mathsf{C}(\Gamma, p^*\pi| K, 1)\, ,
    \end{equation}
    as well as Wilson lines labeled by representations of $K$. 
    There is then a mixed anomaly between the $\Rep(K)$ Wilson lines and the $G$ surfaces, controlled by the extension class of the short exact sequence \eqref{eq:Gamma_SES} \cite{Tachikawa:2017gyf}.

    \newl Secondly, in the IR, gauging the $K$ 0-form symmetry acting on the trivial 3d TFT produces a $(2\!+\!1)$-dimensional Dijkgraaf--Witten TFT with symmetry
    \begin{equation}
        \CIR \simeq \Sigma\cZ(\Vect_K^{\psi|_K})\, .
    \end{equation}
    Notice that since $\mathrm{ker}(p) = K$, the restriction $p^*\pi|_{K}$ vanishes, hence $\psi|_K \in \operatorname{Z}^3(K; U(1))$ is a 3-cocycle.
    
    \item To finalize the construction, we need a tensor 2-functor from $\TwoVec_{G}^{\pi}$ into $\CIR $.
    We note that this Dijkgraaf--Witten theory comes naturally equipped with an action from the UV gauged symmetry $\mathsf{C}(\Gamma, p^*\pi|K, 1)$, so we define a new functor $\cF$ that realizes the $\CUV$ sub-symmetry to be the composition
    \begin{equation}
    \label{eq:generalized F2F from WWW}
        \cF \, : \, \CUV \hookrightarrow \mathsf{C}(\Gamma, p^*\pi|K, 1) \longrightarrow \CIR\, .
    \end{equation}
\end{enumerate}

\newl In order to better illustrate the functor \eqref{eq:generalized F2F from WWW} as a 3-group homomorphism, let us consider the special case where $\Gamma$ is a central extension of $G$, characterized by a Postnikov 2-cocycle 
\begin{equation}
    \varphi \in \operatorname{Z}^2(G; K)\, .
\end{equation}

\newl We can use the Lyndon--Hochschild--Serre (LHS) spectral sequence methods to concretely construct the underlying 3-group homomorphism (see appendix~\ref{app:spec-seq} for details).
The LHS spectral sequence computes $\operatorname{H}^{4}(\Gamma;U(1))$ from the cohomologies of $G$ and $K$.
Its starting page is
\begin{equation}
    E_{2}^{p,q}:=\operatorname{H}^{p}\big(G;\operatorname{H}^{q}(K;U(1))\big)\,,
\end{equation}
where, for a central extension, $G$ acts trivially on $\operatorname{H}^{q}(K;U(1))$.
It carries differentials $d_{r}:E_{r}^{p,q}\to E_{r}^{p+r,q-r+1}$ with each page being the cohomology of the previous one. 

\newl Importantly, the bottom row $E_{2}^{4,0}=\operatorname{H}^{4}(G;U(1))$ maps to $\operatorname{H}^{4}(\Gamma;U(1))$ via the pull-back $p^{*}$. In particular, the class $[p^{*}\pi]$ vanishes if $[\pi]\in E_{2}^{4,0}$ is hit by the differential
\begin{align}
    d_{2} \, :\, E_{2}^{2,1}=\operatorname{H}^{2}(G;\hat{K})&\to E_{2}^{4,0}\, ,\nonumber\\
    [\hat{\varphi}] &\mapsto [\varphi\smile \hat{\varphi}]\, ,
\end{align}
given by the cup product with the extension class, where $\hat K$ denotes the Pontryagin dual.
At the level of cochains, this cup product takes the form
\begin{equation}
    (\varphi \smile \hat{\varphi})(g_1, g_2, g_3, g_4) = \langle \varphi(g_1, g_2) ,\, \hat{\varphi}(g_3, g_4) \rangle\, ,
\end{equation}
where $\langle -,-\rangle$ denotes the evaluation pairing on $K\times\hat{K}$, and the class $[p^{*}\pi]$ is trivial if there exists a 2-cocycle $\hat{\varphi}\in \operatorname{Z}^{2}(G;\hat{K})$ such that
\begin{equation}\label{eq:d2-triv-cond}
    (\varphi\smile \hat{\varphi}) + \delta\phi = \pi\, ,
\end{equation}
where $\phi$ is some $U(1)$-valued 3-cochain on $G$ and $\delta$ denotes the usual group cohomology differential.
Now let us write each element of $\Gamma$ as a pair $(k,g)\in \Gamma$, then $p(k,g) = g$, and we further define $q(k,g) = k$. 
By definition we have
\begin{equation}
    p^*\varphi = -\delta q\, ,
\end{equation}
and hence by pulling back~\eqref{eq:d2-triv-cond} we can see that setting
\begin{equation}\label{eq:Gamma-spt-phase}
    \psi = p^*\phi - (q\smile p^*\hat{\varphi})\, ,
\end{equation}
ensures $\delta\psi = p^*\pi$. Notice that this choice of $\psi$ is special, since $p$ restricted to $K\subset\Gamma$ is trivial, this 3-cochain actually vanishes on $K$. The IR symmetry we are constructing will hence take the form
\begin{equation}
    \CIR \simeq \Sigma\cZ(\Vect_K)\, .
\end{equation}

\ni Now onto the 3-group homomorphism. The first component is the group homomorphism
\begin{equation}
    \rho \, :\, G \to \Aut^{br}(\cZ(\Vect_K))\, ,
\end{equation}
induced by the action of $\Gamma/K \simeq G$ on $K$ which, since we have assumed $\Gamma$ to be a central extension, is trivial. This simplifies matters as it means we do not have to compute $\widetilde{\cO_3}$, because its pull-back by $\rho$ vanishes anyway.

\newl The 1-form symmetry $\Pi_1$ of the IR in this situation is
\begin{equation}
    \Pi_1 \simeq K \times \hat{K}\, .
\end{equation}
The symmetry fractionalization data hence decomposes into two 2-cocycles valued in $K$ and $\hat K$, which will turn out to be given by the 2-cocycles $\varphi$ and $\hat{\varphi}$ that we introduced earlier, respectively.
\begin{figure}[h!]
    \centering
    \begin{tikzpicture}[scale = 0.75]

        \draw[black, thick] (-1, 0) node[anchor = east]{$g$}-- (0, -0.5) -- (1, 1.5) -- (0, 2) -- (-1, 0);
        \draw[black, thick] (1, 1.5) -- (1.75, -0.5) node[anchor = south west]{$g'$} -- (0.35, 0.2);
        \draw[black, thick] (0,2) -- (0, 4) node[anchor = south east]{$gg'$} -- (1, 3.5) -- (1, 1.5);
        \draw[red, ultra thick] (1, 1.5) -- (0, 2);

        \draw[blue, thick] (1, 1.5) -- (3, 3) node[black, anchor = west]{$\varphi(g, g')$} -- (2, 3.5) -- (1, 2.75);

        \draw[black] (4.5, 1.75) node{\huge$\strut\rightsquigarrow$};

        \begin{scope}[shift = {(6.5, 0)}]

        \draw[blue, thick] (0.5, 1.75) ellipse (1 and 1.25);
        \draw[black] (1, 2.5) node[anchor = south west]{$\strut\; \hat{k} \in \hat{K}$};
        
        \filldraw[black, thick, fill = white] (1, 1.5) -- (1.75, -0.5) -- (0.75, 0);
        \draw[black, thick] (0.75, 0) -- (0.35, 0.2);

        \draw[black, thick] (0.5, 1.75) -- (0, 2) -- (-1, 0) -- (-0.5, -0.25);
         \filldraw[black, thick, fill = white] (-0.5, -0.25) -- (0, -0.5) --  (1, 1.5) -- (0.5, 1.75);
        
        \draw[black, thick] (0.5, 1.75) -- (0,2) -- (0, 4) -- (0.5, 3.75);
        \filldraw[black, thick, fill = white] (0.5, 3.75) -- (1, 3.5) -- (1, 1.5) -- (0.5, 1.75);
        
        \draw[red, ultra thick] (1, 1.5) -- (0, 2);
        \end{scope}

        \draw[black] (9, 1.75) node{\huge$\strut =$};

        \begin{scope}[shift = {(11, 0)}]
        \draw[black, thick] (-1, 0) -- (0, -0.5) -- (1, 1.5) -- (0, 2) -- (-1, 0);
        \draw[black, thick] (1, 1.5) -- (1.75, -0.5) -- (0.35, 0.2);
        \draw[black, thick] (0,2) -- (0, 4) -- (1, 3.5) -- (1, 1.5);
        \draw[red, ultra thick] (1, 1.5) -- (0, 2);
        \end{scope}

        \draw[black] (14, 1.75) node{\large$\strut\times\langle\varphi(g,\, g'), \hat{k}\rangle$};
        
    \end{tikzpicture}
    \caption{Before gauging, the fusion of two $G\simeq \Gamma/K$ defects lies in the twisted sector given by $\varphi(g,g') \in K$. After gauging, the same junction carries charge $\varphi(g,g') \in K$ under the dual $\hat{K}$ 1-form symmetry.}
    \label{fig:varphi-diagram}
\end{figure}
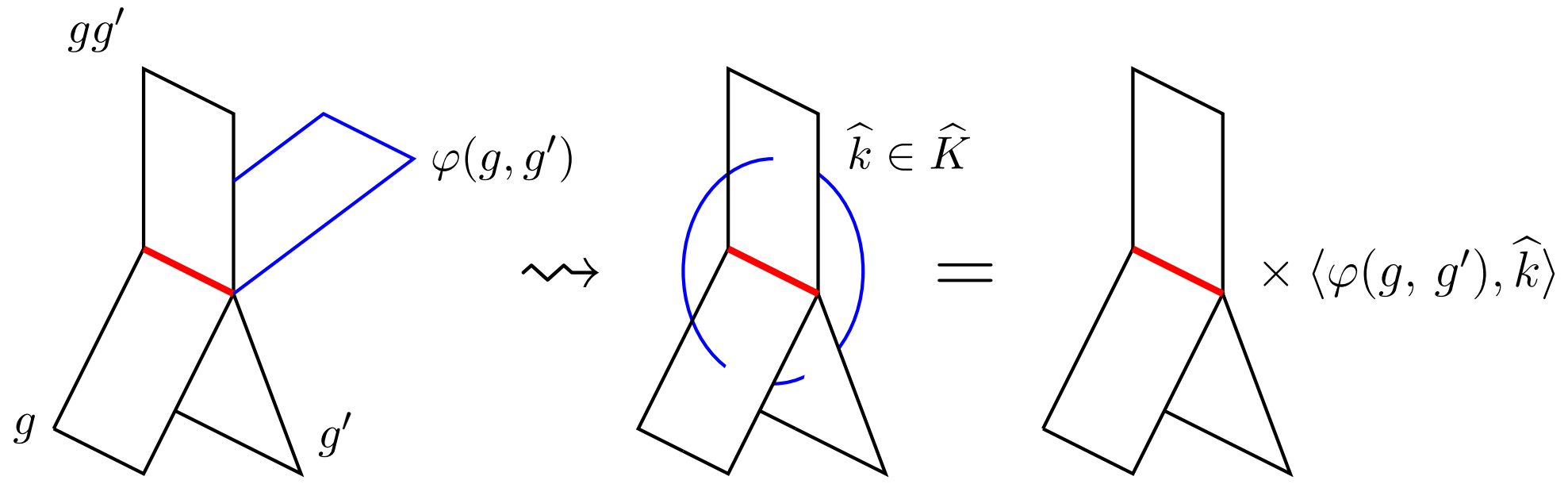

\newl The 2-cocycle data valued in $K$ describes the symmetry fractionalization of $G$ by defects charged under the $\hat K[1]$ symmetry present in the IR. As illustrated above in figure~\ref{fig:varphi-diagram}, the fusion of $G$ defects is twisted by $\varphi\in K$, which means after gauging $K$ the fusion carries charge $\varphi$ under the $\hat K[1]$ symmetry~\cite{Tachikawa:2017gyf}.

\newl To see this more concretely, consider the following field theory argument over a three dimensional manifold $M$. Denote by $\mathbf{a} \in \operatorname{C}^1(M; K)$ and $\hat{\mathbf{b}}\in \operatorname{C}^1(M; \hat K)$ the dynamical fields for the three-dimensional Dijkgraaf--Witten theory, whose Lagrangian is given by
\begin{equation}
    \cL_{\rm DW} = \langle d\mathbf{a} ,\, \hat{\mathbf{b}}\rangle\, .
\end{equation}
We then turn on static background fields $\mathbf{g}_1\in \operatorname{C}^1(M; G)$ for the 0-form $G$ symmetry and $\mathbf{k}_2 \in \operatorname{C}^2(M; K)$ for the 1-form $K$ symmetry, then in the presence of the extension class $\varphi$, the action is shifted to
\begin{equation}
    \cL_{\rm DW} = \langle d\mathbf{a} + \mathbf{k}_2 + \varphi(\mathbf{g}_1) ,\, \hat{\mathbf{b}}\rangle\, .
\end{equation}
As expected, this shift is exactly equivalent to switching on a symmetry fractionalization of the form
\begin{equation}
    \mathbf{k}_2 \mapsto \mathbf{k}_2 + \varphi(\mathbf{g}_1)\, .
\end{equation}
Now switch off $\mathbf{k}_2$ and instead switch on a static background $\hat{\mathbf{k}}_2 \in \operatorname{C}^2(M; \hat K)$ for the $\hat K$ 1-form symmetry. In addition to this, we also couple the $\Gamma$-symmetric SPT phase
\begin{equation}
    \psi(\mathbf{g}_1,\mathbf{a})=\phi(\mathbf{g}_1) -\langle\mathbf{a},\, \hat\varphi(\mathbf{g}_1) \rangle\, ,
\end{equation}
resulting in the modified Lagrangian
\begin{equation}\label{eq:dw-shifted-action}
    \cL_{\rm DW} = \langle d\mathbf{a} + \varphi(\mathbf{g}_1) ,\, \hat{\mathbf{b}}\rangle - \langle \mathbf{a},\, \hat{\mathbf{k}}_2 + \hat\varphi(\mathbf{g}_1)\rangle + \phi(\mathbf{g}_1)\, .
\end{equation}
The shift appearing in the second term corresponds precisely to a symmetry fractionalization 
\begin{equation}
    \hat{\mathbf{k}}_2\mapsto \hat{\mathbf{k}}_2 + \hat{\varphi}(\mathbf{g}_1)\, .
\end{equation}
As a happy consequence of this computation, in the last remaining term of~\eqref{eq:dw-shifted-action} we have also calculated the remaining 3-cochain data to be precisely $\phi$. To better understand this result, we can compute the pull-back of the $\widetilde{\cO_4}$-obstruction given by the braiding
\begin{equation}
    (\rho, \varphi, \hat{\varphi})^*\widetilde{\cO_4} = \varphi\smile \hat{\varphi} = \pi - \delta\phi\, ,
\end{equation}
which, as expected, cancels the anomaly $\pi$ up to the coboundary of $\phi$. Notice that when we fixed $\psi$ in~\eqref{eq:Gamma-spt-phase} we actually had the freedom to shift it by a 3-cocycle, and that freedom is also reflected here: shifting $\phi$ by a 3-cocycle preserves the cancellation condition but moves us to a different equivalence class of generalized fiber 2-functor.

\subsection{Example: $p\, :\, Q_8 \to \bbZ_2^2$}\label{subsec:ex-Q8}
In this subsection we present an example wherein we construct a generalized fiber 2-functor for a three-dimensional $G = \bbZ_2^2$ symmetry with anomaly $[\pi] \in \operatorname{H}^4(\bbZ_2^2;U(1)) \simeq \bbZ_2^2$.
We do this by utilizing the fact that the symmetry extension
\begin{equation}
    K = \bbZ_2 \to Q_8 \xrightarrow{p} \bbZ_2^2\, ,
\end{equation}
with $Q_8 = \langle i, j, k \mid i^2 = j^2 = k^2 = ijk = -1 \rangle$ being the quaternion group, admits a trivialization $\delta\psi = p^*\pi$ for any choice of $\pi$, simply because $\operatorname{H}^4(Q_8;U(1)) = 0$. Following the arguments of the previous subsection, this means we can start from a three-dimensional invertible TFT with $2\Vect^{p^*\pi}_{Q_8}$ symmetry, and then gauge $K = \{1, -1\}$ to obtain a Dijkgraaf--Witten theory with a $2\Vect_{\bbZ_2^2}^\pi$ symmetry. This TFT has a unique vacuum and describes a generalized fiber 2-functor of the form
\begin{equation}
    \cF \; : \; 2\Vect_{\bbZ_2^2}^\pi \to \Sigma\cZ(\Vect_{\bbZ_2})\, .
\end{equation}

\newl To construct a concrete example, we introduce the notation $a = (a_1,{a_2}) \in \bbZ_2\times\bbZ_2$ with $a_1,{a_2} = 0,1$, and specify a mixed anomaly
\begin{equation}\label{eq:q8-anomaly}
    \pi(a, b, c, d) = (-1)^{a_2 b_1 c_1 d_1} \, .
\end{equation}
Here we switch to the multiplicative notation for cochains.

\newl We also introduce the notation $\tilde{a} = (-1)^{a_0}i^{a_1} j^{{a_2}} \in Q_8$ for $a_0 = 0,1$ so that $p(\tilde{a}) = a$, $q(\tilde{a}) = (-1)^{a_0}$, and the extension class is given by
\begin{equation}
    \varphi(a, b) = (-1)^{{a_1}b_1 + {a_2}b_2 + {a_2}b_1}\, .
\end{equation}

\newl Then we can quickly validate that the 2-cocycle
\begin{equation}
    \hat{\varphi}(a, b) = (-1)^{{a_1}b_1}
\end{equation}
and 3-cochain
\begin{equation}
    \phi(a, b, c) = i^{({a_1}+{a_2})b_1c_1}
\end{equation}
taken together satisfy the condition~\eqref{eq:d2-triv-cond}. Hence we can construct a 3-cochain
\begin{equation}
    \psi(\widetilde{a}, \widetilde{b}, \widetilde{c}) = \frac{\phi(a, b,c)}{\langle q(\widetilde{a}), \hat{\varphi}(b, c)\rangle} = i^{({a_1}+{a_2})b_1c_1}(-1)^{a_0b_1c_1}
\end{equation}
over $Q_8$, satisfying $p^*\pi = \delta\psi$.

\newl Putting this all together, we can construct a generalized fiber 2-functor which is summarized by the 3-group homomorphism
\begin{equation}
    \cF^\times \; : \; \bbG_{\pi} \to \mathrm{Pic}(\cZ(\Vect_{\bbZ_2})) \simeq \mathrm{BrPic}(\Vect_{\bbZ_2})
\end{equation}
for $G = \bbZ_2^2$ and $\pi$ as in~\eqref{eq:q8-anomaly}, given by the following data:
\begin{itemize}
    \item[---] A trivial group homomorphism 
    \begin{align}
        \rho \; :\; \bbZ_2^2 &\to \Pi_0 \simeq \bbZ_2\, ,\nonumber\\
        a &\mapsto 1\, .
    \end{align}

    \item[---] Our pair of 2-cocycles 
    \begin{equation}
    \strut (\varphi, \hat{\varphi}) \in \operatorname{Z}^2(\bbZ_2^2; K\times\hat{K})\, .
    \end{equation}

    \item[---] The 3-cochain $\strut\phi \in \operatorname{C}^3(\bbZ_2^2;U(1))$ satisfying
    \begin{equation}
         \strut\delta\phi = ( \varphi \smile \hat{\varphi})^{-1}\,\pi \, .
    \end{equation}
\end{itemize}

\section{The SymTFT and Genuine Fiber 2-Functors}\label{sec:symtft}

\noindent The SymTFT framework provides an economical and systematic way to classify and characterize gapped phases with categorical symmetries.
 Given a $(2\!+\!1)$-dimensional quantum system $\cT$ with fusion 2-categorical symmetry $\cC$, the SymTFT is a fully extended 4d Douglas--Reutter TQFT $\mathrm{DR}[\cC]$, together with a distinguished topological boundary $\fB_\cC$ (called the \emph{symmetry boundary}) on which the $2$-category of defects is $\cC$.

\begin{figure}[H]
    \centering
    \begin{tikzpicture}[scale = 1]
    
        \draw[gray, thick] (0,2) -- (1.5,2) node[anchor = south, black]{$\strut \mathrm{DR}[\cC]$} -- (3, 2);
        \draw[gray, thick] (-3,1) -- (-1, 1);
        \draw[gray, thick] (-3,-1) -- (-1, -1);
        \draw[gray, thick] (-0.8,0) -- (0, 0);
        \draw[gray, thick] (-2,0) -- (-1.2, 0);
        \draw[gray, thick] (-2,2) -- (0, 2);
        \node at (-2,0.5){$ 
    \mathrm{CY}[\cT]\;\strut$};
        \draw[gray, thick] (-1,1) -- (2, 1);
        \draw[gray, thick] (0,0) -- (3, 0);
        \draw[gray, thick] (-1,-1) -- (2, -1);
        \draw[black, thick] (-1, -1) -- (-1, 1)  -- (0, 2) -- (0, 0) -- (-1, -1);
        \node at (-0.5,0.5){$\widetilde{\cT}\;\strut$};
        \node at (2.5,0.5){$\strut\;\fB_\cC$};
        \begin{scope}[shift={(3,0)}]
        \draw[black, thick] (-1, -1) -- (-1, 1) -- (0, 2) -- (0, 1)  -- (0, 0) -- (-1, -1);
        \end{scope}

        \draw[black] (4, 0.5) node{\huge$\strut\rightsquigarrow$};

        \begin{scope}[shift = {(8, 0)}]
            \draw[black, thick] (-1, -1) -- (-1, 1) -- (0, 2)  -- (0, 0) -- (-1, -1);
        \draw[gray, thick] (-3,1) -- (-1, 1);
        \draw[gray, thick] (-3,-1) -- (-1, -1);
        \draw[gray, thick] (-0.8,0) -- (0, 0);
        \draw[gray, thick] (-2,0) -- (-1.2, 0);
        \draw[gray, thick] (-2,2) -- (0, 2);
        \node at (-2,0.5){$ 
    \mathrm{CY}[{\cT}]\;\strut$};
    \node at (-.5,0.5){$\strut\;\cT$};
        \end{scope}
    \end{tikzpicture}
    \caption{The sandwich construction provides a heuristic way to reconstruct the $\cC$-symmetric theory $\cT$ via the interval compactification of the SymTFT $\mathrm{DR}[\cC]$.
    The theory $\cT$ may carry an intrinsic anomaly, in which case the relative boundary $\widetilde\cT$ is an interface to the 4d anomaly theory, which is given by a Crane--Yetter TQFT $\mathrm{CY}[\cT]$.
    }
    \label{fig:symtft-sandwich}
\end{figure}
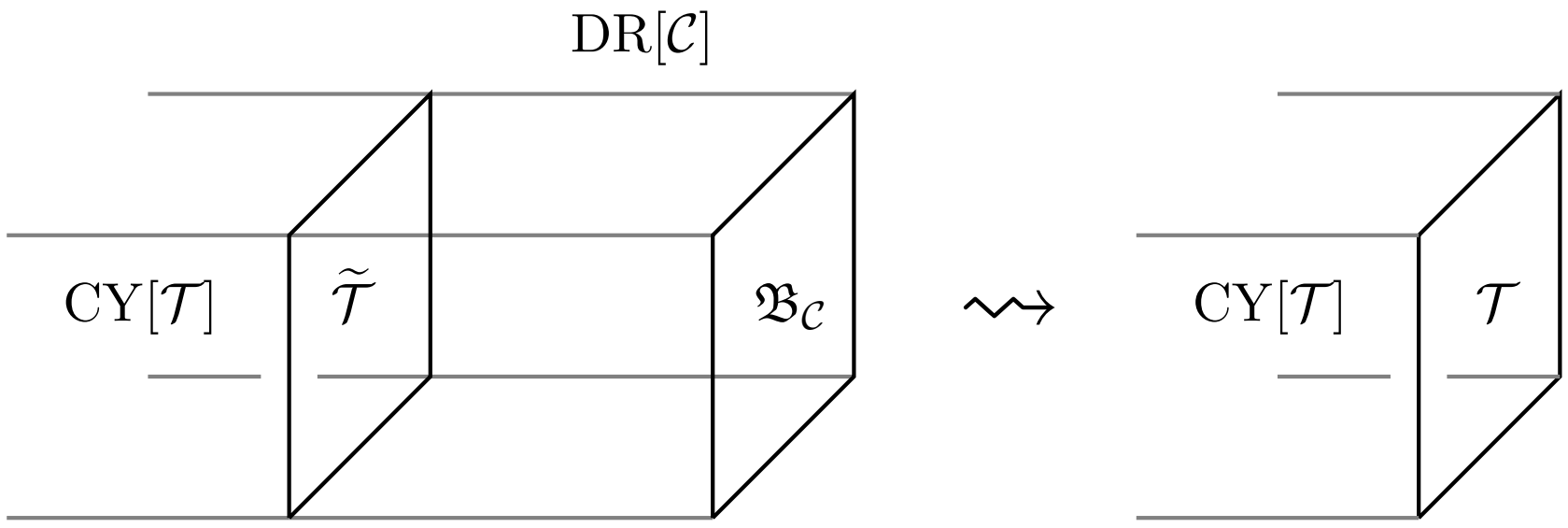
\ni A common interpretation of the SymTFT is the sandwich construction, illustrated in figure~\ref{fig:symtft-sandwich}, where the theory $\cT$ can be obtained from the SymTFT $\mathrm{DR}[\cC]$ supported on an interval, with a (non-topological) boundary $\widetilde{\cT}$ (called the relative boundary) hosting the dynamical degrees of freedom living on one end of the interval, and the symmetry boundary hosting $\cC$ on the other. Compactifying that interval then reproduces the original theory $\cT$.

\newl In general $\mathcal{T}$ may carry its own intrinsic gravitational anomaly, in which case it is not a genuine $(2\!+\!1)$-dimensional theory but rather a relative theory attached to the boundary of a 4d anomaly theory, given by the invertible Crane--Yetter TQFT $\mathrm{CY}[\mathcal{T}]$ \cite{Crane:1993if,Walker:2011mda,Freed:2012bs,Haioun:2025hen}. This intrinsic anomaly is genuine extra data completely independent of the existing symmetry data $\cC$. In the sandwich construction this anomaly is carried entirely by the relative boundary $\widetilde{\mathcal{T}}$, which is then an interface~\footnote{We will continue to use the name ``relative boundary" despite this subtlety, to match the existing literature.} between $\mathrm{DR}[\cC]$ and $\mathrm{CY}[\mathcal{T}]$.

\newl This construction lends itself to both the classification of 2-categorical symmetries, and of $(2\!+\!1)$-dimensional gapped phases exhibiting those symmetries. For instance, exchanging the symmetry boundary for a topological interface to some other invertible 4d theory prior to interval compactification produces a new theory $\cT'$ whose symmetry $\cC'$ is related to $\cC$ by a (generalized) gauging procedure~\cite{Bullimore:2024khm, Bhardwaj:2024qiv, Bhardwaj:2025piv}. If we instead exchange the relative boundary with some choice of topological boundary, then the interval compactification would result in a $3$-dimensional gapped phase for $\cC$~\cite{Bhardwaj:2024qiv, Bhardwaj:2025piv}.

\newl The SymTFT $\mathrm{DR}[\cC]$, together with its spectrum of topological boundary conditions, can be understood very efficiently by studying those bulk topological defects with codimension greater than 1. These surface and line defects organize into a braided fusion 2-category given by the Drinfeld center $\cZ(\cC)$~\cite{kong:2020center,Douglas:2018qfz, Decoppet:2022kcf, Johnson-Freyd:2021chu, Bhardwaj:2025piv}.
The remaining codimension-1 defects and boundary conditions then arise from condensing those surfaces and lines~\footnote{In the case of genuine codimension-1 defects, this is an example of the principle of remote detectability~\cite{kong:2020classification,Johnson-Freyd:2020utr}.}, and carry no independent 
topological data beyond that which is already encoded in $\cZ(\mathcal{C})$.

\newl In the rest of this section we will describe the SymTFT, including its defect content and topological boundary conditions, in more detail. We also present a description of RG flow from the SymTFT perspective, which gives physical intuition on how to construct fiber 2-functors.

\subsection{Bosonic and Fermionic Symmetries}\label{subsec:bosonic-fermionic}
Fusion 2-categorical symmetries $\mathcal{C}$ fall into two families according to the structure of their transparent genuine lines \cite{Xu:2026thesis,Decoppet:2024htz}. 
The transparent lines form the M\"uger center $\mathcal{Z}_2(\Omega\mathcal{C})$, where $\Omega\mathcal{C}$ is the braided fusion category of all genuine lines in $\mathcal{C}$. 
By Deligne's theorem \cite{Deligne2002}, the transparent lines are either all bosonic, forming the fusion category $\Rep(H)$ for a finite group $H$, or include some emergent fermionic lines, and form the fusion category $\Rep(H,z)$, where $(H,z)$ is a finite group $H$
with a central order-2 element $z$ generating fermion parity. 
In the latter case, $\Rep(H,z)$ is equivalent to $\Rep(H)$ as a fusion category, but not as a braided fusion category. Instead the spin of a line defect is given by the map $\Rep(H,z) \to \Rep(\mathbb{Z}_2^f) \simeq \mathsf{sVec}$, and those with non-trivial spin describe emergent fermion lines.
We call these two types of fusion 2-categories all-boson (AB) type and emergent fermion (EF) type fusion 2-categories, respectively.
We should emphasize however, that this nomenclature should not be confused with that of the underlying theory.
It is indeed possible for a bosonic $(2\!+\!1)$-dimensional quantum system (i.e., a system whose spectrum of particles contains only bosons) to exhibit an EF type symmetry. 
These theories can be obtained by bosonizing fermionic theories~\cite{Chen:2017fvr, Kapustin:2017jrc, Gaiotto:2020iye}.


\newl In order to concretely describe the structure of the SymTFT, we now restrict our focus to (AB) type fusion 2-categories.
Following~\cite{Decoppet:2022kcf, Bullimore:2024khm, Decoppet:2024htz}, we begin with the observation that any such $\cC$ is related by a gauging operation (i.e., Morita equivalent) to another (AB) fusion 2-category of the form
\begin{equation}\label{eq:c-prime}
\cC' = 2\Vec_G^\pi \boxtimes \Sigma M\,,
\end{equation}
where $G$ is a finite group, $\pi \in \operatorname{Z}^4(G;U(1))$ is an 't Hooft anomaly for the 0-form symmetry $G$, and $M$ is a non-degenerate braided fusion category.
This Morita equivalence corresponds to gauging the full M\"uger center 
$\mathcal{Z}_2(\Omega\mathcal{C}) \simeq \Rep(H)$ 1-form symmetry in $\cC$.
After gauging one obtains $\cC'$, whose loop space $\Omega\mathcal{C}' = M$ is the (non-degenerate) braided fusion category obtained by taking the modular closure~\footnote{For a braided fusion category $B$ whose M\"uger center is Tannakian, $\mathcal{Z}_2(B)\simeq\Rep(H)$, the \emph{modular closure} is the de-equivariantization $M:=B\boxtimes_{\Rep(H)}\Vec$.
It is a non-degenerate braided fusion category, and $B$ is recovered from it as the $H$-equivariantization, $B\simeq M^H$. 
Physically, the transparent lines in $\Rep(H)$ are the Wilson lines of a gauged $H$ 0-form symmetry.
These are bosonic and braid trivially with everything. 
The modular closure is the category of lines of the theory obtained by un-gauging $H$.} of $\Omega\cC$~\cite{Mueger:1999yb}.
Furthermore, since every braided autoequivalence of the modular category $M$ is implemented by a condensation surface \cite{etingof2009,Gaiotto:2019xmp}, the lines of $M$ decouple, and the remaining surfaces form a fusion 2-category without lines, which is grouplike \cite{Johnson-Freyd:2020twl}, hence $\cC'\simeq\TwoVec^\pi_G\boxtimes\Sigma M$.
%


\newl This Morita equivalence extends to an equivalence of TQFTs: $\mathrm{DR}[\cC] \simeq \mathrm{DR}[\cC']$. We hence see that for any (AB) type fusion 2-category $\cC$, the SymTFT $\mathrm{DR}[\cC]$ is a 4d Dijkgraaf--Witten theory for the group $G$ with topological action $\pi$, stacked with an invertible Crane--Yetter phase corresponding to the Witt class of $M$~\cite{Lan:2018vjb, Johnson-Freyd:2020usu}.

\subsection{Topological Defects and Boundary Conditions}\label{subsec:symtft-defects-and-boundaries}
We now turn to the defect content of the SymTFT of an (AB) fusion 2-categorical symmetry $\cC$. To begin, we note that our two Morita equivalent fusion 2-categories $\mathcal{C}$ and $\mathcal{C}'$, and indeed any two Morita equivalent fusion 2-categories, have corresponding SymTFTs that are equivalent, and in particular~\cite{Decoppet:2022kcf}
\begin{equation}
\label{eq:Morita}
\cZ(\mathcal{C}) \simeq \cZ(\mathcal{C}')\,.
\end{equation}
It therefore suffices to compute $\cZ(\mathcal{C})$ using the canonical representative $\cC'$.
Using the fact that the Drinfeld center distributes over the 2-Deligne tensor product, we find
\begin{equation}
\label{eq:Drinfeld}
\begin{split}
\cZ(2\Vec_G^\pi \boxtimes \Sigma M) \simeq&\, 
\cZ(2\Vec_G^\pi) \boxtimes \cZ(\Sigma M) \\ 
\simeq&\, \cZ(2\Vec_G^\pi)\,.
\end{split}
\end{equation}
Notice that since $M$ is non-degenerate, $\cZ(\Sigma M)$ is trivial, corresponding to the defect content (or lack thereof) of a $4$-dimensional invertible Crane--Yetter phase.

\newl {\bf Bulk Topological Defects.} Due to \eqref{eq:Morita} and \eqref{eq:Drinfeld}, the SymTFT bulk defect content is controlled entirely by 
$\cZ(2\Vec_G^\pi)$ and so the braided fusion 2-category 
$\cZ(\mathcal{C})$ has
\begin{equation}
\pi_0(\cZ(\mathcal{C})) \cong \mathrm{Conj}(G), \qquad 
\Omega(\cZ(\mathcal{C})) \simeq \Rep(G)
\end{equation}
where $\pi_0(\cZ(\mathcal{C}))$ refers to the connected/condensation components of the SymTFT, and $\mathrm{Conj}(G)$ denotes the set of conjugacy classes of $G$.

\newl Surface defects are labeled by finite semisimple $\pi$-twisted $G$-crossed categories \cite{kong:2020center,Xu:2024etale}: these are $G$-graded categories equipped with a ($\pi$-twisted) $G$-action that changes the grading by conjugation.
Surfaces then fuse according to the Deligne tensor product of such categories, where the new grading is determined via the group composition, and the $G$-action acts distributively.
Defects in the connected component of the identity form the symmetric fusion 2-category $2\Rep(G)$, and are condensations of those genuine line defects that are the Wilson lines from the Dijkgraaf--Witten gauge theory, and themselves form the symmetric fusion category $\Rep(G)$.

\newl {\bf Topological Boundaries.} We now describe the SymTFT topological boundaries.
Before describing boundary conditions for a general choice of $\cC$, let us first isolate the Dijkgraaf--Witten component by setting $M$ to be trivial. This fixes $\cC' = 2\Vect_G^\pi$. 
There exists a canonical (Dirichlet) topological boundary that supports the symmetry $\cC'$, that we denote by $\fB_\mathrm{Dir}$.
This boundary is, up to the stacking of a decoupled 3d TQFT, characterised uniquely by the fact that all bulk Wilson lines in $\Rep(G)$ condense on it.

\newl Starting from $\fB_{\rm Dir}$, a large class of boundaries supporting symmetries described by group-theoretical fusion 2-categories~\cite{Etingof:2002vpd}
\begin{equation}
    \cC = \mathsf{C}(G, \pi | H, \psi)\, ,
\end{equation}
are obtained by gauging a subgroup $H\subseteq G$ that is non-anomalous, $[\pi|_H]=0\in \operatorname{H}^4(H;U(1))$, together with a choice of trivialization $\psi\in \operatorname{C}^3(H;U(1))$, $\delta\psi=\pi|_H$, denoted by
\begin{equation}
\label{eq:NeumannH boundary}
    \fB_{(H,\psi)} = \frac{\fB_{\rm Dir}\boxtimes {\rm SPT}_{H}^{\psi}}  {H_{\rm diag}}\,.
\end{equation}
Physically, $\psi$ is the boundary counterterm whose variation cancels the inflow of $\pi|_H$ onto the $H$ subsymmetry of $\fB_{\rm Dir}$, rendering it gaugeable.
Two trivializations $\psi, \psi'$ of $\pi|_H$ differ by 3-cocycles $\psi -  \psi' \in \operatorname{Z}^3(H;U(1))$, and inequivalent gaugings of a fixed $H$ are labeled by the discrete torsion  $\operatorname{H}^3(H;U(1))$. 
When the representative $\pi$ is chosen with $\pi|_H=0$, which is always possible, $\psi$ is a 3-cocycle and $\fB_{(H,\psi)}$ is obtained by stacking the $H$-SPT phase $\psi$ onto $\fB_{\rm Dir}$ and gauging the diagonal $H$.

\newl The boundary $\fB_{(H,\psi)}$ supports the group-theoretical symmetry $\mathsf{C}(G, \pi | H, \psi)$, whose surfaces lie in connected components
\begin{equation}
    \pi_0\mathsf{C}(G, \pi | H, \psi) \simeq H\backslash G / H\, ,
\end{equation}
that fuse according to the ring structure on double cosets. This symmetry also contains genuine lines, which are Wilson lines labeled by $\Rep(H)$. This boundary is characterized by the restriction functor
 \begin{equation}
    \mathrm{Res}^G_H \, :\, \Rep(G) \to \Rep(H)\, ,
\end{equation}
which captures the boundary restriction of Wilson lines living in the SymTFT bulk.

\newl The above only provides a subset of topological boundaries of the SymTFT.
To obtain the full spectrum, we need to employ a \emph{generalized} gauging procedure~\cite{Bullimore:2024khm, Bhardwaj:2024qiv}.
This involves stacking an $H$-symmetric TFT onto $\fB_\mathrm{Dir}$ before gauging. In this construction we no longer require $H$ to be non-anomalous because we can pick a coupling to cancel that anomaly, labeled by a $\pi|_H$-twisted faithfully $H$-graded fusion category $A$~\cite{Xu:2024etale, Bullimore:2024khm, Pearson:2024snm, Inamura:2025cum, Bhardwaj:2025jtf}. After stacking and gauging, we obtain yet more topological boundaries that describe distinct fusion 2-categories that are not necessarily group-theoretical themselves, but are still Morita equivalent to something group-theoretical. 
We denote the new boundaries by
\begin{equation}
    \fB_{(H,\, A)} = \frac{\fB_\mathrm{Dir}\boxtimes \mathrm{TV}[A_1]}{H_{\rm diag}}\, ,
\end{equation}
where $A=\bigoplus_{h\in H}A_h$ is a faithfully $H$-graded, $\pi|_H$-twisted fusion category with trivial component $A_1$.
Concretely, we stack the 3d Turaev--Viro TFT $\mathrm{TV}[A_1]$ whose lines form $\cZ(A_1)$, onto the Dirichlet boundary.
The grading of $A$ equips this TFT with an $H$ symmetry, namely the 3-group homomorphism 
\begin{equation}
    H\to\mathrm{BrPic}(A_1)\simeq\mathrm{Pic}(\cZ(A_1))
\end{equation}
determined by the components $A_h$ \cite{etingof2009}, and the $\pi|_H$-twist of the pentagonator of $A$ means that this symmetry carries the 't Hooft anomaly $\pi|_H^{-1}$.
The Dirichlet boundary $\fB_{\rm Dir}$ carries the subgroup symmetry $H\subseteq G$ with anomaly $\pi|_H$.
The diagonal $H$ symmetry of the stacked theory is therefore anomaly-free and can be gauged.
Notice in particular, that we can reproduce the boundaries $\fB_{(H, \psi)}$ from~\eqref{eq:NeumannH boundary} by picking $A = \Vect_H^\psi$.
Algebraically, gauging $H_{\rm diag}$ is the condensation of $A$ regarded as a connected rigid algebra in $2\Vec^{\pi}_G$, supported in degrees $H\subseteq G$.
Every connected rigid algebra in $2\Vec^\pi_G$ is of this form, so this construction exhausts the boundaries of the SymTFT.

\newl {\bf Interfaces to Invertible 4d Theories.} Beyond topological boundaries, one may also consider topological \emph{interfaces} between $\mathrm{DR}[\cC]$ and invertible 4d (Crane--Yetter) theories. Starting from the Dijkgraaf--Witten TFT, an interface to the invertible theory $\mathrm{CY}[M]^{-1}$ can be folded to produce a boundary for the more general (AB) type SymTFT $\mathrm{DR}[\cC']$ where $\cC'$ is given by~\eqref{eq:c-prime}.

\newl Such interfaces arise from the same stacking-and-gauging procedure as before, once we allow the stacked 3d TFT to be a Reshetikhin--Turaev theory $\mathrm{RT}[M]$ whose input $M$ lies in a non-trivial Witt class.
This is precisely the case that the previous construction could not reach since $\mathrm{RT}[M]$, not being the Turaev--Viro theory of any fusion category, cannot be presented as an algebra internal to $2\Vec^\pi_G$.
Instead it is a boundary of the invertible theory $\mathrm{CY}[M]$ \cite{Walker:2011mda,Haioun:2025hen,Haioun:2023oyb}, which is what turns the result of the construction from a boundary into an interface.

\newl Concretely, we equip $\mathrm{RT}[M]$ with an anomalous $H$ symmetry, 
equivalently a $\pi|_H$-twisted $H$-crossed braided extension $\widehat{M}_H = \bigoplus_{h\in H} \widehat{M}_h$ where $\widehat{M}_1 = M$, whose 't Hooft anomaly cancels that of $H\subseteq G$ on $\fB_{\rm Dir}$.
Stacking and gauging the diagonal $H$ then produces a topological interface between $\mathrm{DR}[\cC]$ and $\mathrm{CY}[M]$ (as depicted in figure~\ref{fig:symtft-sandwich}) denoted as
\begin{equation}
    \fB_{(H,\, \widehat{M}_H)} = \frac{\fB_\mathrm{Dir}\boxtimes \mathrm{RT}[M]}{H_{\rm diag}}\, .
\end{equation}
Taken together, the boundaries and interfaces $\fB_{(H, \widehat{M}_H)}$, enumerated over all possible non-degenerate braided fusion categories $M$, describe every topological interface between the 4d Dijkgraaf--Witten theory and an invertible Crane--Yetter theory. By folding such interfaces, we capture every (AB) fusion 2-categorical symmetry boundary. Moreover, this labeling reproduces the classification of topological boundaries by Lagrangian algebras internal to $\cZ(2\Vect_G^\pi)$~\cite{Decoppet:2022dnz-Morita,Decoppet:2023hkr,Xu:2024etale}.
The Witt class itself is invisible to the Drinfeld center, which is why the interfaces to non-trivial $\mathrm{CY}[M]$ appear there only after forgetting the invertible theory.

\subsection{Gapped Phases and RG 2-Functors}\label{subsec:symtft-rg-flow}
We now move on to the study of $\cC$-symmetric gapped phases, with a particular focus on describing the RG 2-functor from the perspective of the SymTFT.
We begin by constructing the sandwich picture for a gapped effective QFT $\TUV$ with symmetry $\CUV$, which flows to an IR fixed point described by a TQFT $\TIR$ with its own symmetry $\CIR$. This construction together with its RG flow is depicted below in figure~\ref{fig:symtft-rg-functor}.

\begin{figure}[H]
    \centering
    \begin{tikzpicture}[scale = .75]
    
    \begin{scope}[shift={(-8, 0)}]
        \draw[gray, thick] (0,2) -- (1.5,2) node[anchor = south, black]{$\strut \mathrm{DR}[\CUV]$} -- (3, 2);
        \draw[gray, thick] (-1,1) -- (2, 1);
        \draw[gray, thick] (0,0) -- (3, 0);
        \draw[gray, thick] (-1,-1) -- (2, -1);
        
        \draw[black, thick] (-1, -1) -- (-1, 1) node[anchor = east]{$\widetilde{\TUV}\;\strut$} -- (0, 2) -- (0, 0) -- (-1, -1);
        \begin{scope}[shift={(3,0)}]
        \draw[black, thick] (-1, -1) -- (-1, 1) -- (0, 2) -- (0, 1) node[anchor = west]{$\strut\;\fB_\CUV$} -- (0, 0) -- (-1, -1);
        \end{scope}
        \draw[draw=bblue, ultra thick] (2.5, -0.5)  -- (2.5, 1.5);
        \node[text=bblue] at (3.3,-0.6) {\large$\strut\qquad X\in \CUV$};
        \draw[black] (5, 0.5) node{\huge$\strut\rightsquigarrow$};
    \end{scope}
    
    \begin{scope}
        \draw[gray, thick] (0,2) -- (1.5,2) node[anchor = south, black]{$\strut \mathrm{DR}[\CUV]$} -- (3, 2);
        \draw[gray, thick] (-1,1) -- (2, 1);
        \draw[gray, thick] (0,0) -- (3, 0);
        \draw[gray, thick] (-1,-1) -- (2, -1);
        
        \draw[black, thick] (-1, -1) -- (-1, 1) node[anchor = east]{$\widetilde{\TIR}'\;\strut$} -- (0, 2) -- (0, 0) -- (-1, -1);
        \begin{scope}[shift={(3,0)}]
        \draw[black, thick] (-1, -1) -- (-1, 1) -- (0, 2) -- (0, 1) node[anchor = west]{$\strut\;\fB_\CUV$} -- (0, 0) -- (-1, -1);
        \end{scope}
        \draw[draw=bblue, ultra thick] (2.5, -0.5)  -- (2.5, 1.5);
        \node[text=bblue] at (3.3,-0.6) {\large$\strut\qquad X\in \CUV$};
        \draw[black] (5, 0.5) node{\huge$\strut\rightsquigarrow$};
    \end{scope}
        \begin{scope}[shift = {(8, 0)}]
        \draw[draw=bblue, ultra thick] (-0.5, -0.5)  -- (-0.5, 1.5);
            \draw[black, thick] (-1, -1) -- (-1, 1) -- (0, 2) node[anchor = west]{$\strut\;\TIR$} -- (0, 0) -- (-1, -1);
        \node[text=bblue] at (0.8,-0.6) {\large$\strut\qquad \cF(X)\in \CIR$};
        \end{scope}
    \end{tikzpicture}
    \caption{The RG flow sends the (non-topological) relative boundary $\widetilde{\TUV}$ to the (topological) boundary $\widetilde{\TIR}'$. Interval compactification then maps a UV defect $X \in \CUV$ to $\cF(X) \in \CIR$. In an effort to keep this diagram readable, we have omitted the intrinsic anomaly theory $\mathrm{CY}[\TUV]$. }
    \label{fig:symtft-rg-functor}
\end{figure}
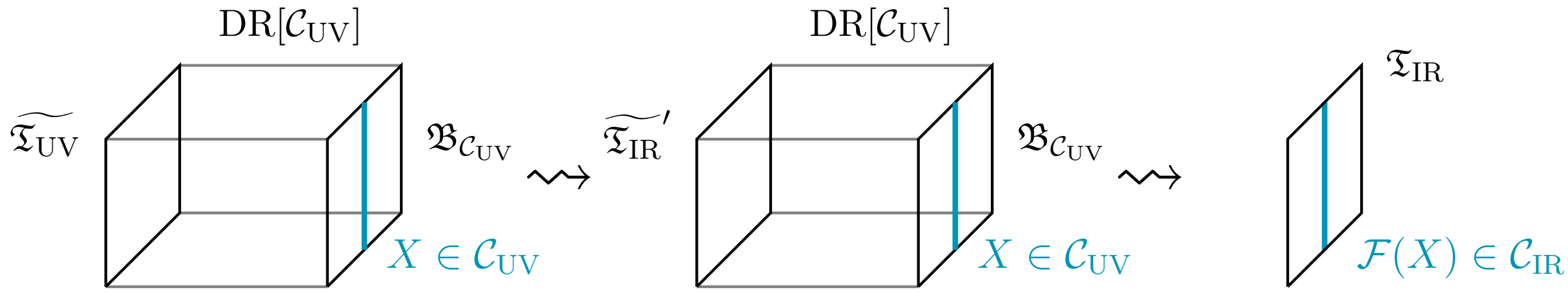

\ni The first arrow appearing in figure~\ref{fig:symtft-rg-functor} represents the RG flow. The SymTFT $\mathrm{DR}[\CUV]$, intrinsic anomaly theory $\mathrm{CY}[\TUV]$ and symmetry boundary $\fB_\CUV$ are all unaffected by this flow, since they are topological and thus completely independent of scale. Meanwhile the relative boundary $\widetilde{\TUV}$ flows to a topological boundary that we denote $\widetilde{\TIR}'$, to distinguish it from the relative boundary $\widetilde{\TIR}$ associated to the SymTFT of $\CIR$.

\newl The second arrow in figure~\ref{fig:symtft-rg-functor} represents the interval compactification, which in this setting produces the TQFT $\TIR$. The benefit of doing this from the point of view of the UV SymTFT is that we can now directly construct the RG 2-functor
\begin{equation}
    \cF \, :\, \CUV \to \CIR\, ,
\end{equation}
which takes UV topological defects to their IR counterparts. As we illustrate above in figure~\ref{fig:symtft-rg-functor}, this functor is implemented by inserting a UV defect on the symmetry boundary prior to compactifying the interval in the IR.

\newl We can use this construction to determine which symmetries in $(2\!+\!1)$ dimensions admit a fiber 2-functor, that is, an RG 2-functor into the trivial TQFT with $\CIR = 2\Vect$.
It was shown in \cite{Decoppet:2023bay} that this is possible if and only if $\CUV$ is group-theoretical and of the form 
\begin{equation}
\CUV \simeq \mathsf{C}(G, \pi| H,\psi)\, , 
\end{equation}
such that $H$ admits a complement $K\subseteq G$, i.e.\ $G=H\bowtie K$ is an exact factorization, $HK=G$ and $H\cap K=\{1\}$, on which $\pi$ is trivializable, $\pi|_K=\delta\omega$. Fiber 2-functors are then classified by such pairs $(K,\omega)$ up to conjugation by $G$ and coboundaries. 

\begin{figure}[H]
    \begin{minipage}[t]{0.45\textwidth}
        \centering
         \begin{tikzpicture}[scale = 1]
            \draw[gray, thick] (0,2) -- (1.5,2) node[anchor = south, black]{$\strut \mathrm{DR}[\CUV]$} -- (3, 2);
            \draw[gray, thick] (-1,1) -- (2, 1);
            \draw[gray, thick] (0,0) -- (3, 0);
            \draw[gray, thick] (-1,-1) -- (2, -1);
            
            \draw[black, thick] (-1, -1) -- (-1, 1) node[anchor = east]{$\widetilde{\TIR}'\;\strut$} -- (0, 2) -- (0, 0) -- (-1, -1);
            \draw[blue, opacity = 0.25, ultra thick] (-0.5, 0.5) -- (2.5, 0.5);
            \filldraw[red] (-0.5, 0.5) circle (1.5pt);
            \begin{scope}[shift={(3,0)}]
            \draw[black, thick] (-1, -1) -- (-1, 1) -- (0, 2) -- (0, 1) node[anchor = west]{$\strut\;\fB_\CUV$} -- (0, 0) -- (-1, -1);
            \draw[blue, ultra thick] (-.5, 0.5) -- (-.5, 1.5);
            \filldraw[red] (-0.5, 0.5) circle (1.5pt);
            \end{scope}
        \end{tikzpicture}
        \caption{A UV symmetry defect is endable in the IR if in the UV SymTFT it can be extended to a bulk line that ends on the IR relative boundary.}
    \label{fig:symtft-twisted-sectors}
    \end{minipage}%
    \hfill%
    \begin{minipage}[t]{0.45\textwidth}
        \centering
         \begin{tikzpicture}[scale = 1]
            \draw[gray, thick] (0,2) -- (1.5,2) node[anchor = south, black]{$\strut \mathrm{DR}[\CUV]$} -- (3, 2);
            \draw[gray, thick] (-1,1) -- (2, 1);
            \draw[gray, thick] (0,0) -- (3, 0);
            \draw[gray, thick] (-1,-1) -- (2, -1);
            
            \draw[black, thick] (-1, -1) -- (-1, 1) node[anchor = east]{$\widetilde{\TIR}'\;\strut$} -- (0, 2) -- (0, 0) -- (-1, -1);
            \draw[blue, opacity = 0.25, ultra thick] (-0.5, 0.5) -- (2.5, 0.5);
            \filldraw[red] (-0.5, 0.5) circle (1.5pt);
            \begin{scope}[shift={(3,0)}]
            \draw[black, thick] (-1, -1) -- (-1, 1) -- (0, 2) -- (0, 1) node[anchor = west]{$\strut\;\fB_\CUV$} -- (0, 0) -- (-1, -1);
            \filldraw[red] (-0.5, 0.5) circle (1.5pt);
            \end{scope}
        \end{tikzpicture}
        \caption{IR topological local operators emerge as a result of bulk lines that can end on the UV symmetry and IR relative boundaries simultaneously.}
    \label{fig:symtft-local-ops}
    \end{minipage}
\end{figure}

\ni In the SymTFT picture, the UV defects that vanish in the IR are precisely those that can be connected to line defects in the SymTFT bulk, that can then end on the IR relative boundary $\widetilde{\TIR}'$ as illustrated in figure~\ref{fig:symtft-twisted-sectors}.
However, for $\CUV$ an (AB) type fusion 2-category, the modular closure $M$ of $\Omega\CUV$ consists of topological lines that are \emph{intrinsic} to the symmetry boundary, and cannot be connected to some line in the bulk SymTFT: this forces $M$ to be trivial and confirms the statement that $\CUV$ needs to be group-theoretical if it is to admit a fiber 2-functor.

\newl For the remainder of this subsection we will assume this is the case, and hence that the SymTFT is a pure Dijkgraaf--Witten theory with finite gauge group $G$ and action $\pi\in \operatorname{Z}^4(G; U(1))$. The same argument goes through for line defects living on the relative $\widetilde{\TIR}'$ boundary, which means the two boundaries we wish to consider are of the form
\begin{equation}
\begin{split}
    \mathfrak B_{\CUV}= \fB_{(H, \psi)}\, , \qquad
    \widetilde{\TIR}' = \fB_{(K, \omega)}\, , 
\end{split}
\end{equation}
where $H,K\subseteq G$ are two choices of non-anomalous subgroups, and $\psi, \omega$ are the corresponding trivializing 3-cochains. Now to construct a fiber 2-functor we must ensure the 3d gapped phase produced by sandwiching these boundaries together is trivial, and so we must ask for two things:
\begin{enumerate}
    \item There are no non-trivial topological local operators, and hence a single vacuum. In the sandwich construction, this translates to ensuring that all non-trivial $G$-Wilson lines living in the 4d bulk can only end on one of the two boundaries \emph{at most}.

    \item There are no non-trivial topological lines, which is enough to say the underlying 3d TFT is trivial. In SymTFT language, this means every boundary Wilson line $\lambda$ on $\fB_\CUV$ must end on $\widetilde{\TIR}'$ in exactly $\dim\lambda$ independent ways, so that it acts by its dimension in the IR.
\end{enumerate}

\ni To answer these questions we recall a version of Frobenius reciprocity, which states that for $H\subseteq G$ a subgroup,
\begin{equation}
    \Hom_{\Rep(G)}(\mathrm{Ind}_H^G(\lambda), \lambda') \simeq \Hom_{\Rep(H)}(\lambda, \mathrm{Res}_H^G(\lambda'))\, ,
\end{equation}
for any two $\lambda \in \Rep(H)$ and $\lambda' \in \Rep(G)$. For our purposes, $\lambda$ is a boundary Wilson line living on $\fB_{\CUV}$, $\lambda'$ is a bulk Wilson line of the SymTFT, and $\mathrm{Res}_H^G$ is the functor that sends bulk Wilson lines to boundary Wilson lines.
This result tells us that the bulk Wilson line corresponding to $\mathrm{Ind}_H^G(\lambda)$ is a direct sum of all those irreducible bulk Wilson lines, whose restriction to $\fB_{\CUV}$ contains $\lambda$ as a sub-representation. For instance, $\mathrm{Ind}_H^G(1_H)$ with $1_H$ being the trivial $H$-representation is a direct sum of all those bulk Wilson lines that can end on $\fB_\CUV$. The same is also true of course for $K\subseteq G$ and $\widetilde{\TIR}'$.

\begin{figure}[H]
    \begin{minipage}[t]{0.45\textwidth}
        \centering
         \begin{tikzpicture}[scale = 1]
            \draw[gray, thick] (0,2) -- (1.5,2) node[anchor = south, black]{$\strut \mathrm{DR}[\CUV]$} -- (3, 2);
            \draw[gray, thick] (-1,1) -- (2, 1);
            \draw[gray, thick] (0,0) -- (3, 0);
            \draw[gray, thick] (-1,-1) -- (2, -1);
            
            \draw[black, thick] (-1, -1) -- (-1, 1) node[anchor = east]{$\widetilde{\TIR}'\;\strut$} -- (0, 2) -- (0, 0) -- (-1, -1);
            \draw[blue, opacity = 0.25, ultra thick] (-0.5, 0.5) -- (2.5, 0.5);
            \filldraw[red] (-0.5, 0.5) circle (1.5pt);
            \filldraw[opacity = 0.5, red] (1, 0.5) circle (1.5pt);
            \begin{scope}[shift={(3,0)}]
            \draw[black, thick] (-1, -1) -- (-1, 1) -- (0, 2) -- (0, 1) node[anchor = west]{$\strut\;\fB_\CUV$} -- (0, 0) -- (-1, -1);
            \draw[blue, ultra thick] (-.5, 0.5) -- (-.5, 1.5);
            \filldraw[red] (-0.5, 0.5) circle (1.5pt);
            \end{scope}
        \end{tikzpicture}
        \caption{The ways the line $\lambda \in \Rep(H)$ can end in the IR are labeled by junctions between $\mathrm{Ind}_H^G(\lambda)$ and $\mathrm{Ind}_K^G(1_K)$ in the SymTFT bulk; $\lambda$ becomes trivial in the IR if there are exactly $\dim\lambda$ of them.}
    \label{fig:symtft-twisted-sectors-2}
    \end{minipage}%
    \hfill%
    \begin{minipage}[t]{0.45\textwidth}
        \centering
         \begin{tikzpicture}[scale = 1]
            \draw[gray, thick] (0,2) -- (1.5,2) node[anchor = south, black]{$\strut \mathrm{DR}[\CUV]$} -- (3, 2);
            \draw[gray, thick] (-1,1) -- (2, 1);
            \draw[gray, thick] (0,0) -- (3, 0);
            \draw[gray, thick] (-1,-1) -- (2, -1);
            
            \draw[black, thick] (-1, -1) -- (-1, 1) node[anchor = east]{$\widetilde{\TIR}'\;\strut$} -- (0, 2) -- (0, 0) -- (-1, -1);
            \draw[blue, opacity = 0.25, ultra thick] (-0.5, 0.5) -- (2.5, 0.5);
            \filldraw[red] (-0.5, 0.5) circle (1.5pt);
            \filldraw[opacity = 0.5, red] (1, 0.5) circle (1.5pt);
            \begin{scope}[shift={(3,0)}]
            \draw[black, thick] (-1, -1) -- (-1, 1) -- (0, 2) -- (0, 1) node[anchor = west]{$\strut\;\fB_\CUV$} -- (0, 0) -- (-1, -1);
            \filldraw[red] (-0.5, 0.5) circle (1.5pt);
            \end{scope}
        \end{tikzpicture}
        \caption{Local operators in the IR are labeled by morphisms from $\mathrm{Ind}_H^G(1_H)$ to $\mathrm{Ind}_K^G(1_K)$ in the SymTFT bulk. The trivial one takes $1_G$ to itself.}
    \label{fig:symtft-local-ops-2}
    \end{minipage}
\end{figure}

\ni To satisfy the first of our two conditions, we are essentially asking that the only bulk Wilson line that appears in both $\mathrm{Ind}_H^G(1_H)$ and $\mathrm{Ind}_K^G(1_K)$ is the trivial one, or in other words:
\begin{equation}\label{eq:cond-1}
    \Hom_{\Rep(G)}(\mathrm{Ind}_H^G(1_H), \mathrm{Ind}_K^G(1_K)) \simeq \bbC\, ,
\end{equation}
as illustrated in figure~\ref{fig:symtft-local-ops-2}. This ensures that no line ending on the $H$-boundary can end on the $K$-boundary, except for the trivial $G$-representation of course, which appears in this permutation representation only once. Now we employ Mackey's formula:
\begin{equation}
    \mathrm{Res}^G_K \mathrm{Ind}^G_H(1_H) \simeq \bigoplus_{g \in K\backslash G / H}\mathrm{Ind}^K_{K\cap {}^{g}H}(1_{K\cap {}^{g}H})\, ,
\end{equation}
and we can immediately see that to satisfy the condition~\eqref{eq:cond-1}, we need $|K\backslash G/ H| = 1 \iff G = HK$.

\newl For the second of our two conditions, we ask that every boundary Wilson line $\lambda\in\Rep(H)$ living on $\fB_{\CUV}$ ends on $\widetilde{\TIR}'$ in exactly $\dim\lambda$ independent ways, as it must if the IR theory is trivial. Equivalently, the fiber 2-functor induces a fiber functor $\Rep(H)\to\Vect$ on loops, which must preserve dimensions. As we illustrate in figure~\ref{fig:symtft-twisted-sectors-2}, this reads
\begin{equation}\label{eq:cond-2}
    \dim \, \Hom_{\Rep(G)}(\mathrm{Ind}_H^G(\lambda), \mathrm{Ind}_K^G(1_K)) = \dim \, \lambda \, .
\end{equation}
Using Mackey's formula together with $G = HK$ and Frobenius reciprocity, the left-hand side equals
\begin{equation}
    \dim \, \Hom_{\Rep(K)}\big(1_K,\mathrm{Ind}^K_{K\cap H}\mathrm{Res}^H_{K\cap H}(\lambda)\big) = \dim \, \lambda^{K\cap H}\,,
\end{equation}
the dimension of the subspace of $\lambda$ fixed by $K\cap H$. Condition~\eqref{eq:cond-2} for all simple $\lambda$ therefore says that $K\cap H$ acts trivially in every irreducible representation of $H$, which happens if and only if $K\cap H=\{1\}$.

\newl Hence a necessary condition for the existence of a fiber 2-functor for $\cC \simeq \mathsf{C}(G, \pi |H, \psi)$ is that there exists a subgroup $K\subseteq G$, such that $\pi|_K$ trivializes, $G = HK$ and $K\cap H = \{1\}$, i.e., $G=H\bowtie K$. Conversely, for such $K$ the sandwich of $\fB_{(H,\psi)}$ and $\fB_{(K,\omega)}$ is the trivial 3d TQFT \cite{Decoppet:2023bay}, as we also verify in section~\ref{subsec:ir-symmetry}, so the condition is also sufficient.

\newl {\bf Comparison with fusion 1-categories.} The trivializations $\psi$ and $\omega$ play no role in \eqref{eq:cond-1} and \eqref{eq:cond-2}. The bulk lines of the 4d Dijkgraaf--Witten theory are Wilson lines, and which of them end on $\fB_{(H,\psi)}$ and $\fB_{(K,\omega)}$, and in how many ways, depends only on $H$ and $K$; the 3-cochains only enter the junctions of surfaces. This is where the criterion departs from its 1-categorical counterpart \cite{Ostrikmodule,Natale:2003}. There, the trivializations are 2-cochains that twist the lines themselves, and $H\cap K\neq\{1\}$ is allowed whenever the induced 2-cocycle on it is non-degenerate. For instance, $\Rep(\bbZ_2^2)$ has two fiber functors, from $K=\{1\}$ and from $K=\bbZ_2^2$ with a non-trivial twist, whereas $2\Rep(\bbZ_2^2)$ has only the one from $K=\{1\}$. Physically, gauging $\bbZ_2^2$ in the non-trivial 2d SPT phase gives a trivially gapped theory, while gauging it in any 3d SPT phase leaves a Dijkgraaf--Witten theory. See section~\ref{subsec:warm-up-ir} and example~\ref{ex:Z2Z2-twisted}.

\section{Warm-up: 3d SymTFT and \texorpdfstring{$(1\!+\!1)$}{(1+1)}-Dimensional Gapped Phases}\label{sec:warm-up}
Before turning to fusion 2-categories, we illustrate the subsequent constructions of section~\ref{sec:gen-fib-fun-symtft} one dimension lower within the domain of 1-categories.
We consider $(1\!+\!1)$-dimensional theories with group-theoretical symmetries, which for simplicity we take to be gauging-related to $\Vec_{G}$, i.e., to a finite group symmetry with trivial anomaly.
Its SymTFT is the $(2\!+\!1)$-dimensional Dijkgraaf--Witten theory of $G$. 
Gapped phases are obtained by compactifying the SymTFT on an interval with a symmetry boundary and a chosen topological physical boundary.

\newl We ask three questions: (i) what is the symmetry $\CUV$ on a given topological symmetry boundary, (ii) what is the symmetry $\CIR$ of the resulting 2d TQFT obtained upon compactifying the SymTFT, and (iii) what is the RG functor $\cF:\CUV\to\CIR$.
We answer them first with module categories, which make the stacking-and-gauging construction of the boundaries explicit, and then with Lagrangian algebras, which describe the same boundaries through the bulk lines that condense on them.
Every statement below has a 2-categorical counterpart in section~\ref{sec:gen-fib-fun-symtft}, which we point out along the way.
Throughout, boxed examples develop the case $G=S_3$.
 
\subsection{The SymTFT}\label{subsec:warm-up-symtft}
The SymTFT of a $(1\!+\!1)$d theory with $\Vec_{G}$ symmetry is the $(2\!+\!1)$d Dijkgraaf--Witten theory of $G$, whose topological lines form the Drinfeld center $\cZ(\Vec_{G})$, a non-degenerate braided fusion category. 
%
 
\newl {\bf Topological lines in the SymTFT.} A topological line defect in the SymTFT is an object of $\cZ(\Vec_{G})$, i.e., a $G$-graded vector space $V=\bigoplus_{x\in G}V_x$ together with a \emph{half-braiding}, i.e.\ isomorphisms $g\otimes V\simeq V\otimes g$ for all $g\in G$, compatible with the fusion of the lines $g$. Since fusion in $\Vec_G$ multiplies degrees,
\begin{equation}
(g\otimes V)_y=V_{g^{-1}y}\,,\qquad (V\otimes g)_y=V_{yg^{-1}}\,,
\end{equation}
and setting $x=g^{-1}y$, a half-braiding is a family of isomorphisms
\begin{equation}
u_g:\ V_x\ \xrightarrow{\ \sim\ }\ V_{gxg^{-1}}\,,\qquad u_g\,u_h=u_{gh}\,,
\label{eq:S3-half-braiding}
\end{equation}
where the second condition is the compatibility with $g\otimes h\simeq gh$. 
Thus $\cZ(\Vec_G)$ is the category of $G$-equivariant vector bundles over $G$ with $G$ acting via conjugation. 
Physically, $x$ is the flux of the line, i.e., the holonomy of the gauge field around it, and $u_g$ implements a gauge transformation, which conjugates the flux.
 
\newl In what follows, we will repeatedly make use of the {\bf orbit-stabilizer principle}, which we now describe.
Let $X$ be a finite set with an action of $G$. 
A $G$-equivariant vector bundle over $X$ is a collection of vector spaces $V_x$, $x\in X$, together with isomorphisms 
\begin{equation}
    u_g:V_x\to V_{g\cdot x}
\end{equation} 
satisfying $u_gu_{g'}=u_{gg'}$. Such a bundle is determined, up to isomorphism, by its fibers at one base point $x_O$ in each orbit $O\subseteq X$, each of which is a representation of the stabilizer $\mathrm{Stab}(x_O)$.
Conversely, any representation of $\mathrm{Stab}(x_O)$ can be transported over $O$. 
Therefore
\begin{equation}
\Big\{G\text{-equivariant vector bundles over }X\Big\}\ \simeq\ \bigoplus_{O\subseteq X}\Rep\big(\mathrm{Stab}(x_O)\big)\,,
\label{eq:S3-orbit-stabiliser}
\end{equation}
so that a simple object is a pair $(O,\rho)$ of an orbit $O$ and an irreducible representation $\rho$ of the stabilizer, of total dimension $\sum_{x\in O}\dim V_x=|O|\dim\rho$.
%
 
\newl {\bf Simple lines in $\cZ(\Vec_{G})$.} 
By the orbit-stabilizer principle \eqref{eq:S3-orbit-stabiliser}, applied to $G$ acting on itself by conjugation, a simple line is specified by an orbit, i.e.\ a conjugacy class $[g]$, and an irreducible representation $\pi$ of the stabilizer of $g$, i.e.\ of its centralizer $C_G(g)$, carried by $V_g$. 
We denote these lines by $\Qc{g}{\pi}$; their dimension is $|[g]|\dim\pi$.
In the gauge theory, $[g]$ is the flux label and $\pi$ the electric charge label. For $g=\mathrm{id}$ the line is a pure Wilson line, and $u_y=R(y)$ is the representation matrix.
 
\begin{examplebox}[Topological lines in $\cZ(\Vec_{S_3})$][label=eq:Z(S3) lines]
We denote the $S_3$ group elements as
\begin{equation}
S_3=\{\mathrm{id},a,a^2,b,ab,a^2b\}\,,\qquad a^3=b^2=\mathrm{id}\,,\qquad ab=ba^2\,,
\label{eq:S3-group}
\end{equation}
with conjugacy classes and centralizers
\begin{equation}
\begin{aligned}
&[\mathrm{id}]=\{\mathrm{id}\}\,,&\quad &[a]=\{a,a^2\}\,,&\quad &[b]=\{b,ab,a^2b\}\,,\\
&C_{S_3}(\mathrm{id})=S_3\,,&\quad &C_{S_3}(a)=\Z_3=\{\mathrm{id},a,a^2\}\,,&\quad &C_{S_3}(b)=\Z_2=\{\mathrm{id},b\}\,.
\end{aligned}
\end{equation}
Unless stated otherwise $\Z_2$ and $\Z_3$ denote these two subgroups. The irreducible representations of $S_3$ are the trivial representation $1$, the sign representation $P$ and the two-dimensional representation $E$, with
\begin{equation}
E(a)=\begin{pmatrix}\omega&0\\0&\omega^2\end{pmatrix},\qquad E(b)=\begin{pmatrix}0&1\\1&0\end{pmatrix},\qquad \omega=e^{2\pi i/3}\,.
\label{eq:S3-E}
\end{equation}
The irreducible representations of $\Z_3$ are labeled $\omega^k$, $k\in\Z_3$ (with $a\mapsto\omega^k$, and $1:=\omega^0$), and those of $\Z_2$ are labeled $\pm$ (with $b\mapsto\pm1$). The characters of $S_3$ are
\begin{equation}
\begin{array}{c|ccc}
 & [\mathrm{id}] & [a] & [b]\\\hline
\chi_1 & 1 & 1 & 1\\
\chi_P & 1 & 1 & -1\\
\chi_E & 2 & -1 & 0
\end{array}
\label{eq:S3-characters}
\end{equation}
The simple lines are the eight lines
\begin{equation}
  \{\Qc{\mathrm{id}}{1}\,,  \Qc{\mathrm{id}}{P}\,,  \Qc{\mathrm{id}}{E}\,,  \Qc{a}{\omega^{k}}\,,  \Qc{b}{\pm}\}\,,\qquad k\in\Z_3\,,
\end{equation}
of dimensions $1,1,2,2,2,2,3,3$, so that $\sum_Q(\dim Q)^2=36=|S_3|^2$.
\end{examplebox}
 
\newl {\bf Fusion.} Fusing two lines multiplies their fluxes and tensors their internal states, with $G$ acting on both factors:
\begin{equation}
(V\otimes W)_x=\bigoplus_{yz=x}V_y\otimes W_z\,,\qquad u_g\big|_{V_y\otimes W_z}=u_g\otimes u_g\,.
\label{eq:S3-center-fusion}
\end{equation}
To decompose the result, one reads off, on each conjugacy class $[x]$, the representation of $C_G(x)$ on the fiber at $x$. 
 
\begin{examplebox}[Fusion rules in $\cZ(\Vec_{S_3})$][label=eq:fusions in Z(S3)]
Here we present three representative computations.
The rest follow analogously.
\begin{itemize}
\item $\Qc{\mathrm{id}}{E}\otimes\Qc{a}{\omega^k}$. The fiber at $a$ is $E\otimes V_a$, on which $a$ acts as $E(a)\,\omega^k=\mathrm{diag}(\omega^{k+1},\omega^{k+2})$. Hence
\begin{equation}
    \Qc{\mathrm{id}}{E}\otimes\Qc{a}{\omega^k}=\Qc{a}{\omega^{k+1}}\oplus\Qc{a}{\omega^{k+2}}\,.
\end{equation}
\item $\Qc{a}{1}\otimes\Qc{a}{1}$. At $x=\mathrm{id}$ the fiber is $V_a\otimes V_{a^2}\oplus V_{a^2}\otimes V_a$. Here $a$ acts trivially and $u_b$ exchanges the two summands, so with $u_b^2=1$ the fiber is $1\oplus P$. At $x=a$ the fiber is $V_{a^2}\otimes V_{a^2}$, on which the centralizer $C_{S_3}(a)=\Z_3$ acts trivially. Hence
\begin{equation}
\Qc{a}{1}\otimes\Qc{a}{1}=\Qc{\mathrm{id}}{1}\oplus\Qc{\mathrm{id}}{P}\oplus\Qc{a}{1}\,.
\end{equation}
\item $\Qc{b}{+}\otimes\Qc{b}{+}$. At $x=\mathrm{id}$ the fiber is $\bigoplus_{y\in[b]}V_y\otimes V_y$, three lines permuted by $S_3$ as the three transpositions, i.e.\ the permutation representation $1\oplus E$. At $x=a$ there are three pairs $(y,z)\in[b]^2$ with $yz=a$, permuted freely by the centralizer $\Z_3$, which gives the regular representation $1\oplus\omega\oplus\omega^2$. Hence
\begin{equation}
\Qc{b}{+}\otimes\Qc{b}{+}=\Qc{\mathrm{id}}{1}\oplus\Qc{\mathrm{id}}{E}\oplus\Qc{a}{1}\oplus\Qc{a}{\omega}\oplus\Qc{a}{\omega^2}\,.
\end{equation}
\end{itemize}    
\end{examplebox}
 
\newl {\bf Braiding.} Moving a line $V$ of flux $y$ past a line $W$ of flux $z$ conjugates the flux of $W$:
\begin{equation}
c_{V,W}:\ V_y\otimes W_z\ \longrightarrow\ W_{yzy^{-1}}\otimes V_y\,,\qquad v\otimes w\longmapsto u_y(w)\otimes v\,.
\label{eq:S3-center-braiding}
\end{equation}
This is the non-abelian Aharonov--Bohm effect.
Three consequences give the modular data used in this section:
\begin{itemize}
\item \emph{Spin.} The twist of a line acts on $V_x$ by $u_x$, i.e., a line is rotated by its own flux. For $\Qc{g}{\pi}$, $u_g$ acts on $V_g$ by $\pi(g)=\chi_\pi(g)/\dim\pi$, which is a scalar by Schur's lemma since $g$ is central in $C_G(g)$.
\item \emph{Wilson lines around fluxes.} For a Wilson line $R$ (flux $\mathrm{id}$) and a line of flux $g$, the double braiding acts on $R$ by the holonomy $R(g)$. A Wilson loop linking a flux of class $[g]$ therefore evaluates to $\chi_R(g)$, compared with $\dim R$ when unlinked; equivalently $S_{R,Q}/S_{1,Q}=\chi_R(g)$ for $Q=\Qc{g}{\pi}$, up to complex conjugation depending on orientation conventions.
\item \emph{Wilson lines among themselves} have trivial double braiding, since $u_{\mathrm{id}}=1$: they form the symmetric, in fact Tannakian, subcategory $\Rep(G)\subset\cZ(\Vec_G)$. These are the lines that condense on $\fB_{\rm Dir}$.
\end{itemize}
 
\subsection{Module categories}\label{subsec:warm-up-module}
There is a canonical Dirichlet boundary $\fB_{\rm Dir}$ of the SymTFT $\cZ(\cC)$, on which the fusion category of (symmetry) lines is $\cC=\Vec_G$.
Consider another topological boundary $\fB$, and let $\cM$ denote the category whose objects are the topological line interfaces between $\fB_{\rm Dir}$ and $\fB$ and whose morphisms are topological junctions between them.
A topological line $g\in\cC$ of $\fB_{\rm Dir}$ can be fused onto an interface $m\in \cM$ from the $\fB_{\rm Dir}$ side, producing a new interface $g\triangleright m$.
Fusing two lines successively is equivalent to first fusing them and then fusing the result onto $m$, up to a topological junction: $(g\otimes h)\triangleright m\simeq g\triangleright(h\triangleright m)$.
This structure, together with the coherence of these junctions, is precisely the definition of a left $\cC$-module category. 
Topological boundaries of $\cZ(\cC)$ are in one-to-one correspondence with indecomposable $\cC$-module categories \cite{Davydov:2010wpa,KitaevKong2012,FuchsSchweigertValentino2013}, the 1-categorical counterpart of the statement used in section~\ref{sec:gen-fib-fun-symtft}.
For $\cC=\Vec_G$ these are classified by pairs $(H,\psi)$ with $H\subseteq G$ a subgroup up to conjugation and $\psi\in \operatorname{H}^2(H;U(1))$ \cite{Ostrikmodule}; we denote the corresponding boundary by $\fB_{(H,\psi)}$, or $\fB_H$ when $\psi$ is trivial. 
 
\begin{examplebox}[Topological boundaries of $\cZ(\Vec_{S_3})$][label=ex:topo boundaries]
Every subgroup of $S_3$ is conjugate to one of $\{\mathrm{id}\}$, $\Z_2$, $\Z_3$ or $S_3$, all of which have vanishing $\operatorname{H}^2(H;U(1))$.
Hence the $S_3$ Dijkgraaf--Witten theory has exactly four topological boundaries $\fB_H$, with $\fB_{\{\mathrm{id}\}}=\fB_{\rm Dir}$ the Dirichlet boundary and $\fB_{S_3}=\fB_{\rm Neu}$ the Neumann boundary.    
\end{examplebox}
 
\noindent To derive the category of interfaces between $\fB_{\rm Dir}$ and $\fB_{(H,\psi)}$, we require some facts about 2d TFTs for which we make a short digression.
 
\newl {\bf An aside on 2d TFTs.} 
A fully extended framed 2d TFT is a symmetric monoidal functor from framed bordisms $\mathrm{Bord}_2$ to the Morita 2-category $\mathrm{Alg}$,
\begin{equation}
    \cT:\mathrm{Bord}_2 \rightarrow \mathrm{Alg}\,.
\end{equation}
The target is the 2-category $\mathrm{Alg}$ whose objects are algebras (the fully dualizable ones being the finite-dimensional semisimple algebras), whose 1-morphisms are bimodules and whose 2-morphisms are bimodule intertwiners.
By the cobordism hypothesis, the 2d TFT $\cT$ is determined by the algebra it assigns to a point, $A=\cT(\mathrm{pt})$ \cite{Lurie:2009keu,schommer2009classification}.
Physically, $A$ is the algebra of topological operators living on a reference boundary condition $\beta$ that contains every vacuum, i.e.\ $A=\End(\beta)$.
All the other data of $\cT$ is constructible from $A$:
\begin{itemize}
    \item {\it Boundary conditions.} The space of boundary-changing operators between $\beta$ and another boundary condition $\gamma$ is acted on by $A$, by fusing operators in $\beta$ onto the junction
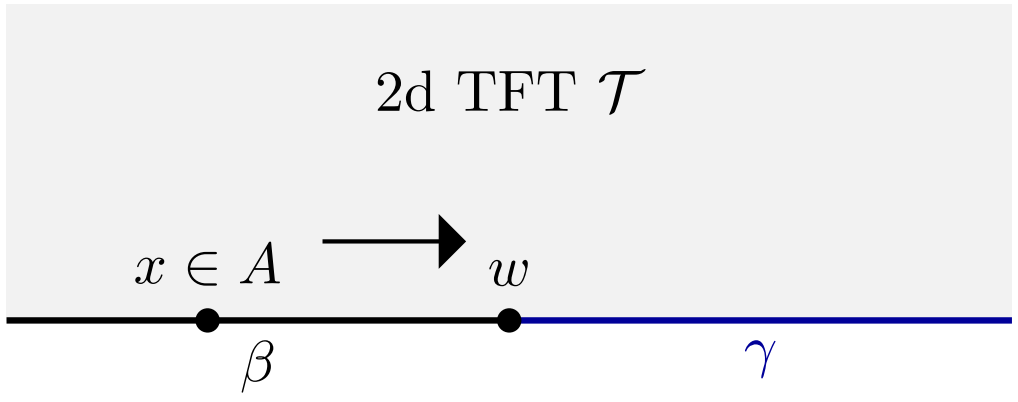
\begin{eqs}
    \begin{tikzpicture}[scale=0.95]
\fill[gray!10] (0,0) rectangle (7,2.2);
\node at (3.5,1.6) {2d TFT $\mathcal{T}$};
\draw[very thick] (0,0) -- (3.5,0);
\draw[very thick,blue!60!black] (3.5,0) -- (7,0);
\node[below] at (1.75,0) {$\beta$};
\node[below,blue!60!black] at (5.25,0) {$\gamma$};
\filldraw (3.5,0) circle (2.2pt);
\node[above] at (3.5,0.1) {$w$};
\filldraw (1.4,0) circle (2.2pt);
\node[above] at (1.4,0.1) {$x\in A$};
\draw[->,thick] (2.2,0.55) -- (3.2,0.55);
\end{tikzpicture}
\end{eqs}
    This identifies boundary conditions with the category $\mathsf{LMod}(A)$, whose objects are $A$-modules and whose morphisms are module maps.
    \item {\it Vacua.} The vacua are in one-to-one correspondence with simple $A$-modules. 
    Specifically, the algebra $A=\bigoplus_{k=1}^{n}\mathrm{Mat}_{r_{k}}(\bbC)$ has $n$ simple modules and $\mathsf{LMod}(A)\simeq\Vec^{\oplus n}$.
    Each vacuum carries exactly one simple boundary condition.
    \item {\it Local operators.} In a TFT, topological local operators and states on a circle are the same thing. Given a local operator $O$, the path integral on a disc with $O$ inserted at its center produces a state $|O\rangle$ on the boundary circle. Conversely, every state on the circle arises in this way: a neighbourhood of the circle is a cylinder, which in a TFT is the identity and can be shrunk to a point. The product of local operators corresponds to the pair of pants, which glues two discs into one.

    \newl A state on the circle can be obtained by gluing the two ends of an interval with the boundary condition $\beta$ at both ends, whose states are $A=\End(\beta)$. An operator $x\in A$ acting at one end of the interval can, after gluing, be moved around the circle to the other end, so the interval states $x\,w$ and $w\,x$ give the same circle state.
    Hence the space of bulk topological local operators is
\begin{equation}
\cT(S^1)=A/[A,A]\,,
\label{eq:2d-local-ops}
\end{equation}
where $[A,A]$ is the linear span of commutators. Here $A/[A,A]$ is the quotient by a subspace: for instance, $\mathrm{Mat}_r(\bbC)/[\mathrm{Mat}_r,\mathrm{Mat}_r]\simeq\bbC$ via the trace. 
For $A=\bigoplus_k\mathrm{Mat}_{r_k}(\bbC)$ one trace survives per block, so the space of local operators is $\bbC^n$. With the product given by the pair of pants, this is the commutative algebra $\cZ(A)\simeq\bbC^n$, spanned by the projectors onto the $n$ vacua.
    \item {\it Topological lines.} A topological line is a linear endofunctor of the category of boundary conditions $\Vec^{\oplus n}$, i.e.\ an $n\times n$ matrix of vector spaces, with fusion given by matrix multiplication:
\begin{equation}
\mathrm{Lines}(\mathcal{T})\simeq\End\big(\Vec^{\oplus n}\big)\simeq\mathrm{Mat}_n(\Vec)\,.
\label{eq:Matn}
\end{equation}
The simple topological lines in $\cT$, denoted $1_{ij}$, are the elementary domain walls between vacua $i$ and $j$, with $1_{ij}$ mapping the boundary condition of vacuum $j$ to that of vacuum $i$.
They satisfy the fusion rules 
\begin{equation}
    1_{ij}\otimes 1_{kl}=\delta_{jk}1_{il}\,,
\end{equation}
and $1_{ii}$ is the identity line within vacuum $i$. 
The identity line $\bigoplus_i1_{ii}$ is not simple for $n>1$, so $\mathrm{Mat}_n(\Vec)$ is multifusion, and $\End(1)\simeq\bbC^n$ is the algebra of local operators.
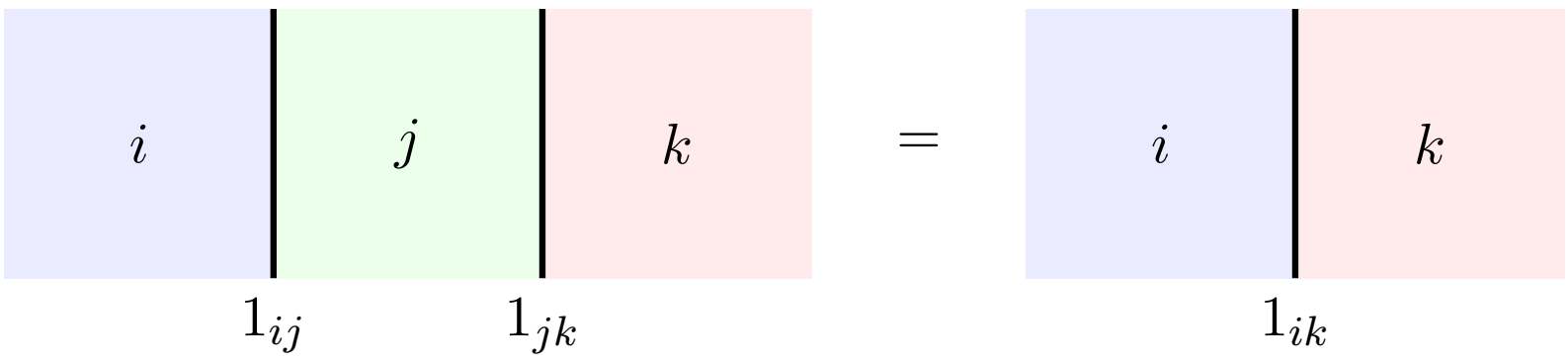
\begin{equation}
\begin{split}
\begin{tikzpicture}[scale=0.9]
\fill[blue!8] (0,0) rectangle (2,2); \fill[green!8] (2,0) rectangle (4,2); \fill[red!8] (4,0) rectangle (6,2);
\node at (1,1) {$i$}; \node at (3,1) {$j$}; \node at (5,1) {$k$};
\draw[very thick] (2,0) -- (2,2); \draw[very thick] (4,0) -- (4,2);
\node[below] at (2,0) {$1_{ij}$}; \node[below] at (4,0) {$1_{jk}$};
\node at (6.8,1) {$=$};
\begin{scope}[xshift=7.6cm]
\fill[blue!8] (0,0) rectangle (2,2); \fill[red!8] (2,0) rectangle (4,2);
\node at (1,1) {$i$}; \node at (3,1) {$k$};
\draw[very thick] (2,0) -- (2,2);
\node[below] at (2,0) {$1_{ik}$};
\end{scope}
\end{tikzpicture}
\end{split}
\end{equation}
\item {\it Morita invariance.} Morita equivalent algebras define the same theory. In particular $A=\bigoplus_{k =1}^n\mathrm{Mat}_{r_k}(\bbC)$ is Morita equivalent to $\bigoplus_{k=1}^n\bbC$, i.e.\ $\mathsf{LMod}(A)\simeq\mathsf{LMod}(\bbC^n)$, and the integers $r_k$, which are Chan--Paton multiplicities of the reference boundary condition, are unphysical. 
For an \emph{oriented} theory, $A$ also carries a symmetric Frobenius trace, which contributes one Euler counterterm $\lambda_k\in\bbC^\times$ per vacuum, $\cT(\Sigma_g)=\sum_k\lambda_k^{1-g}$ \cite{schommer2009classification}. 
This datum is invisible in $\mathsf{LMod}(A)$ and plays no role below.
\end{itemize}
 
\newl {\bf Interfaces between $\fB_{\rm Dir}$ and $\fB_{(H,\psi)}$.}
We now derive $\cM_{(H,\psi)}$, the category of interfaces between $\fB_{\rm Dir}$ and $\fB_{(H,\psi)}$.
The procedure involves two steps: (i) stack a 2d TFT $\cT$ with an $H$ symmetry, $H\subseteq G$, on half of $\fB_{\rm Dir}$, and (ii) gauge the diagonal $H$ on the stacked half.
This produces the boundary condition $\fB_{(H,\psi)}$ on the stacked half, and the line where it meets the untouched half is an interface between $\fB_{\rm Dir}$ and $\fB_{(H,\psi)}$:
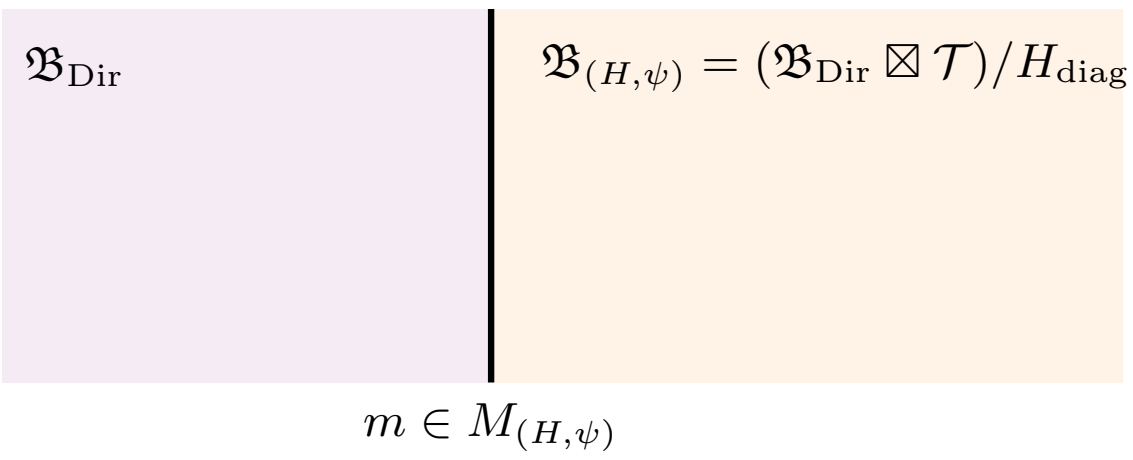
\begin{equation}
\begin{split}
\begin{tikzpicture}[scale=0.95]
\fill[violet!8] (0,0) rectangle (3.4,2.6);
\fill[orange!10] (3.4,0) rectangle (7.8,2.6);
\node at (.5,2.2) {\footnotesize$\fB_{\rm Dir}$};
\node at (5.8,2.2) {\footnotesize$\fB_{(H,\psi)}=(\fB_{\rm Dir}\boxtimes\cT)/H_{\rm diag}$};
\draw[very thick] (3.4,0) -- (3.4,2.6);
\node[below] at (3.4,0) {\footnotesize $m\in M_{(H,\psi)}$};
\end{tikzpicture}
\end{split}
\end{equation}
\begin{enumerate}
\item {\it Stacking an $H$-symmetric TFT.} 
One stacks an auxiliary TFT $\cT$ on the right half of $\fB_{\rm Dir}$:
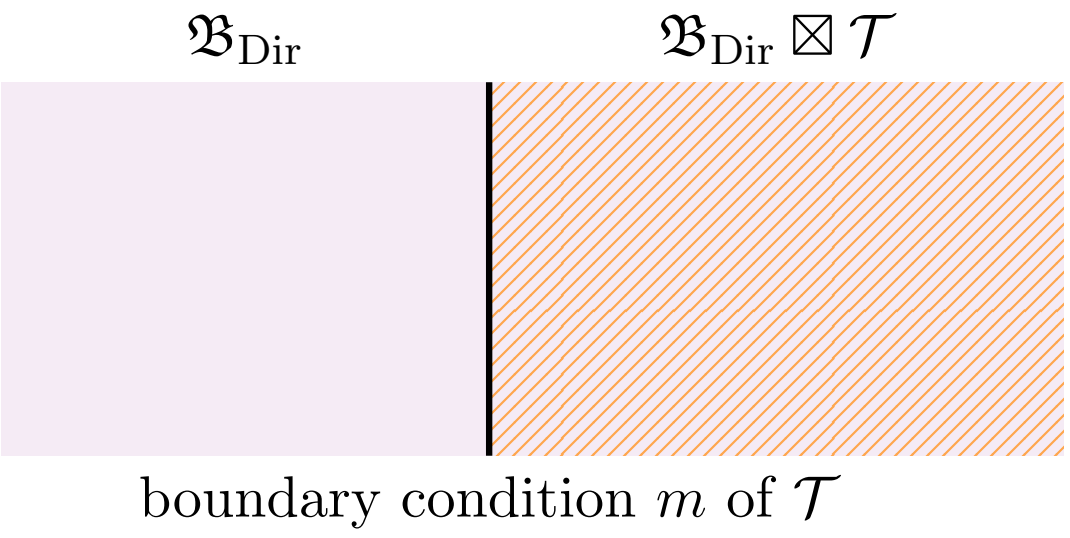
\begin{equation}
    \begin{split}
        \begin{tikzpicture}[scale=0.95]
\fill[violet!8] (0,0) rectangle (7.4,2.6);
\fill[pattern=north east lines,pattern color=orange!70] (3.4,0) rectangle (7.4,2.6);
\node at (1.7,2.9) {$\fB_{\rm Dir}$};
\node at (5.4,2.9) {$\fB_{\rm Dir}\boxtimes\cT$};
\draw[very thick] (3.4,0) -- (3.4,2.6);
\node[below] at (3.4,0) {boundary condition $m$ of $\cT$};
\end{tikzpicture}
    \end{split}
\end{equation}
Since $\fB_{\rm Dir}$ continues across it, an interface between $\fB_{\rm Dir}$ and $\fB_{\rm Dir}\boxtimes\cT$ is specified by a line $g\in G$ of $\fB_{\rm Dir}$ fused onto it together with a boundary condition $m$ of $\cT$, with $\cT$ presented by an algebra $A_1=\bigoplus_{k=1}^{n}\mathrm{Mat}_{r_k}(\bbC)$. We denote the category of boundary conditions of $\cT$ by
\begin{equation}
\wt \cM
=\mathsf{LMod}(A_1)\simeq\Vec^{\oplus n}\,,
\label{eq:Mtilde}
\end{equation}
so that the interfaces between $\fB_{\rm Dir}$ and $\fB_{\rm Dir}\boxtimes\cT$ form the $\cC$-module category $\Vec_G\boxtimes\wt \cM$, with $\cC=\Vec_G$ acting on the first factor.

\newl An $H$ symmetry of $\cT$ is specified by a monoidal functor 
\begin{equation}
    \Lambda: \Vec_{H} \longrightarrow \mathrm{Mat}_{n}(\Vec)\,.
\end{equation}
Since $\mathrm{Mat}_{n}(\Vec)=\End(\Vec^{\oplus n})$, this is equivalent to an action of $H$ on the boundary conditions, $h\triangleright m = \Lambda(h)(m)$, which makes $\wt \cM$ a $\Vec_H$-module category that we denote by $\wt \cM_\Lambda$.
For a single vacuum, $\mathrm{Mat}_1(\Vec)= \Vec$ and $\Lambda$ is a fiber functor.
Fusing the line $h$ onto the unique simple boundary condition $1_\cM$ leaves it invariant up to an invertible boundary junction $a_h:1_\cM\to h\triangleright 1_\cM$, at which the $h$-line emerges from the boundary.

\begin{eqs}
\label{eq:a_h}
    \begin{tikzpicture}[scale=0.9]
\begin{scope}[xshift=5.4cm]
\fill[pattern=north east lines,pattern color=orange!70] (0,0) rectangle (3.2,2.6);
\draw[very thick] (0,0) -- (0,2.6);
\draw[thick,orange!80!black] (3.2,1.3) -- (0,1.3); \node[right,orange!80!black] at (3.2,1.3) {$h$};
\filldraw (0,1.3) circle (2.2pt); \node[left] at (0,1.3) {$a_h$};
\node[left] at (0,0.4) {$1_\cM$}; \node[left] at (0,2.2) {$h\triangleright1_\cM$};
\end{scope}
\end{tikzpicture}
\end{eqs}
Since $\Hom_{\wt{\cM}}(1_{\cM},h\triangleright 1_{\cM})\cong \bbC$, $a_h$ is unique up to a non-zero number $a_h\mapsto \lambda(h)a_h$.
Two lines $h_1$ and $h_2$ can end on $1_{\cM}$ in two ways that are related by $\psi (h_1,h_2)\in U(1)$
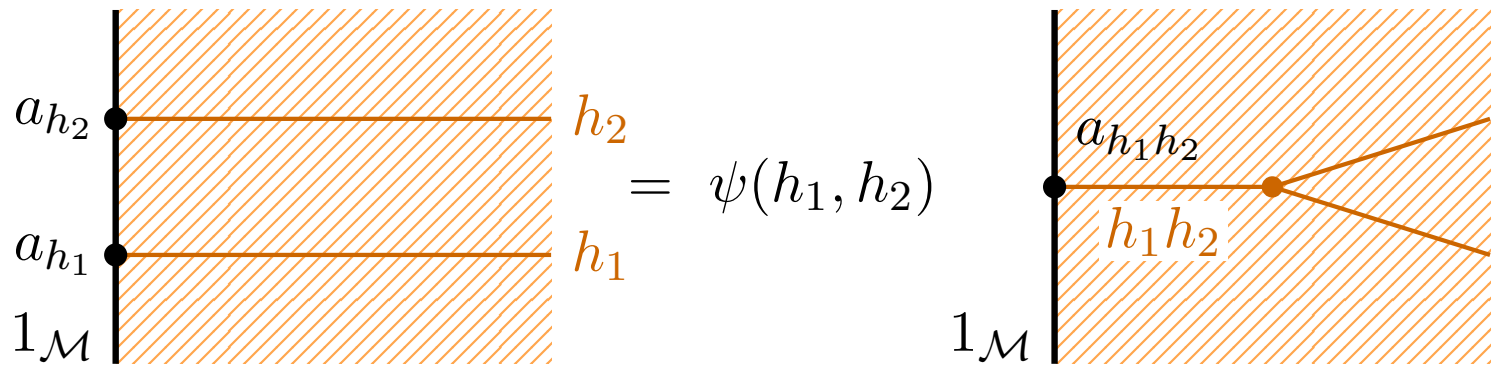
\begin{eqs}
    \begin{tikzpicture}[scale=0.9]
\fill[pattern=north east lines,pattern color=orange!70] (0,0) rectangle (3.2,2.6);
\draw[very thick] (0,0) -- (0,2.6);
\draw[thick,orange!80!black] (3.2,0.8) -- (0,0.8); \node[right,orange!80!black] at (3.2,0.8) {$h_1$};
\draw[thick,orange!80!black] (3.2,1.8) -- (0,1.8); \node[right,orange!80!black] at (3.2,1.8) {$h_2$};
\filldraw (0,0.8) circle (2.2pt); \node[left] at (0,0.8) {$a_{h_1}$};
\filldraw (0,1.8) circle (2.2pt); \node[left] at (0,1.8) {$a_{h_2}$};
\node[left] at (0,0.2) {$1_\cM$};
\node at (4.9,1.3) {$=\ \psi(h_1,h_2)$};
\begin{scope}[xshift=6.9cm]
\fill[pattern=north east lines,pattern color=orange!70] (0,0) rectangle (3.2,2.6);
\draw[very thick] (0,0) -- (0,2.6);
\draw[thick,orange!80!black] (3.2,0.8) -- (1.6,1.3); \draw[thick,orange!80!black] (3.2,1.8) -- (1.6,1.3);
\draw[thick,orange!80!black] (1.6,1.3) -- (0,1.3);
\filldraw[orange!80!black] (1.6,1.3) circle (2pt);
\node[below,orange!80!black,fill=white,inner sep=1pt] at (0.8,1.2) {$h_1h_2$};
\filldraw (0,1.3) circle (2.2pt); \node[above right] at (0,1.35) {$a_{h_1h_2}$};
\node[left] at (0,0.2) {$1_\cM$};
\end{scope}
\end{tikzpicture}
\label{eq:fiber_functor_associator}
\end{eqs}
where $m_{h_1,h_2}:h_1\triangleright(h_2\triangleright 1_\cM)\to(h_1h_2)\triangleright1_\cM$ fuses the two lines.
Associativity of fusion in $\Vec_{H}$ requires $\delta\psi=1$, i.e.\ $\psi$ is a 2-cocycle, and rescaling the $a_h$ changes $\psi$ by a coboundary. 
We recover the fact that fiber functors of $\Vec_H$ are classified by $\operatorname{H}^2(H;U(1))$: a single-vacuum $H$-symmetric TFT is an $H$-SPT.
The junctions $a_h$ form an $H$-graded algebra $A$: for a boundary condition $\beta$ of $\cT$, its degree-$h$ piece $A_h=\Hom(\beta,h\triangleright\beta)$ is the space of junctions at which $h$ emerges from $\beta$, and its product is the composite in \eqref{eq:fiber_functor_associator}.
For a single vacuum and $\beta=1_\cM$, $A_h=\bbC\,a_h$ and \eqref{eq:fiber_functor_associator} is the multiplication $a_{h_1}a_{h_2}=\psi(h_1,h_2)\,a_{h_1h_2}$, so the twisted group algebra $A=\bbC^\psi[H]$ encodes exactly the same data as the fiber functor. 
This is the 1-categorical counterpart of the graded fusion category $A=\bigoplus_hA_h$ of section~\ref{sec:gen-fib-fun-symtft}.
 
\newl More generally, for an $n$-vacua TFT each $\Lambda(h)$ is an invertible object of $\mathrm{Mat}_{n}(\Vec)$, i.e.\ a permutation of the vacua.\footnote{When the source category contains non-invertible lines, $\Lambda(x)$ acts on the vacua by a matrix of non-negative integers.}
If $\wt \cM_\Lambda$ is indecomposable, the vacua form a single $H$-orbit; a representative vacuum is stabilized by a subgroup $L\subseteq H$, $n=|H/L|$, and $L$ acts on it by a fiber functor with class $[\psi]\in \operatorname{H}^{2}(L;U(1))$, so that $\wt \cM_\Lambda\simeq\Vec(H/L)^{\psi}$.
We will see at the end of step 2 that stacking such a TFT and gauging $H$ gives the same boundary as stacking the single-vacuum $L$-SPT $\psi$ and gauging $L$, so it suffices to consider single-vacuum TFTs.
 
\item {\it Gauging the diagonal $H$ symmetry.} 
As a preparation, consider first gauging $H$ in $\cT$ alone. The boundary conditions of $\cT/H$ form the category $\wt \cM^{H}$ of $H$-equivariant objects of $\wt \cM_\Lambda$: an object $m\in\wt \cM$ together with isomorphisms $\rho(h):h\triangleright m \to m$ for all $h\in H$, such that
\begin{equation}
\begin{tikzcd}[column sep=large,row sep=large]
h\triangleright(h'\triangleright m) \arrow[r,"h\triangleright\rho(h')"] \arrow[d,"m_{h,h'}"'] & h\triangleright m \arrow[d,"\rho(h)"] \\
(hh')\triangleright m \arrow[r,"\rho(hh')"'] & m
\end{tikzcd}
\label{eq:equiv-condition}
\end{equation}
commutes, where by functoriality $h\triangleright \rho(h'): h\triangleright (h' \triangleright m) \to h \triangleright m$.
A morphism from $(m,\rho)$ to $(m',\rho')$ is a morphism $f:m\to m'$ in $\wt \cM$ that intertwines the isomorphisms, $f\circ\rho(h)=\rho'(h)\circ(h\triangleright f)$.
Physically, $\wt \cM^H$ consists of the $H$-invariant boundaries together with a choice of how the $H$-lines, which after gauging are summed over, end on them topologically.
For a single vacuum, $\wt \cM=\Vec$, an $H$-equivariant object is a vector space $V$ together with isomorphisms $\rho(h):h\triangleright V\to V$ satisfying \eqref{eq:equiv-condition}, and using the junctions \eqref{eq:a_h} to identify $h\triangleright V$ with $V$, the matrices $\hat\rho(h):=\rho(h) \circ a_{h}\in \End(V)$ satisfy $\hat{\rho}(h_1)\hat{\rho}(h_2)=\psi(h_1,h_2)\hat{\rho}(h_1h_2)$. Therefore
\begin{equation}
\wt \cM^H=\Big\{(V,\hat\rho)\ :\ \hat\rho(h)\hat\rho(h')=\psi(h,h')\,\hat\rho(hh')\Big\}=\Rep^\psi(H)\simeq\mathsf{LMod}\big(\bbC^\psi[H]\big)\,,
\label{eq:MtildeH}
\end{equation}
whose simple objects are the irreducible $\psi$-projective representations of $H$, which need not be one-dimensional. This is the counterpart of \eqref{eq:h-eq-vs-graded-ext}.
 
\newl We now gauge the diagonal $H$ on $\fB_{\rm Dir}\boxtimes \cT$.
A topological line interface between $\fB_{\rm Dir}$ and $\fB_{(H,\psi)}$ is a topological line on the boundary of spacetime across which the boundary condition changes from $\fB_{\rm Dir}$ to $\fB_{(H,\psi)}$.
Since $\fB_{\rm Dir}$ continues across it, it is labeled by a pair $[g,m]$ of a symmetry line $g\in G$ of $\fB_{\rm Dir}$ fused onto it and a boundary condition $m\in\wt \cM$ of $\cT$, on which $G$ acts by $g'\triangleright[g,m]=[g'g,m]$.
A line $h\in H$ moved across the interface can be traded for the $h$-line of $\cT$, the combination of the two being a gauge line and hence invisible, and absorbed into $m$:
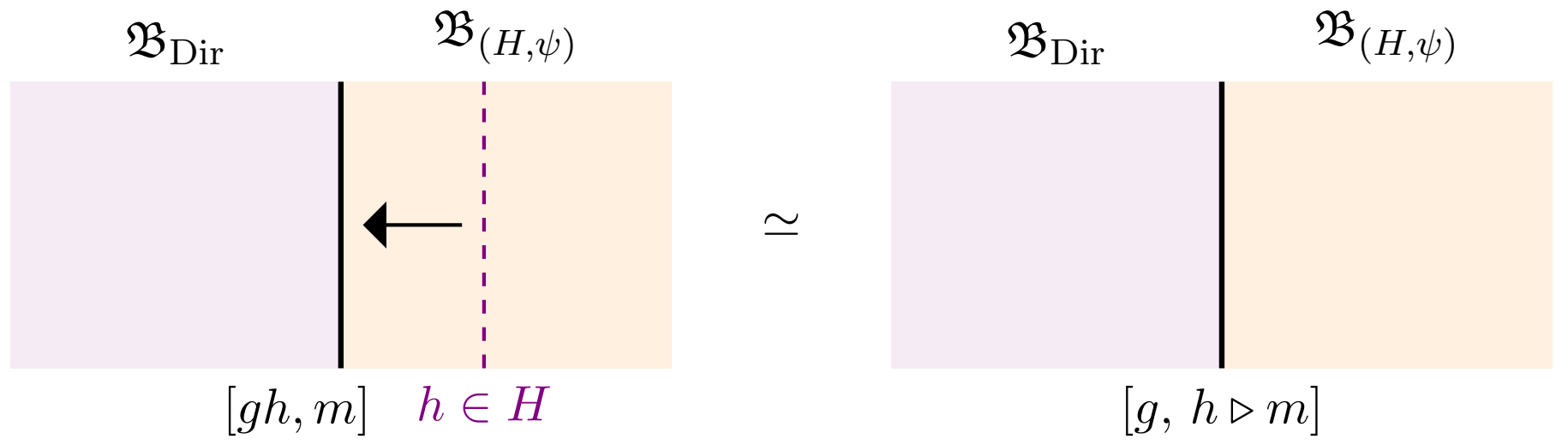
\begin{eqs}
    \begin{tikzpicture}[scale=0.85]
\fill[violet!8] (0,0) rectangle (3.0,2.6);
\fill[orange!12] (3.0,0) rectangle (6.0,2.6);
\draw[very thick] (3.0,0) -- (3.0,2.6);
\node[above] at (1.5,2.6) {$\fB_{\rm Dir}$}; \node[above] at (4.5,2.6) {$\fB_{(H,\psi)}$};
\draw[thick,violet,dashed] (4.3,0) -- (4.3,2.6); \node[below,violet] at (4.3,0) {$h\in H$};
\draw[->,thick] (4.1,1.3) -- (3.2,1.3);
\node[below] at (2.6,0) {$[gh,m]$};
\node at (7.0,1.3) {$\simeq$};
\begin{scope}[xshift=8.0cm]
\fill[violet!8] (0,0) rectangle (3.0,2.6);
\fill[orange!12] (3.0,0) rectangle (6.0,2.6);
\draw[very thick] (3.0,0) -- (3.0,2.6);
\node[above] at (1.5,2.6) {$\fB_{\rm Dir}$}; \node[above] at (4.5,2.6) {$\fB_{(H,\psi)}$};
\node[below] at (3.0,0) {$[g,\,h\triangleright m]$};
\end{scope}
\end{tikzpicture}
\end{eqs}
so that $[gh,m]\simeq[g,h\triangleright m]$ and the interfaces form the balanced tensor product
\begin{equation}
\cM_{(H,\psi)}=\Vec_{G}\boxtimes_{\Vec_H}\wt \cM_\Lambda\,,
\label{eq:induced}
\end{equation}
i.e.\ the module category induced from $H$ to $G$ \cite{etingof2009,DSPS2014}.
Choosing representatives $g_i$ of the cosets $i\in G/H$ and defining little-group elements by $g\,g_i=g_{g\cdot i}\,l_i(g)$, $l_i(g)\in H$, this gives $\cM_{(H,\psi)}\simeq\bigoplus_{i\in G/H}\wt \cM_i$, with simple objects $(m,i)$, $m\in \wt \cM$ simple, and module action
\begin{equation}
g\triangleright(m,i)=\big(l_i(g)\triangleright m,\ g\cdot i\big)\,,
\label{eq:induced-action}
\end{equation}
the counterpart of the induced module 2-category of section~\ref{sec:gen-fib-fun-symtft}.
For a single-vacuum $\cT$, $\cM_{(H,\psi)}\simeq\Vec(G/H)^{\psi}$: one simple interface $\delta_i$ per coset, permuted as $g\triangleright\delta_i=\delta_{g\cdot i}$.
The cocycle $\psi$ does not change which interface is produced; it enters only through the associator of the module action, $\psi\big(l_{g'\cdot i}(g),l_i(g')\big)$.
Finally, inducing from $L$ to $H$ and then from $H$ to $G$ is the same as inducing from $L$ to $G$, $\Vec_G\boxtimes_{\Vec_H}\Vec(H/L)^{\psi}\simeq\Vec_G\boxtimes_{\Vec_L}\Vec^{\psi}$, which shows the claim made at the end of step 1.
\end{enumerate}
 
\newl {\bf Right modules.} The physical boundary will be placed on the other side of the SymTFT, so we also need the right-module counterpart. The interfaces between $\fB_{(K,\psi_K)}$ and $\fB_{\rm Dir}$, with $\fB_{\rm Dir}$ now on the right, form the right $\cC$-module category
\begin{equation}
\cN_{(K,\psi_K)}\simeq\Vec(K\backslash G)^{\psi_K}\,,\qquad \delta_{\bar\imath}\triangleleft g=\delta_{\bar\imath\cdot g}\,,
\label{eq:right-module}
\end{equation}
built from the right cosets $\bar\imath\in K\backslash G$ in the same way, and obtained from $\cM_{(K,\psi_K)}$ by reversing the orientation of the interfaces.
 
\newl We fix a symmetry boundary $\fB_{(H,\psi_H)}$ with interfaces $\cM=\cM_{(H,\psi_H)}$, and a physical boundary $\fB_{(K,\psi_K)}$ with interfaces $\cN=\cN_{(K,\psi_K)}$, and answer the three questions in the following subsections.
To keep formulas readable we take $\psi_H$ and $\psi_K$ trivial in what follows, which covers $G=S_3$, and indicate the modifications for non-trivial cocycles in words.
We then write $\cM_H=\Vec(G/H)$ and $\cN_K=\Vec(K\backslash G)$, with simple interfaces $\delta_i$, $i\in G/H$, and $\delta_{\bar\imath}$, $\bar\imath\in K\backslash G$.
 
\subsubsection{The UV symmetry \texorpdfstring{$\CUV$}{CUV}}\label{subsec:warm-up-uv}
A topological line $X$ of $\fB_H$ can end on an interface $m\in \cM_H$ from the $\fB_H$ side, and beyond the junction the interface is a new interface $X(m)$; similarly a line $g\in G$ of $\fB_{\rm Dir}$ ends on it from the other side, producing $g\triangleright m$:
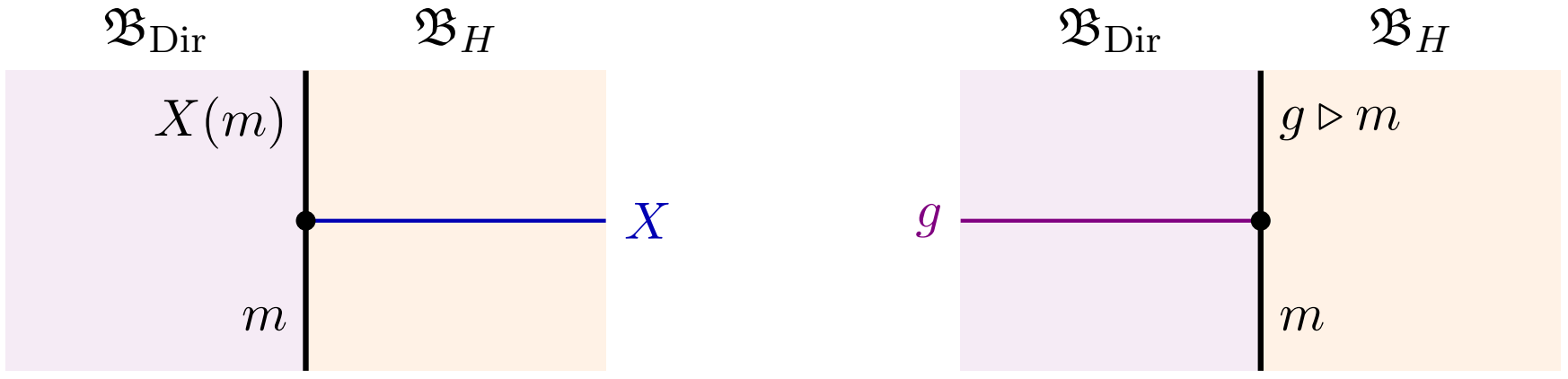
\begin{eqs}
\begin{tikzpicture}[scale=0.85,baseline=(c)]
\coordinate (c) at (0,1.2);
\fill[violet!8] (0,0) rectangle (2.6,2.6);
\fill[orange!10] (2.6,0) rectangle (5.2,2.6);
\node[above] at (1.3,2.6) {$\fB_{\rm Dir}$}; \node[above] at (3.9,2.6) {$\fB_H$};
\draw[very thick] (2.6,0) -- (2.6,2.6);
\draw[thick,blue!70!black] (5.2,1.3) -- (2.6,1.3); \node[right,blue!70!black] at (5.2,1.3) {$X$};
\filldraw (2.6,1.3) circle (2.2pt);
\node[left] at (2.6,0.45) {$m$}; \node[left] at (2.6,2.15) {$X(m)$};
\end{tikzpicture}
\qquad\qquad
\begin{tikzpicture}[scale=0.85,baseline=(c)]
\coordinate (c) at (0,1.2);
\fill[violet!8] (0,0) rectangle (2.6,2.6);
\fill[orange!10] (2.6,0) rectangle (5.2,2.6);
\node[above] at (1.3,2.6) {$\fB_{\rm Dir}$}; \node[above] at (3.9,2.6) {$\fB_H$};
\draw[very thick] (2.6,0) -- (2.6,2.6);
\draw[thick,violet] (0,1.3) -- (2.6,1.3); \node[left,violet] at (0,1.3) {$g$};
\filldraw (2.6,1.3) circle (2.2pt);
\node[right] at (2.6,0.45) {$m$}; \node[right] at (2.6,2.15) {$g\triangleright m$};
\end{tikzpicture}
\label{eq:CH-actions}
\end{eqs}
The two endpoints can be moved past each other along the interface, which gives topological junctions
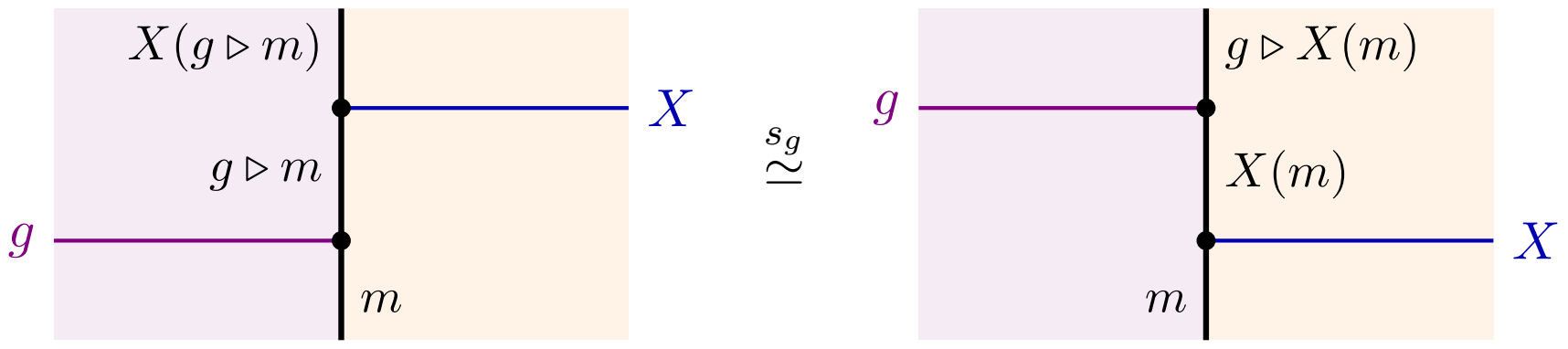
\begin{eqs}
\begin{tikzpicture}[scale=0.8,baseline=(c)]
\coordinate (c) at (0,1.4);
\fill[violet!8] (0,0) rectangle (2.6,3.0);
\fill[orange!10] (2.6,0) rectangle (5.2,3.0);
\draw[very thick] (2.6,0) -- (2.6,3.0);
\draw[thick,violet] (0,0.9) -- (2.6,0.9); \node[left,violet] at (0,0.9) {$g$};
\draw[thick,blue!70!black] (5.2,2.1) -- (2.6,2.1); \node[right,blue!70!black] at (5.2,2.1) {$X$};
\filldraw (2.6,0.9) circle (2.2pt); \filldraw (2.6,2.1) circle (2.2pt);
\node[right] at (2.6,0.35) {\small$m$};
\node[left] at (2.6,1.5) {\small$g\triangleright m$};
\node[left] at (2.6,2.65) {\small$X(g\triangleright m)$};
\end{tikzpicture}
\ \ \overset{s_g}{\simeq}\ \ 
\begin{tikzpicture}[scale=0.8,baseline=(c)]
\coordinate (c) at (0,1.4);
\fill[violet!8] (0,0) rectangle (2.6,3.0);
\fill[orange!10] (2.6,0) rectangle (5.2,3.0);
\draw[very thick] (2.6,0) -- (2.6,3.0);
\draw[thick,blue!70!black] (5.2,0.9) -- (2.6,0.9); \node[right,blue!70!black] at (5.2,0.9) {$X$};
\draw[thick,violet] (0,2.1) -- (2.6,2.1); \node[left,violet] at (0,2.1) {$g$};
\filldraw (2.6,0.9) circle (2.2pt); \filldraw (2.6,2.1) circle (2.2pt);
\node[left] at (2.6,0.35) {\small$m$};
\node[right] at (2.6,1.5) {\small$X(m)$};
\node[right] at (2.6,2.65) {\small$g\triangleright X(m)$};
\end{tikzpicture}
\label{eq:CH-middle}
\end{eqs}
Writing the action of $X$ on the right, $m\triangleleft X:=X(m)$, the interfaces form a $(\cC,\CUV)$-bimodule category, and $s_g:(g\triangleright m)\triangleleft X\simeq g\triangleright(m\triangleleft X)$ is its middle associator. Its compatibility with the fusion of the lines of $\fB_{\rm Dir}$, the mixed pentagon of the bimodule category, reads $s_{gg'}=(g\triangleright s_{g'})\circ s_g$ for trivial module associator.
Equivalently, $X$ is a $\cC$-module functor of $\cM_H$.
Conversely, every module functor arises from a line on $\fB_H$, since the boundary is determined by its interfaces with $\fB_{\rm Dir}$, and fusing two lines of $\fB_H$ composes the corresponding functors.
Hence the symmetry supported on $\fB_H$ is the dual fusion category
\begin{equation}
\CUV=\cC_H:=\End_{\cC}(\cM_H)\,,
\label{eq:CH}
\end{equation}
the counterpart of \eqref{eq:uv-equivalence} \cite{Ostrikmodule,EGNO}.
It is Morita equivalent to $\cC$, so that $\cZ(\cC_H)\simeq\cZ(\cC)$ and all topological boundaries bound the same SymTFT.
Equivalently, $\cM_H$ is the category of right modules over the algebra $\bbC[H]=\bigoplus_{h\in H}h$ in $\Vec_G$, and $\cC_H$ is the category of $\bbC[H]$--$\bbC[H]$ bimodules in $\Vec_G$: the lines on which the mesh of $H$ gauge lines can end from both sides, i.e., the symmetry lines of the theory obtained by gauging $H$ \cite{Bhardwaj:2017xup}.
 
\newl {\bf A recipe for $\CUV$.} The structure of $\cC_H$ follows from \eqref{eq:CH} by the following steps, each of which has a direct physical meaning.
\begin{enumerate}
\item \emph{Interfaces are cosets.} The simple interfaces are $\delta_i$, $i\in G/H$, permuted by the lines of $\fB_{\rm Dir}$ as $g\triangleright\delta_i=\delta_{g\cdot i}$.
\item \emph{Lines are matrices of junction spaces.} A line $U$ is determined by what it does to each simple interface,
\begin{equation}
U(\delta_j)=\bigoplus_{i\in G/H}U_{ij}\otimes\delta_i\,,\qquad U_{ij}:=\Hom\big(\delta_i,U(\delta_j)\big)\,,
\label{eq:CH-matrix}
\end{equation}
where $U_{ij}$ is the space of junctions at which $U$, ending on $\delta_j$, turns it into $\delta_i$:
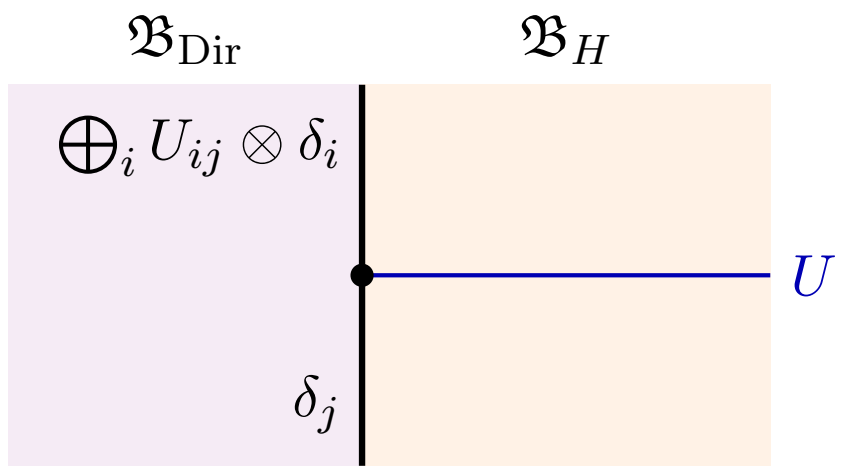
\begin{eqs}
\begin{tikzpicture}[scale=0.9,baseline=(c)]
\coordinate (c) at (0,1.3);
\fill[violet!8] (0,0) rectangle (2.6,2.8);
\fill[orange!10] (2.6,0) rectangle (5.6,2.8);
\draw[very thick] (2.6,0) -- (2.6,2.8);
\draw[thick,blue!70!black] (5.6,1.4) -- (2.6,1.4); \node[right,blue!70!black] at (5.6,1.4) {$U$};
\filldraw (2.6,1.4) circle (2.2pt);
\node[left] at (2.6,0.45) {$\delta_j$}; \node[left] at (2.6,2.35) {$\bigoplus_iU_{ij}\otimes\delta_i$};
\node[above] at (1.3,2.8) {$\fB_{\rm Dir}$}; \node[above] at (4.1,2.8) {$\fB_H$};
\end{tikzpicture}
\label{eq:CH-column}
\end{eqs}
Fusion of lines is matrix multiplication,
\begin{equation}
(U\circ V)_{ik}=\bigoplus_{j}U_{ij}\otimes V_{jk}\,,
\label{eq:CH-composition}
\end{equation}
as for the lines $\mathrm{Mat}_n(\Vec)$ of a 2d TFT \eqref{eq:Matn}, and a morphism of lines is a matrix of linear maps $U_{ij}\to U'_{ij}$.
\item \emph{Lines commute with the symmetry of $\fB_{\rm Dir}$.} Comparing $U(g\triangleright\delta_j)=\bigoplus_{i}U_{i,\,g\cdot j}\otimes\delta_{i}$ with $g\triangleright U(\delta_j)=\bigoplus_iU_{ij}\otimes\delta_{g\cdot i}$, the middle associator $s_g$ of \eqref{eq:CH-middle} is the same as a family of isomorphisms
\begin{equation}
u_g:\ U_{ij}\ \xrightarrow{\ \sim\ }\ U_{g\cdot i,\,g\cdot j}\,,\qquad u_g\,u_{g'}=u_{gg'}\,,
\label{eq:CH-equivariance}
\end{equation}
where $u_g$ is the inverse of the component of $s_g$ and the second condition is the mixed pentagon. Moving the endpoint of $g$ past that of $U$ relabels both interfaces by $g$, and $u_g$ identifies the junction spaces before and after. A line on $\fB_H$ is therefore a $G$-equivariant vector bundle over $G/H\times G/H$, with $G$ acting diagonally.
Morphisms commute with the $u_g$, and \eqref{eq:CH-composition} carries the equivariant structure $u_g\otimes u_g$. 
This is the counterpart of the $\cC$-linearity condition \eqref{eq:c-linearity-0}.

\item \emph{Double cosets.} By \eqref{eq:CH-equivariance} all entries in an orbit $\{(g\cdot i,g\cdot j)\}_{g\in G}$ are isomorphic, so a simple line is supported on a single orbit. The orbits are labeled by the double cosets $H\backslash G/H$, via $(g_iH,g_jH)\mapsto Hg_i^{-1}g_jH$, and every orbit contains an entry $p_0=(H,gH)$ in the first row. Its stabilizer consists of the $k\in G$ with $kH=H$ and $kgH=gH$, i.e.,
\begin{equation}
\mathrm{Stab}(H,gH)=H\cap gHg^{-1}\,.
\end{equation}
\item \emph{A simple line is a double coset with a residual charge.} By the orbit-stabilizer principle \eqref{eq:S3-orbit-stabiliser}, a simple line is a pair
\begin{equation}
\Big(HgH\in H\backslash G/H\,,\ \ \rho\in\mathrm{Irr}\big(H\cap gHg^{-1}\big)\Big)\,,\qquad \dim=\frac{|HgH|}{|H|}\,\dim\rho\,,
\label{eq:CH-simples}
\end{equation}
where $|HgH|/|H|$ is the number of non-zero entries in a column. To build it explicitly, choose for each entry $p$ of the orbit an element $k_p\in G$ with $k_p\cdot p_0=p$ (and $k_{p_0}=\mathrm{id}$), put a copy of $\rho$ in every entry, and set
\begin{equation}
u_g\big|_p=\rho\big(k_{g\cdot p}^{-1}\,g\,k_p\big)\,:\ U_p\to U_{g\cdot p}\,.
\label{eq:CH-transport}
\end{equation}
The element $k_{g\cdot p}^{-1}gk_p$ maps $p_0\mapsto p\mapsto g\cdot p\mapsto p_0$, so it lies in the stabilizer: this is Wigner's little-group construction, with $k_p$ playing the role of a standard boost.
\item \emph{Fusion is read off at one entry per orbit.} By \eqref{eq:CH-composition}, the entry $(U\circ V)_{p_0}$, $p_0=(i,k)$, is a sum over intermediate interfaces $j$, and the stabilizer of $p_0$ acts on it by permuting $j\mapsto s\cdot j$ with the matrices $u_s\otimes u_s$. The multiplicity of the simple line $(HgH,\rho)$ in $U\otimes V$ is the multiplicity of $\rho$ in $(U\circ V)_{p_0}$, viewed as a representation of $\mathrm{Stab}(p_0)$.
\item \emph{The associator.} Composition of functors is strictly associative, and the associator only appears once products are identified with simple lines. Two facts fix it in practice: the lines on the trivial double coset form $\Rep(H)$ with its standard associator, and if $H$ has a complement $K$, i.e.\ $KH=G$ and $K\cap H=\{\mathrm{id}\}$, then reading off a single column, $U\mapsto\bigoplus_iU_{i,H}$, is a fiber functor on $\cC_H$, because $K$ moves every column to the first one by a unique element.
\end{enumerate}
\noindent Physically, the lines on the trivial double coset are supported on the diagonal and form $\Rep(H)$.
These are the Wilson lines of the $H$ gauge field that survives on $\fB_H$, and generate the symmetry dual to the gauged $H$. 
The lines on a non-trivial double coset are what remains of the symmetry lines $x\notin H$.
Such a line does not commute with the gauged $H$ by itself, and only combinations of $x$ with its $H$-conjugates survive, possibly dressed with a residual charge $\rho$ of $H\cap xHx^{-1}$.
For non-trivial $\psi_H$ the steps are the same, except that the module associator of $\Vec(G/H)^{\psi_H}$ makes the $u_g$ compose up to phases, and the stabilizer representations $\rho$ become projective, with a 2-cocycle determined by $\psi_H$ \cite{Ostrik2003double}.
%
 
\begin{examplebox}[The symmetries $\cC_H$ of the four boundaries of $\cZ(\Vec_{S_3})$][label=CH on S3]
We denote by $1_{ij}$ the elementary matrices of $\mathrm{Mat}_n(\Vec)\simeq\End(\cM_H)$, which describe the underlying matrix of a line once its equivariant structure is forgotten.
\begin{enumerate}
\item {\bf $H=\{\mathrm{id}\}$.} The double cosets are the group elements, all with trivial stabilizer, so the simple lines are labeled by $g\in S_3$, with $U_g(\delta_z)=\delta_{zg^{-1}}$ and $U_g\circ U_{g'}=U_{gg'}$. Hence $\cC_{\{\mathrm{id}\}}\simeq\Vec_{S_3}$.
\item {\bf $H=S_3$.} There is a single entry, with stabilizer $S_3$, so a line is a representation and $\cC_{S_3}\simeq\Rep(S_3)$, with simple lines $1$, $P$, $E$.
\item {\bf $H=\Z_2$.} The cosets $i\in\{0,1,2\}$, $i\leftrightarrow a^i\Z_2$, are permuted as $a\cdot i=i+1$ and $b\cdot i=-i$ mod $3$. There are two double cosets, $\Z_2=\{\mathrm{id},b\}$ and $\Z_2a\Z_2=\{a,a^2,ab,a^2b\}$.
The first is the diagonal orbit $\{(i,i)\}$, with stabilizer $\Z_2$ at $(0,0)$. Choosing $k_{(i,i)}=a^i$ in \eqref{eq:CH-transport}, $u_a$ acts trivially and $u_b$ by $\rho(b)$ on every entry. It carries two lines, $1$ and $P$, both with underlying matrix $\bigoplus_i1_{ii}$, on which $b$ acts by $+1$ and $-1$ respectively.
The second is the orbit of all six off-diagonal entries, with trivial stabilizer $\Z_2\cap a\Z_2a^{-1}=\{\mathrm{id},b\}\cap\{\mathrm{id},a^2b\}=\{\mathrm{id}\}$. It carries a single line $E$, with underlying matrix $\bigoplus_{i\neq j}1_{ij}$, on which $u_g$ simply permutes the entries, and dimension $|\Z_2a\Z_2|/|\Z_2|=2$.
Matrix multiplication gives $P\otimes P=1$, $P\otimes E=E$ and $E\otimes E=2\bigoplus_i1_{ii}\oplus E$ at the level of underlying matrices. In the entry $(0,0)$ of $E\circ E$, the two paths $E_{01}\otimes E_{10}$ and $E_{02}\otimes E_{20}$ are exchanged by $u_b$, since $b\cdot1=2$, so the stabilizer acts with eigenvalues $\pm1$ and the diagonal part is $1\oplus P$:
\begin{equation*}
E\otimes E=1\oplus P\oplus E\,.
\end{equation*}
Physically, $P$ is the Wilson line of the gauged $\Z_2$, and $E$ is the gauge-invariant combination $a\oplus a^2$, which cannot distinguish $a$ from $a^2$ since they are exchanged by the gauged $b$.
\item {\bf $H=\Z_3$.} Since $\Z_3$ is normal, the two double cosets are the cosets $\Z_3$ and $b\Z_3$, and both orbits of entries, $\{(0,0),(1,1)\}$ and $\{(0,1),(1,0)\}$, have stabilizer $\Z_3$. All six simple lines are invertible: $\eta^k$ with underlying matrix $1_{00}\oplus1_{11}$ and $\sigma_k$ with underlying matrix $1_{01}\oplus1_{10}$, on which $a$ acts by $\omega^k$ on the entry $(0,0)$, respectively $(0,1)$. By \eqref{eq:CH-transport} with $k_{(1,1)}=k_{(1,0)}=b$, $a$ acts on the second entry of each orbit by $\rho(bab)=\rho(a^2)=\omega^{-k}$.
Reading off the eigenvalue of $u_a$ along the single path through each base entry,
\begin{equation*}
\eta^k\otimes\eta^l=\eta^{k+l}\,,\qquad \eta^k\otimes\sigma_l=\sigma_{k+l}\,,\qquad \sigma_l\otimes\eta^k=\sigma_{l-k}\,,\qquad \sigma_k\otimes\sigma_l=\eta^{k-l}\,,
\end{equation*}
where the asymmetry between the second and third rules comes from the conjugate eigenvalue on the entry $(1,1)$ of $\eta^k$. These are the relations of $S_3$ under $\eta^k\mapsto a^k$, $\sigma_k\mapsto a^kb$.
Physically, $\eta$ is the Wilson line of the gauged $\Z_3$ and $\sigma_0$ the surviving image of $b$, so that the boundary symmetry is $\widehat{\Z}_3\rtimes\Z_2\simeq S_3$.
\end{enumerate}
\newl Associators: in both cases $H$ has a complement $K$, namely $K=\Z_3$ for $H=\Z_2$ and $K=\Z_2$ for $H=\Z_3$, so step 7 provides a fiber functor. This fixes the associators: $\cC_{\Z_2}\simeq\Rep(S_3)$, since semisimple Hopf algebras of dimension $6$ are group algebras or their duals \cite{Masuoka1995}, and $\cC_{\Z_3}\simeq\Vec_{S_3}$, since $\Vec^\omega_{S_3}$ admits a fiber functor only for trivial $\omega$. In summary, $\cC_{\{\mathrm{id}\}}\simeq\cC_{\Z_3}\simeq\Vec_{S_3}$ and $\cC_{\Z_2}\simeq\cC_{S_3}\simeq\Rep(S_3)$.
\end{examplebox}
 
\subsubsection{The IR symmetry \texorpdfstring{$\CIR$}{CIR}}\label{subsec:warm-up-ir}
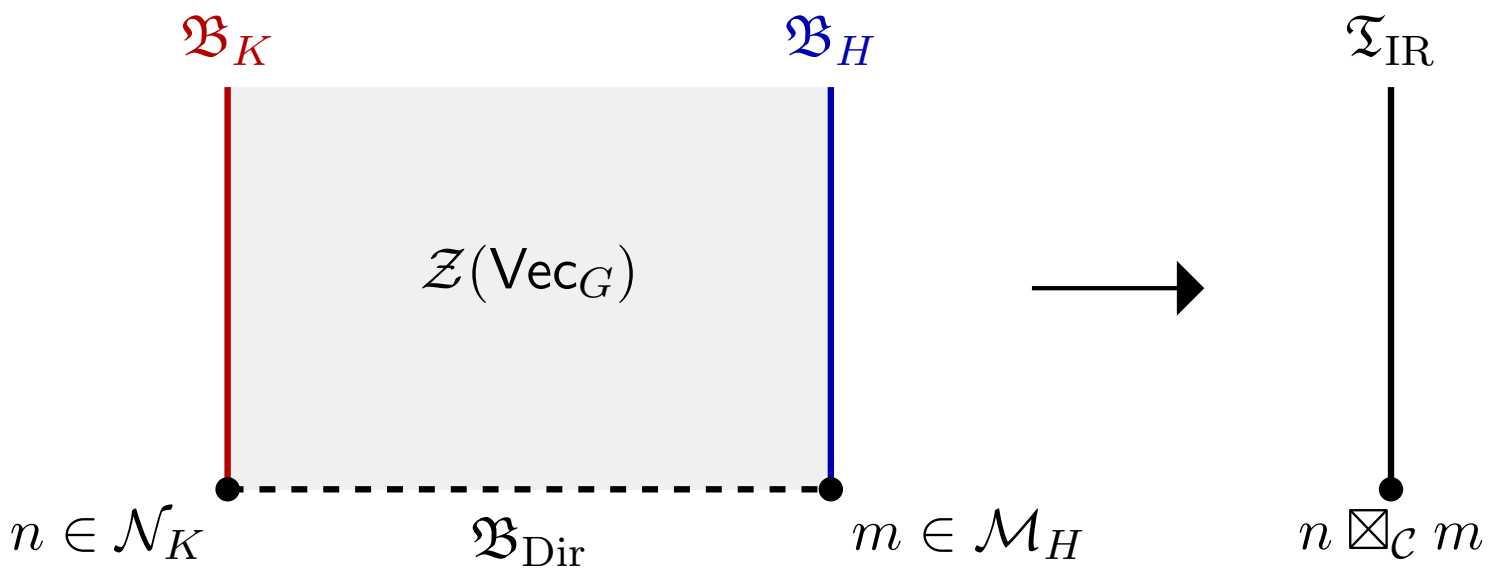
\begin{figure}[ht]
\centering
\begin{tikzpicture}[scale=0.95]
\fill[gray!12] (0,0) rectangle (4.2,2.8);
\draw[very thick,red!70!black] (0,0) -- (0,2.8) node[above] {$\fB_K$};
\draw[very thick,blue!70!black] (4.2,0) -- (4.2,2.8) node[above] {$\fB_H$};
\draw[very thick,dashed] (0,0) -- (4.2,0);
\node[below] at (2.1,-0.05) {$\fB_{\rm Dir}$};
\filldraw (0,0) circle (2.2pt) node[below left] {$n\in \cN_K$};
\filldraw (4.2,0) circle (2.2pt) node[below right] {$m\in \cM_H$};
\node at (2.1,1.5) {$\cZ(\Vec_G)$};
\draw[->,thick] (5.6,1.4) -- (6.8,1.4);
\draw[very thick] (8.1,0) -- (8.1,2.8) node[above] {$\TIR$};
\filldraw (8.1,0) circle (2.2pt) node[below] {$n\boxtimes_\cC m$};
\end{tikzpicture}
\caption{A strip of $\fB_{\rm Dir}$ joins the physical boundary $\fB_K$ to the symmetry boundary $\fB_H$, with interfaces $n\in \cN_K$ and $m\in \cM_H$ at its corners. After interval compactification this becomes a boundary condition $n\boxtimes_\cC m$ of the IR 2d TQFT.}
\label{fig:warm-up-sandwich}
\end{figure}

To read off the IR theory, we insert a strip of the Dirichlet boundary between the physical boundary $\fB_K$ and the symmetry boundary $\fB_H$, as in figure~\ref{fig:warm-up-sandwich}.
After interval compactification, a boundary condition of the IR theory is specified by an interface $n\in \cN_K$ at one corner of the strip and an interface $m\in \cM_H$ at the other (figure~\ref{fig:warm-up-sandwich}).
A symmetry line $g$ of the strip can be moved from one corner to the other, so $(n\triangleleft g)\boxtimes m$ and $n\boxtimes(g\triangleright m)$ describe the same boundary condition.
The boundary conditions of the IR theory therefore form the balanced product $\cN_K\boxtimes_\cC \cM_H$, whose objects are objects of $\cN_K\boxtimes \cM_H$ together with coherent isomorphisms implementing this move, and as in \eqref{eq:cir}
\begin{equation}
\CIR\simeq\End_{\Vec}\big(\cN_K\boxtimes_\cC \cM_H\big)\,.
\label{eq:warm-up-cir}
\end{equation}
 
\newl {\bf Computing the balanced product.} An object of $\cN_K\boxtimes \cM_H$ is a vector bundle $W$ over the set $K\backslash G\times G/H$ of pairs of interfaces. Moving a line $g$ across the strip acts on these pairs by
\begin{equation}
g:\ (\bar r,\,i)\ \longmapsto\ (\bar r\cdot g^{-1},\ g\cdot i)\,, \quad \bar r\in K\backslash G\,,  \ i \in G/H\,,
\label{eq:S3-combined}
\end{equation}
and the balancing isomorphisms are exactly a $G$-equivariant structure on $W$ for this action. The orbits of \eqref{eq:S3-combined} are the double cosets $K\backslash G/H$, via $(Ky,zH)\mapsto KyzH$, and the stabilizer of the pair $(Kg,H)$ is $H\cap g^{-1}Kg$. The orbit-stabilizer principle \eqref{eq:S3-orbit-stabiliser} then gives
\begin{equation}
\cN_K\boxtimes_\cC \cM_H\ \simeq\ \bigoplus_{KgH\in K\backslash G/H}\Rep\big(H\cap g^{-1}Kg\big)\,,
\label{eq:S3-balanced}
\end{equation}
the counterpart of \eqref{eq:boundary-cat}: the $H\cap K^g$-graded fusion category $[\bar B\boxtimes A]_{H\cap K^g}$ there is replaced here by the group algebra of $H\cap g^{-1}Kg$. For non-trivial cocycles, the summand $\Rep(L_g)$, $L_g:=H\cap g^{-1}Kg$, is replaced by the category $\Rep^{\alpha_g}(L_g)\simeq\mathsf{LMod}\big(\bbC^{\alpha_g}[L_g]\big)$ of $\alpha_g$-projective representations, where
\begin{equation}
\alpha_g(h,h')=\psi_H(h,h')\,\psi_K\big(ghg^{-1},gh'g^{-1}\big)^{-1}\,,\qquad h,h'\in L_g\,,
\label{eq:alpha-g}
\end{equation}
is a 2-cocycle on $L_g$. Its overall orientation depends on conventions, but replacing $\alpha_g$ by $\alpha_g^{-1}$ affects none of the statements below.
 
\newl {\bf Two readings of the result.} The simple objects of \eqref{eq:S3-balanced} have two interpretations.
\begin{itemize}
\item \emph{Interfaces in the SymTFT.} Shrinking the Dirichlet strip, they are the simple topological line interfaces between $\fB_K$ and $\fB_H$. The double coset records how the reduction of the gauge field to $K$ on one side is glued to its reduction to $H$ on the other. The residual gauge group on the interface is $H\cap g^{-1}Kg$, and the interface can carry a charge $\sigma\in\mathrm{Irr}(H\cap g^{-1}Kg)$ under it.
\item \emph{Vacua of the IR theory.} After interval compactification, they are the simple boundary conditions of the IR 2d TQFT, one for each vacuum. The number of vacua and the IR symmetry are therefore
\begin{equation}
n(H,K)=\sum_{KgH\in K\backslash G/H}\big|\mathrm{Irr}(H\cap g^{-1}Kg)\big|\,,\qquad \CIR\simeq\mathrm{Mat}_{n(H,K)}(\Vec)\,,
\label{eq:S3-vacua}
\end{equation}
by \eqref{eq:Matn}. 
Equivalently, \eqref{eq:S3-balanced} says that the IR theory is presented by the algebra $\bigoplus_{KgH}\bbC[H\cap g^{-1}Kg]$, i.e., it is a direct sum of 2d gauge theories, one with gauge group $H\cap g^{-1}Kg$ for each double coset. By \eqref{eq:2d-local-ops}, its local operators form 
\begin{equation}
\bigoplus_{KgH \in K\backslash G/H}\cZ\big(\bbC[H\cap g^{-1}Kg]\big)\simeq\bbC^{n(H,K)}\,.
\end{equation}
In each summand, one basis is given by the class sums, i.e.\ the gauge-invariant holonomies, and another by the central idempotents, which are the vacuum projectors.
For non-trivial cocycles, $\bbC[L_g]$ is replaced by $\bbC^{\alpha_g}[L_g]$, and $|\mathrm{Irr}(L_g)|$ in \eqref{eq:S3-vacua} by the number of irreducible $\alpha_g$-projective representations of $L_g$.
\end{itemize}
This is the main structural difference from section~\ref{subsec:ir-symmetry}. There, the summand of a double coset in \eqref{eq:boundary-cat} is a 2-category of module categories over a fusion category, which is connected, so each double coset contributes exactly one vacuum. Here the summand $\Rep(H\cap g^{-1}Kg)$ is a semisimple 1-category, whose distinct simple objects admit no non-zero morphisms between them, so each irreducible representation of the stabilizer is a separate vacuum. The representations that label lines of a single vacuum one dimension higher label different vacua here.
The cocycles sharpen this difference. Here, a non-degenerate $\alpha_g$ makes $\bbC^{\alpha_g}[L_g]$ a full matrix algebra, so a non-trivial stabilizer can still contribute a single vacuum, with no residual gauge theory. One dimension higher, the stabilizer instead contributes the fusion category $[\bar B\boxtimes A]_{H\cap K^g}$, of dimension $|H\cap K^g|\dim(A_1)\dim(B_1)$, and the lines $\cZ\big([\bar B\boxtimes A]_{H\cap K^g}\big)$ it leaves in the IR are trivial only if this dimension is one. No choice of twist can therefore compensate a non-trivial stabilizer in $(2\!+\!1)$ dimensions.
 
\newl {\bf Single vacuum and fiber functors.} By \eqref{eq:S3-vacua}, there is a single vacuum if and only if
\begin{enumerate}[label=(\roman*)]
\item there is a single double coset, $KH=G$, and
\item its stabilizer $L=H\cap K$ contributes a single vacuum, i.e.\ $\bbC^{\alpha}[L]$ with $\alpha=\psi_H|_L\,\psi_K|_L^{-1}$ is a simple algebra.
\end{enumerate}
Condition (ii) says that $\alpha$ is non-degenerate on $L$, so that $L$ is a group of central type and $|L|$ is a perfect square. For trivial cocycles it reduces to $H\cap K=\{\id\}$, i.e.\ $K$ is a complement of $H$ and $G=H\bowtie K$ is an exact factorization. In general, however, $H\cap K$ can be non-trivial; the smallest example is $G=H=K=\Z_2^2$ with $[\psi_K]\neq0$, see example~\ref{ex:Z2Z2-twisted}.
Every indecomposable $\cC_H$-module category is of the form $\cN_{(K,\psi_K)}\boxtimes_\cC \cM_{(H,\psi_H)}$ \cite{Ostrikmodule}, and fiber functors of $\cC_H$ are its module categories of rank one. Conditions (i) and (ii) are therefore exactly the criterion for the group-theoretical fusion category $\cC_H$ to admit a fiber functor \cite{Ostrikmodule,Natale:2003}.
This is the 1-categorical counterpart of the criterion of section~\ref{subsec:symtft-rg-flow}, with one difference: there, a twist cannot compensate a non-trivial intersection, and $H\cap K=\{\id\}$ is required (see also the comparison at the end of that section).
Compared with \eqref{eq:single-vacuum}, both conditions are needed already for a single vacuum, and a single vacuum forces the IR theory to be trivial, $\CIR\simeq\Vec$. In two dimensions there is therefore no analogue of a generalized fiber 2-functor.
 
\begin{examplebox}[Vacua of the $S_3$ sandwiches][label=ex:S3 vacua]
For $(H,K)=(\Z_2,\Z_2)$ there are two double cosets, $\{\mathrm{id},b\}$ and $\{a,a^2,ab,a^2b\}$, with stabilizers $\Z_2$ and $\Z_2\cap a^{-1}\Z_2a=\{\mathrm{id},b\}\cap\{\mathrm{id},ab\}=\{\mathrm{id}\}$, so there are $2+1=3$ vacua. For $(H,K)=(\Z_3,\Z_3)$ the two double cosets $\Z_3$ and $b\Z_3$ both have stabilizer $\Z_3$, giving $3+3=6$ vacua. For $(H,K)=(\Z_2,\Z_3)$ there is a single double coset, since $\Z_3\Z_2=S_3$, with trivial stabilizer, giving a single vacuum. All sixteen cases are
\begin{center}
\small
\renewcommand{\arraystretch}{1.3}
\begin{tabular}{|c||c|c|c|c|}
\hline
$\fB_{\rm sym}$ $\backslash$ $\fB_{\rm phys}$ & $\fB_{\{\mathrm{id}\}}$ & $\fB_{\Z_2}$ & $\fB_{\Z_3}$ & $\fB_{S_3}$ \\
\hline\hline
$\fB_{\{\mathrm{id}\}}$ $(\Vec_{S_3})$ & 6 & 3 & 2 & 1 \\
$\fB_{\Z_2}$ $(\Rep(S_3))$ & 3 & 3 & 1 & 2 \\
$\fB_{\Z_3}$ $(\Vec_{S_3})$ & 2 & 1 & 6 & 3 \\
$\fB_{S_3}$ $(\Rep(S_3))$ & 1 & 2 & 3 & 3 \\
\hline
\end{tabular}
\end{center}
When $KH=S_3$, \eqref{eq:S3-balanced} reduces to $\Rep(H\cap K)$, and the IR theory is the 2d pure gauge theory with gauge group $H\cap K$: its vacua are labeled by the irreducible representations of $H\cap K$, and its local operators are the gauge-invariant holonomies, i.e.\ the center $\cZ(\bbC[H\cap K])$ of the group algebra, whose central idempotents are the vacuum projectors. One dimension higher, the same pair of boundaries gives a single vacuum carrying the lines $\cZ(\Vec_{H\cap K})$ of the 3d Dijkgraaf--Witten theory of $H\cap K$; for instance $(H,K)=(\Z_2,S_3)$ gives the toric code there, and the two vacua of the $\Z_2$ gauge theory here.
There are four single-vacuum pairs: $(\{\mathrm{id}\},S_3)$ and $(S_3,\{\mathrm{id}\})$ are the trivial phases of $\Vec_{S_3}$ and $\Rep(S_3)$ in their standard presentations, while $(\Z_2,\Z_3)$ and $(\Z_3,\Z_2)$ express that the exact factorization $S_3=\Z_2\bowtie\Z_3$ provides fiber functors for both $\cC_{\Z_2}\simeq\Rep(S_3)$ and $\cC_{\Z_3}\simeq\Vec_{S_3}$.
Since $\operatorname{H}^2(L;U(1))=0$ for every subgroup $L\subseteq S_3$, no non-trivial cocycles can occur, and condition (ii) always reduces to $H\cap K=\{\mathrm{id}\}$. Accordingly, $\Rep(S_3)$ and $\Vec_{S_3}$ each have a unique fiber functor up to equivalence, realized in both of their presentations.
\end{examplebox}

\begin{examplebox}[A fiber functor without a complement][label=ex:Z2Z2-twisted]
Take $G=\Z_2^2$, the symmetry boundary $\fB_{G}$ with trivial $\psi_H$, so that $\CUV=\cC_G\simeq\Rep(\Z_2^2)$, and the physical boundary $\fB_{(G,\psi_K)}$ with $[\psi_K]$ the generator of $\operatorname{H}^2(\Z_2^2;U(1))\simeq\Z_2$, e.g.\ $\psi_K(a,b)=(-1)^{a_1b_2}$. There is a single double coset, with stabilizer $L=G$, and $\alpha=\psi_K^{-1}$ is non-degenerate: in $\bbC^{\alpha}[\Z_2^2]$ the generators $u_{(1,0)}$ and $u_{(0,1)}$ square to one and anticommute, so that $\bbC^{\alpha}[\Z_2^2]\simeq\mathrm{Mat}_2(\bbC)$, generated by the Pauli matrices $X$ and $Z$. Hence $\cN\boxtimes_\cC\cM\simeq\Vec$, there is a single vacuum, and every line of $\Rep(\Z_2^2)$ acts by its dimension. This is the second fiber functor of $\Rep(\Z_2^2)$, besides the one given by the complement $K=\{1\}$, although here $H\cap K=G$. Physically, it describes gauging $\Z_2^2$ in the non-trivial $\Z_2\times\Z_2$ SPT phase, which leaves a unique gapped ground state.
One dimension higher, pairing $\fB_G$ with $\fB_{(G,\omega)}$ leaves the 3d Dijkgraaf--Witten theory with lines $\cZ(\Vec^{\omega}_{\Z_2^2})$, of global dimension $16$, for every $\omega\in\operatorname{H}^3(\Z_2^2;U(1))$.
\end{examplebox}
 
\subsubsection{The RG functor \texorpdfstring{$\cF:\CUV\to\CIR$}{F : CUV -> CIR}}\label{subsec:warm-up-rg}

Inserting a line $X$ of $\CUV$ on the symmetry boundary and compactifying the slab produces a line of the IR theory. 
The line only acts on the interface at its own corner of the strip,
\begin{equation}
\cF:\ \cC_H\longrightarrow\CIR\simeq\mathrm{Mat}_n(\Vec)\,,\qquad X\longmapsto\Id_{\cN_K}\boxtimes_\cC X\,,
\label{eq:S3-RG}
\end{equation}
which is well defined because $X$ is $\cC$-linear. 
Concretely, $X$ acts on a ($G$-equivariant) bundle $W$ over $K\backslash G\times G/H$ through its matrix \eqref{eq:CH-matrix} in the second label,
\begin{equation}
\big(\cF(X)W\big)_{(\bar r,i)}=\bigoplus_{j\in G/H}X_{ij}\otimes W_{(\bar r,j)}\,,\qquad u_g=u_g^X\otimes u_g^W\,,
\label{eq:S3-RG-matrix}
\end{equation}
and since $X$ never touches the first label, $\cF(X\circ Y)=\cF(X)\circ\cF(Y)$. Physically, $\cF$ is the action of the symmetry on the vacua; therefore specifying it is the same as specifying the gapped phase as a $\CUV$-module category.

\newl {\bf A recipe for $\cF$.} Recall that a vacuum $v$ is a simple equivariant bundle, i.e.,  it is supported on one orbit $O$ of pairs $p=(\bar r,j)$, with stabilizer $S$ at a base point $p_0$, and it carries an irreducible representation $\rho$ of $S$.
Its fiber at every $p\in O$ is a copy of $\rho$, and $u_g$ acts from $p$ to $g\cdot p$ by $\rho(k_{g\cdot p}^{-1}gk_p)$, with $k_p\cdot p_0=p$. To compute $\cF(X)v$:
\begin{enumerate}
\item \emph{Move the support.} By \eqref{eq:S3-RG-matrix}, the fiber of $\cF(X)v$ at $(\bar r,i)$ is $\bigoplus_jX_{ij}\otimes v_{(\bar r,j)}$.
The second label of each point of $O$ is moved along the non-zero entries of the matrix of $X$, and the first label is untouched.
The result may be supported on several orbits $O'$.
\item \emph{Read off one fiber per orbit.} For each such $O'$, take a base point $p'_0=(\bar r',i')$ with stabilizer $S'$. An element $s\in S'$ maps the summand $j$ of the fiber at $p'_0$ to the summand $s\cdot j$, so only the summands it fixes contribute to its character.
\item \emph{Decompose.} The multiplicity of the vacuum $v'=(O',\sigma)$  for some  $\sigma \in \Rep({\rm Stab}(p_0'))$ is determined by the multiplicity of the $\sigma$-irrep at the fiber at $s$.
\end{enumerate}
Writing $1_{v'v}$ for the elementary IR line from $v$ to $v'$, the result is $\cF(X)\simeq\bigoplus_{v,v'}\cF(X)_{v'v}\otimes1_{v'v}$, and the integer matrices $\dim\cF(X)_{v'v}$ multiply according to the fusion rules of $\CUV$.

\newl {\bf No emergent symmetry.} Every IR line $1_{v'v}$ appears in some $\cF(X)$, because the vacua of an indecomposable physical boundary form an indecomposable $\CUV$-module category, whose simple objects are all connected by the action of $\CUV$. So no gapped phase obtained from a sandwich has an emergent symmetry in this sense, and the question of minimality addressed in section~\ref{subsec:minimality} is genuinely 2-categorical.

\begin{examplebox}[The RG functor for the symmetry boundary $\fB_{\Z_2}$][label=ex:S3-RG]
On $\fB_{\Z_2}$, the interfaces are $\delta_i$, $i\in\{0,1,2\}$, with $a\cdot i=i+1$ and $b\cdot i=-i$, and $\CUV\simeq\Rep(S_3)$ has the simple lines of example~\ref{CH on S3}: $P$, the identity matrix on which $g$ acts by $\mathrm{sgn}(g)$, and $E$, the off-diagonal matrix on which $g$ only permutes the entries.

\newl {\bf (a) $K=\Z_3$: a fiber functor.} The right cosets are $\Z_3$ and $\Z_3b$, labeled $r\in\{0,1\}$, so the pairs are $(r,i)$, with $a\cdot(r,i)=(r,i+1)$ and $b\cdot(r,i)=(1-r,-i)$. All six pairs form a single orbit with trivial stabilizer, since $\Z_2\cap\Z_3=\{\mathrm{id}\}$, so there is a single vacuum $v$ with one-dimensional fibers, and $\CIR\simeq\Vec$.
\begin{itemize}
\item $\cF(P)v$: by step 1 its fiber at $(r,i)$ is $P_{ii}\otimes v_{(r,i)}$, one-dimensional on the same orbit. The stabilizer is trivial, so by step 3 this is one copy of $v$; the signs $\mathrm{sgn}(g)$ are absorbed by rescaling the fiber at $(r,i)$ by $(-1)^r$. Hence $\cF(P)=1$.
\item $\cF(E)v$: its fiber at $(r,i)$ is $\bigoplus_{j\neq i}E_{ij}\otimes v_{(r,j)}$, two-dimensional, so $\cF(E)=2$.
\end{itemize}
Every line acts by its quantum dimension: $\cF:\Rep(S_3)\to\Vec$ is the fiber functor, and the IR theory is trivially gapped. It exists because $\Z_3$ is a complement of $\Z_2$, i.e.\ $S_3=\Z_2\Z_3$ with $\Z_2\cap\Z_3=\{\mathrm{id}\}$, so that $S_3$ is the Zappa--Sz\'ep product of $\Z_2$ and $\Z_3$, here the semidirect product $\Z_3\rtimes\Z_2$. This is the 1-categorical counterpart of the exact factorizations of section~\ref{subsec:symtft-rg-flow}.

\newl {\bf (b) $K=S_3$: spontaneous breaking of $\Z_2$.} Now $S_3\backslash S_3$ is a point, so the pairs are just $i\in\{0,1,2\}$, forming a single orbit with stabilizer $\{\mathrm{id},b\}$ at $i=0$. There are two vacua $v_\pm$, on which $b$ acts at $i=0$ by $\pm1$, and $\CIR\simeq\mathrm{Mat}_2(\Vec)$.
\begin{itemize}
\item $\cF(P)v_\pm$: its fiber at $0$ is $P_{00}\otimes(v_\pm)_0$, on which $b$ acts by the product of its actions on the two factors, $(-1)(\pm1)=\mp1$. Hence $\cF(P):v_\pm\mapsto v_\mp$, i.e., 
\begin{eqs}
    \cF(P)=1_{+-}\oplus 1_{-+}
\end{eqs}
\item $\cF(E)v_\pm$: its fiber at $0$ is $E_{01}\otimes(v_\pm)_1\oplus E_{02}\otimes(v_\pm)_2$. Since $b\cdot1=2$, $b$ exchanges the two summands and fixes neither, so $\chi(b)=0$ and $\chi(\mathrm{id})=2$, and the multiplicities of $\pm$ are 1 each. Hence $\cF(E):v_\pm\mapsto v_+\oplus v_-$.
\begin{eqs}
    \cF(E)=1_{++}\oplus 1_{+-}\oplus 1_{-+} \oplus 1_{--}\,.
\end{eqs}
\end{itemize}
The line $P$ exchanges the two vacua, so the $\Z_2$ symmetry it generates is spontaneously broken. 
The IR theory is the 2d $\Z_2$ gauge theory, with gauge group $H\cap K=\Z_2$, in agreement with example~\ref{ex:S3 vacua}.
\end{examplebox}

\subsection{Lagrangian algebras}\label{subsec:warm-up-lagrangian}
The same boundaries can be described from the bulk, by the lines of $\cZ(\Vec_G)$ that condense on them.
This is the 1-categorical counterpart of section~\ref{subsec:lagrangian-algebras}, and we follow the same structure.
We first describe the Lagrangian algebra of each boundary explicitly, and then recover $\CUV$, $\CIR$ and the RG functor from it. Each example below is repeated one dimension higher in section~\ref{subsec:lagrangian-algebras}, and table~\ref{tab:examples} there lists the corresponding examples side by side.

\newl A {\bf Lagrangian algebra} in $\cZ(\Vec_G)$ is a line $\cL=\bigoplus_{x\in G}\cL_x$ with a unital and associative product that commutes with the half-braiding and is commutative with respect to the braiding.
Additionally, it is  connected, separable and has dimension $\dim\cL=\sqrt{\dim\cZ(\Vec_G)}=|G|$ which in turn implies that it has trivial category of local modules.
Physically, $\cL$ collects the bulk lines that can end on the boundary: the product fuses their endpoints; commutativity lets the endpoints move around each other; connectedness makes the boundary indecomposable; and the dimension condition makes the condensation maximal \cite{Kong:2013aya}. Lagrangian algebras in $\cZ(\cC)$ are in one-to-one correspondence with indecomposable $\cC$-module categories \cite{Davydov:2010wpa}, and hence with topological boundaries \cite{KitaevKong2012,FuchsSchweigertValentino2013}.

\newl {\bf The Lagrangian algebra of $\fB_{(H,\psi)}$} can be read off from its construction. 
On $\fB_{\rm Dir}$ any non-trivial bulk flux line cannot end.
It continues along the boundary as the symmetry line $x\in \Vec_{G}$.
After stacking $\cT$ and gauging the diagonal $H$, for $x\in H$ this line is identified with the $x$-line of $\cT$, their combination being a gauge line, and the latter ends at the junction $a_x$ of \eqref{eq:a_h}. So a bulk flux $x\in H$ ends on $\fB_{(H,\psi)}$, and we denote its endpoint by $e_x$. Fusing two endpoints amounts to fusing the two boundary lines before they end, so by \eqref{eq:fiber_functor_associator} the endpoints form the boundary algebra $\bbC^\psi[H]$, $e_xe_y=\psi(x,y)\,e_{xy}$, with $\psi$ normalized. A gauge transformation $h\in H$ conjugates both the flux and the junction, $u_he_x=e_he_xe_h^{-1}$. A gauge transformation $g\notin H$ instead produces the endpoint of the conjugate flux $gxg^{-1}$, which we denote $[g,e_x]:=u_g(e_x)$. Since $u_{gh}(e_x)=u_g(e_he_xe_h^{-1})$, only the coset $gH$ matters, and the bulk lines that end on $\fB_{(H,\psi)}$ form the induced algebra
\begin{equation}
\cL_{(H,\psi)}=G\times_H\bbC^\psi[H]\,,\qquad [gh,e_x]=[g,\,e_he_xe_h^{-1}]\quad(h\in H)\,,
\label{eq:Lagrangian-induced}
\end{equation}
spanned by the gauge transforms $[g,e_x]$ of the endpoints of fluxes in $H$. Its structure is
\begin{eqs}
\deg[g,e_x]=gxg^{-1}\,,\qquad u_k[g,e_x]=[kg,e_x]\,,\qquad \eta=\sum_{gH\in G/H}[g,e_{\mathrm{id}}]\,,    
\end{eqs}
where $\eta$ is the algebra unit.
The product structure is 
\begin{equation}
\begin{gathered}
[g,e_x]\cdot[gh,e_y]=[g,\,e_xe_he_ye_h^{-1}]\quad(h\in H)\,,\qquad [g,e_x]\cdot[g',e_y]=0\quad(g'H\neq gH)\,,
\end{gathered}
\label{eq:Lagrangian-structure}
\end{equation}
i.e., endpoints transported by the same coset are multiplied in $\bbC^\psi[H]$ after bringing them to the same representative, while those transported by different cosets multiply to zero, so that the units $[g,e_{\mathrm{id}}]$ of the different cosets are orthogonal idempotents whose sum is the gauge-invariant unit $\eta$.

\newl {\bf Block structure.} Choose representatives $g_i$ of the cosets $i\in G/H$, with $g_0=\mathrm{id}$. Every $g\in G$ is uniquely $g=g_ih$ with $h\in H$, so by \eqref{eq:Lagrangian-induced} every symbol is a multiple of one of the $|G|$ elements $[g_i,e_x]$, which form a basis. In this basis the product is
\begin{equation}
[g_i,e_x]\cdot[g_j,e_y]=\delta_{ij}\,\psi(x,y)\,[g_i,e_{xy}]\,.
\end{equation}
So, as an ordinary algebra, $\cL_{(H,\psi)}$ is the direct sum of $|G/H|$ blocks
\begin{equation}
B_i:=\mathrm{span}\big\{[g_i,e_x]\big\}_{x\in H}\,\simeq\,\bbC^\psi[H]\,,\qquad B_i\cdot B_j=0\quad(i\neq j)\,,
\end{equation}
one copy of the boundary algebra for each coset, i.e.\ for each simple interface $\delta_i$. The block $B_i$ is graded by the stabilizer $H_i=g_iHg_i^{-1}$ of the coset $i$, since $[g_i,e_x]$ has flux $g_ixg_i^{-1}$, and its unit $p_i:=[g_i,e_{\mathrm{id}}]$ is an idempotent, with $\eta=\sum_ip_i$. Writing $kg_i=g_{k\cdot i}\,l_i(k)$ with $l_i(k)\in H$, the half-braiding acts as
\begin{equation}
u_k[g_i,e_x]=[g_{k\cdot i},\,e_le_xe_l^{-1}]\,,\qquad l=l_i(k)\,,
\end{equation}
mapping the block $B_i$ to $B_{k\cdot i}$: $G$ permutes the blocks as it permutes the cosets, and the little-group element conjugates the label.

\newl {\bf Which lines condense.} A simple line $\Qc{g}{\pi}$ is determined by its fiber at the single flux $g$, which carries the representation $\pi$ of the centralizer $C_G(g)$. A morphism $\Qc{g}{\pi}\to\cL_{(H,\psi)}$ must preserve the grading and commute with the half-braiding, so it is determined by a $C_G(g)$-equivariant map from $\pi$ to the flux-$g$ part of $\cL_{(H,\psi)}$. The flux-$g$ elements are the $[g_i,e_x]$ with $g_ixg_i^{-1}=g$. Such an $x\in H$ exists if and only if $g\in H_i$, i.e.\ $g$ fixes the coset $i$, and it is then unique. So the flux-$g$ part has one basis element for each coset fixed by $g$, and the centralizer $C_G(g)$ permutes these cosets. Hence, for $\psi$ trivial,
\begin{equation}
n_H\big(\Qc{g}{\pi}\big)=\dim\Hom\big(\Qc{g}{\pi},\cL_H\big)=\Big[\,\bbC\big[(G/H)^g\big]:\pi\,\Big]_{C_G(g)}\,,
\label{eq:CH-condensation}
\end{equation}
the multiplicity of $\pi$ in the permutation representation of $C_G(g)$ on the set $(G/H)^g$ of cosets fixed by $g$. In words: a flux line $g$ can end on $\fB_H$ only if $g$ is conjugate into $H$, and its charge $\pi$ must appear in how the centralizer permutes the cosets that $g$ fixes. For non-trivial $\psi$, an element $k\in C_G(g)$ that fixes the coset $i$ acts on the basis element $[g_i,e_x]$, $x=g_i^{-1}gg_i$, by the discrete-torsion phase
\begin{equation}
\varepsilon_x(l)=\frac{\psi(l,x)}{\psi(x,l)}\,,\qquad l=g_i^{-1}kg_i\in C_H(x)\,,
\label{eq:charges}
\end{equation}
which follows from $u_k[g_i,e_x]=[g_i,e_le_xe_l^{-1}]$. The permutation representation is then twisted by these phases, so fluxes condense together with a compensating charge and the condensing lines become dyonic. In the two extreme cases: for the Dirichlet boundary only the identity fixes a coset, and $\cL_{\{\mathrm{id}\}}=\Fun(G)=\bigoplus_R\dim R\,\Qc{\mathrm{id}}{R}$ is the regular representation, so only Wilson lines end, each $\dim R$ times; for the Neumann boundary every flux fixes the single coset with trivial action, so $\cL_G=\bbC[G]=\bigoplus_{[g]}\Qc{g}{1}$ and every flux ends once, with trivial charge.

\begin{examplebox}[The Lagrangian algebras of $\cZ(\Vec_{S_3})$][label=ex:S3-Lagrangians]
Take $H=\Z_2=\{\mathrm{id},b\}$, $\psi=1$, and representatives $g_i=a^i$ of the cosets $i\in\{0,1,2\}$. Since $a\cdot a^i=a^{i+1}$ and $b\cdot a^i=a^{-i}b$, the cosets are permuted as $a\cdot i=i+1$ and $b\cdot i=-i$, with little-group elements $l_i(a)=\mathrm{id}$ and $l_i(b)=b$.

\newl {\it Basis and grading.} Since $\Z_2$ is abelian, the identification \eqref{eq:Lagrangian-induced} reads $[gh,e_x]=[g,e_x]$; for instance $[ab,e_b]=[a,e_b]$. The basis consists of
\begin{equation*}
p_i:=[a^i,e_{\mathrm{id}}]\ \ (\deg\ \mathrm{id})\,,\qquad f_i:=[a^i,e_b]\ \ (\deg\ x_i)\,,\qquad x_0=b\,,\ x_1=a^2b\,,\ x_2=ab\,,
\end{equation*}
where $x_i=a^iba^{-i}$ is the transposition fixing the coset $i$.

\newl {\it Half-braiding.} Conjugation by $e_b$ is trivial, so $u_kp_i=p_{k\cdot i}$ and $u_kf_i=f_{k\cdot i}$; for instance $u_a$ maps $f_0$ (flux $b$) to $f_1$ (flux $aba^{-1}=a^2b$).

\newl {\it Product and blocks.} $p_ip_j=\delta_{ij}p_i$, $p_if_j=f_jp_i=\delta_{ij}f_i$ and $f_if_j=\delta_{ij}p_i$, with unit $\eta=p_0+p_1+p_2$. So $\cL_{\Z_2}=B_0\oplus B_1\oplus B_2$ with $B_i=\mathrm{span}\{p_i,f_i\}\simeq\bbC[H_i]$, i.e.\ $\cL_{\Z_2}\simeq\bbC[\Z_2]^{\oplus3}$ as an ordinary algebra.

\newl {\it Decomposition.} By \eqref{eq:CH-condensation}, one flux class at a time:
\begin{itemize}
\item $\mathrm{id}$ fixes all three cosets, and $C_{S_3}(\mathrm{id})=S_3$ permutes them: $\mathrm{span}\{p_0,p_1,p_2\}$ is the permutation representation $1\oplus E$, spanned by $p_0+p_1+p_2$ and by $\sum_i\omega^{\mp i}p_i$;
\item $a$ fixes no coset, since $a\cdot i=i+1$: no line $\Qc{a}{\omega^k}$ condenses;
\item $b$ fixes only the coset $0$, and $C_{S_3}(b)=\{\mathrm{id},b\}$ acts on $f_0$ trivially, $u_bf_0=f_0$: one copy of $\Qc{b}{+}$, whose fibers at $a^2b$ and $ab$ are $f_1$ and $f_2$.
\end{itemize}
The other three Lagrangian algebras are computed in the same way. Altogether,
\begin{equation}
\begin{aligned}
\cL_{\{\mathrm{id}\}}&=\Qc{\mathrm{id}}{1}\oplus\Qc{\mathrm{id}}{P}\oplus2\,\Qc{\mathrm{id}}{E}&&\simeq\Fun(S_3)\,,\\
\cL_{\Z_2}&=\Qc{\mathrm{id}}{1}\oplus\Qc{\mathrm{id}}{E}\oplus\Qc{b}{+}&&\simeq\bbC[\Z_2]^{\oplus3}\,,\\
\cL_{\Z_3}&=\Qc{\mathrm{id}}{1}\oplus\Qc{\mathrm{id}}{P}\oplus2\,\Qc{a}{1}&&\simeq\bbC[\Z_3]^{\oplus2}\,,\\
\cL_{S_3}&=\Qc{\mathrm{id}}{1}\oplus\Qc{a}{1}\oplus\Qc{b}{+}&&\simeq\bbC[S_3]\,,
\end{aligned}
\label{eq:S3-Lagrangians}
\end{equation}
where the right-hand sides are the underlying ordinary algebras. Each has dimension $6=|S_3|$, and all summands are bosons. For instance, the multiplicity $2$ in $\cL_{\Z_3}$ arises because $a$ fixes both cosets of $\Z_3$ and its centralizer $\Z_3$ acts trivially on them.

\newl The 2-categorical counterpart of $\cL_{\Z_2}$ is the Lagrangian algebra $\cL(A)$ of the boundary $\fB_{(\bbZ_2,\Ising)}$, computed in the same steps in example~\ref{ex:S3-2lag-algebras}. The idempotents $p_i$ become three copies of $\cZ(\Vect_{\bbZ_2})$, and the elements $f_i$ become the twist defects $\sigma_\pm^{(i)}$.
\end{examplebox}

\begin{examplebox}[The Lagrangian algebra $\cL_{\Z_3}$][label=ex:S3-LZ3]
The single-vacuum sandwich of example~\ref{ex:S3-RG}(a) pairs $\fB_{\Z_2}$ with $\fB_{\Z_3}$, so we spell out $\cL_{\Z_3}$ in the same steps as $\cL_{\Z_2}$ in example~\ref{ex:S3-Lagrangians}. Its 2-categorical counterpart is the Lagrangian algebra $\cL(B)$ of the physical boundary of the running example of section~\ref{sec:gen-fib-fun-symtft}, described in example~\ref{ex:S3-2lag-LB}.

\newl {\it Cosets.} $K=\Z_3=\{\mathrm{id},a,a^2\}$ is normal, with cosets $K$ and $bK$, labeled $c\in\{0,1\}$, with representatives $g_0=\mathrm{id}$ and $g_1=b$. Since $ab=ba^2$, the element $a$ fixes both cosets, with little-group elements $l_0(a)=a$ and $l_1(a)=a^2$, while $b$ exchanges them, $b\cdot c=1-c$, with $l_c(b)=\mathrm{id}$.

\newl {\it Basis and grading.} By \eqref{eq:Lagrangian-induced} the basis consists of the six elements
\begin{equation*}
q_c^k:=[g_c,e_{a^k}]\,,\qquad \deg q_0^k=a^k\,,\qquad \deg q_1^k=ba^kb^{-1}=a^{-k}\,,\qquad k\in\Z_3\,.
\end{equation*}

\newl {\it Half-braiding.} Conjugation by $e_{a^l}$ is trivial, since $K$ is abelian, so $u_aq_c^k=q_c^k$ and $u_bq_c^k=q_{1-c}^k$; for instance $u_b$ maps $q_0^1$ (flux $a$) to $q_1^1$ (flux $a^2$).

\newl {\it Product and blocks.} $q_c^kq_{c'}^{k'}=\delta_{cc'}\,q_c^{k+k'}$, with unit $\eta=q_0^0+q_1^0$. So $\cL_{\Z_3}=B_0\oplus B_1$ with $B_c=\mathrm{span}\{q_c^k\}_{k\in\Z_3}\simeq\bbC[\Z_3]$, one block for each coset.

\newl {\it Decomposition.} By \eqref{eq:CH-condensation}, one flux class at a time:
\begin{itemize}
\item $\mathrm{id}$ fixes both cosets, which $S_3$ permutes: $\mathrm{span}\{q_0^0,q_1^0\}$ is the permutation representation $1\oplus P$, spanned by $q_0^0+q_1^0$ and $q_0^0-q_1^0$;
\item $a$ fixes both cosets: the flux-$a$ elements are $q_0^1$ and $q_1^2$, on which $C_{S_3}(a)=\Z_3$ acts trivially, giving two copies of $\Qc{a}{1}$. They are labeled by the two elements $a,a^2\in K$ through which the flux $a$ is seen from the two cosets;
\item $b$ fixes no coset: no line $\Qc{b}{\pm}$ condenses.
\end{itemize}
This reproduces $\cL_{\Z_3}=\Qc{\mathrm{id}}{1}\oplus\Qc{\mathrm{id}}{P}\oplus2\,\Qc{a}{1}$ in \eqref{eq:S3-Lagrangians}. The two copies of $\Qc{a}{1}$ will reappear as the two twisted-sector operators of $E$ in example~\ref{ex:S3-lag-rg-Z2Z3}.

\newl {\it No discrete torsion.} The boundaries with $K=\Z_3$ are labeled by $\psi\in \operatorname{H}^2(\Z_3;U(1))=0$, so $\cL_{\Z_3}$ is the only Lagrangian algebra of this type. One dimension up, they are labeled by $\omega\in \operatorname{H}^3(\Z_3;U(1))\simeq\Z_3$, and a non-trivial $\omega$ twists the product of $\cL(B)$ on its flux-$[a]$ part (example~\ref{ex:S3-2lag-LB}). This twist has no 1-categorical counterpart.
\end{examplebox}
\subsubsection{The UV symmetry \texorpdfstring{$\CUV$}{CUV}}\label{subsec:warm-up-lag-uv}
A line on the boundary $\fB_{(H,\psi)}$ is a bulk line on which the condensate can end from one side, i.e.\ a left $\cL_{(H,\psi)}$-module in $\cZ(\Vec_G)$. The symmetry of the boundary is therefore
\begin{equation}
\CUV\simeq{}_{\cL_{(H,\psi)}}\cZ(\Vec_G)\,,
\label{eq:warm-up-lag-uv}
\end{equation}
with tensor product $\otimes_{\cL}$ and unit $\cL$ \cite{Davydov:2010wpa}, the counterpart of the 2-category of left $\cL(A)$-modules in section~\ref{subsec:lag-uv}, \eqref{eq:lag-CUV}.

\newl {\bf Free modules.} The simplest boundary lines are obtained by bending a bulk line $Q$ onto the boundary, which gives the free module
\begin{equation}
\Phi_H:\ \cZ(\Vec_G)\longrightarrow{}_{\cL_H}\cZ(\Vec_G)\,,\qquad Q\longmapsto\cL_H\otimes Q\,,
\label{eq:warm-up-free-module}
\end{equation}
the bulk-to-boundary map. It is monoidal, and a module map out of $\cL_H\otimes Q$ is determined by its restriction to $Q$, so that
\begin{equation}
\Hom_{{}_{\cL_H} \cZ(\Vec_G)}\big(\Phi_H(Q),V\big)\simeq\Hom_{\cZ(\Vec_G)}(Q,V)
\label{eq:warm-up-reciprocity}
\end{equation}
for every module $V$. Two consequences make $\CUV$ computable.
\begin{itemize}
\item \emph{Endpoints.} Taking $V=\cL_H$, the junctions at which $Q$ ends on the boundary form $\Hom_{\cL_H}(\Phi_H(Q),\cL_H)\simeq\Hom(Q,\cL_H)$, which recovers the multiplicities $n_H(Q)$ of \eqref{eq:CH-condensation}:
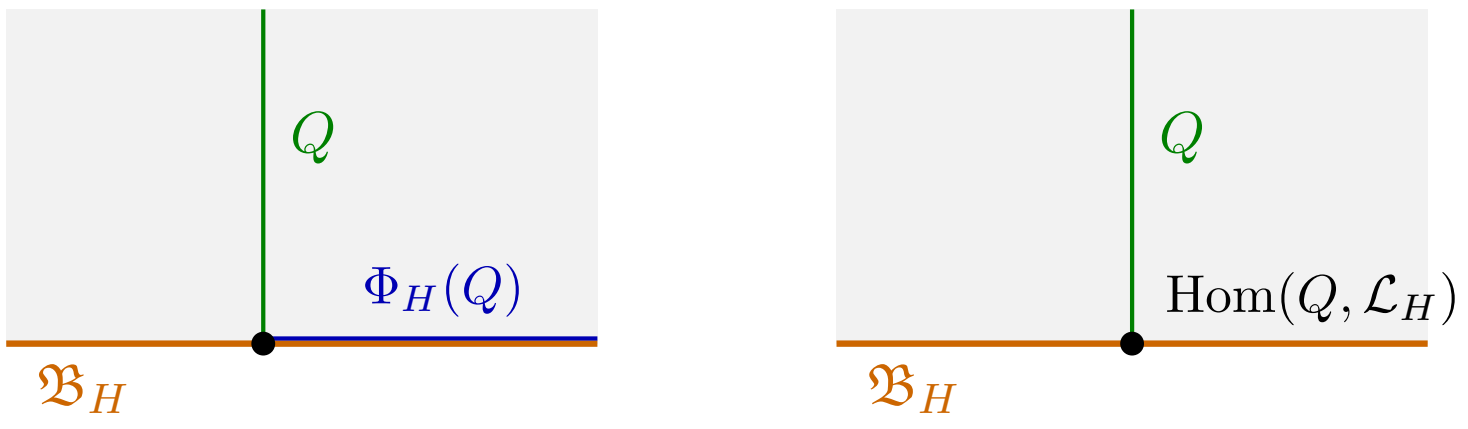
\begin{eqs}
\begin{tikzpicture}[scale=0.85,baseline=(c)]
\coordinate (c) at (0,1.0);
\fill[gray!10] (0,0) rectangle (4.6,2.6);
\draw[very thick,orange!80!black] (0,0) -- (4.6,0); \node[below,orange!80!black] at (0.6,0) {$\fB_H$};
\draw[thick,green!50!black] (2.0,2.6) -- (2.0,0);
\draw[thick,blue!70!black] (2.0,0.04) -- (4.6,0.04);
\node[right,green!50!black] at (2.05,1.6) {$Q$};
\filldraw (2.0,0) circle (2.4pt);
\node[above,blue!70!black] at (3.4,0.05) {\small$\Phi_H(Q)$};
\end{tikzpicture}
\qquad\qquad
\begin{tikzpicture}[scale=0.85,baseline=(c)]
\coordinate (c) at (0,1.0);
\fill[gray!10] (0,0) rectangle (4.6,2.6);
\draw[very thick,orange!80!black] (0,0) -- (4.6,0); \node[below,orange!80!black] at (0.6,0) {$\fB_H$};
\draw[thick,green!50!black] (2.3,2.6) -- (2.3,0); \node[right,green!50!black] at (2.35,1.6) {$Q$};
\filldraw (2.3,0) circle (2.4pt);
\node[right] at (2.4,0.35) {\small$\Hom(Q,\cL_H)$};
\end{tikzpicture}
\label{eq:CH-bulk-boundary}
\end{eqs}
\item \emph{Simplicity.} Taking $V=\Phi_H(Q')$ gives $\Hom_{\cL_H}(\Phi_H(Q),\Phi_H(Q'))\simeq\Hom(Q\otimes\bar Q',\cL_H)$. In particular, $\Phi_H(Q)$ is simple if and only if the only summand of $\cL_H$ contained in $Q\otimes\bar Q$ is the unit, and its dimension as a boundary line is $\dim Q$.
\end{itemize}
The multiplicity of a simple boundary line $X$ in $\Phi_H(Q)$ is then the multiplicity of $Q$ in $X$, viewed as a bulk line. In the module-category description of section~\ref{subsec:warm-up-uv}, $\Phi_H(Q)$ is the matrix with entries $\Phi_H(Q)_{ji}=\bigoplus_{x\cdot i=j}V_x$, acting on the interfaces through the flux of $Q$; on Wilson lines it is the restriction functor $\Rep(G)\to\Rep(H)$.

\begin{examplebox}[The symmetry of $\fB_{\Z_2}$ from $\cL_{\Z_2}$][label=ex:S3-lag-uv]
Let $\cL=\cL_{\Z_2}=1\oplus E\oplus\Qc{b}{+}$, where we abbreviate $\Qc{\mathrm{id}}{R}=R$. The global dimension of $\CUV$ is $\dim\cZ(\Vec_{S_3})/\dim\cL=36/6=6$.
\begin{itemize}
\item \emph{Simple lines.} $P\otimes P=1$ and $\Qc{a}{1}\otimes\Qc{a}{1}=1\oplus P\oplus\Qc{a}{1}$ each meet $\cL$ only in $1$, so $P:=\cL\otimes P$ and $E:=\cL\otimes\Qc{a}{1}$ are simple, of dimensions $1$ and $2$. Neither is the unit, since $P,\Qc{a}{1}\not\subset\cL$. As $1^2+1^2+2^2=6$, the simple lines are $1,P,E$.
\item \emph{Fusion.} Since $\Phi_{\Z_2}$ is monoidal, $P\otimes P=\Phi(P\otimes P)=1$, $P\otimes E=\Phi(P\otimes\Qc{a}{1})=E$ and $E\otimes E=\Phi(1\oplus P\oplus\Qc{a}{1})=1\oplus P\oplus E$.
\item \emph{Associator.} The lines $1,P,\Qc{a}{1}$ span a fusion subcategory $D\subset\cZ(\Vec_{S_3})$, on which $\Phi_{\Z_2}$ is fully faithful, since $\Hom(Q\otimes Q',\cL)=\delta_{QQ'}\bbC$ for $Q,Q'\in D$. It is therefore an equivalence $D\simeq\CUV$. All lines of $D$ are bosons with trivial mutual braiding, so $D$ is Tannakian, and hence $\CUV\simeq D\simeq\Rep(S_3)$.
\item \emph{Bulk-to-boundary map.} As bulk lines, $1=1\oplus E\oplus\Qc{b}{+}$, $P=P\oplus E\oplus\Qc{b}{-}$ and $E=\Qc{a}{1}\oplus\Qc{a}{\omega}\oplus\Qc{a}{\omega^2}\oplus\Qc{b}{+}\oplus\Qc{b}{-}$. Reading off the multiplicities,
\begin{equation*}
\begin{gathered}
\Phi_{\Z_2}:\quad P\mapsto P\,,\quad E\mapsto1\oplus P\,,\quad \Qc{a}{\omega^k}\mapsto E\,,\\
\Qc{b}{+}\mapsto1\oplus E\,,\quad \Qc{b}{-}\mapsto P\oplus E\,.
\end{gathered}
\end{equation*}
\end{itemize}
The other boundaries are treated in the same way, giving $\cC_{\{\mathrm{id}\}}\simeq\cC_{\Z_3}\simeq\Vec_{S_3}$ and $\cC_{\Z_2}\simeq\cC_{S_3}\simeq\Rep(S_3)$. The coincidences are explained by the braided autoequivalence
\begin{equation}
\varphi:\ \Qc{\mathrm{id}}{E}\ \leftrightarrow\ \Qc{a}{1}\,,\qquad\text{all other lines fixed}\,,
\label{eq:S3-duality}
\end{equation}
which exchanges $\cL_{\{\mathrm{id}\}}\leftrightarrow\cL_{\Z_3}$ and $\cL_{\Z_2}\leftrightarrow\cL_{S_3}$; indeed $D=\varphi(\Rep(S_3))$.

\newl Example~\ref{ex:S3-2lag-uv} repeats the first two steps one dimension higher: free modules over $\cL(A)$ give the components of $\CUV$, and the lines out of the unit give $\Omega\CUV\simeq\cZ(\Vect_{\bbZ_2})^{\bbZ_2}$, the counterpart of the Wilson lines $1,P$ and of the line $E$.
\end{examplebox}

\begin{examplebox}[The symmetry of $\fB_{\Z_2}$ as bimodules in $\Vec_{S_3}$][label=ex:S3-lag-uv-bimod]
The Lagrangian algebra $\cL_{\Z_2}$ is the full center of the boundary algebra $A:=\bbC[\Z_2]=\bbC e_{\mathrm{id}}\oplus\bbC e_b$ of section~\ref{subsec:warm-up-module}, viewed as an algebra in $\Vec_{S_3}$ supported on $\Z_2$. Left $\cL_{\Z_2}$-modules in $\cZ(\Vec_{S_3})$ are equivalent to $A$-bimodules in $\Vec_{S_3}$ \cite{Kong:2009inh,davydov2010centre}, which gives a second, more explicit, description of the symmetry, anticipated in section~\ref{subsec:warm-up-uv}, where $\cC_H$ was identified with the $\bbC[H]$--$\bbC[H]$ bimodules in $\Vec_G$:
\begin{equation}
\CUV\simeq{}_{\cL_{\Z_2}}\cZ(\Vec_{S_3})\simeq{}_A(\Vec_{S_3})_A\,.
\label{eq:S3-lag-uv-bimod}
\end{equation}
\begin{itemize}
\item \emph{Objects.} A line is an $S_3$-graded vector space $V=\bigoplus_gV_g$ with commuting left and right $A$-actions compatible with the grading, $e_bV_g\subseteq V_{bg}$ and $V_ge_b\subseteq V_{gb}$. Fusion is $\otimes_A$, with unit $A$.
\item \emph{Double cosets.} The actions only move the degree within a double coset, so every bimodule splits over $\Z_2\backslash S_3/\Z_2$, as in example~\ref{CH on S3}, and on each double coset it is determined by its component at a representative.
\item \emph{Trivial double coset.} Here $V=V_{\mathrm{id}}\oplus e_bV_{\mathrm{id}}$, and $V_{\mathrm{id}}$ carries the action $v\mapsto e_bve_b$ of $\Z_2$. The simple objects are $1=A$, on which this action is trivial, and $P$, on which it is $-1$. The free bimodule $A\otimes A$ has $(A\otimes A)_{\mathrm{id}}=\mathrm{span}\{e_{\mathrm{id}}\otimes e_{\mathrm{id}},e_b\otimes e_b\}$, whose two basis vectors are exchanged, so $A\otimes A\simeq1\oplus P$.
\item \emph{Double coset $\Z_2a\Z_2=\{a,a^2,ab,a^2b\}$.} The stabilizer $\Z_2\cap a\Z_2a^{-1}$ is trivial, so $V\simeq A\otimes V_a\otimes A$ is freely generated by its component $V_a$, and the simple object is $E:=A\otimes\bbC_a\otimes A$, where $\bbC_a$ is $\bbC$ in degree $a$. It has components $\bbC$ in each of the four degrees, and dimension $4/\dim A=2$.
\item \emph{Fusion.} Splitting the middle $A=\bbC e_{\mathrm{id}}\oplus\bbC e_b$,
\begin{equation}
\begin{split}
E\otimes_AE&=A\otimes\bbC_a\otimes A\otimes\bbC_a\otimes A\\
&\simeq\big(A\otimes\bbC_{a^2}\otimes A\big)\oplus\big(A\otimes\bbC_{aba}\otimes A\big)\simeq E\oplus1\oplus P\,,
\end{split}
\label{eq:S3-EE-bimod}
\end{equation}
since $a^2=bab$, so that the two factors $e_b$ are absorbed into the outer copies of $A$, and $aba=b\in\Z_2$, so that $A\otimes\bbC_b\otimes A\simeq A\otimes A$. This is a third derivation of $E\otimes E=1\oplus P\oplus E$, after examples~\ref{CH on S3} and~\ref{ex:S3-lag-uv}.
\end{itemize}
The 2-categorical counterpart is example~\ref{ex:S3-2lag-uv}, with $\bbC$ replaced by $\Vec$, $A=\bbC[\Z_2]$ by $\Ising$ and $\bbC_a$ by $(\Vec_{\Z_2})_a$. There the same manipulation gives $X\otimes X\simeq\chi\oplus X$. The one qualitative difference is that the counterpart $\chi$ of $A\otimes A\simeq1\oplus P$ is a single simple surface: the condensation of the Wilson line $P$ is one surface, while in two dimensions $1$ and $P$ remain distinct lines.
\end{examplebox}

\subsubsection{The IR symmetry \texorpdfstring{$\CIR$}{CIR}}\label{subsec:warm-up-lag-ir}
In the sandwich with symmetry boundary $\fB_H$ and physical boundary $\fB_K$, a line of the IR theory is a bulk line in the slab on which the condensate of $\fB_H$ can end from one side and that of $\fB_K$ from the other. As a category,
\begin{equation}
\CIR\simeq{}_{\cL_H}\cZ(\Vec_G)_{\cL_K}\,,
\label{eq:warm-up-lag-ir}
\end{equation}
the $(\cL_H,\cL_K)$-bimodules in $\cZ(\Vec_G)$, with fusion given by the fusion of parallel lines in the slab \cite{Kong:2013aya}; this is the counterpart of ${}_{\cL(A)}\fZ_{\cL(B)}$ in \eqref{eq:lag-CIR-bimod} from section~\ref{subsec:lag-ir}. The IR identity line is the free bimodule $\cL_H\otimes\cL_K$, and, as in \eqref{eq:warm-up-reciprocity}, its endomorphisms are
\begin{equation}
\End_{\CIR}(\cL_H\otimes\cL_K)\simeq\Hom_{\cZ(\Vec_G)}\big(1,\cL_H\otimes\cL_K\big)\simeq\bigoplus_Q\Hom(Q,\cL_H)\otimes\Hom(\bar Q,\cL_K)\,.
\label{eq:warm-up-local-ops}
\end{equation}
An IR local operator is thus a bulk line stretched across the slab, with one endpoint on each boundary. Since the local operators of a 2d TQFT with $n$ vacua form $\bbC^n$, and Lagrangian algebras are self-dual,
\begin{equation}
n(H,K)=\sum_Q n_H(Q)\,n_K(Q)=\dim\Hom_{\cZ(\Vec_G)}\big(\cL_H,\cL_K\big)\,,
\label{eq:warm-up-vacua-lag}
\end{equation}
which reproduces \eqref{eq:S3-vacua}.

\newl {\bf Order parameters.} The decomposition \eqref{eq:warm-up-local-ops} organizes the IR local operators by their generalized charges \cite{Bhardwaj:2023idu}. A line $Q$ common to $\cL_H$ and $\cL_K$ ends on $\fB_K$ in $n_K(Q)$ ways. On $\fB_H$ its endpoint can sit at the end of any boundary line $X\subseteq\Phi_H(Q)$: for $X=1$ this gives a genuine local operator, and otherwise an operator in the twisted sector of $X$. The lines in $\cL_K$ are therefore the generalized charges of the order parameters of the phase, and the restriction $\Phi_H(Q)$ tells whether an order parameter is a genuine local operator, a string operator, or comes in a multiplet containing both.

\begin{examplebox}[IR theories for the symmetry boundary $\fB_{\Z_2}$][label=ex:S3-lag-ir]
Here $\cL_{\Z_2}=1\oplus E\oplus\Qc{b}{+}$ and $\CUV\simeq\Rep(S_3)$.
\begin{itemize}
\item \emph{$K=S_3$}, with $\cL_{S_3}=1\oplus\Qc{a}{1}\oplus\Qc{b}{+}$. The common lines are $1$ and $\Qc{b}{+}$, so by \eqref{eq:warm-up-vacua-lag} there are two vacua $v_\pm$, and $\CIR\simeq\mathrm{Mat}_2(\Vec)$. The line $\Qc{b}{+}$ stretched across the slab is the order parameter: the holonomy $b$ measured from one boundary to the other, $O=v_+-v_-$. Since $\Phi_{\Z_2}(\Qc{b}{+})=1\oplus E$, it comes in a multiplet with an operator in the $E$-twisted sector. Under $P$, which links the flux $b$ with $\chi_P(b)=-1$, it is odd, $O\mapsto-O$: this is the $\Z_2$ SSB phase of $\Rep(S_3)$.
\item \emph{$K=\Z_3$}, with $\cL_{\Z_3}=1\oplus P\oplus2\,\Qc{a}{1}$. The only common line is $1$, so there is a single vacuum, $\CIR\simeq\Vec$, and no order parameter: the phase is trivially gapped. This is the case carried over to $(2\!+\!1)$ dimensions in examples~\ref{ex:S3-2mod-ir} and~\ref{ex:S3-2lag-ir}.
\end{itemize}
\end{examplebox}

\begin{examplebox}[The single-vacuum sandwich $(\fB_{\Z_2},\fB_{\Z_3})$ without decomposing][label=ex:S3-lag-ir-Z2Z3]
We can also reproduce example~\ref{ex:S3-lag-ir} without decomposing the Lagrangian algebras into simple lines. This is the form of the computation that carries over to one dimension higher, in example~\ref{ex:S3-2lag-ir}.
\begin{itemize}
\item \emph{Local operators as invariants.} A morphism $1\to V$ in $\cZ(\Vec_{S_3})$ picks a vector of flux $\mathrm{id}$ that is invariant under the half-braiding, so $\Hom(1,V)=(V_{\mathrm{id}})^{S_3}$. By \eqref{eq:S3-center-fusion}, the flux-$\mathrm{id}$ part of a product is $(V\otimes W)_{\mathrm{id}}=\bigoplus_xV_x\otimes W_{x^{-1}}$. The local operators \eqref{eq:warm-up-local-ops} are therefore
\begin{equation}
\End_{\CIR}(\cL_{\Z_2} \otimes \cL_{\Z_3})\simeq\Big(\bigoplus_{x\in S_3}(\cL_{\Z_2})_x\otimes(\cL_{\Z_3})_{x^{-1}}\Big)^{S_3}\,.
\label{eq:S3-local-ops-invariants}
\end{equation}
\item \emph{Supports.} $\cL_{\Z_2}$ is supported on $\{\mathrm{id}\}\cup[b]$ (example~\ref{ex:S3-Lagrangians}) and $\cL_{\Z_3}$ on $\{\mathrm{id}\}\cup[a]$ (example~\ref{ex:S3-LZ3}). The supports meet only at the identity, which is the statement that $1$ is the only line common to both algebras. Only $x=\mathrm{id}$ contributes:
\begin{equation*}
(\cL_{\Z_2})_{\mathrm{id}}\otimes(\cL_{\Z_3})_{\mathrm{id}}=\mathrm{span}\{p_i\otimes q_c^0\}\simeq\Fun\big(S_3/\Z_2\times S_3/\Z_3\big)\,.
\end{equation*}
\item \emph{Orbits.} $S_3$ permutes the six pairs $(i,c)$ with stabilizer $\Z_2\cap\Z_3=\{\mathrm{id}\}$, i.e.\ freely and transitively. The invariants are one-dimensional, spanned by $\sum_{i,c}p_i\otimes q_c^0$, so there is a single vacuum. This is the orbit count $\Z_2\backslash S_3/\Z_3=\{S_3\}$ of \eqref{eq:S3-combined} once more.
\item \emph{Bimodules.} Equivalently, the boundary conditions of the IR theory are the $(\bbC[\Z_2],\bbC[\Z_3])$-bimodules in $\Vec_{S_3}$. Every $g\in S_3$ factorizes uniquely as $g=hk$ with $h\in\Z_2$ and $k\in\Z_3$, so such a bimodule is $V\simeq\bbC[\Z_2]\otimes V_{\mathrm{id}}\otimes\bbC[\Z_3]$, with $V_{\mathrm{id}}$ an arbitrary vector space. The category of boundary conditions is $\Vec$, and $\CIR\simeq\Vec$.
\end{itemize}
In example~\ref{ex:S3-2lag-ir} of section~\ref{sec:gen-fib-fun-symtft} the same steps are applied, with $\cL_{\Z_2}$ and $\cL_{\Z_3}$ replaced by the Lagrangian algebras $\cL(A)$ and $\cL(B)$ in $\cZ_1(2\Vect_{S_3})$ of the boundaries $\fB_{(\bbZ_2,\Ising)}$ and $\fB_{(\bbZ_3,B)}$. The flux-$\mathrm{id}$ part of $\cL(A)\otimes\cL(B)$ consists of six copies of $\cZ(\Vec_{\Z_2})$ instead of six copies of $\bbC$, the free and transitive $S_3$-action again gives a single vacuum, and the invariants are now the IR lines $\cZ(\Vec_{\Z_2})$ rather than the scalars. Likewise the category of boundary conditions becomes $\mathsf{LMod}(\Vec_{\Z_2})$ instead of $\Vec$.
\end{examplebox}

\subsubsection{The RG functor \texorpdfstring{$\cF:\CUV\to\CIR$}{F : CUV -> CIR}}\label{subsec:warm-up-lag-rg}
A line of the symmetry boundary is brought through the slab and the condensate of the physical boundary is allowed to end on it. The RG functor is therefore the free module functor
\begin{equation}
\cF:\ {}_{\cL_H}\cZ(\Vec_G)\longrightarrow{}_{\cL_H}\cZ(\Vec_G)_{\cL_K}\,,\qquad X\longmapsto X\otimes\cL_K\,,
\label{eq:warm-up-free}
\end{equation}
with the left $\cL_H$-action of $X$ and the regular right $\cL_K$-action, the counterpart of the free module functor of section~\ref{subsec:lagrangian-algebras}. In particular, $\cF(\cL_H)=\cL_H\otimes\cL_K$ is the IR identity line.

\newl {\bf Twisted sectors.} An operator in the twisted sector of a UV line $X$ lives at the end of $X$ on the symmetry boundary, and therefore in the IR at the end of $\cF(X)=\bigoplus_{v,v'}\cF(X)_{v'v}\otimes1_{v'v}$. Only the diagonal components $1_{vv}$ can end, on the vacuum $v$, so twisted-sector operators of $X$ are supported on the vacua that $X$ maps to themselves, and their number is $\Tr\cF(X):=\sum_v\dim\cF(X)_{vv}$. Counting the same operators in the SymTFT, a line $Q$ ends on $\fB_K$ in $n_K(Q)$ ways and at the end of $X$ on $\fB_H$ in $[\Phi_H(Q):X]$ ways, so that
\begin{equation}
\operatorname{tr}\cF(X)=\sum_Qn_K(Q)\,\big[\Phi_H(Q):X\big]=\big[\Phi_H(\cL_K):X\big]\,.
\label{eq:warm-up-trace}
\end{equation}
For $X=1$ this is the number of vacua. The traces $\Tr\cF(X)$ are the characters of the action of $\CUV$ on the vacua, and are determined by the Lagrangian algebras alone.

\begin{examplebox}[The RG functor for the symmetry boundary $\fB_{\Z_2}$][label=ex:S3-lag-rg]
Using the bulk-to-boundary map of example~\ref{ex:S3-lag-uv}:
\begin{itemize}
\item \emph{$K=S_3$}: $\Phi_{\Z_2}(\cL_{S_3})=\Phi(1)\oplus\Phi(\Qc{a}{1})\oplus\Phi(\Qc{b}{+})=2\cdot1\oplus2E$, so $\big(\Tr\cF(1),\Tr\cF(P),\Tr\cF(E)\big)=(2,0,2)$. This matches $\cF(P)=\left(\begin{smallmatrix}0&1\\1&0\end{smallmatrix}\right)$ and $\cF(E)=\left(\begin{smallmatrix}1&1\\1&1\end{smallmatrix}\right)$ of example~\ref{ex:S3-RG}: $P$ fixes no vacuum and has no twisted-sector operators, while the two $E$-twisted operators come from $\Qc{a}{1}$ (a string operator, since $\Phi_{\Z_2}(\Qc{a}{1})=E$) and from the multiplet of the order parameter $\Qc{b}{+}$.
\item \emph{$K=\Z_3$}: $\Phi_{\Z_2}(\cL_{\Z_3})=\Phi(1)\oplus\Phi(P)\oplus2\,\Phi(\Qc{a}{1})=1\oplus P\oplus2E$, so the traces are $(1,1,2)$. With a single vacuum, this is $\cF(P)=1$ and $\cF(E)=2$: every line acts by its quantum dimension, and $\cF$ is the fiber functor of $\Rep(S_3)$. Its 2-categorical counterpart is the generalized fiber 2-functor of examples~\ref{ex:S3-2mod-rg} and~\ref{ex:S3-2lag-rg}.
\end{itemize}
\end{examplebox}

\begin{examplebox}[The fiber functor of $\Rep(S_3)$ from bimodules][label=ex:S3-lag-rg-Z2Z3]
In the bimodule description of examples~\ref{ex:S3-lag-uv-bimod} and~\ref{ex:S3-lag-ir-Z2Z3}, with $A=\bbC[\Z_2]$ and $B=\bbC[\Z_3]$, a line $Y\in\CUV$ acts on the boundary conditions of the IR theory by $M\mapsto Y\otimes_AM$, the bimodule form of \eqref{eq:S3-RG}. Since $M\simeq A\otimes M_{\mathrm{id}}\otimes B$, it suffices to read off the flux-$\mathrm{id}$ component of $Y\otimes_AM$.
\begin{itemize}
\item \emph{The line $E$.} $E\otimes_AM\simeq A\otimes\bbC_a\otimes A\otimes M_{\mathrm{id}}\otimes B$, whose degree-$\mathrm{id}$ part collects the triples $(h,h',k)\in\Z_2\times\Z_2\times\Z_3$ with $h\,a\,h'\,k=\mathrm{id}$. There are exactly two, $(\mathrm{id},\mathrm{id},a^2)$ and $(b,b,a)$, each contributing one copy of $M_{\mathrm{id}}$:
\begin{equation}
(E\otimes_AM)_{\mathrm{id}}\simeq M_{\mathrm{id}}\oplus M_{\mathrm{id}}\,,\qquad\text{i.e.}\qquad\cF(E)=2\,.
\label{eq:S3-FE-bimod}
\end{equation}
\item \emph{The lines $1$ and $P$.} $(P\otimes_AM)_{\mathrm{id}}\simeq M_{\mathrm{id}}$, so $\cF(P)=\cF(1)=1$.
\end{itemize}
This is the fiber functor of example~\ref{ex:S3-RG}(a), now obtained by reading off a single component, the bimodule version of the column read-off in step 7 of section~\ref{subsec:warm-up-uv}.

\newl {\it Twisted sector.} The two summands of $\cF(E)$ are labeled by the elements $k=a^2,a$ of $K$ through which the flux carried by $E$ is absorbed by the physical boundary. Correspondingly, the two $E$-twisted operators counted by \eqref{eq:warm-up-trace}, $\Tr\cF(E)=2$, are the two copies of $\Qc{a}{1}$ in $\cL_{\Z_3}$ (example~\ref{ex:S3-LZ3}), stretched across the slab and ending on $E$, since $\Phi_{\Z_2}(\Qc{a}{1})=E$ (example~\ref{ex:S3-lag-uv}).

\newl {\it Uniqueness.} Fiber functors of this type with $K=\Z_3$ are labeled by $\psi\in \operatorname{H}^2(\Z_3;U(1))=0$ \cite{Ostrikmodule}, so this one is unique.

\newl One dimension up, example~\ref{ex:S3-2lag-rg} repeats this computation. There $X\mapsto1\oplus1$, the two summands are again labeled by $a^2$ and $a$, and the twisted sector of $X$ again comes from the two copies of the flux-$[a]$ surface in $\cL(B)$. The new feature is that these labels now carry the 3-cocycle $\omega\in \operatorname{H}^3(\Z_3;U(1))\simeq\Z_3$, which has no counterpart here.
\end{examplebox}

\section{Generalized Fiber 2-Functors From The SymTFT}\label{sec:gen-fib-fun-symtft}
In section~\ref{sec:intro}, we introduced the notion of a generalized fiber 2-functor, motivated by the question:
\begin{center}
\it
    Given a UV fusion 2-categorical symmetry, can one realize a gapped phase in which it acts on an IR $(2\!+\!1)$-dimensional TQFT with a unique vacuum state?
\end{center}
In section~\ref{sec:sym-ext} we then supplemented that discussion with an explicit example constructed using symmetry extensions, and pointed out how that same construction extends to \emph{any} invertible symmetry in $(2\!+\!1)$ dimensions; every fusion 2-category of the form $\CUV  = 2\Vect_G^\pi$ admits a generalized fiber 2-functor
\begin{equation}
    \cF \, : \, 2\Vect_G^\pi \to \Sigma\cB\, ,
\end{equation}
with $\cB$ the braided fusion category of line defects of some Dijkgraaf--Witten TQFT determined by symmetry extension of $(G, \pi)$.

\newl In this section, which constitutes the main contribution of this work, we bring the two previous threads together: the SymTFT picture of boundaries and RG 2-functors from section~\ref{sec:symtft}, and the explicit constructions of generalized fiber 2-functors from section~\ref{sec:sym-ext}.
%
We seek to demonstrate that the above result extends to all (AB)-type symmetries in $(2\!+\!1)$ dimensions, including non-invertible ones, by leveraging the fact that these symmetries will nonetheless always be related to some invertible symmetry by a (generalized) gauging operation \cite{Decoppet:2024htz, Bullimore:2024khm, Bhardwaj:2024qiv} (see also theorem \ref{thm:AB-center} in the appendix).

\newl We will then present two constructions for building RG 2-functors, and in particular, generalized fiber 2-functors. The first is a method using module 2-categories, which is applicable whenever $\CUV$ is Morita equivalent to an invertible fusion 2-category. The second uses Lagrangian algebras and is more general, being applicable to any (AB)-type fusion 2-category; we work it out in detail in the same Morita class, where it can be compared with the first, and explain how it extends to all (AB)-type fusion 2-categories in remark~\ref{rmk:general-lagrangian}. Throughout we freely use the criterion for genuine fiber 2-functors established in section~\ref{subsec:symtft-rg-flow}, and refer to section~\ref{sec:symtft} for the SymTFT conventions. The $2$-categorical notions used below including algebras and their modules, condensation and full centers, module 2-categories and Morita equivalence are collected in appendix~\ref{app:2cat}, to which we now refer rather than restating all such definitions here.

Table~\ref{tab:examples} lists the examples of the running $S_3$ example in sections~\ref{sec:warm-up} and~\ref{sec:gen-fib-fun-symtft} side by side, together with a dictionary between the 1- and 2-categorical objects.

\begin{table}[htbp]
\centering
\small
\renewcommand{\arraystretch}{1.2}
\begin{tabular}{@{}lcccc@{}}
\toprule
& \multicolumn{2}{c}{$(1\!+\!1)$d, $\cZ(\Vec_{S_3})$, section~\ref{sec:warm-up}} & \multicolumn{2}{c}{$(2\!+\!1)$d, $\cZ_1(2\Vect_{S_3})$, section~\ref{sec:gen-fib-fun-symtft}}\\
\cmidrule(lr){2-3}\cmidrule(lr){4-5}
& modules (\ref{subsec:warm-up-module}) & Lagrangian (\ref{subsec:warm-up-lagrangian}) & modules (\ref{subsec:module-2-cats}) & Lagrangian (\ref{subsec:lagrangian-algebras})\\
\midrule \midrule
boundaries & example~\ref{ex:topo boundaries} & examples~\ref{ex:S3-Lagrangians}, \ref{ex:S3-LZ3} & examples~\ref{ex:Ising from VecZ2 ModCat}, \ref{ex:S3-2mod-interfaces} & examples~\ref{ex:S3-2lag-algebras}, \ref{ex:S3-2lag-LB}\\ \graymidrule
$\CUV$ & example~\ref{CH on S3} & examples~\ref{ex:S3-lag-uv}, \ref{ex:S3-lag-uv-bimod} & example~\ref{ex:S3-2mod-uv} & example~\ref{ex:S3-2lag-uv}\\ \graymidrule
$\CIR$ & example~\ref{ex:S3 vacua} & examples~\ref{ex:S3-lag-ir}, \ref{ex:S3-lag-ir-Z2Z3} & example~\ref{ex:S3-2mod-ir} & example~\ref{ex:S3-2lag-ir}\\ \graymidrule
$\cF$ & example~\ref{ex:S3-RG} & examples~\ref{ex:S3-lag-rg}, \ref{ex:S3-lag-rg-Z2Z3} & example~\ref{ex:S3-2mod-rg} & example~\ref{ex:S3-2lag-rg}\\
\bottomrule
\end{tabular}

\vspace{3em}
\begin{tabularx}{\textwidth}{@{}l>{\raggedright\arraybackslash}X>{\raggedright\arraybackslash}X@{}}
\toprule
& $(1\!+\!1)$d & $(2\!+\!1)$d\\
\midrule \midrule
symmetry boundary & $\fB_{\Z_2}$, $A=\bbC[\Z_2]$, $\cL_{\Z_2}$ & $\fB_{(\bbZ_2,\Ising)}$, $A=\Ising$, $\cL(A)$\\ \graymidrule
physical boundary & $\fB_{\Z_3}$, $B=\bbC[\Z_3]$, $\cL_{\Z_3}$ & $\fB_{(\bbZ_3,\Vect^\omega_{\bbZ_3})}$, $B=\Vect^\omega_{\bbZ_3}$, $\cL(B)$\\ \graymidrule
UV symmetry & $\Rep(S_3)\simeq{}_A(\Vec_{S_3})_A$ & ${}_{\Ising}(2\Vect_{S_3})_{\Ising}$, with $\Omega\CUV\simeq\cZ(\Vect_{\bbZ_2})^{\bbZ_2}$\\ \graymidrule
trivial double coset & $1$ and $P$, with $A\otimes A\simeq1\oplus P$ & $1$ and the simple object $\chi=\Ising\boxtimes_{A_1}\Ising$\\ \graymidrule
other double coset & $E=A\otimes\bbC_a\otimes A$, $E\otimes E\simeq1\oplus P\oplus E$ & $X=\Ising\boxtimes_{A_1}(A_1)_a\boxtimes_{A_1}\Ising$, $X\otimes X\simeq\chi\oplus X$\\ \graymidrule
IR theory & trivially gapped, $\CIR\simeq\Vec$ & 3d $\bbZ_2$ gauge theory, $\CIR\simeq\Sigma\cZ(\Vect_{\bbZ_2})$\\ \graymidrule
RG functor & $\cF(E)=2$ & $\cF(X)\simeq1\oplus1$, $\Omega\cF$ forgets $\bbZ_2$-equivariance\\ \graymidrule
twisted sector of $E$, $X$ & two copies of $\Qc{a}{1}$ in $\cL_{\Z_3}$ & two copies of $S_{[a],\bbZ_3}$ in $\cL(B)$\\ \graymidrule
discrete torsion of $\fB_{\Z_3}$ & $\operatorname{H}^2(\Z_3;U(1))=0$ & $[\omega]\in \operatorname{H}^3(\bbZ_3;U(1))\simeq\bbZ_3$\\
\bottomrule
\end{tabularx}
\caption{The running $S_3$ example. Top: where each step is computed, in the module and Lagrangian-algebra models, in $(1\!+\!1)$ and $(2\!+\!1)$ dimensions. Bottom: dictionary between the single-vacuum sandwiches $(\fB_{\Z_2},\fB_{\Z_3})$ of section~\ref{sec:warm-up} and $(\fB_{(\bbZ_2,\Ising)},\fB_{(\bbZ_3,\Vect^\omega_{\bbZ_3})})$ of section~\ref{sec:gen-fib-fun-symtft}.}
\label{tab:examples}
\end{table}
\subsection{Single-Vacuum Gapped Phases}\label{subsec:single-vacuum}
We now demonstrate the core result of this work, that all AB-type fusion 2-categories admit generalized fiber 2-functors. First recall from section~\ref{subsec:symtft-defects-and-boundaries}, that the four-dimensional SymTFT for an AB-type symmetry $\CUV$ in $(2\!+\!1)$ dimensions is of the form
\begin{equation}
    \mathrm{DW}(G, \pi) \boxtimes \mathrm{CY}[M]\, ,
\end{equation}
where $\mathrm{DW}(G,\pi)$ denotes a Dijkgraaf--Witten theory with finite gauge group $G$ and action $\pi \in \operatorname{Z}^4(G; U(1))$, and $\mathrm{CY}[M]$ denotes the invertible Crane--Yetter phase associated to the Witt class of $M$, a non-degenerate braided fusion category given by the modular closure of $\Omega\CUV$.

\newl Recall from section~\ref{sec:sym-ext} also, that any pair $(G, \pi)$ admits a symmetry extension
\begin{equation}
    K \longrightarrow \Gamma \overset{p}\longrightarrow G\,,
\end{equation}
and 3-cochain $\psi\in \operatorname{C}^3(\Gamma; U(1))$ such that $p^*\pi = \delta\psi$, and that for each such extension we can construct a canonical generalized fiber 2-functor
\begin{equation}\label{eq:sym-ext-fun}
    \cF \, : \, 2\Vect_G^\pi \to \Sigma\cZ(\Vect_K^{\psi|_K})\, ,
\end{equation}
completing the proof that any invertible fusion 2-category admits a generalized fiber 2-functor.
Translated into the SymTFT picture, this canonical 2-functor can be obtained from SymTFT RG flow with the IR relative boundary
\begin{equation}
    \widetilde{\TIR}' = \fB_{(G,\, B)}\, , \qquad B = \Vect^\psi_\Gamma\, ,\qquad B_1 = \Vect_K^{\psi|_K}\, ,
\end{equation}
where the $G$-grading on $B$ is given by $p : \Gamma \to G$. Notice that this boundary has the property that no non-trivial SymTFT (bulk) Wilson line can condense on it.

\newl Finally, recall that in section~\ref{subsec:symtft-rg-flow}, to construct a (genuine) fiber 2-functor, we had to ensure two conditions:
\begin{enumerate}[label=(\roman*)]
    \item No non-trivial bulk Wilson line ends on both boundaries, \eqref{eq:cond-1}, so that the IR has no non-trivial local operators and hence a single vacuum; and 
    
    \item Every boundary Wilson line $\lambda$ on $\fB_\CUV$ ends on $\widetilde{\TIR}'$ in exactly $\dim\lambda$ independent ways, \eqref{eq:cond-2}, so that the UV lines act by their dimensions and $\Omega\cF$ is a fiber functor.
\end{enumerate}
To construct a generalized fiber 2-functor we now drop condition (ii) and keep only (i).
Then, for any AB-type fusion 2-category $\CUV$, we can always pick the boundary
\begin{equation}
    \widetilde{\TIR}' = \fB_{(G,\, B)} \boxtimes \mathrm{RT}[M]\, ,
\end{equation}
obtained by stacking the above symmetry extension boundary for $\mathrm{DW}(G, \pi)$ onto the boundary of $\mathrm{CY}[M]$ that supports the Reshetikhin--Turaev theory $\mathrm{RT}[M]$, whose fusion 2-category of surface defects is $\Sigma M$.
No non-trivial bulk line of the SymTFT ends on this choice of boundary.
It follows then that the compactification of the SymTFT with this choice of IR relative boundary produces no non-trivial topological local operators, and thus has a single vacuum. 
This completes the proof that all AB-type fusion 2-categorical symmetries admit generalized fiber 2-functors.

\subsection{Module 2-Categories}\label{subsec:module-2-cats}
Given a four-dimensional SymTFT associated to some fusion 2-categorical symmetry $\cC$, topological boundaries are characterized in their entirety by $\cC$-module 2-categories~\footnote{Viewed as a fully extended 4d TFT in $4\Vect$ (the Morita 4-category of multifusion 2-categories), this is a natural consequence of the cobordism hypothesis. A lax monoidal natural transformation from the SymTFT $\mathrm{DR}[\cC]$ to the trivial theory assigns the point to a $(\cC, 2\Vect)$-bimodule.}. 
We refer the reader to~\cite{Decoppet:2021ssm, Pearson:2024snm} for details on module 2-categories.
Physically this module catalogues two-dimensional topological interfaces back to the boundary supporting $\cC$, as illustrated below in figure~\ref{fig:symtft-corners}.

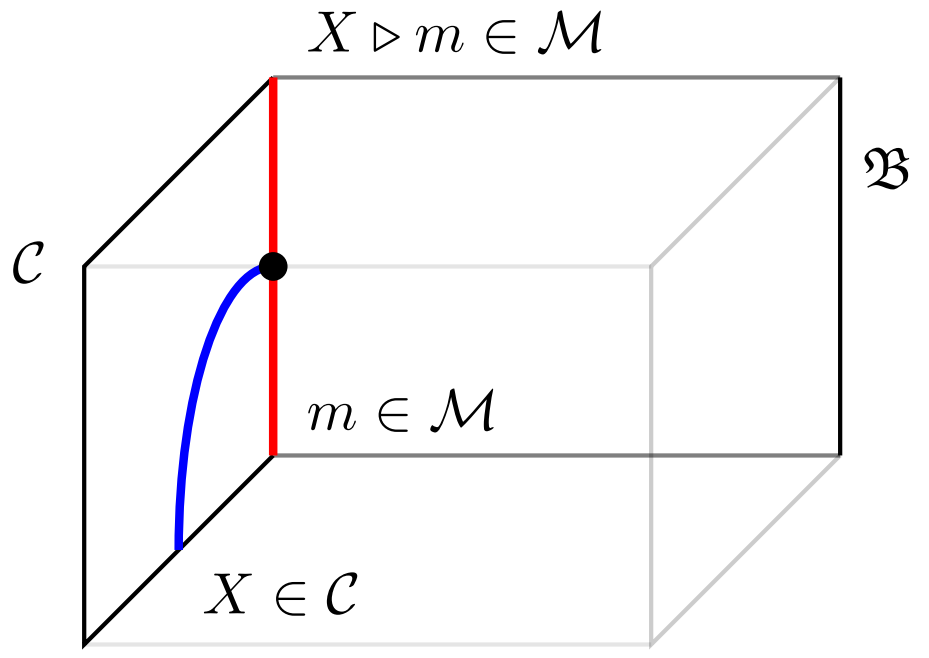
\begin{figure}[H]
    \centering
    \begin{tikzpicture}[scale = 1.25]
        \draw[gray, thick] (0,2) -- (3, 2);
        \draw[gray, thick, opacity = 0.2] (-1,1) -- (2, 1);
        \draw[gray, thick] (0,0) -- (3, 0);
        \draw[gray, thick, opacity = 0.2] (-1,-1) -- (2, -1);
        
        \draw[black, thick] (0, 0) -- (-1, -1) -- (-1, 1) node[anchor = east]{$\cC\;\strut$} -- (0, 2);
        \draw[blue, ultra thick] (-0.5, -0.5) node[anchor = north west, black]{$X \in \cC$} arc[x radius = 0.5, y radius = 1.5, start angle = 180, end angle = 90];
        \draw[red, ultra thick] (0,2) node[anchor = south west, black]{$\, X \triangleright m \in \cM$} -- (0,0) node[anchor = south west, black]{$\, m \in \cM$};
        \filldraw[black] (0,1) circle (2pt);
        \begin{scope}[shift={(3,0)}]
        \draw[black, thick, opacity = 0.2] (0,0) -- (-1, -1) -- (-1, 1) -- (0,2);
        \draw[black, thick] (0, 2) -- (0, 1.5) node[anchor = west]{$\fB\strut$} -- (0, 0);
        \end{scope}
    \end{tikzpicture}
    \caption{The boundary $\fB$ is characterized by a 2-category of 2d interfaces $m\in \cM$ back to $\cC$. $\cM$ then inherits a (left) $\cC$-module action from colliding defects $X\in \cC$.}
    \label{fig:symtft-corners}
\end{figure}

\newl In the case that $\cC = 2\Vect_G^\pi$ is invertible and the SymTFT is given by a pure 4d Dijkgraaf--Witten theory, this labeling of boundaries is equivalent to $\pi$-projective 3-representations of $G$~\cite{Decoppet:2023bay,Xu:2024etale, Bullimore:2024khm}, and the corresponding boundaries support those fusion 2-categories that are Morita equivalent to $\cC$.
Irreducible $\pi$-projective 3-representations of $G$ are labeled (up to equivalence) by:
\begin{itemize}
    \item A (conjugacy class of) subgroup $H\subseteq G$,
    \item A (Morita class of) $\pi|_H$-twisted faithfully $H$-graded fusion category
    \begin{equation}
        A \simeq \bigoplus_{h \in H} A_h\, .
    \end{equation}
\end{itemize}
This coincides with our previously mentioned classification of boundaries $\fB_{(H, A)}$ from section~\ref{sec:symtft}.

\newl Fixing the data as above, we can extract the corresponding (left) $\cC$-module following~\cite{etingof2009,Xu:2024etale,Pearson:2024snm}. First we define a module 2-category over $\widetilde{\cC} = 2\Vect_H^{\pi|_H}$ whose underlying linear 2-category is
\begin{equation}
    \widetilde{\cM} = \mathsf{LMod}(A_1)\, .
\end{equation}
This 2-category catalogues boundary conditions for the 3d Turaev--Viro TQFT we are stacking on the boundary, with line defects forming the braided fusion category $\cZ(A_1)$.
We then give a $\widetilde{\cC}$-module action on $\widetilde{\cM}$ using the 3-group homomorphism encoded by $A$:
\begin{align}
    A_\bullet \, :\, H \rtimes^{\pi|_H} \bbC^\times[2]  &\to {\rm BrPic}(A_1)\, ,\nonumber\\
    h &\mapsto A_h\, .
\end{align}
Under this action a simple object $h\in H$ acts on $m\in \widetilde{\cM}$ via the balanced tensor product
\begin{equation}
    h \triangleright m = A_h \boxtimes_{A_1} m\, ,
\end{equation}
and the remaining higher data is all encoded in the monoidal structure of $A_\bullet$. Before moving on, we point that the $H$-equivariant data on objects of $\widetilde{\cM}$ with respect to this action can be neatly repackaged as
\begin{equation}\label{eq:h-eq-vs-graded-ext} 
    \widetilde{\cM}^H \simeq \mathsf{LMod}(A)\, .
\end{equation}
This 2-category catalogues choices of boundary conditions for a 3d TQFT with lines given by $\cZ(A)$, which is obtained by gauging the $H$ symmetry on the Turaev--Viro TQFT for $A_1$~\footnote{This interpretation is literal when $\pi|_H$ is trivial, so that $A$ is an ordinary fusion category. For $[\pi|_H]\neq0$, $A$ is only a $\pi|_H$-twisted fusion category: the $H$ symmetry of $\mathrm{TV}[A_1]$ is anomalous and can only be gauged together with $\fB_{\rm Dir}$, whose anomaly it cancels. The equivalence \eqref{eq:h-eq-vs-graded-ext} then holds for $\pi|_H$-twisted equivariant objects and modules over the twisted fusion category $A$, and there is no stand-alone 3d TQFT with lines $\cZ(A)$.}.

\begin{examplebox}[
Ising module categories from $\Vec_{\Z_2}$ module categories][label=ex:Ising from VecZ2 ModCat]    
    {\bf The 3d $\bbZ_2$ gauge theory.} As an example, take the 3d Turaev--Viro TQFT associated to $A_1 = \Vect_{\bbZ_2}$, which describes Dijkgraaf--Witten theory with gauge group $\bbZ_2$. 
    This theory has four topological lines generated via fusion by the $e$ and $m$ lines, which braid non-trivially with one another.
    
    \newl {\bf Its boundaries.} Topological boundaries are labeled by 2-representations
    \begin{align}
        \widetilde{\cM} &= \mathsf{LMod}(\Vect_{\bbZ_2}) \nonumber \\
        &\simeq 2\Rep(\bbZ_2)\, .
    \end{align}
    There are two simple objects, corresponding to a pair of boundaries that condense the $e$ and $m$ lines, respectively. They both support a global 0-form $\bbZ_2$ symmetry, and are related to one another by gauging.

    \newl {\bf The $e$--$m$ symmetry.} The TQFT bulk has a 0-form $\bbZ_2$ symmetry that permutes the $e$ and $m$ lines, and hence permutes the two boundaries mentioned above. This symmetry is encoded by the faithfully $\bbZ_2$-graded fusion category
\begin{equation}\label{eq:S3-Ising-grading}
    A = \Ising\, , \qquad A_1 = \Vect_{\bbZ_2}\, ,\qquad A_{-1} = \Vect\, ,
\end{equation}
whose simple objects we denote $1$, $g$ and $\epsilon$, which fuse as
\begin{equation}
    g\otimes g = 1\, ,\qquad g\otimes\epsilon = \epsilon\, ,\qquad \epsilon \otimes\epsilon = 1 \oplus g\, .
\end{equation}
Thought of as an $A_1$-bimodule, $A_{-1}$ describes a non-trivial surface defect on which $1 \oplus em$ condenses, and bringing this defect to either of the two boundaries then acts as gauging.

\newl {\bf Gauging.} Gauging this bulk $\bbZ_2$ symmetry produces the 3d Turaev--Viro TQFT for $\Ising$, whose boundaries are labeled, by \eqref{eq:h-eq-vs-graded-ext}, by
    \begin{equation}
        \widetilde{\cM}^{\bbZ_2}  = \mathsf{LMod}(\Ising)\, .
    \end{equation}
Since the $\bbZ_2$ exchanges the $e$- and $m$-condensing boundaries, a $\bbZ_2$-equivariant boundary condition must contain both. Indeed, $\mathsf{LMod}(\Ising)$ has a single simple object, the regular module $\Ising$. Its underlying $A_1$-module is $A_1\oplus A_{-1}\simeq\Vect_{\bbZ_2}\oplus\Vect$, the sum of the two simple objects of $\widetilde{\cM}$, and its equivariant structure exchanges them. After gauging, the two boundaries of the $\bbZ_2$ gauge theory are therefore identified into the unique topological boundary of $\mathrm{TV}[\Ising]$. This $\bbZ_2$-graded fusion category $A=\Ising$ is the one used to build the symmetry boundary $\fB_{(\bbZ_2,\Ising)}$ of the running example in example~\ref{ex:S3-2mod-interfaces}.
\end{examplebox}

\ni The full $\cC$-module 2-category $\cM$ is then obtained by induction on $G$, and has the underlying linear 2-category
\begin{equation}
    \cM = \bigoplus_{i \in G/H} (\mathsf{LMod}(A_1))_i\, ,
\end{equation}
which includes a grading by left $H$-cosets. We denote the transitive action of $g\in G$ by $g : i \mapsto g\cdot i$, and we then designate representatives $i = g_iH$ and little group elements
\begin{equation}
    g g_i = g_{g\cdot i} l_i(g)\, .
\end{equation}
The action of $g\in G$ on an object $(M, i) \in (\mathsf{LMod}(A_1))_i$ is then given:
\begin{equation}
    g\triangleright (M, i) := (A_{l_i(g)} \boxtimes_{A_1}M,\,{g\cdot i})\, .
\end{equation}

\begin{examplebox}[Topological interfaces between boundaries of 4d $S_3$ SymTFT][label=ex:S3-2mod-interfaces]
    Consider a theory whose SymTFT is the 4d Dijkgraaf--Witten theory with gauge group $G = S_3$. This TQFT admits a canonical Dirichlet boundary condition $\fB_{\rm Dir}$ hosting the 0-form $S_3$ symmetry.
    Continuing the earlier example, we will consider the symmetry boundary
    \begin{equation}
        \fB_{(\bbZ_2, A)} = \frac{\fB_{\rm Dir}\boxtimes \mathrm{TV}[\Vect_{\bbZ_2}]}{(\bbZ_2)_{\rm diag}}\, ,
    \end{equation}
    obtained by coupling the 3d $\bbZ_2$ Dijkgraaf--Witten TQFT to the boundary and gauging. Since they are all related by conjugation, $\bbZ_2\subset S_3$ can be chosen to be any order-2 subgroup. The $\bbZ_2$-action on the 3d TQFT $\mathrm{TV}[\Vect_{\bbZ_2}]$ is given by $A = \Ising$ as a $\bbZ_2$-graded fusion category.
    
    \newl Interfaces between $\fB_{\rm Dir}$ and $\fB_{(\bbZ_2, A)}$ are labeled by objects of the induced $2\Vect_{S_3}$-module
    \begin{equation}
         \cM = \bigoplus_{i \in S_3/\bbZ_2} (2\Rep(\bbZ_2))_i\, .
    \end{equation}
    This 2-category has three connected components that are permuted by $S_3$, each containing two objects that are permuted by its stabilizing $\bbZ_2$ subgroup.
\end{examplebox}

\ni Later we will also need the corresponding construction for a right $\cC$-module 2-category, for which we replace $\mathsf{LMod}(A_1)$ for $\mathsf{RMod}(A_1)$, and act from the right with $A_\bullet$ instead. The induction in this case is instead on right cosets $\bar\imath \in H\backslash G$ with right action $g : \bar\imath \mapsto \bar\imath\cdot g$ and little group elements
\begin{equation}
    g_{\,\bar\imath}\, g = r_{\,\bar\imath }(g)g_{\,\bar\imath\cdot g}\, .
\end{equation}
Notice that this classification is ultimately the same, because we can flip a boundary from being on the left to being on the right by employing the equivalences
\begin{equation}
    \mathsf{LMod}(A_1) \simeq \mathsf{RMod}(\bar{A_1})\, , \qquad \mathsf{Bimod}(A_1) \simeq \mathsf{Bimod}(\bar{A_1})\, ,
\end{equation}
where $\bar{A_1}$ denotes the orientation reversal of $A_1$. The action from $G$ is then also flipped by taking the reversal $\bar{A}$, which describes an $H$-graded extension of $\bar{A_1}$ with grading inverse to that of $A$.

\newl In this subsection we restrict ourselves to the above case, where $\CUV$ is Morita equivalent to the invertible fusion 2-category $\cC = 2\Vect_G^\pi$ and corresponds to a (left) $\cC$-module $\cM$. In addition to this, we also ask that the theory $\TUV$ we work in has no intrinsic anomaly, so that the IR relative boundary $\widetilde{\TIR}'$ is a genuine boundary of the SymTFT that we identify with a (right) $\cC$-module $\cN$.

\begin{figure}[H]
    \centering
    \begin{tikzpicture}
        
        \draw[black, thick] (-1, -1) -- (-1, 0.5) node[anchor = east]{$\widetilde{\TIR}' = \fB_{(K, B)}\;\strut$} -- (-1, 1) -- (0, 2) -- (0, 0) -- (-1, -1);
        \draw[red, ultra thick] (0,2) node[anchor =  east, black]{$\strut \cN\;$} -- (0,0);
        \begin{scope}[shift={(3,0)}]
        \draw[black, thick] (-1, -1) -- (-1, 1) -- (0, 2) -- (0, 0.5) node[anchor = west]{$\strut\;\CUV \sim\fB_{(H, A)}$} -- (0,1) -- (0, 0) -- (-1, -1);
        \draw[blue, ultra thick] (0,2) node[anchor =  west, black]{$\;\strut \cM$} -- (0,0);
        \end{scope}

        \draw[gray, thick] (0,2) -- (1.5,2) node[anchor = south, black]{$\strut \cC \sim \fB_{\rm Dir}$} -- (3, 2);
        \draw[gray, thick, opacity = 0.2] (-1,1) -- (2, 1);
        \draw[gray, thick] (0,0) -- (3, 0);
        \draw[gray, thick, opacity = 0.2] (-1,-1) -- (2, -1);
    \end{tikzpicture}
    \caption{The symmetry boundary supporting $\CUV$ is $\fB_{(H,A)}$ and blue interfaces with $\fB_{\rm Dir}$ are objects of $\cM$. The IR relative boundary $\widetilde{\TIR}'$ is $\fB_{(K, B)}$ and red interfaces with $\fB_{\rm Dir}$ are objects of $\cN$. }
    \label{fig:symtft-rg-functor-2}
\end{figure}
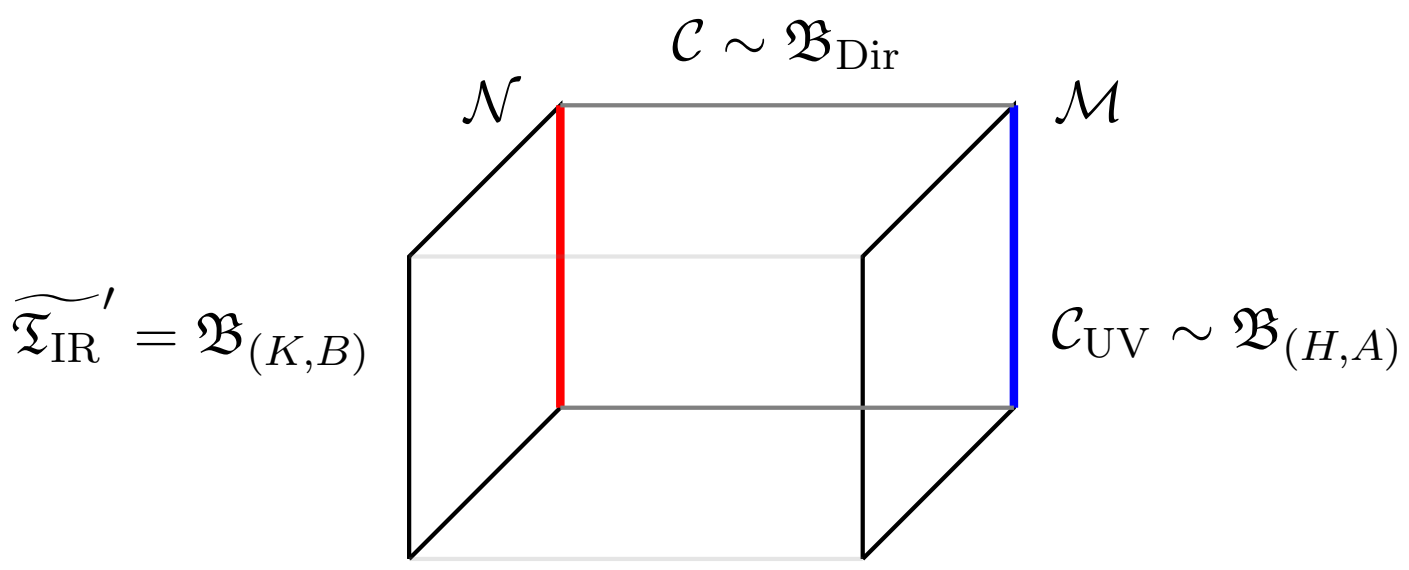
\ni Moving forward, we will fix $\cM$ and $\cN$ by setting the symmetry boundary supporting $\CUV$ to be $\fB_{(H, A)}$, and the IR relative boundary $\widetilde{\TIR}'$ to be $\fB_{(K, B)}$, these identifications are summarized and illustrated above in figure~\ref{fig:symtft-rg-functor-2}. Our goal is then to answer the three following questions:
\begin{enumerate}
    \item What is the symmetry $\CUV$?
    \item What is the symmetry $\CIR$?
    \item What is the RG functor $\cF : \CUV \to \CIR$?
\end{enumerate}
To produce a generalized fiber 2-functor, we will need to select $\cM$ and $\cN$ such that $\CIR$ is fusion, rather than multifusion. This selection should reflect the argument from section~\ref{subsec:symtft-rg-flow}, that the IR will contain no non-trivial topological local operators if and only if 
\begin{equation}
    H\backslash G/ K = \{G\}\, ,
\end{equation}
or equivalently, $G = HK$.

\subsubsection{The UV Symmetry \texorpdfstring{$\CUV$}{CUV}}\label{subsec:uv-symmetry}
Recall that the symmetry $\CUV$ living on $\fB_{(H, A)}$ can be obtained by starting from $\fB_{\rm Dir}$, coupling an $H$-symmetric TFT with anyon lines $\cZ(A_1)$, and gauging the subgroup $H$ diagonally. All surface defects in that TFT are condensation, and so after gauging we expect (connected components of) objects in $\CUV$ to be labeled by double $H$-cosets
\begin{equation}
    \pi_0\CUV \simeq H\backslash G/ H\, ,
\end{equation}
and loop objects to be labeled by
\begin{equation}
    \Omega\CUV \simeq \cZ(A_1)^H\, ,
\end{equation}
corresponding to $H$-invariant combinations of anyons, stacked with Wilson lines which comprise the M\"uger center
\begin{equation}
    \cZ_2(\Omega\CUV) \simeq \Rep(H)\, .
\end{equation}

\ni Given the (left) $\cC$-module 2-category $\cM$ corresponding to $\fB_{(H, A)}$, we can compute $\CUV$ directly using~\footnote{This also follows from the cobordism hypothesis. The modification of that natural transformation which assigns the point to $\cM$, assigns the point to a choice of $\cC$-linear endomorphism of $\cM$. }
\begin{equation}\label{eq:uv-equivalence}
    \CUV \simeq \End_{\cC}(\cM)\, .
\end{equation}
This follows from our earlier discussion about interfaces between boundaries.
$\cM$ should admit a (right) $\CUV$-module action that commutes with the existing $\cC$-module action.
The 2-category of all such actions is precisely the 2-category of $\cC$-linear endomorphisms of $\cM$ appearing above.

\newl First we must compute the 2-category of ordinary (i.e., only $2\Vect$-linear) endomorphisms of $\cM$, and to do so we pick a convention where endomorphisms act from the left. Linear endomorphisms of $\cM$ are constructed from twice-categorified matrix elements
\begin{equation}
    (i,j;\, U) \in \End_{2\Vect}(\cM)\, ,
\end{equation}
$\cM$-bigraded by $i,j \in G/H$, where $U\in \mathsf{Bimod}(A_1)$ is an $A_1$-bimodule, which acts on an object $(M,k)\in \cM$ as
\begin{equation}
    (i,j;\, U)\triangleright(M, k) = 
    \begin{cases}
        \left(U\boxtimes_{A_1} M, i\right)\quad &\text{if } j=k\, ,\\
        \emptyset &\text{otherwise}.
    \end{cases}
\end{equation}
More general linear endomorphisms are then matrices, obtained by taking direct sums of the form
\begin{equation}\label{eq:u-expansion}
    \underline{U} = \bigoplus_{i,j}(i,j;\,U_{i,j})\, .
\end{equation}
The 1-morphisms between such objects preserve the $G/H$-bigrading and, together with the 2-morphisms, are given by those morphisms in $\mathsf{Bimod}(A_1)$.

\newl For a linear endomorphism $\underline{U}\in \End_{2\Vect}(\cM)$ to be $\cC$-linear it must come equipped with 1-isomorphisms
\begin{equation}\label{eq:c-linearity}
g\triangleright\left(\underline{U}\triangleright(M,k)\right) \simeq \underline{U}\triangleright \left(g\triangleright(M,k)\right)
\end{equation}
for each $(M,k)\in \cM$ and $g\in G$, which satisfy conditions of their own up to higher data.
For now we will focus on the condition imposed above, as expanding each side of that equivalence using~\eqref{eq:u-expansion} yields
\begin{equation}\label{eq:c-linearity-0}
    A_{l_i(g)}\boxtimes_{A_1} U_{i,j} \simeq U_{g\cdot i, g\cdot j}\boxtimes_{A_1} A_{l_j(g)}\, .
\end{equation}
Notice that this implies it is not possible in general for a simple linear endomorphism to satisfy this condition all by itself; whenever the component $U_{i,j}$ is non-vanishing, the components $U_{g\cdot i, g\cdot j}$ for each $g\in G$ must also be non-vanishing. Since the $G$-action on $G/H$ is transitive, we can always pick a representative bimodule $U_{H, j}$. Such a representative is not unique however, and since the trivial coset is stabilized by $H$, the various gradings $j \in G/H$ that appear in such representatives form disconnected components of a single $H$-double coset.

\newl To label simple $\cC$-linear endomorphisms then, first fix a double coset with representative $g\in G$ stabilized by $H_{g} = H \cap {}^{g} H$, where ${}^{g} H = gHg^{-1}$. Then the remaining data is an $A_1$-bimodule $U$ equipped with 1-isomorphisms
\begin{equation}\label{eq:c-linearity-1}
    u_h \, :\, A_h\boxtimes_{A_1} U \to U\boxtimes_{A_1} A_{h^g}\, ,
\end{equation}
that commute $U$ with the $H_{ g}$ action, and 2-isomorphisms~\footnote{The conditions appearing here have been put into a schematic form to make them more readable: we suppress the associators of $A$ and the whiskerings, but not the 3-cochain $\xi_g$ in \eqref{eq:c-linearity-3}, whose cohomology class cannot be absorbed.}
\begin{equation}\label{eq:c-linearity-2}
    u_{h,h'}\, :\, u_h \circ u_{h'}\to u_{hh'}\, ,
\end{equation}
that compose the 1-isomorphisms, satisfying an associativity condition
\begin{equation}\label{eq:c-linearity-3}
    u_{h, h'h''}\, u_{h',h''} = \xi_g(h,h',h'')\; u_{hh',h''}\, u_{h,h'}\, ,
\end{equation}
for all $h,h',h'' \in H_{ g}$.
The 3-cochain $\xi_g\in\operatorname{C}^3(H_g;U(1))$ appearing here is determined by $\pi$ and $g$. The two sides of \eqref{eq:c-linearity-1} involve $A$ through the two embeddings $h\mapsto h$ and $h\mapsto h^g$ of $H_g$ into $H$, so they are twisted by $\pi|_{H_g}$ and by its conjugate $\pi^g(h_1,\ldots,h_4):=\pi(h_1^g,\ldots,h_4^g)$ respectively. Conjugation acts trivially on the cohomology of $G$, so these two cocycles differ by a coboundary,
\begin{equation}\label{eq:xi-g}
    \pi|_{H_g} = \pi^g\cdot\delta\xi_g\, ,
\end{equation}
and $\xi_g$ is the canonical such 3-cochain built from $\pi$ and $g$; an explicit formula is given in \cite{Decoppet:2023bay}. For $g=1$ it is trivial, and so it is whenever $\pi$ is trivial, but in general its class is non-trivial even when $\pi|_H$ is trivial. For example, for $G=\bbZ_2^a\times\bbZ_2^b$, $H=\bbZ_2^a$ and $\pi(g_1,g_2,g_3,g_4)=(-1)^{x_1x_2x_3y_4}$ with $g_i=(x_i,y_i)$, one has $\pi|_H=1$, while for the double coset of $g=b$ the cochain $\xi_b(a^r,a^s,a^t)=(-1)^{rst}$ represents the non-trivial class of $\operatorname{H}^3(\bbZ_2;U(1))$.

\newl More generally, when $g$ centralizes $H_g$, as in this example, the two embeddings agree, $\pi^g=\pi|_{H_g}$, and $\xi_g$ is a 3-cocycle: it is the restriction to $H_g\subseteq C_G(g)$ of the transgression $\tau_g(\pi)$ of $\pi$. For instance, the Drinfeld center $\cZ_1(2\Vect_G^\pi)$ is the dual of $2\Vect_{G\times G}^{\pi\boxtimes\pi^{-1}}$ with respect to the diagonal subgroup $G\subseteq G\times G$, whose double cosets are the conjugacy classes $[x]$ of $G$, with representatives $(x,1)$ and stabilizers $C_G(x)$. In this case $\xi_{(x,1)}=\tau_x(\pi)$, which recovers the decomposition $\cZ_1(2\Vect_G^\pi)\simeq\bigoplus_{[x]}2\Rep(C_G(x),\tau_x(\pi))$ of \cite{Kong:2019brm}. In general $g$ does not centralize $H_g$, and $\xi_g$ is only a cochain satisfying \eqref{eq:xi-g}.

\newl Thus we see that using the equivalence~\eqref{eq:uv-equivalence}, simple objects of $\CUV$ are labeled by:
\begin{itemize}
    \item $HgH \in  H\backslash G/H$ a double $H$-coset.
    \item An $H_g$-equivariant $A_1$-bimodule $U$ under the action
    \begin{equation}
       h\; :\; U \mapsto A_h \boxtimes_{A_1} U \boxtimes_{A_1} A_{h^g}^{op}\, ,
    \end{equation}
    where the equivariant structure is twisted by $\xi_g$ as in \eqref{eq:c-linearity-3}.
\end{itemize}
Equivalently, by the same repackaging as in \eqref{eq:h-eq-vs-graded-ext}, the component of $\CUV$ on the double coset $HgH$ consists of the modules over the faithfully $H_g$-graded fusion category $\bigoplus_{h\in H_g}A_h\boxtimes\bar{A_{h^g}}$, whose associator is corrected by $\xi_g$; by \eqref{eq:xi-g} this is exactly what is needed for it to satisfy the ordinary pentagon equation. For $A=\Vect_H^\psi$ this fusion category is pointed, $\Vect_{H_g}^{\xi}$ with $\xi$ the 3-cocycle built from $\xi_g$, $\psi|_{H_g}$ and its conjugate $\psi^g$, and one recovers the components $\mathsf{Mod}(\Vect_{H_g}^{\xi})$ of the group-theoretical fusion 2-category $\mathsf{C}(G,\pi|H,\psi)$ \cite{Decoppet:2023bay}; in the example above, the component of $b$ is $\mathsf{Mod}(\Vect^{\xi_b}_{\bbZ_2})$, with a single simple object, rather than $\mathsf{Mod}(\Vect_{\bbZ_2})$.
The latter labeling captures $H_g$-invariant surface defects of the Turaev--Viro TFT associated to $A_1$, stacked with condensations of the Wilson lines of $H$. These defects are all condensation, so the connected components of simple objects are labeled only by double cosets.

\newl The monoidal structure of $\CUV$ is the composition of $\cC$-linear endomorphisms of $\cM$. On the matrices \eqref{eq:u-expansion} it is matrix multiplication, with the relative tensor product $\boxtimes_{A_1}$ in place of the product,
\begin{equation}\label{eq:CUV-fusion-matrix}
    (\underline{U}\circ\underline{U}')_{i,k}=\bigoplus_{j\in G/H}U_{i,j}\boxtimes_{A_1}U'_{j,k}\, ,
\end{equation}
and the $\cC$-linear structure \eqref{eq:c-linearity} of the composite is the composite of the $\cC$-linear structures, so the composite is again $\cC$-linear. The associator and unitors of $\CUV$ are inherited from those of $\boxtimes_{A_1}$ in $\mathsf{Bimod}(A_1)$. For simple objects supported on the double cosets $HgH$ and $Hg'H$, the composite is supported on the double cosets contained in $HgH\cdot Hg'H$, which categorifies the fusion rules of a group-theoretical fusion category. Equivalently, under $\CUV\simeq{}_A(2\Vect_G^\pi)_A$, fusion is the relative Deligne tensor product $\boxtimes_A$ of $A$-bimodule categories; see \eqref{eq:lag-CUV} and the explicit computation \eqref{eq:S3-XX-bimod} for the running example.

\newl We can also compute the category $\Omega\CUV$ of topological lines comprising the UV 1-form symmetry. These correspond to $\cC$-linear natural transformations of the identity endomorphism, which we associate to the trivial double coset with trivial $A_1$-bimodule $U = A_1$ and trivial equivariantization data. Concretely these are bimodule functors
\begin{equation}
    f : A_1 \to A_1\, ,
\end{equation}
that commute with the 1-isomorphisms $u_h$ for each $h \in H$, up to 2-isomorphisms satisfying additional higher conditions. Since the category of bimodule functors $f$ is precisely the Drinfeld center $\cZ(A_1)$, this is just equivalent to the $H$-equivariantization
\begin{equation}
    \Omega\CUV \simeq \cZ(A_1)^H\, .
\end{equation}
Notice, this braided fusion category is not non-degenerate, having M\"uger center given precisely by $\mathsf{Rep}(H)$. Later we will return to this point in more detail, but it is worth mentioning that the M\"uger center describes lines that do not carry charge under any 1-form symmetry, and as such represents the largest sub-category that is trivializable in the IR. One way to characterize a ``minimal" IR phase of $\CUV$ then, is to condense the M\"uger center in the IR producing the modular closure of $\Omega\CUV$, given simply by $\cZ(A_1)$.

\begin{examplebox}[The UV symmetry of $\fB_{(\bbZ_2,\Ising)}$][label=ex:S3-2mod-uv]
    Continuing with our running example (example~\ref{ex:S3-2mod-interfaces}), we now compute the UV symmetry $\CUV$ for $G = S_3$, $H = \bbZ_2$ and $A = \Ising$. This is the symmetry living on the boundary $\fB_{(\bbZ_2, A)}$ of the 4d $S_3$ Dijkgraaf--Witten TQFT. Its 1-categorical counterpart is the symmetry $\Rep(S_3)$ of $\fB_{\Z_2}$ in example~\ref{CH on S3}, and the same computation with Lagrangian algebras is example~\ref{ex:S3-2lag-uv}.

    \newl Connected components of $\CUV$ are labeled by double cosets
    \begin{equation}\label{eq:S3-double-cosets}
        \bbZ_2\backslash S_3 / \bbZ_2\, ,
    \end{equation}
    of which there are two.
    Objects of the identity-connected component are all condensation, and are labeled by $\bbZ_2$-equivariant bimodules over $A_1 = \Vect_{\bbZ_2}$.
    In particular, the identity bimodule $A_1$ gives two simple objects. The unit $1$ is $A_1$ with its canonical $\bbZ_2$-equivariant structure. The object $\chi$ is induced from the trivial subgroup of $\bbZ_2$: its underlying bimodule is $A_1\oplus A_1$, with the two summands exchanged by $\bbZ_2$. Physically, $\chi$ is the condensation of the Wilson line $1^-$ on the trivial surface, so it is not invertible, $\chi\otimes\chi\simeq2\chi$. As $\Ising$-bimodules, $1=\Ising$ and $\chi=\Ising\boxtimes_{A_1}\Ising$ (example~\ref{ex:S3-2lag-uv}).

    \newl Objects of the other connected component, with representative $g=a$ and $H_a=\{\id\}$, are labeled by bimodules over $A_1$ with no equivariant data. The surface defect $X \in \CUV$ corresponding to the identity bimodule in this component is non-invertible with fusion
    \begin{equation}
        X\otimes X \simeq \chi \oplus X \, .
    \end{equation}
    This is the 2-categorical counterpart of $E\otimes E=1\oplus P\oplus E$ in $\Rep(S_3)$, and it is derived from the composition \eqref{eq:CUV-fusion-matrix} in \eqref{eq:S3-XX-bimod}.

    \newl Loop objects of $\CUV$ corresponding to genuine topological lines are labeled by
    \begin{equation}\label{eq:S3-OmegaCUV}
        \Omega\CUV \simeq \cZ(\Vect_{\bbZ_2})^{\bbZ_2}\, ,
    \end{equation}
    which contains 5 simple objects:
    \begin{itemize}
    \item[--] The objects $1$ and $em$, which are $\bbZ_2$-invariants, together with a choice of irreducible $\bbZ_2$-representation, which we denote $1^{\pm}$ and $em^{\pm}$, and which fuse as
    \begin{align}
        1^{\pm}\otimes 1^{\pm} = 1^{+}\, ,&\qquad 
        1^{\pm}\otimes 1^{\mp} = 1^{-}\, ,\nonumber\\
        1^{\pm}\otimes em^{\pm} = em^{+}\, ,&\qquad 1^{\pm}\otimes em^{\mp} = em^{-}\, ,\nonumber\\
        em^{\pm}\otimes em^{\pm} = 1^{+}\, ,&\qquad em^{\pm}\otimes em^{\mp} = 1^{-}\, .
    \end{align}
    In particular, $1^{+}$ is the identity object.
    
    \item[--] The non-invertible composite object $\eta = e \oplus m$, which forms a $\bbZ_2$-invariant and fuses as
    \begin{gather}
        1^{\pm}\otimes\eta \; =\; em^{\pm}\otimes \eta \; = \; \eta\, ,\\
        \eta \otimes \eta \; =\;  1^+ \oplus 1^-\oplus em^+ \oplus em^-\, .
    \end{gather}
\end{itemize}
The M\"uger center $\cZ_2\Omega\CUV \simeq \Rep(\bbZ_2)$ is given by the symmetric subcategory generated by $\{1^+, 1^-\}$, and writing $\cZ(\Ising) \simeq \Ising\boxtimes\bar{\Ising}$, we can see that the inclusion $\Omega\CUV \subseteq \cZ(\Ising)$ sends
\begin{equation}\label{eq:S3-OmegaCUV-in-ZIsing}
    1^- \mapsto g\bar{g}\, ,\qquad em^+ \mapsto g\, ,\qquad em^- \mapsto \bar{g}\, ,\qquad \eta \mapsto \epsilon\bar{\epsilon}\, .
\end{equation}
\end{examplebox}

\subsubsection{The IR Symmetry \texorpdfstring{$\CIR$}{CIR}}\label{subsec:ir-symmetry}
Recall that the IR TFT $\TIR$ supporting the symmetry $\CIR$ is obtained from the SymTFT by sandwiching the symmetry boundary $\fB_{(H,A)}$, corresponding to the left $\cC$-module $\cM$, with the IR relative boundary $\fB_{(K, B)}$, corresponding to some other right $\cC$-module $\cN$.

\newl Setting $(H,A) = (1, \Vect)$ results in $\fB_{(H, A)} = \fB_{\rm Dir}$ and $\CUV \simeq \cC$. Sandwiching with $\fB_{(K, B)}$ produces an IR phase for $\cC$ with $G/K$ vacua, each hosting anyon lines characterized by $\cZ(\bar{B_1})$. Switching $(H, A)$ back on then has the effect of coupling anyon lines from $\cZ(A_1)$ and gauging a diagonal $H$ subgroup. Gauging the right action of $H$ on $G/K$ results in vacua labeled by $(K,H)$-double cosets. In the case that $G = HK$ and there is only one vacuum, it will host anyons living in the $(H\cap K)$-equivariantization of some $(H\cap K)$-crossed braided extension of $\cZ(\bar{B_1} \boxtimes A_1)$, equivalent to
\begin{equation}
    \Omega\CIR \simeq \cZ\left( \,\bigoplus_{h \in H\cap K} \bar{B_{h^{-1}}}\boxtimes A_h \right)\, .
\end{equation}

\begin{figure}[H]
    \centering
    \begin{tikzpicture}[scale = 1.25]
        \draw[gray, thick] (0,2) -- (1.5, 2) node[anchor = south, black]{$\fB_{\rm Dir}$} -- (3, 2);
        \draw[gray, thick, opacity = 0.2] (-1,1) -- (2, 1);
        \draw[gray, thick] (0,0) -- (3, 0);
        \draw[gray, thick, opacity = 0.2] (-1,-1) -- (2, -1);
        
        \draw[black, thick] (0, 0) -- (-1, -1) -- (-1, 0.5) node[anchor = east]{$\fB_{(K, B)}\;\strut$} -- (-1,1) -- (0, 2);
        \draw[red, ultra thick] (0,2) node[anchor = south east, black]{$\cN$} -- (0,0);
        \begin{scope}[shift={(3,0)}]
        \draw[black, thick] (0,0) -- (-1, -1) -- (-1, 1) -- (0,2);
        \draw[blue, ultra thick] (0, 2) node[anchor = south west, black]{$\cM$} -- (0, 0.5) node[anchor = west, black]{$\;\fB_{(H, A)}\strut$} -- (0, 0);
        \end{scope}

         \draw[black] (5, 0.5) node{\huge$\strut\rightsquigarrow$};

        \begin{scope}[shift = {(7, 0)}]
            \draw[black, thick] (-1, -1) -- (-1, 1) -- (0, 2) -- (0,0.5) node[anchor = west]{$\strut\;\TIR$} -- (0,1) -- (0, 0) -- (-1, -1);
            \draw[violet, ultra thick] (0,2) node[anchor = west, black]{$\;\strut\cN\boxtimes_\cC\cM$} -- (0,0);
        \end{scope}
    \end{tikzpicture}
    \caption{Sandwiching the 2-categories of interfaces $\cN$ and $\cM$ together balances their shared $\cC$-module action, producing a 2-category of topological boundaries in the IR given by $\cN\boxtimes_\cC\cM$.}
    \label{fig:symtft-3-boundaries}
\end{figure}
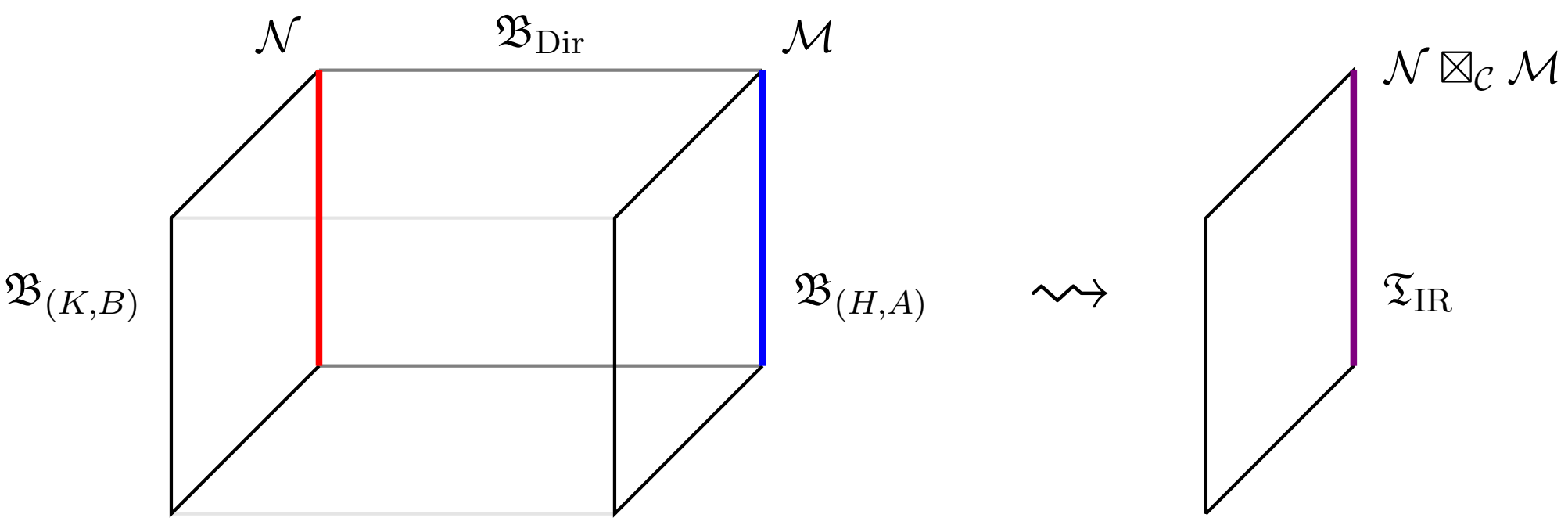

\ni One way to understand what TFT this process describes is to insert a copy of the Dirichlet boundary that connects the symmetry and relative boundaries, as illustrated in figure~\ref{fig:symtft-3-boundaries}. This way when we close the interval, the resulting TFT now has a 2-category of boundaries labeled by pairs of objects in $\cN$ and $\cM$. Now notice that given a defect $X\in \cC$ inserted on $\fB_{\rm Dir}$, we have the choice to collide it with either $\cN$ from the right, or $\cM$ from the left. It hence follows that the linear 2-category of boundary conditions is equivalent to the \emph{balanced} tensor product $\cN\boxtimes_\cC\cM$~\footnote{This is yet another result obtainable from the cobordism hypothesis: the composition of boundaries thought of as 1-morphisms in $4\Vect$ is given by the balanced bimodule product.}.
This 2-category of boundaries should be thought of as a module 2-category over the $2\Vect$ symmetry appearing in the trivial 3d TFT. Following similar logic to before then, we can hence identify $\CIR$ with the category of linear endomorphisms
\begin{equation}\label{eq:cir}
    \CIR \simeq \End_{2\Vect}(\cN\boxtimes_\cC\cM)\, .
\end{equation}
To compute the balanced product $\cN\boxtimes_\cC \cM$, we must first build $\cN\boxtimes \cM$, which is bigraded by $K\backslash G\times G/H$ and of the form
\begin{equation}
    \cN\boxtimes \cM  = \bigoplus_{\bar\imath,\, i}\left( \mathsf{RMod}(B_1)\right)_{\,\bar\imath} \boxtimes\left(\mathsf{LMod}(A_1)\right)_i\simeq \bigoplus_{\bar\imath,\, i}\left(\mathsf{LMod}(\bar{B_1}\boxtimes A_1)\right)_{\,\bar\imath,\, i}\, .
\end{equation}
We denote objects with bigrading $(\bar\imath, i)\in K\backslash G\times G/H$ by
\begin{equation}
    \left(M, \bar\imath, i\right) \in \cN\boxtimes \cM\, ,
\end{equation}
where $M \in \mathsf{LMod}(\bar{B_1}\boxtimes A_1)$. There is then a combined action from $g \in G$ given by
\begin{equation}
    g \triangleright\left(M, \bar\imath, i\right) = \left((\bar{B_{r_{\,\bar\imath}(g^{-1})}}\boxtimes A_{l_i(g)})\boxtimes_{\,\bar{B_1}\boxtimes A_1} M, \,\bar\imath\triangleleft g^{-1},\, g\triangleright i\right)\, ,
\end{equation}
where $l_i(g)$ and $r_{\bar\imath}(g)$ are the little group elements associated to cosets $i$ of $H$ and $\bar\imath$ of $K$ respectively. 
This corresponds physically to acting simultaneously on $\cN$ from the right with $g^{-1}$ and $\cM$ from the left with $g$, which should leave the boundary of $\TIR$ invariant. 
Thus we see that objects of $\cN\boxtimes_\cC\cM$ are hence given by $G$-equivariant objects of $\cN\boxtimes\cM$ with respect to this combined action.

\newl Since the action of $G$ on $G/H$ is transitive, inside any $G$-orbit of simple objects in $\cN\boxtimes\cM$ we can pick a representative object with the trivial coset $i = H$. 
This representative is not unique however, since we can still act with the stabilizer $H$, which identifies all the gradings by $\bar\imath$ that lie in the same $(K,H)$-double coset. 
Once we fix a double coset with representative $g \in G$, the remaining data is a $H\cap K^g$-equivariant (left) module category over $\bar{B_1}\boxtimes A_1$.
Notice the action of $H\cap K^g$ on this choice of representative is encoded by the faithfully graded fusion category
\begin{equation}
    [\bar{B}\boxtimes A]_{H\cap K^g} = \bigoplus_{h \in H\cap K^{g}} \bar{B_{{}^{g}(h)^{-1}}}\boxtimes A_{h}\, ,
\end{equation}
and so it follows from~\eqref{eq:h-eq-vs-graded-ext} that this remaining data is equivalent to a choice of (left) module category over $[\bar{B}\boxtimes A]_{H\cap K^g}$.
As for the components of $\CUV$ in section~\ref{subsec:uv-symmetry}, for $g\neq1$ and non-trivial $\pi$ the two factors are twisted by $\pi$ restricted along two different embeddings of $H\cap K^g$ into $G$, and the associator of $[\bar{B}\boxtimes A]_{H\cap K^g}$ is understood to be corrected by the corresponding 3-cochain $\xi_g$ of \eqref{eq:xi-g}. For $g=1$ the twists $\pi|_{H\cap K}$ of $A$ and $\pi|_{H\cap K}^{-1}$ of $\bar{B}$ cancel on the nose, so $[\bar{B}\boxtimes A]_{H\cap K}$ is an ordinary fusion category with no correction. In particular, none of the single-vacuum formulas below are affected.

\newl To get the full spectrum of boundaries, we enumerate over all choices of $(K,H)$-double coset to obtain
\begin{equation}\label{eq:boundary-cat}
    \cN\boxtimes_\cC\cM \simeq \bigoplus_{HgK \in H\backslash G/ K} \mathsf{LMod}\left([\bar{B}\boxtimes A]_{H\cap K^{ g}}\right)\, .
\end{equation}
Notice that because $\cN\boxtimes_\cC\cM$ labels the boundaries of $\TIR$, its connected components labeled by double cosets in $H\backslash G/K$ are in 1-1 correspondence with the vacua of the IR theory. To obtain our case of interest where we have a single vacuum in the IR, we must restrict to pairs $(H,K)$ admitting a single double coset,
\begin{equation}\label{eq:single-vacuum}
    H\backslash G/K = \{G\} \quad\Longleftrightarrow\quad HK = G\, .
\end{equation}
We emphasize, in keeping with the discussion in the introduction, that \eqref{eq:single-vacuum} is a statement about \emph{connectedness}---a single sector of genuine local operators---and nothing more.
In particular it does not by itself assert that the IR carries no non-trivial topological lines; determining the surviving line category requires the separate computation of $\Omega\CIR$ below.

\newl Restricting ourselves to the single vacuum condition of $G= HK$, identity~\eqref{eq:boundary-cat} reduces to
\begin{equation}
    \cN\boxtimes_\cC\cM \simeq \mathsf{LMod}\left([\bar{B}\boxtimes A]_{H\cap K}\right)\, ,
\end{equation}
and from our earlier assertion~\eqref{eq:cir} we obtain
\begin{equation}
    \CIR \simeq \mathsf{Bimod}\left([\bar{B}\boxtimes A]_{H\cap K}\right)\, , \qquad \Omega\CIR \simeq \cZ\left([\bar{B}\boxtimes A]_{H\cap K}\right)\, .
\end{equation}
Notice that because $[\bar{B}\boxtimes A]_{H\cap K}$ is (faithfully) graded by $H\cap K$, the IR 1-form symmetry $\Omega\CIR$ carries a grading by conjugacy classes of $H\cap K$, and the component with grading $[1]$ is precisely
\begin{equation}\label{eq:OmegaCIR-neutral}
    \Omega\CIR_{[1]} \simeq \left(\cZ(\bar{B_1})\boxtimes\cZ(A_1)\right)^{H\cap K}\, ,
\end{equation}
since the neutral component of $[\bar{B}\boxtimes A]_{H\cap K}$ is $\bar{B_1}\boxtimes A_1$. It contains $\cZ(A_1)^{H\cap K}$ as the subcategory of objects whose $\cZ(\bar{B_1})$ factor is the unit.
Notice also, that when $K = G$ and $B_1\simeq\Vect$, this matches $\Omega\CUV$. This is not accidental, since this describes a phase in which all the UV $\Rep(H)$ lines survive to the IR, and so the RG functor is just inclusion; for $B_1\not\simeq\Vect$ the physical boundary supplies additional neutral lines $\cZ(\bar{B_1})$ in the IR.
Before moving on to computing the RG 2-functor for general $H$ and $K$, we make one final remark: the computation of $\Omega\CIR$ above agrees with the findings of section~\ref{subsec:symtft-rg-flow}, in that to describe a genuine fiber 2-functor we require $\Omega\CIR \simeq \Vect$ which, since we have already assumed $G=HK$, holds if and only if $A_1 \simeq B_1\simeq \Vect$ and $H\cap K= \{1\}$.

\begin{examplebox}[The single-vacuum IR symmetry $\CIR$ with $\fB_{(\bbZ_3, \omega)}$][label=ex:S3-2mod-ir]
   Earlier, in example~\ref{ex:S3-2mod-uv}, we produced a UV symmetry $\CUV$ living on the boundary $\fB_{(\bbZ_2, A)}$ to the 4d $S_3$ Dijkgraaf--Witten TQFT. We now return to this example to compute $\CIR$ for a phase that has a single vacuum.
   
   \newl For this example, it is sufficient to pick the topological physical boundary $\fB_{(K,B)}$ with
\begin{equation}\label{eq:S3-IR-choice}
    K = \bbZ_3 \subset S_3\, ,\qquad B =\Vect^\omega_{\bbZ_3}\, ,
\end{equation}
where $\omega$ is a 3-cocycle representing a non-trivial class $[\omega]\in \operatorname{H}^3(\bbZ_3;U(1))\simeq\bbZ_3$. This is the 2-categorical counterpart of the boundary $\fB_{\Z_3}$ paired with $\fB_{\Z_2}$ in example~\ref{ex:S3 vacua}.

\newl With $H = \bbZ_2$ and $K = \bbZ_3$, we have that $S_3 = HK$ and thus an IR symmetry $\CIR$ whose surfaces are all condensation. Moreover, this choice of subgroups also has that $H\cap K = \{1\}$, meaning lines in this TQFT are given by
\begin{equation}\label{eq:S3-OmegaCIR}
    \Omega\CIR \simeq \cZ(\Vect_{\bbZ_2})\, ,
\end{equation}
identifying it with the 3d $\bbZ_2$ Dijkgraaf--Witten theory whose simple lines we continue to denote by $\{1, e, m, em\}$. The same result is derived from the Lagrangian algebras of the two boundaries in example~\ref{ex:S3-2lag-ir}.
\end{examplebox}

\subsubsection{The RG 2-Functor \texorpdfstring{$\cF : \CUV \to \CIR$}{F : CUV -> CIR}}\label{subsec:rg-functor}
With the UV and IR symmetries identified, the RG 2-functor itself is now essentially forced.
Recall that the RG 2-functor $\cF : \CUV \to \CIR$ is obtained by inserting a $\CUV$ defect on the symmetry boundary $\fB_{(H, A)}$, and then sandwiching with IR relative boundary $\fB_{(K, B)}$ to produce a defect in $\CIR$. This is illustrated below in figure~\ref{fig:symtft-rg-functor-3}.

\newl Starting with the simpler case of $H = \{1\}$, we have a UV 1-form symmetry $\Omega\CUV \simeq\cZ(A_1)$, and an IR phase with $G/K$ vacua, each containing genuine lines labeled by objects of $\cZ(\bar{B_1}\boxtimes A_1)$. 
In this example the UV 1-form symmetry is non-degenerate and as such there are no Wilson lines. The RG 2-functor at the level of the 1-form symmetry is simply the braided monoidal inclusion
\begin{equation}   
\Omega\cF_{H = \{1\}} \, :\, \cZ(A_1) \hookrightarrow \cZ(\bar{B_1}\boxtimes A_1)\, .
\end{equation}
As we have now discussed at length, switching $H$ back on has the effect of gauging. The UV 1-form symmetry becomes $\Omega\CUV \simeq \cZ(A_1)^H$. Assuming $G = HK$ so that the IR has a single vacuum; the IR 1-form symmetry becomes
\begin{equation}
    \Omega\CIR \simeq  \cZ\left([\bar{B}\boxtimes A]_{H\cap K}\right)\, .
\end{equation}
Matching equivariance data from the UV and IR, we can see that the RG functor $\Omega\cF$ comes from a composition of braided monoidal functors
\begin{equation}
    \cZ(A_1)^H \longrightarrow \cZ(A_1)^{H\cap K} \hookrightarrow \left(\cZ(\bar{B_1})\boxtimes\cZ(A_1)\right)^{H\cap K} \hookrightarrow \cZ\left( [\bar{B}\boxtimes A]_{H\cap K} \right)\, ,
\end{equation}
where the first step forgets all but the $H\cap K$-equivariant data, the second tensors with the unit of $\cZ(\bar{B_1})$, and the third is just the inclusion of $\Omega\CIR_{[1]}$ from \eqref{eq:OmegaCIR-neutral}, with grading concentrated on the conjugacy class of the identity.

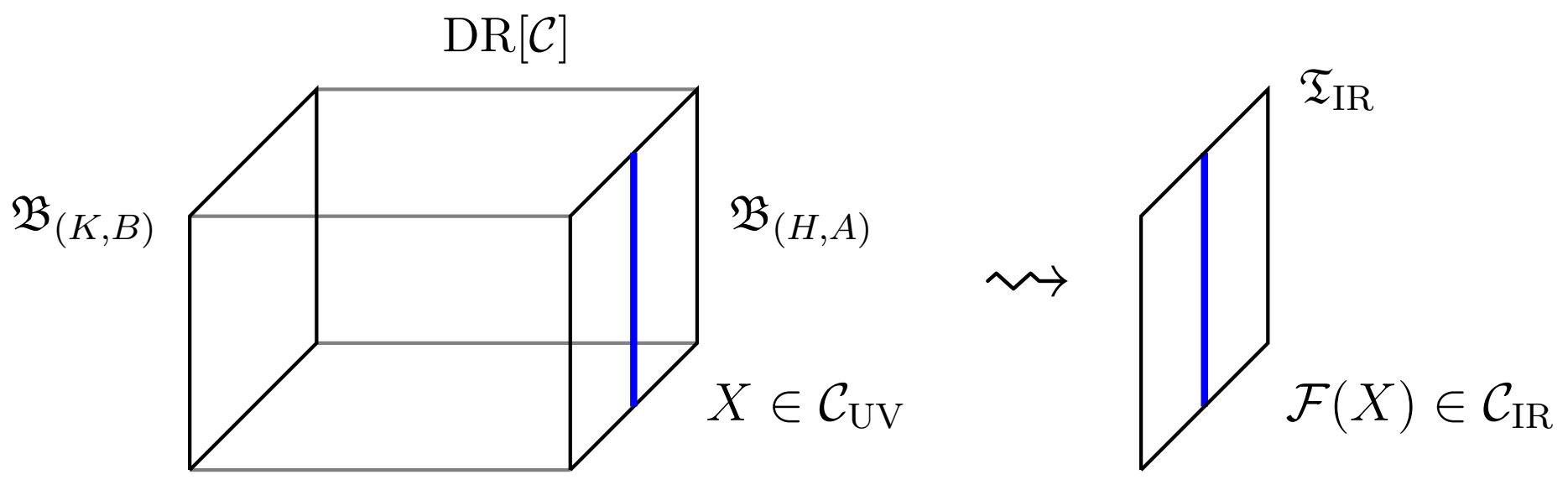
\begin{figure}[H]
    \centering
    \begin{tikzpicture}[scale = 1]
    
        \draw[gray, thick] (0,2) -- (1.5,2) node[anchor = south, black]{$\strut \mathrm{DR}[\cC]$} -- (3, 2);
        \draw[gray, thick] (-1,1) -- (2, 1);
        \draw[gray, thick] (0,0) -- (3, 0);
        \draw[gray, thick] (-1,-1) -- (2, -1);
        
        \draw[black, thick] (-1, -1) -- (-1, 1) node[anchor = east]{$\fB_{(K, B)}\;\strut$} -- (0, 2) -- (0, 0) -- (-1, -1);
        \begin{scope}[shift={(3,0)}]
        \draw[black, thick] (-1, -1) -- (-1, 1) -- (0, 2) -- (0, 1) node[anchor = west]{$\strut\;\fB_{(H, A)}$} -- (0, 0) -- (-1, -1);
        \end{scope}
        \draw[blue, ultra thick] (2.5, -0.5) node[black, anchor = west]{\large$\strut\quad X\in \CUV$} -- (2.5, 1.5);

        \draw[black] (5.5, 0.5) node{\huge$\strut\rightsquigarrow$};

        \begin{scope}[shift = {(7.5, 0)}]
        \draw[blue, ultra thick] (-0.5, -0.5) node[black, anchor = west]{\large$\strut\quad\,\cF(X)\in \CIR$} -- (-0.5, 1.5);
            \draw[black, thick] (-1, -1) -- (-1, 1) -- (0, 2) node[anchor = west]{$\strut\;\TIR$} -- (0, 0) -- (-1, -1);
        \end{scope}
    \end{tikzpicture}
    \caption{Interval compactification maps a UV defect $X \in \CUV$ to $\cF(X) \in \CIR$.}
    \label{fig:symtft-rg-functor-3}
\end{figure}
\ni In our module 2-category formalism, the left hand side of figure~\ref{fig:symtft-rg-functor-3} corresponds to constructing a linear endomorphism of $\cN\boxtimes\cM$ by pairing an endomorphism $X\in\CUV$ of $\cM$, with the identity on $\cN$. To get the right picture, we must balance that product of endomorphisms, giving the 2-functor
\begin{equation}\label{eq:RG-functor}
\begin{split}
    \cF \; : \; \End_{\cC}(\cM) &\longrightarrow \End_{2\Vect}\big(\cN \boxtimes_{\cC}{\cM}\big)\, ,\\
    X &\longmapsto \Id_{\cN} \boxtimes_\cC X\, ,
\end{split}
\end{equation}
which is well-defined precisely because $X$ is $\cC$-linear. This 2-functor is identified with the RG 2-functor via our previous identifications $\CUV \simeq \End_\cC(\cM)$ and $\CIR \simeq \End_{2\Vect}(\cN\boxtimes_\cC\cM)$, so we continue to denote it by $\cF$.
Monoidality is transparent in this presentation.
Since the balanced 2-Deligne tensor product is functorial in each variable and $\Id_{\cN}$ is strictly idempotent under composition, we have
\begin{equation}
\begin{split}
    \cF(X\circ Y) &= \Id_\cN\boxtimes_{\cC}(X\circ Y)\\
    &\simeq \big(\Id_{\cN}\boxtimes_{\cC} X\big)\circ\big(\Id_{\cN}\boxtimes_{\cC} Y\big)\\
    &= \cF(X)\circ\cF(Y)\, ,
\end{split}
\end{equation}
together with $\cF(\Id_\cM) \simeq \Id_{\cN\boxtimes_\cC\cM}$, and the coherence data of these equivalences is inherited from the associator of the balanced product. Hence $\cF$ is a tensor 2-functor, which is exactly the RG 2-functor \eqref{eq:RG functor} of the introduction. Under the single-vacuum condition $HK = G$ of \eqref{eq:single-vacuum}, the target is a connected fusion 2-category and $\cF$ is a generalized fiber 2-functor in the sense fixed there.

\newl Translating this equivalence into a computation at the level of objects is straightforward.
First take $\underline{U}\in \CUV$ a $\cC$-linear endomorphism of $\cM$, and $\underline{M}\in \cN\boxtimes_\cC\cM$ a boundary for $\TIR$. 
Acting on $\underline{M}$ with $\underline{U}$ will produce a new boundary, and incorporating the $\cC$-linear structure of $\underline{U}$ will allow us to extract the IR defect $\cF(\underline{U})$ it corresponds to in the IR.

\newl Recall that a simple object in $\CUV$ is labeled by an $H$-double coset $HgH$ together with an irreducible $H_g$-equivariant $A_1$-bimodule $U$, where $H_g=H\cap{}^{g}H$ is the stabilizer of the pair of cosets $(H,gH)$. 
The underlying linear endomorphism of $\cM$ is then given by
\begin{equation}
    \underline{U} = \bigoplus_{\alpha \in G/H_g} (g_\alpha H, g_\alpha gH; U_\alpha)\, ,
\end{equation}
where we define
\begin{equation}
     U_{H_g} := U\, ,\qquad U_\alpha := A_{l_H(g_\alpha)} \boxtimes_{A_1} U \boxtimes_{A_1} A^{op}_{l_{gH}(g_\alpha)}\, ,
\end{equation}
to ensure that $\underline{U}$ admits a $\cC$-linear structure.
Recall also that when $G=HK$, a simple object of $\cN\boxtimes_\cC\cM$ is labeled by an irreducible (left) module $M$ over $[\bar{B}\boxtimes A]_{H\cap K}$. 
The underlying object in $\cN\boxtimes\cM$ is then given by
\begin{equation}
    \underline{M} = \bigoplus_{\beta \in G/H\cap K} (M_\beta,\, Kg_\beta^{-1},\, g_\beta H)\, ,
\end{equation}
where we define
\begin{equation}
      M_{H\cap K} := M\, ,\qquad M_\beta := \left( \bar{B_{r_{K}(g_\beta^{-1})}}\boxtimes A_{l_{H}(g_\beta)} \right) \boxtimes_{\bar{B_1}\boxtimes A_1} M\, ,
\end{equation}
to ensure $\underline{M}$ admits a $\cC$-balanced structure.
Tensoring $\underline{U}$ with the identity on $\cN$ and subsequently evaluating that endomorphism on $\underline{M}$, we find
\begin{equation}
    \left(\Id_\cN\boxtimes\underline{U}\right)(\underline{M}) = \bigoplus_{\alpha\, ,\beta|\, g_\alpha g H = g_\beta H} \left((\bar{B_1}\boxtimes U_\alpha)\boxtimes_{\bar{B_1}\boxtimes A_1} M_\beta,\,\,Kg_\beta^{-1},\,\, g_\alpha H\right)\, .
\end{equation}
To understand what object in $\cN\boxtimes_\cC\cM$ this must descend to, it is sufficient to restrict our focus to a representative object with trivial bigrading $(K, H)$.
Expanding each of the $M_\beta$ and factoring out $M$ yields the $(\bar{B_1}\boxtimes A_1)$-bimodule
\begin{equation}
    \bigoplus_{\substack{g_\alpha \in H\, ,g_\beta \in K\\ g_\alpha g H = g_\beta H}}\left(\bar{B_{g_\beta^{-1}}}\boxtimes \left(U_\alpha \boxtimes_{A_1} A_{l_{H}(g_\beta)} \right)\right)\, ,
\end{equation}
where we used $r_K(g_\beta^{-1})=g_\beta^{-1}$ for $g_\beta\in K$, choosing the representative $g_K=\id$ of the trivial coset.
This is not enough to identify an IR defect however, since we may recall that objects of $\CIR$ are actually labeled by $[\bar{B}\boxtimes A]_{H\cap K}$-bimodules. This apparent discrepancy is because so far we have only treated $M$ as a $(\bar{B_1}\boxtimes A_1)$-module, and is easily resolved by pre-composing with the forgetful functor
\begin{equation}
    [\bar{B}\boxtimes A]_{H\cap K} \boxtimes_{[\bar{B}\boxtimes A]_{H\cap K}} (-) \, :\, \mathsf{LMod}([\bar{B}\boxtimes A]_{H\cap K}) \longrightarrow \mathsf{LMod}(\bar{B_1}\boxtimes A_1)\, ,
\end{equation}
and using that the tensor product of $\bar{B}$ is reversed, $\bar{B_x}\boxtimes_{\bar{B_1}}\bar{B_y}\simeq\bar{B_{yx}}$, results in an object $\cF(\underline{U})\in\CIR$ given by
\begin{equation}
\begin{split}
    \cF(\underline{U}) &\simeq \bigoplus_{\substack{h\in H\cap K\\g_\alpha \in H\, ,g_\beta \in K\\ g_\alpha g H = g_\beta H}}\left( \bar{B_{(g_\beta h)^{-1}}} \boxtimes \left( U_\alpha \boxtimes_{A_1} A_{l_H(g_\beta h)} \right) \right)\\
    &\simeq \bigoplus_{\substack{g_\alpha \in H\, ,k \in K\\ g_\alpha g H = k H}}\left( \bar{B_{k^{-1}}} \boxtimes \left( U_\alpha \boxtimes_{A_1} A_{l_H(k)} \right) \right)\, ,
\end{split}
\end{equation}
where in the second step we used that $k=g_\beta h$ runs over $K$ exactly once as $g_\beta$ runs over the representatives of $K/(H\cap K)$ and $h$ over $H\cap K$.
This naturally has a right $[\bar{B}\boxtimes A]_{H\cap K}$-module action, given by the right action on itself: $\bar{B_{h^{-1}}}\boxtimes A_h$ sends the summand labeled by $(\alpha,k)$ to the one labeled by $(\alpha,kh)$.

\newl To obtain the left module action, pick $h \in H\cap K$ and act with $\bar{B_{h^{-1}}}\boxtimes A_h$. On the $\bar{B}$ factor this gives $\bar{B_{h^{-1}}}\boxtimes_{\bar{B_1}}\bar{B_{k^{-1}}}\simeq\bar{B_{(hk)^{-1}}}$. On the $A$ factor we use the $\cC$-linear structure \eqref{eq:c-linearity-0} of $\underline{U}$, with $i=H$ and $j=kH$, together with the identity $l_{kH}(h)\,l_H(k)=l_H(hk)$, to obtain
\begin{equation}
\begin{split}
    A_h \boxtimes_{A_1} U_\alpha \boxtimes_{A_1} A_{l_{H}(k)} &\simeq U_{h\cdot \alpha} \boxtimes_{A_1} A_{l_{kH}(h)} \boxtimes_{A_1} A_{l_{H}(k)}\\
    &\simeq U_{h\cdot \alpha} \boxtimes_{A_1} A_{l_{H}(hk)}\, .
\end{split}
\end{equation}
Since $hg_\alpha gH=hkH$, the left action of $\bar{B_{h^{-1}}}\boxtimes A_h$ sends the summand labeled by $(\alpha,k)$ to the one labeled by $(h\cdot\alpha,hk)$. It therefore permutes the summands and leaves the total object invariant, completing the $[\bar{B}\boxtimes A]_{H\cap K}$-bimodule structure on $\cF(\underline{U})$.

\begin{examplebox}[The RG 2-functor from the 4d $S_3$ SymTFT][label=ex:S3-2mod-rg]
    We have now constructed both a UV symmetry (example~\ref{ex:S3-2mod-uv}) and an IR TQFT with a single vacuum (example~\ref{ex:S3-2mod-ir}), starting from 4d $S_3$ Dijkgraaf--Witten theory. The RG 2-functor that connects these two points gives a generalized fiber 2-functor, which we will now describe.

    \newl For these choices, the RG 2-functor
    \begin{equation}\label{eq:S3-2RG}
        \cF \; :\; \CUV \to \CIR
    \end{equation}
    sends both connected components of $\CUV$ to the single connected component of $\CIR\simeq\mathsf{Bimod}(\Vect_{\bbZ_2})$. More specifically, since $H\cap K=\{\id\}$, the formula for $\cF(\underline{U})$ above forgets the $H_g$-equivariant data of a simple object $(HgH,U)$ and sums the components $U_\alpha$ of $\underline{U}$ that are reached from $K$, i.e.\ those with $g_\alpha\in\bbZ_2$ and $g_\alpha gH=g_\beta H$ for some $g_\beta\in\bbZ_3$. For the objects of example~\ref{ex:S3-2mod-uv}, 
    \begin{equation}
        \cF(1)\simeq1\, ,\qquad \cF(\chi)\simeq1\oplus1\, ,\qquad \cF(X)\simeq1\oplus1\, ,
    \end{equation}
    and more generally $\cF(X_U)\simeq U\oplus\big(A_{-1}\boxtimes_{A_1}U\boxtimes_{A_1}A_{-1}\big)$ for the object $X_U$ with $A_1$-bimodule $U$ in the component of $X$. These images are computed from the Lagrangian algebras in example~\ref{ex:S3-2lag-rg}, see \eqref{eq:S3-FX} and \eqref{eq:S3-FXU}, where the twisted sector of $X$ and the role of $\omega$ are also discussed. The 1-categorical counterpart is the fiber functor of $\Rep(S_3)$ in example~\ref{ex:S3-RG}, with $\cF(E)=2$.
    
    \newl Evaluating $\cF$ on loop objects yields the braided tensor functor
\begin{equation}\label{eq:S3-OmegaF}
    \Omega\cF \; :\; \cZ(\Vect_{\bbZ_2})^{\bbZ_2} \to \cZ(\Vect_{\bbZ_2})\, ,
\end{equation}
that forgets the $\bbZ_2$-equivariant data, sending
\begin{equation}
    1^\pm \mapsto 1\, ,\qquad
    em^\pm \mapsto em\, ,\qquad
    \eta \mapsto e \oplus m\, .
\end{equation}
It is surjective, so this phase is minimal in the sense of section~\ref{subsec:minimality} below.
\end{examplebox}


\subsubsection{Minimality}\label{subsec:minimality}
We now return to one of the questions laid out in the introduction concerning the existence of a \emph{minimal} IR phase, by which we mean a gapped phase with a single vacuum, and the smallest amount of emergent symmetry.

\newl To begin, we first ask what it would mean for an IR phase to have \emph{no} emergent 1-form symmetry. In that case, every line appearing in $\Omega\CIR$ should be accessible via the RG functor on lines
\begin{equation}
    \Omega\cF \; : \; \Omega\CUV \longrightarrow \Omega\CIR\, .
\end{equation}
That is to say, $\Omega\cF$ is \emph{surjective}: every object $Y\in \Omega\CIR$ appears as a sub-object of $\cF(X)\in \Omega\CIR$, where $X\in \Omega\CUV$. Recall that the M\"uger center of $\Omega\CUV$ is given by
\begin{equation}
    \cZ_2(\Omega\CUV) \simeq \Rep(H)\, ,
\end{equation}
and its modular closure by
\begin{equation}\label{eq:modular-closure}
    \Omega\CUV \boxtimes_{\Rep(H)} \Vect \simeq \cZ(A_1)\, .
\end{equation}
We now use the description of surjective braided tensor functors~\cite{DGNO, Davydov:2010wpa}. 
If $F:\cD\to\cE$ is a surjective braided tensor functor between braided fusion categories, then $\cE\simeq\cD_{\cQ}$ is the category of (right) modules over the connected \'etale algebra $\cQ=F^R(1)$, where $F$ is the free-module functor and $F^R$ is its right adjoint. 
Since $\cE$ is braided, every $\cQ$-module is local, which forces $\cQ\in\cZ_2(\cD)$, so $F$ is a de-equivariantization. 
For $\cD=\Omega\CUV$, the connected \'etale algebras in $\cZ_2(\cD)\simeq\Rep(H)$ are the function algebras $\Fun(H/H')$ over $H$-cosets, for subgroups $H'\subseteq H$, and the corresponding de-equivariantization is $\cZ(A_1)^{H'}$. 
This tells us
\begin{equation}\label{eq:surjective-br-mon-fun}
  \Omega\cF \text{ is surjective} \,\Rightarrow \, \Omega\CIR \simeq \cZ(A_1)^{H'}\, , 
\end{equation}
where $H'\subseteq H$ is a subgroup, and also that in such cases, $\Omega\cF$ is (up to the equivalence~\eqref{eq:surjective-br-mon-fun}) equivalent to the forgetful functor
\begin{equation}
    \cZ(A_1)^{H} \to \cZ(A_1)^{H'}\, .
\end{equation}
This places a strong condition on when the RG functor can be surjective: it implies that $\cZ_2(\Omega\CIR) \simeq \Rep(H')$, but because $\Omega\CIR$ is non-degenerate, this forces $H' = \{1\}$ and $\Omega\CIR\simeq \cZ(A_1)$.
Comparing this condition with the general form
\begin{equation}
    \Omega\CIR \simeq \cZ\left( [\bar{B}\boxtimes A]_{H\cap K} \right)\, ,
\end{equation}
we see this is only possible when, in addition to our initial assumption that $G=HK$, we also need $B_1= \Vect$, $K$ is non-anomalous, and $H\cap K = \{1\}$.
Indeed, comparing Frobenius--Perron dimensions,
\begin{equation}
\begin{split}
    \operatorname{FPdim}\,\cZ\left( [\bar{B}\boxtimes A]_{H\cap K} \right) &= \big(|H\cap K|\,\operatorname{FPdim}(A_1)\operatorname{FPdim}(B_1)\big)^2\, ,\\
    \operatorname{FPdim}\,\cZ(A_1) &= \operatorname{FPdim}(A_1)^2\, ,
\end{split}
\end{equation}
so the two can only agree if $|H\cap K|\operatorname{FPdim}(B_1) = 1$, i.e.\ $H\cap K=\{1\}$ and $B_1\simeq\Vect$. Then $B\simeq\Vect_K^\psi$ is pointed, and its associator $\psi$ trivializes $\pi|_K$. Conversely, under these conditions $[\bar{B}\boxtimes A]_{H\cap K}\simeq A_1$ and $\Omega\cF$ is the forgetful functor $\cZ(A_1)^H\to\cZ(A_1)$.
Notice that when $H$ is non-anomalous and $A_1 = \Vect$, this precisely reproduces the conditions for the existence of a genuine fiber 2-functor discussed in section~\ref{subsec:symtft-rg-flow}.

\newl When the above conditions can be satisfied, the IR phase described above is the minimal one, but when it is not possible to satisfy them, the problem of finding a minimal IR phase becomes much more complicated in general because there are multiple directions to check for the next best candidate:
\begin{itemize}
    \item If non-anomalous $K$ satisfying $G=HK$ exist, we could focus on maintaining $B_1 = \Vect$ and minimizing $H\cap K$.
    
    \item If subgroups $K$ exist such that $G=HK$ factorizes, we could focus on maintaining $H\cap K = \{1\}$ and minimizing $B_1$.
\end{itemize}
When both of these strategies are available, it is not clear in general which we should prioritize, and even if it were the case that only one of the above strategies is possible, it is not clear that it will produce a unique minimal IR.

\subsection{Lagrangian Algebras}\label{subsec:lagrangian-algebras}
The boundaries $\fB_{(H,A)}$ of section~\ref{subsec:module-2-cats} can also be described from the bulk, by the surfaces of the SymTFT that condense on them.
This is the 2-categorical counterpart of section~\ref{subsec:warm-up-lagrangian}, and we follow the same steps: we describe the bulk surfaces, read off the Lagrangian algebra of each boundary from its construction, determine which bulk surfaces end on it, and then recover $\CUV$, $\CIR$ and the RG 2-functor.
The payoff is a simple form of the latter: it sends a surface on the symmetry boundary to the free module it generates over the Lagrangian algebra of the physical boundary.
Each example below has a 1-categorical counterpart in section~\ref{subsec:warm-up-lagrangian}, to which we point along the way, and the precise definitions of algebras, modules and Lagrangian algebras in 2-categories are collected in appendix~\ref{app:2cat}.
\newl {\bf Topological surfaces and lines of the SymTFT.} Set
\begin{equation}
    \fZ := \cZ_1(2\Vect_G^\pi)\, ,
\end{equation}
the braided fusion 2-category of topological surface defects of the 4d Dijkgraaf--Witten SymTFT with finite group symmetry $G$ and a 4-cocycle twist $\pi$.
We first make the description of the bulk surfaces in section~\ref{subsec:symtft-defects-and-boundaries} explicit, in parallel with the lines of $\cZ(\Vec_G)$ in section~\ref{subsec:warm-up-symtft}. 
For simplicity we take $\pi$ trivial, which covers the running example for $G = S_3$; a non-trivial $\pi$ twisting the coherence data below has been discussed in \cite{Kong:2019brm}.
\begin{itemize}
\item \emph{Objects.} A surface is a $G$-crossed finite semisimple category, i.e., a $G$-graded finite semisimple category $X=\bigoplus_{x\in G}X_x$ together with a $G$-action by functors $T_g:X_x\to X_{gxg^{-1}}$ and coherent isomorphisms 
\begin{equation}
\begin{gathered}
T_gT_h\simeq T_{gh} \, ,
\\
T_1\simeq \Id\,,
\end{gathered}
\end{equation} which satisfy higher associativity and unitality coherence conditions.
This categorifies the half-braiding \eqref{eq:S3-half-braiding}: $\fZ$ is the 2-category of $G$-equivariant 2-vector bundles over $G$, with $G$ acting by conjugation, and $x$ is the flux of the surface.
\item \emph{Morphisms.} A 1-morphism, i.e.\ a line between two surfaces, is a $G$-crossed functor, and a 2-morphism, i.e.\ a point-like junction between two such lines, is a $G$-crossed natural transformation.
\item \emph{Fusion and braiding.} As in \eqref{eq:S3-center-fusion} and \eqref{eq:S3-center-braiding},
\begin{equation}\label{eq:Z-fusion}
\begin{gathered}
(X\otimes Y)_z=\bigoplus_{xy=z}X_x\boxtimes Y_y\,,\\
\beta_{X,Y}:\ X_x\boxtimes Y_y\to Y_{xyx^{-1}}\boxtimes X_x\,,\quad s\boxtimes t\mapsto T_x(t)\boxtimes s\,,
\end{gathered}
\end{equation}
with the diagonal $G$-action. The unit $\bone$ is $\Vect$ neutrally graded, with the trivial action.
\item \emph{Lines out of the unit.} A 1-morphism $\bone\to X$ picks an object of flux $1 \in G$ together with a $G$-equivariant structure, so
\begin{equation}\label{eq:Z-hom-from-unit}
\Hom_\fZ(\bone,X)\simeq(X_1)^G\, .
\end{equation}
For $X=\bone$ this gives the Wilson lines, $\Omega\fZ\simeq (\Vect)^G \simeq \Rep(G)$.
\item \emph{Simple surfaces.} By the orbit-stabilizer principle \eqref{eq:S3-orbit-stabiliser} one categorical level up, a simple surface is a conjugacy class $[x]$ together with a simple 2-representation of the centralizer $C_G(x)$ on the fiber $X_x$, i.e.\ a subgroup $L\subseteq C_G(x)$ up to conjugation and a class in $\operatorname{H}^2(L;U(1))$ \cite{Kong:2019brm}. We write $S_{[x],L}$ when this class is trivial: its fiber over $x$ is the permutation 2-representation $\Vect(C_G(x)/L)$. Physically, $S_{[x],L}$ is the magnetic surface of flux $[x]$ on which the gauge group is broken down to $L$. In particular $S_{[\id],L}$ is the condensation of the Wilson lines $\Fun(G/L)=\mathrm{Ind}_L^G(1)$, and $S_{[\id],G}=\bone$. By 2-categorical Frobenius reciprocity,
\begin{equation}\label{eq:Z-hom-from-simple}
\Hom_\fZ\big(S_{[x],L},X\big)\simeq(X_x)^L\,,
\end{equation}
the category of $L$-equivariant objects of the fiber $X_x$.
\end{itemize}

\newl {\bf Lagrangian algebras.} In section~\ref{subsec:warm-up-lagrangian}, a topological boundary of $\cZ(\Vec_G)$ was described by the line $\cL$ collecting the bulk lines that end on it, with a product that fuses their endpoints.
One dimension higher, a boundary is obtained by condensing bulk surfaces, and a single surface does not condense on its own: what proliferates is a \emph{network} of surfaces, glued along lines and points \cite{Gaiotto:2019xmp,Roumpedakis:2022aik,Zhao:2022yaw}.
The data that make such a network consistent are the structure of an algebra in $\fZ$. Compared with section~\ref{subsec:warm-up-lagrangian}, the properties of the product are promoted to structure, carried by the lines and points of the network:
\begin{itemize}
\item \emph{Surface.} The surface $\cL\in\fZ$ is the sum of the bulk surfaces that condense. As a 2-vector bundle, $\cL=\bigoplus_{x\in G}\cL_x$ with a $G$-action $T_g$. 

\item \emph{Fusing two sheets.} Where two sheets of the network meet, they merge back into the network along a line defect located at the seam, which gives a multiplication 1-morphism $m:\cL\otimes\cL\to\cL$, i.e.\ by \eqref{eq:Z-fusion} a family of crossed functors $\cL_x\boxtimes\cL_y\to\cL_{xy}$, $s\boxtimes t\mapsto s\otimes t$. It is the counterpart of the product that fuses endpoints in section~\ref{subsec:warm-up-lagrangian}.

\item \emph{Creating the network.} The network can open up out of the trivial surface. This is a unit 1-morphism $i:\bone\to\cL$, i.e.\ by \eqref{eq:Z-hom-from-unit} an object $1_\cL\in(\cL_1)^G$, the counterpart of the unit $\eta$ one dimension lower.

\item \emph{Local moves.} Three sheets can be merged in two orders, and a sheet can be merged with a freshly created one. In section~\ref{subsec:warm-up-lagrangian} the two results had to be equal. Here they are two different seams, and the network must provide point-like junctions between them, $(s\otimes t)\otimes u\simeq s\otimes(t\otimes u)$ and $1_\cL\otimes s\simeq s\simeq s\otimes1_\cL$. Consistency under further local moves of the network imposes the pentagon and triangle conditions, so that $\cL$ becomes a $G$-crossed monoidal category.

\item \emph{Moving sheets past each other.} Two sheets of the network can be exchanged. A sheet $t$ passing through a sheet $s$ of flux $x$ becomes $T_x(t)$, by the braiding in \eqref{eq:Z-fusion}, so the exchange is a point-like junction $s\otimes t\simeq T_x(t)\otimes s$, i.e.\ a $G$-crossed braiding on $\cL$. It categorifies the notion of commutativity, and makes $\cL$ a braided algebra in $\fZ$ \cite{Decoppet:2023hkr,Xu:2024etale}.
\end{itemize}
Two conditions then single out the networks that produce a topological boundary.
\begin{itemize}
    \item The algebra $\cL$ is \emph{connected} if $1_\cL$ is simple in $(\cL_1)^G$, so that the network can be created from the vacuum in a single way and the boundary is indecomposable; it is also \emph{separable}, so that small bubbles of the network can be shrunk away without creating new topological sectors. These conditions imply $\cL$ is a $G$-crossed braided multifusion category, essentially induced from an $H$-crossed braided fusion subcategory for some subgroup $H\subseteq G$; we will explicitly see this structure in just a moment.
    
    \item It is \emph{Lagrangian} if the condensation is maximal: no non-trivial surface or line of the SymTFT survives in the condensed phase, i.e.\ the 2-category of local $\cL$-modules is trivial \cite{Zhao:2022yaw,Decoppet:2023hkr}. This replaces the dimension condition $\dim\cL=|G|$ of section~\ref{subsec:warm-up-lagrangian}.
\end{itemize}
Separable braided algebras are called \emph{étale}, generalizing anyon condensation in $(2\!+\!1)$ dimensions \cite{Kong:2013aya}, and Lagrangian algebras in $\fZ$ correspond to topological boundaries of the SymTFT, when the bulk can be recovered from the boundary as a Drinfeld center \cite{Zhao:2022yaw,Decoppet:2023hkr,Xu:2024etale}. For a precise definition see definition~\ref{def:lagrangian-algebra}.

\newl {\bf The Lagrangian algebra of $\fB_{(H,A)}$} can be read off from its construction, in the same steps as $\cL_{(H,\psi)}$ in section~\ref{subsec:warm-up-lagrangian}, with the endpoints of lines replaced by the end lines of surfaces.
\begin{enumerate}[label=(\arabic*)]
\item \emph{Dirichlet boundary.} On $\fB_{\rm Dir}$ no bulk surface of non-trivial flux can end: pushed onto the boundary, a surface of flux $x$ continues along it as the symmetry surface $x\in2\Vect_G^\pi$.
\item \emph{Fluxes in $H$ can end.} After stacking $\mathrm{TV}[A_1]$ and gauging the diagonal $H$, for $x\in H$ this surface is identified with the symmetry surface of $\mathrm{TV}[A_1]$ that implements $x$, i.e.\ the invertible surface labeled by the $A_1$-bimodule category $A_x$. Their combination is a gauge surface, so the bulk surface of flux $x$ can now end on $\fB_{(H,A)}$.
\item \emph{End lines.} The surface ends along a line, and the lines along which the surface $A_x$ of $\mathrm{TV}[A_1]$ can end are its twisted lines: the objects of $A_x$ that commute with $A_1$, which form the \emph{relative center}
\begin{equation}
\cZ_{A_1}(A_x) := \Fun_{A_1|A_1}(A_1, A_x) \, .
\end{equation}
More explicitly, an object of $\cZ_{A_1}(A_x)$ is a pair $(s,\gamma^s)$ where $s\in A_x$ and $\gamma^s$ is a natural isomorphism $\gamma^s_a: s\otimes a \simeq a\otimes s$ for all $a\in A_1$, satisfying one of the hexagon conditions.
This ensures that $s$ can consistently end the surface $A_x$ along a line that commutes with the $A_1$-action.
These end lines are the counterpart of the endpoints $e_x$ one dimension higher.
%
%
\item \emph{Fusing end lines.} Fusing two end lines amounts to fusing the two surfaces before they end, so the end lines form the $H$-crossed (relative) Drinfeld center \cite{Gelaki:2009graded}
\begin{equation}\label{eq:ZH-A}
    \cZ_H(A)=\bigoplus_{h\in H}\cZ_{A_1}(A_h)\,,
\end{equation}
with the tensor product induced from $A$. 
It is the counterpart of the boundary algebra $\bbC^\psi[H]$: for $A=\Vect_H^\omega$ one finds $\cZ_H(A)\simeq\Vect_H^\omega$ with $H$ acting by conjugation.
\item \emph{Gauge transformations in $H$.} A gauge transformation $h\in H$ conjugates both the flux and its end line, $T_h:\cZ_{A_1}(A_x)\to\cZ_{A_1}(A_{hxh^{-1}})$, the counterpart of $u_he_x=e_he_xe_h^{-1}$. 
This makes $\cZ_H(A)$ an $H$-crossed braided fusion category. 
Equivalently, 
\begin{equation}
    \cZ_H(A)\simeq\cZ(A)\boxtimes_{\Rep H}\Vect
\end{equation} 
is the de-equivariantization of $\cZ(A)$ along the Tannakian subcategory $\Rep(H)\subset\cZ(A)$ defined by the grading, and its trivial component is 
\begin{equation}
\cZ_{2}(\Rep H,\cZ(A))\boxtimes_{\Rep H}\Vect\simeq\cZ(A_1) \, .
\end{equation}
\item \emph{Gauge transformations outside $H$.} A gauge transformation $g\notin H$ instead produces the end line of the conjugate flux $gxg^{-1}$, which we denote $[g,s]$ for $s\in\cZ_{A_1}(A_x)$. Only the coset $gH$ matters, $[gh,s]\simeq[g,T_h(s)]$, so the bulk surfaces that end on $\fB_{(H,A)}$ form the induced $G$-crossed braided multifusion category
\begin{equation}\label{eq:LA-general}
    \cL(A)\simeq\Fun_H\big(G,\cZ_H(A)\big)\simeq\bigoplus_{i\in G/H}\cZ_H(A)^{(i)}\,,
\end{equation}
the counterpart of the induced algebra \eqref{eq:Lagrangian-induced}.
\item \emph{Block structure.} As in section~\ref{subsec:warm-up-lagrangian}, choose representatives $g_i$ of the cosets $i\in G/H$ and write $kg_i=g_{k\cdot i}\,l_i(k)$. The copy $\cZ_H(A)^{(i)}$ consists of the end lines $[g_i,s]$ and is graded by $g_iHg_i^{-1}$, since $[g_i,s]$ has flux $g_ixg_i^{-1}$. The $G$-action permutes the copies,
\begin{equation}\label{eq:LA-action}
    T_k[g_i,s]\simeq[g_{k\cdot i},\,T_{l_i(k)}(s)]\,,
\end{equation}
and end lines in different copies have vanishing product, $[g_i,s]\otimes[g_j,t]\simeq\delta_{ij}\,[g_i,s\otimes t]$, with unit $\bigoplus_i[g_i,1]$. 
This captures the block structure of section~\ref{subsec:warm-up-lagrangian}, with one copy of $\cZ_H(A)$ for each connected component $i\in G/H$ of the interfaces $\cM$ between $\fB_{\rm Dir}$ and $\fB_{(H,A)}$. 
%
\end{enumerate}
Mathematically, $\cL(A)$ is the full center of $A$, regarded as an algebra in $2\Vect_G^\pi$, and it is a Lagrangian algebra in $\fZ$ (with details in appendix~\ref{app:condensation}).

\begin{remark}\label{rmk:general-lagrangian}
    The algebras $\cL(A)$ are not the most general Lagrangian algebras in $\fZ$. The steps above only used that $\cZ_H(A)$ is an $H$-crossed braided fusion category whose neutral component $\cZ(A_1)$ is non-degenerate. Replacing it by an arbitrary ($\pi|_H$-twisted) $H$-crossed braided extension $\widehat{M}_H=\bigoplus_{h\in H}\widehat{M}_h$ of a non-degenerate braided fusion category $M=\widehat{M}_1$, as in section~\ref{subsec:symtft-defects-and-boundaries}, the same induction gives
    \begin{equation}\label{eq:L-general-Witt}
        \cL\big(H,\widehat{M}_H\big):=\Fun_H\big(G,\widehat{M}_H\big)\simeq\bigoplus_{i\in G/H}\widehat{M}_H^{(i)}\, ,
    \end{equation}
    and every Lagrangian algebra in $\fZ$ is of this form \cite{Decoppet:2023hkr,Xu:2024etale}. It describes the boundary $\fB_{(H,\widehat{M}_H)}$, and ${}_{\cL(H,\widehat{M}_H)}\fZ$ is the corresponding (AB)-type fusion 2-category, with lines $M^H$; by theorems~\ref{thm:AB-center} and~\ref{thm:canonical-lagrangian}, every (AB)-type fusion 2-category arises in this way. The full centers $\cL(A)=\cL(H,\cZ_H(A))$ are exactly the case where $M\simeq\cZ(A_1)$ is Witt trivial, i.e.\ where the symmetry is Morita equivalent to $2\Vect_G^\pi$ (warning~\ref{warn:center-does-not-determine-morita}). 
    
    \newl In what follows, the descriptions of $\CUV$, $\CIR$ and $\cF$ as modules, bimodules and free modules over Lagrangian algebras in $\fZ$, namely the first equivalence in \eqref{eq:lag-CUV}, \eqref{eq:lag-OmegaCIR} and \eqref{eq:lag-free}, apply to $\cL(H,\widehat{M}_H)$ in place of $\cL(A)$. Only the comparison with bimodules in $2\Vect_G^\pi$, and hence with section~\ref{subsec:module-2-cats}, requires $M$ to be Witt trivial. We keep to the algebras $\cL(A)$ in order to make this comparison, and because they cover all of our examples.
\end{remark}

\newl {\bf Which surfaces end.} A bulk surface $S$ ends on $\fB_{(H,A)}$ with topological lines and junctions, which form the category $\Hom_\fZ(S,\cL(A))$, the 2-categorical counterpart of the multiplicity $n_H(Q)=\dim\Hom(Q,\cL_H)$ of \eqref{eq:CH-condensation}.
\begin{enumerate}[label=(\arabic*)]
\item \emph{Reduction to one fiber.} For a simple surface $S_{[x],L}$, Frobenius reciprocity \eqref{eq:Z-hom-from-simple} reduces this to the fiber of $\cL(A)$ at the single flux $x$. 
By the block structure, the copy $i$ contributes to this fiber if and only if $x\in g_iHg_i^{-1}$, i.e.\ $x$ fixes the coset $i$, and it then contributes the end lines $\cZ_{A_1}(A_{x_i})$ of the flux $x_i:=g_i^{-1}xg_i\in H$.
\item \emph{General formula.} Hence
\begin{equation}\label{eq:LA-endpoints}
    \Hom_\fZ\big(S_{[x],L},\cL(A)\big)\simeq\Big(\bigoplus_{i\in(G/H)^x}\cZ_{A_1}(A_{x_i})\Big)^L\,,
\end{equation}
the category of $L$-equivariant objects, where $L\subseteq C_G(x)$ permutes the cosets fixed by $x$ and acts on the end lines through \eqref{eq:LA-action}.
\item \emph{Line permutation.} A surface of flux $[x]$ can end on $\fB_{(H,A)}$ only if $x$ is conjugate into $H$, and it ends along the twisted lines of $\mathrm{TV}[A_1]$ seen from the cosets that $x$ fixes, symmetrized over $L$.
This is \eqref{eq:CH-condensation} one categorical level up: the permutation representation $\bbC[(G/H)^x]$ of the centralizer is replaced by a permutation 2-representation, and the discrete-torsion phases $\varepsilon_x$ by the action of $l_i(k)$ on the twisted lines, which can permute them, as $e\leftrightarrow m$ in example~\ref{ex:S3-2lag-algebras}, and twist their equivariant structure through the associator of $A$, as in example~\ref{ex:S3-2lag-LB}.
\item \emph{Extreme cases.} For the Dirichlet boundary, $H=\{1\}$ and $A=\Vect$, only $x= 1$ fixes a coset, and $\cL(A)\simeq S_{[1],\{1\}}$ is the condensation of all Wilson lines, the counterpart of $\cL_{\{1\}}=\Fun(G)$. For the Neumann boundary, $H=G$ and $A=\Vect_G$, every flux fixes the single coset and acts trivially on its end line, so $\cL(A)\simeq\bigoplus_{[x]}S_{[x],C_G(x)}$ and every flux ends exactly once, the counterpart of $\cL_G=\bbC[G]$.
\item \emph{Lines out of the unit.} For the unit $\bone=S_{[1],G}$, \eqref{eq:LA-endpoints} gives the lines out of the unit, $\Hom_\fZ(\bone,\cL(A))\simeq(\cL(A)_1)^G\simeq\cZ(A_1)^H$. These are the lines on the boundary $\fB_{(H,A)}$, i.e.\ $\Omega\CUV$, as we will see in section~\ref{subsec:lag-uv}.
\end{enumerate}

\begin{examplebox}[The Lagrangian algebra $\cL(A)$ of the symmetry boundary][label=ex:S3-2lag-algebras]
We now describe, in terms of the surfaces of the 4d $S_3$ Dijkgraaf--Witten SymTFT $\fZ=\cZ_1(2\Vect_{S_3})$, the Lagrangian algebra of the symmetry boundary $\fB_{(\bbZ_2,\Ising)}$ of the running example of section~\ref{subsec:module-2-cats}, with $H=\bbZ_2=\{\id,b\}$ and $A=\Ising$ graded as in \eqref{eq:S3-Ising-grading}. We follow example~\ref{ex:S3-Lagrangians}, which treats the boundary $\fB_{\Z_2}$ of $\cZ(\Vec_{S_3})$ in the same steps, and use the same cosets, representatives and little-group elements. The physical boundary is treated in example~\ref{ex:S3-2lag-LB}.

\newl {\bf Simple surfaces of $\fZ$.} All groups $\operatorname{H}^2(L;U(1))$ with $L\subseteq S_3$ vanish, so by the classification of simple surfaces above, the simple surfaces of $\fZ$ are
\begin{equation}\label{eq:S3-simple-surfaces}
\begin{gathered}
S_{[\id],L}\ \ (L=\{\id\},\bbZ_2,\bbZ_3,S_3)\,,\\
S_{[a],L}\ \ (L=\{\id\},\bbZ_3)\,,\qquad S_{[b],L}\ \ (L=\{\id\},\bbZ_2)\,,
\end{gathered}
\end{equation}
with unit $\bone=S_{[\id],S_3}$. They are the 2-categorical counterparts of the eight lines $\Qc{g}{\pi}$ of example~\ref{eq:Z(S3) lines}, with the representations $\pi$ of the centralizer replaced by its subgroups $L$.

\newl {\bf $\cL(A)$: the crossed center.} The Wilson line $1^-\mapsto g\bar g$ of \eqref{eq:S3-OmegaCUV-in-ZIsing} generates the Tannakian subcategory $\Rep(\bbZ_2)\subset\cZ(\Ising)\simeq\Ising\boxtimes\bar{\Ising}$ defined by the grading, and
\begin{equation}\label{eq:S3-Z2-crossed-center}
  \cZ_{\bbZ_2}(\Ising)=\cZ(\Ising)\boxtimes_{\Rep(\bbZ_2)}\Vect
  \;\simeq\;\underbrace{\{1,\,e,\,m,\,em\}}_{\text{degree }\id}\;\oplus\;\underbrace{\{\sigma_+,\,\sigma_-\}}_{\text{degree }b}\,.
\end{equation}
The neutral component is $\cZ(A_1)=\cZ(\Vect_{\bbZ_2})$. It comes from the centralizer of $\Rep(\bbZ_2)$, with $\epsilon\bar\epsilon\mapsto e\oplus m$ and $1\bar g,\,g\bar1\mapsto em$. The twisted component contains the two $e$--$m$ twist defects, of quantum dimension $\sqrt2$, which come from the pairs $\{1\bar\epsilon,\,g\bar\epsilon\}$ and $\{\epsilon\bar1,\,\epsilon\bar g\}$. Their fusion is
\begin{equation}
  \sigma_\pm\otimes\sigma_\pm=1\oplus em\,,\qquad
  \sigma_\pm\otimes\sigma_\mp=e\oplus m\,,\qquad
  e\otimes\sigma_\pm=m\otimes\sigma_\pm=\sigma_\mp\,.
\end{equation}
The residual $\bbZ_2$-action $Y\mapsto b\cdot Y$ exchanges $e\leftrightarrow m$ and fixes all other simple objects.

\newl {\bf $\cL(A)$: objects and grading.} By \eqref{eq:LA-general}, $\cL(A)\simeq\bigoplus_{i=0}^2\cZ_{\bbZ_2}(\Ising)^{(i)}$ has one copy for each coset $i=a^i\bbZ_2$, i.e.\ for each order-two subgroup $\{\id,x_i\}$, with $x_i=a^iba^{-i}$ as in example~\ref{ex:S3-Lagrangians}. Copy $i$ has its neutral part in degree $\id$ and its twist defects in degree $x_i$:
\begin{equation}
\begin{gathered}
  \cL(A)_{\id}\simeq\bigoplus_{i}\big\{1^{(i)},e^{(i)},m^{(i)},em^{(i)}\big\}\,,\\
  \cL(A)_{x_i}\simeq\big\{\sigma_+^{(i)},\sigma_-^{(i)}\big\}\,,\qquad
  \cL(A)_{a}=\cL(A)_{a^2}=0\,.
\end{gathered}
\end{equation}
This categorifies the basis of example~\ref{ex:S3-Lagrangians}: the idempotents $p_i$ become the three copies of $\cZ(\Vect_{\bbZ_2})$, and the elements $f_i$ of flux $x_i$ become the twist defects $\sigma_\pm^{(i)}$.

\newl {\bf $\cL(A)$: the $S_3$-action.} With the little-group elements $l_i(a)=\id$ and $l_i(b)=b$ of example~\ref{ex:S3-Lagrangians},
\begin{equation}
T_a\big(Y^{(i)}\big)=Y^{(i+1)}\,,\qquad T_b\big(Y^{(i)}\big)=(b\cdot Y)^{(-i)}\,,
\end{equation}
so that, for instance, $T_b(e^{(0)})=m^{(0)}$ while $T_b(\sigma_\pm^{(0)})=\sigma_\pm^{(0)}$. This categorifies the half-braiding $u_k$ there.

\newl {\bf $\cL(A)$: product, unit and braiding.} The multiplication 1-morphism $m:\cL(A)\otimes\cL(A)\to\cL(A)$ is the fusion within each copy and vanishes between different copies,
\begin{equation}
m\big(Y^{(i)}\boxtimes Y'^{(j)}\big)=\delta_{ij}\,(Y\otimes Y')^{(i)}\,,\qquad\text{e.g.}\quad m\big(\sigma_+^{(i)}\boxtimes\sigma_+^{(i)}\big)=1^{(i)}\oplus em^{(i)}\,,
\end{equation}
the counterpart of \eqref{eq:Lagrangian-structure}. The unit 1-morphism $i:\bone\to\cL(A)$ is the equivariant object $1^{(0)}\oplus1^{(1)}\oplus1^{(2)}\in(\cL(A)_\id)^{S_3}$, the counterpart of $\eta=p_0+p_1+p_2$. The braiding 2-morphism $m\circ\beta\Rightarrow m$ is the $\bbZ_2$-crossed braiding in each copy, $s\otimes t\simeq T_{x}(t)\otimes s$ for $s$ of degree $x$. For instance $\sigma_+^{(0)}\otimes e^{(0)}\simeq m^{(0)}\otimes\sigma_+^{(0)}$: the line $e$ becomes $m$ when it crosses the twist defect.

\newl {\bf $\cL(A)$: decomposition into bulk surfaces.} Viewing $\cL(A)$ as an $S_3$-equivariant 2-vector bundle over $S_3$, we read off its decomposition fiber by fiber, as in the last step of example~\ref{ex:S3-Lagrangians}:
\begin{itemize}
\item over $\id$, the twelve simple objects form three $S_3$-orbits: $\{1^{(i)}\}$ and $\{em^{(i)}\}$, each with stabilizer $\bbZ_2$, and $\{e^{(i)},m^{(i)}\}$, with trivial stabilizer since $T_b(e^{(0)})=m^{(0)}$. So $\cL(A)_\id\simeq\Vect(S_3/\bbZ_2)^{\oplus2}\oplus\Vect(S_3)$, which gives $2\,S_{[\id],\bbZ_2}\oplus S_{[\id],\{\id\}}$;
\item over $a$, $\cL(A)_a=0$: no surface of flux $[a]$ ends on $\fB_{(\bbZ_2,\Ising)}$;
\item over $b$, $\cL(A)_b=\{\sigma_\pm^{(0)}\}$, on which $C_{S_3}(b)=\{\id,b\}$ acts trivially, which gives $2\,S_{[b],\bbZ_2}$.
\end{itemize}
Altogether,
\begin{equation}\label{eq:S3-LA-surfaces}
\cL(A)\simeq2\,S_{[\id],\bbZ_2}\oplus S_{[\id],\{\id\}}\oplus2\,S_{[b],\bbZ_2}\,.
\end{equation}
Compare $\cL_{\Z_2}=\Qc{\id}{1}\oplus\Qc{\id}{E}\oplus\Qc{b}{+}$ in \eqref{eq:S3-Lagrangians}. For $A=\Vect_{\bbZ_2}$, i.e.\ without the factor $\mathrm{TV}[\Vect_{\bbZ_2}]$, the same steps give $\cL(A)\simeq S_{[\id],\bbZ_2}\oplus S_{[b],\bbZ_2}$, the direct categorification: $\Fun(S_3/\bbZ_2)=1\oplus E$ becomes $S_{[\id],\bbZ_2}$, and $\Qc{b}{+}$ becomes $S_{[b],\bbZ_2}$. For $A=\Ising$, the anyons $1$ and $em$, which are fixed by the $e$--$m$ symmetry, each give a copy of $S_{[\id],\bbZ_2}$, the doublet $\{e,m\}$ gives $S_{[\id],\{\id\}}$, and the two twist defects give the two copies of $S_{[b],\bbZ_2}$.

\newl {\bf $\cL(A)$: endpoints of bulk surfaces.} By \eqref{eq:LA-endpoints}, the surface $S_{[b],\bbZ_2}$ of transposition flux ends along the lines of $(\cL(A)_b)^{\bbZ_2}$, i.e.\ along a twist defect $\sigma_\pm$ carrying a $\bbZ_2$-charge. These four lines are the lines $1\bar\epsilon$, $g\bar\epsilon$, $\epsilon\bar1$ and $\epsilon\bar g$ of $\cZ(\Ising)$, of dimension $\sqrt2$, which are absent from $\Omega\CUV$. Physically, transposition fluxes end on $\fB_{(\bbZ_2,\Ising)}$ because $\bbZ_2$ is gauged there, and the end line is dressed by a twist defect of $\mathrm{TV}[\Vect_{\bbZ_2}]$ because the gauged $\bbZ_2$ acts by $e\leftrightarrow m$. Surfaces of flux $[a]$ do not end: $\Hom_\fZ(S_{[a],L},\cL(A))=0$.
\end{examplebox}

\begin{examplebox}[The Lagrangian algebra $\cL(B)$ of the physical boundary][label=ex:S3-2lag-LB]
We now turn to the physical boundary $\fB_{(\bbZ_3,\Vect^\omega_{\bbZ_3})}$ of the running example, \eqref{eq:S3-IR-choice}, following example~\ref{ex:S3-LZ3} and the steps of example~\ref{ex:S3-2lag-algebras}. Here $K=\bbZ_3=\{\id,a,a^2\}$ and $B=\Vect^\omega_{\bbZ_3}$, with $[\omega]\neq0$ in $\operatorname{H}^3(\bbZ_3;U(1))\simeq\bbZ_3$. Since $B_1=\Vect$, the crossed center is $\cZ_{\bbZ_3}(B)\simeq\Vect^\omega_{\bbZ_3}$, with one simple object $k$ in each degree $k\in\bbZ_3$, trivial $\bbZ_3$-action on objects, and associator $\omega$. The same steps give:
\begin{itemize}
\item \emph{Objects and grading:} $\cL(B)\simeq\Vect^{\omega\,(0)}_{\bbZ_3}\oplus\Vect^{\omega\,(1)}_{\bbZ_3}$, with one copy for each coset $c\in\{K,bK\}=\{0,1\}$ of example~\ref{ex:S3-LZ3}, and with $k^{(0)}$ in degree $a^k$ and $k^{(1)}$ in degree $ba^kb^{-1}=a^{-k}$. So $\cL(B)$ is supported on $\{\id\}\cup[a]$.
\item \emph{The $S_3$-action:} $T_a$ fixes every object and $T_b$ exchanges $k^{(0)}\leftrightarrow k^{(1)}$. Both copies carry the same $\omega$, since $b$ acts on $\bbZ_3$ by inversion, which acts trivially on $\operatorname{H}^3(\bbZ_3;U(1))\simeq \operatorname{H}^4(\bbZ_3;\bbZ)$.
\item \emph{Product and unit:} $m(k^{(c)}\boxtimes k'^{(c')})=\delta_{cc'}(k+k')^{(c)}$, with associator $\omega(k,k',k'')$ on each copy, and unit $0^{(0)}\oplus0^{(1)}$. This categorifies the blocks $B_c\simeq\bbC[\Z_3]$ of example~\ref{ex:S3-LZ3}, now with the 3-cocycle $\omega$ as associator.
\item \emph{Decomposition:} over $\id$, the two units are exchanged by $b$, with stabilizer $\bbZ_3$, which gives $S_{[\id],\bbZ_3}$. Over $a$, the objects $1^{(0)}$ and $2^{(1)}$ are fixed by $C_{S_3}(a)=\bbZ_3$, which gives two copies of $S_{[a],\bbZ_3}$. Nothing lies over $b$. Hence
\begin{equation}\label{eq:S3-LB-surfaces}
\cL(B)\simeq S_{[\id],\bbZ_3}\oplus2\,S_{[a],\bbZ_3}\,,
\end{equation}
the categorification of $\cL_{\Z_3}=\big(\Qc{\id}{1}\oplus\Qc{\id}{P}\big)\oplus2\,\Qc{a}{1}$, with the two copies of $S_{[a],\bbZ_3}$ again labeled by $a,a^2\in K$.
\item \emph{Lines and endpoints:} the boundary lines are $\Hom_\fZ(\bone,\cL(B))\simeq(\cL(B)_\id)^{S_3}\simeq\Rep(\bbZ_3)$, onto which the bulk Wilson lines restrict; in particular $P$ ends on $\fB_{(\bbZ_3,\Vect^\omega_{\bbZ_3})}$. The 1-categorical discrete-torsion charges $\varepsilon_x$ of \eqref{eq:charges} are replaced here by the slant product $i_a\omega\in \operatorname{H}^2(\bbZ_3;U(1))=0$, so the flux-$[a]$ surfaces end without an extra charge. The twist $\omega$ is therefore invisible in \eqref{eq:S3-LB-surfaces}, and only enters through the associator of the product of $\cL(B)$ on its flux-$[a]$ part.
\end{itemize}
\end{examplebox}

\subsubsection{The UV Symmetry \texorpdfstring{$\CUV$}{CUV}}\label{subsec:lag-uv}
We write ${}_{\cL}\fZ$, $\fZ_{\cL}$ and ${}_{\cL}\fZ_{\cL'}$ for the 2-categories of left $\cL$-modules, right $\cL$-modules and $(\cL,\cL')$-bimodules in $\fZ$, i.e.\ $\mathsf{LMod}_\fZ(\cL)$, $\mathsf{RMod}_\fZ(\cL)$ and $\mathsf{Bimod}_\fZ(\cL,\cL')$ (appendix~\ref{app:algebras-modules}).

\newl {\bf The UV symmetry.} The symmetry on $\fB_{(H,A)}$ is the 2-category of left $\cL(A)$-modules,
\begin{equation}\label{eq:lag-CUV}
\CUV\simeq{}_{\cL(A)}\fZ\simeq{}_A(2\Vect_G^\pi)_A\,,
\end{equation}
the counterpart of \eqref{eq:warm-up-lag-uv} and \eqref{eq:S3-lag-uv-bimod}. Every left $\cL(A)$-module is a condensation of a free module $\cL(A)\otimes Y$ with $Y\in\fZ$, and, as in \eqref{eq:warm-up-reciprocity},
\begin{equation}\label{eq:free-adjunction}
\Hom_{\CUV}\big(\cL(A)\otimes Y,\,V\big)\simeq\Hom_{\fZ}\big(Y,\,V\big)
\end{equation}
for every module $V$. In particular, by \eqref{eq:Z-hom-from-unit}, $\Omega\CUV\simeq\Hom_\fZ(\bone,\cL(A))\simeq(\cL(A)_1)^G\simeq\cZ(A_1)^H$, reproducing the result of section~\ref{subsec:uv-symmetry}.

\begin{examplebox}[The UV symmetry from $\cL(A)$][label=ex:S3-2lag-uv]
We compute $\CUV={}_{\cL(A)}\fZ$ for the Lagrangian algebra of example~\ref{ex:S3-2lag-algebras}, in parallel with examples~\ref{ex:S3-lag-uv} and~\ref{ex:S3-lag-uv-bimod}, and recover the results of example~\ref{ex:S3-2mod-uv}.

\newl {\bf Components from free modules.} By \eqref{eq:free-adjunction}, the free module functor $Y\mapsto\cL(A)\otimes Y$ is the 2-categorical counterpart of the bulk-to-boundary map $\Phi_H$ of \eqref{eq:warm-up-free-module}. A bulk surface of flux $x$ pushed onto the boundary moves the interfaces in $M$ by $i\mapsto x\cdot i$. As for the matrix $\Phi_H(Q)_{ji}=\bigoplus_{x\cdot i=j}V_x$ of section~\ref{subsec:warm-up-lag-uv}, its free module is therefore supported on the double cosets $Hg_i^{-1}xg_iH$, i.e.\ on those double cosets that meet the conjugacy class of $x$. For the simple surfaces \eqref{eq:S3-simple-surfaces}:
\begin{itemize}
  \item $S_{[\id],L}$ lands in the component $\bbZ_2$ of the unit;
  \item $S_{[a],L}$ lands in $\bbZ_2a\bbZ_2$;
  \item $S_{[b],L}$ lands in both, since $b\in\bbZ_2$ and $ab\in\bbZ_2a\bbZ_2$.
\end{itemize}
Hence $\pi_0\CUV\simeq\bbZ_2\backslash S_3/\bbZ_2$, reproducing \eqref{eq:S3-double-cosets}.

\newl {\bf Lines.} By \eqref{eq:free-adjunction} and \eqref{eq:Z-hom-from-unit},
\begin{equation}
  \Omega\CUV=\End_{\CUV}\big(\cL(A)\big)\simeq\Hom_{\fZ}\big(\bone,\cL(A)\big)\simeq\big(\cL(A)_{\id}\big)^{S_3}\simeq\cZ(\Vect_{\bbZ_2})^{\bbZ_2}\,.
\end{equation}
The last step is the orbit decomposition over $\id$ in example~\ref{ex:S3-2lag-algebras}: the orbits $\{1^{(i)}\}$ and $\{em^{(i)}\}$, with stabilizer $\bbZ_2$, give $1^\pm$ and $em^\pm$, and the free orbit $\{e^{(i)},m^{(i)}\}$ gives $\eta$. This reproduces \eqref{eq:S3-OmegaCUV}. The twist defects sit in non-trivial degree and do not contribute. A bulk Wilson line $R$ maps to $R\boxtimes1_{\cL(A)}$, i.e.\ to its restriction to $\Rep(\bbZ_2)=\langle1^-\rangle$, the M\"uger center.

\newl {\bf Explicit description: bimodules in $2\Vect_{S_3}$.} By \eqref{eq:lag-CUV}, $\CUV\simeq{}_{\Ising}(2\Vect_{S_3})_{\Ising}$, the 2-categorical counterpart of ${}_{\bbC[\Z_2]}(\Vec_{S_3})_{\bbC[\Z_2]}$ in example~\ref{ex:S3-lag-uv-bimod}. We write $\Ising=A_1\oplus A_{-1}$, with $A_1=\Vect_{\bbZ_2}$ in degree $\id$ and $A_{-1}=\Vect$ in degree $b$, as in \eqref{eq:S3-Ising-grading}.
\begin{itemize}
\item \emph{Objects and morphisms.} An object is an $S_3$-graded finite semisimple category $M=\bigoplus_gM_g$ with commuting left and right $\Ising$-actions compatible with the grading, $\Ising_h\boxtimes M_g\to M_{hg}$ and $M_g\boxtimes\Ising_h\to M_{gh}$, where $\Ising_\id=A_1$ and $\Ising_b=A_{-1}$. 1-morphisms are graded bimodule functors and 2-morphisms are bimodule natural transformations. Fusion is $\boxtimes_{\Ising}$, with unit $\Ising$.
\item \emph{Double cosets.} As in example~\ref{ex:S3-lag-uv-bimod}, every bimodule splits over $\bbZ_2\backslash S_3/\bbZ_2$. On each double coset $HgH$ it is determined by its component $M_g$, an $A_1$-bimodule category that is equivariant under $H_g=H\cap gHg^{-1}$. These are exactly the labels of the simple objects found in section~\ref{subsec:uv-symmetry}.
\item \emph{Trivial double coset.} Here we find the unit $1=\Ising$, with components $A_1$ and $A_{-1}$ in degrees $\id$ and $b$, and
\begin{equation}
\chi:=\Ising\boxtimes_{A_1}\Ising\,,\qquad \chi_\id\simeq A_1\oplus\big(A_{-1}\boxtimes_{A_1}A_{-1}\big)\simeq A_1\oplus A_1\,,
\end{equation}
whose two summands are exchanged by the stabilizer $\bbZ_2$. These are the objects $1$ and $\chi$ of example~\ref{ex:S3-2mod-uv}: $\chi$ is the condensation of the Wilson line $1^-$ on the trivial surface. It is the counterpart of $A\otimes A\simeq1\oplus P$ in example~\ref{ex:S3-lag-uv-bimod}, but unlike $A\otimes A$ it is a single simple object.
\item \emph{Double coset $\bbZ_2a\bbZ_2$.} The stabilizer $H_a$ is trivial, so every bimodule is freely generated by its component in degree $a$, which can be an arbitrary $A_1$-bimodule category $U$:
\begin{equation}\label{eq:S3-XU}
X_U:=\Ising\boxtimes_{A_1}U_a\boxtimes_{A_1}\Ising\,,\qquad U\in\mathsf{Bimod}(A_1)\simeq\Sigma\cZ(\Vect_{\bbZ_2})\,.
\end{equation}
The object $X$ of example~\ref{ex:S3-2mod-uv} is $X:=X_{A_1}$, the counterpart of $E=A\otimes\bbC_a\otimes A$. Explicitly, $X$ has components $A_1=\Vect_{\bbZ_2}$ in degrees $a$ and $a^2=bab$, and $A_{-1}=\Vect$ in degrees $ab$ and $a^2b=ba$.
\item \emph{Fusion.} Splitting the middle $\Ising=A_1\oplus A_{-1}$ exactly as in \eqref{eq:S3-EE-bimod},
\begin{equation}\label{eq:S3-XX-bimod}
\begin{split}
X\boxtimes_{\Ising}X&\simeq\big(\Ising\boxtimes_{A_1}(A_1)_{a^2}\boxtimes_{A_1}\Ising\big)\\
&\quad\oplus\big(\Ising\boxtimes_{A_1}(A_{-1})_{aba}\boxtimes_{A_1}\Ising\big)\simeq X\oplus\chi\,.
\end{split}
\end{equation}
Here $a^2=bab$, so that the two factors $A_{-1}$ are absorbed into the outer copies of $\Ising$, and $aba=b\in\bbZ_2$. This reproduces $X\otimes X\simeq\chi\oplus X$ of example~\ref{ex:S3-2mod-uv}, now derived from the composition \eqref{eq:CUV-fusion-matrix}.
\end{itemize}
\end{examplebox}

\subsubsection{The IR Symmetry \texorpdfstring{$\CIR$}{CIR}}\label{subsec:lag-ir}
Suppose now we fix two topological boundary conditions of the 4d Dijkgraaf--Witten SymTFT classified by $(H,A)$ and $(K,B)$ respectively. Let $\cL(A)$ and $\cL(B)$ be their full centers as Lagrangian algebras in $\fZ$.
We choose the interval orientation so that topological defects in the left boundary are modeled by left $\cL(A)$-modules and those in the right boundary by right $\cL(B)$-modules; reversing the interval exchanges left and right.
Concretely, the corresponding boundary defect 2-categories are
\begin{equation}
    \CUV = {}_{\cL(A)} {\fZ} \simeq {}_{A}(2\Vect^\pi_G)_A,
    \qquad
    \fZ_{\cL(B)} \simeq {}_{B}(2\Vect^\pi_G)_B.
\end{equation}
The 3d theory obtained by compactifying the interval with these two boundary conditions has defect 2-category
\begin{equation}\label{eq:lag-CIR-bimod}
    \CIR = {}_{\cL(A)} {\fZ} \boxtimes_{\fZ} {\fZ}_{\cL(B)}
    \simeq
    {}_{\cL(A)} {\fZ}_{\cL(B)}
    \simeq
    \End({}_{A}(2\Vect^\pi_G)_B),
\end{equation}
where ${}_{A}(2\Vect^\pi_G)_B$ is the 2-category of boundary conditions of the IR theory, the counterpart of $\cN\boxtimes_\cC\cM$ in \eqref{eq:boundary-cat}. 

\newl Let us be explicit about the monoidal structure in \eqref{eq:lag-CIR-bimod}, which is \emph{not} the composition of bimodules (the latter does not even make sense for $\cL(A)\neq\cL(B)$). Since $\cL(A)$ and $\cL(B)$ are braided algebras, ${}_{\cL(A)}\fZ$ and $\fZ_{\cL(B)}$ are monoidal under the relative tensor products $\otimes_{\cL(A)}$ and $\otimes_{\cL(B)}$, and the free-module 2-functors from $\fZ$ are monoidal and central. This induces a monoidal structure on $\,{}_{\cL(A)}\fZ\boxtimes_\fZ\fZ_{\cL(B)}$, corresponding to the fusion of surfaces in the slab. In the bimodule presentation, it is the relative tensor product over both algebras at once,
\begin{equation}\label{eq:lag-CIR-fusion}
    X\otimes_{\cL(A),\cL(B)}Y\, ,
\end{equation}
which coequalizes the left $\cL(A)$-actions and the right $\cL(B)$-actions on $X\otimes Y$, after using the braiding of $\fZ$ to bring $\cL(A)$ to the left of $X$ and its inverse to bring $\cL(B)$ to the right of $Y$. Under the last equivalence in \eqref{eq:lag-CIR-bimod}, it becomes the composition of endofunctors of ${}_{A}(2\Vect^\pi_G)_B$. This is the 2-categorical counterpart of the structure used in section~\ref{subsec:warm-up-lagrangian}; we take the compatibility of the equivalences \eqref{eq:lag-CIR-bimod} with these monoidal structures as the mathematical formulation of the sandwich construction, and do not attempt to prove it here.

\newl With this monoidal structure the unit of $\CIR$ is the free bimodule $\cL(A)\otimes\cL(B)$, rather than either algebra alone, so by the bimodule version of \eqref{eq:free-adjunction} its lines are
\begin{equation}\label{eq:lag-OmegaCIR}
    \Omega\CIR\simeq\Hom_\fZ\big(\bone,\cL(A)\otimes\cL(B)\big)\simeq\big((\cL(A)\otimes\cL(B))_1\big)^G\,,
\end{equation}
the counterpart of the local operators \eqref{eq:warm-up-local-ops}: an IR line is a bulk line out of the trivial surface into the product of the two condensates.
Note the orientation: since the inverse braiding is used for the right boundary in \eqref{eq:lag-CIR-fusion}, the factor $\cL(B)$ contributes to $\Omega\CIR$ with its braiding reversed. For instance, for $G=\{1\}$ and $A=\Vect$, \eqref{eq:lag-OmegaCIR} gives the category $\cZ(B)$ with the reverse braiding, $\cZ(B)^{\mathrm{rev}}\simeq\cZ(\bar{B})$, in agreement with $\Omega\CIR\simeq\cZ(\bar{B_1}\boxtimes A_1)$ in section~\ref{subsec:ir-symmetry}.
In the single-vacuum situations considered below, this fusion 2-category is of the form
\begin{equation}
    \CIR = {}_{\cL(A)} {\fZ}_{\cL(B)}
    \simeq \Sigma M
\end{equation}
for a non-degenerate braided fusion category $M$.

The bimodule 2-category ${}_{\cL(A)} {\fZ}_{\cL(B)}$ is exactly the target for a generalized fiber 2-functor, when we fix one of the relative boundaries in the sandwich to be the fusion 2-category symmetry.

\begin{figure}[H]
    \centering
    \begin{tikzpicture}[scale=.72, transform shape, >=Latex, every node/.style={align=center,font=\small}]
        \node[draw, rounded corners, inner sep=3pt] (bulk) {$\fZ = \cZ(2\Vect_G^\pi)$\\ 4d DW bulk};
        \node[draw, rounded corners, below left=.8cm and .55cm of bulk, inner sep=3pt] (A) {$\CUV = {}_{\cL(A)} {\fZ}$\\ $(H,A)$ boundary};
        \node[draw, rounded corners, below right=.8cm and .55cm of bulk, inner sep=3pt] (B) {${\fZ}_{\cL(B)}$\\ $(K,B)$ boundary};
        \node[draw, rounded corners, below=2.05cm of bulk, inner sep=3pt] (tensor) {$\CIR = {}_{\cL(A)} {\fZ}_{\cL(B)}$\\ 3d compactified theory};
        \node[draw, rounded corners, below=.75cm of tensor, inner sep=3pt] (target) {single vacuum $\Sigma M$};
        \draw[->] (bulk) -- (A);
        \draw[->] (bulk) -- (B);
        \draw[->] (A) -- (tensor);
        \draw[->] (B) -- (tensor);
        \draw[->] (tensor) -- (target);
    \end{tikzpicture}
    \caption{The bimodule model for a generalized fiber $2$-functor.  Two Lagrangian algebras in the bulk Drinfeld center define two boundary fusion $2$-categories.  Their relative tensor product over the bulk is the $2$-category of bimodules, identified in the single-vacuum case with $\Sigma M$.}
    \label{fig:bimodule-fiber-functor}
\end{figure}
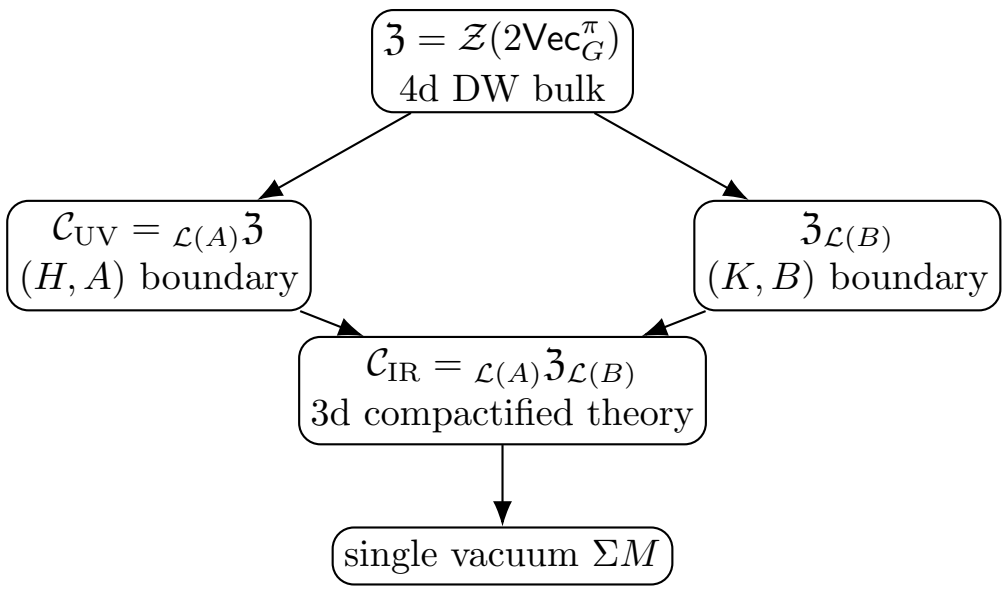

\begin{examplebox}[The IR symmetry from $\cL(A)$ and $\cL(B)$][label=ex:S3-2lag-ir]
We compute $\CIR={}_{\cL(A)}\fZ_{\cL(B)}$ for the Lagrangian algebras \eqref{eq:S3-LA-surfaces} and \eqref{eq:S3-LB-surfaces} of examples~\ref{ex:S3-2lag-algebras} and~\ref{ex:S3-2lag-LB}, step by step as in example~\ref{ex:S3-lag-ir-Z2Z3}, and recover example~\ref{ex:S3-2mod-ir}.
\begin{enumerate}
\item \emph{Lines as lines out of the unit.} By \eqref{eq:lag-OmegaCIR} and \eqref{eq:Z-hom-from-unit},
\begin{equation}
  \Omega\CIR\simeq\Hom_{\fZ}\big(\bone,\cL(A)\otimes\cL(B)\big)\simeq\big((\cL(A)\otimes\cL(B))_\id\big)^{S_3}\,,
\end{equation}
the counterpart of \eqref{eq:S3-local-ops-invariants}.
\item \emph{Grading.} By \eqref{eq:Z-fusion}, $(\cL(A)\otimes\cL(B))_\id=\bigoplus_{x\in S_3}\cL(A)_x\boxtimes\cL(B)_{x^{-1}}$, with the diagonal $S_3$-action.
\item \emph{Supports.} $\cL(A)$ is supported on $\{\id\}\cup[b]$ and $\cL(B)$ on $\{\id\}\cup[a]$, so only $x=\id$ contributes: the twist defects $\sigma_\pm^{(i)}$ of $\cL(A)$ and the flux-$[a]$ objects of $\cL(B)$ never pair up. In terms of \eqref{eq:S3-LA-surfaces} and \eqref{eq:S3-LB-surfaces}, no surface of non-trivial flux is common to both algebras, the counterpart of $1$ being the only line common to $\cL_{\Z_2}$ and $\cL_{\Z_3}$. Hence
\begin{equation}
\begin{split}
  (\cL(A)\otimes\cL(B))_\id&=\cL(A)_{\id}\boxtimes\cL(B)_{\id}\\
  &\simeq\bigoplus_{(i,c)\,\in\,S_3/\bbZ_2\times S_3/\bbZ_3}\cZ(\Vect_{\bbZ_2})^{(i,c)}\,.
\end{split}
\end{equation}
\item \emph{Orbits.} $S_3$ acts by $T_g\big(Y^{(i,c)}\big)=(l_i(g)\cdot Y)^{(g\cdot i,\,g\cdot c)}$, permuting the six labels $(i,c)$ with stabilizer $\bbZ_2\cap\bbZ_3=\{\id\}$, i.e.\ freely and transitively, exactly as the pairs $(i,c)$ of example~\ref{ex:S3-lag-ir-Z2Z3}. An $S_3$-equivariant object is therefore determined, with no further constraint, by its component on the copy $(0,0)$, and the tensor product and braiding can be computed there:
\begin{equation}
  \Omega\CIR\simeq\Big(\bigoplus_{(i,c)}\cZ(\Vect_{\bbZ_2})^{(i,c)}\Big)^{S_3}\simeq\cZ(\Vect_{\bbZ_2})\,.
\end{equation}
\item \emph{Vacua.} The unit object $\bigoplus_{(i,c)}1^{(i,c)}$ lies in a single orbit, so it is simple and the IR has a single vacuum. This is the counterpart of the one-dimensional invariants in example~\ref{ex:S3-lag-ir-Z2Z3}, and of $H\backslash G/K=\{G\}$ in \eqref{eq:single-vacuum}, and it agrees with \eqref{eq:S3-OmegaCIR}. The twist $\omega$ does not enter, since it lives on the flux-$[a]$ part of $\cL(B)$.
\item \emph{The 2-category.} By \eqref{eq:lag-CIR-bimod}, $\CIR\simeq\End\big({}_{\Ising}(2\Vect_{S_3})_{B}\big)$. Every $g\in S_3$ factorizes uniquely as $g=hk$ with $h\in\bbZ_2$ and $k\in\bbZ_3$, so an $(\Ising,B)$-bimodule is determined by its identity component, $M\simeq\Ising\boxtimes_{A_1}M_\id\boxtimes B$, where $M_\id$ is an arbitrary left $A_1$-module category. Hence
\begin{equation}\label{eq:S3-IR-boundaries}
\begin{gathered}
{}_{\Ising}(2\Vect_{S_3})_{B}\simeq\mathsf{LMod}(\Vect_{\bbZ_2})\,,\\
\CIR\simeq\mathsf{Bimod}(\Vect_{\bbZ_2})\simeq\Sigma\cZ(\Vect_{\bbZ_2})\,.
\end{gathered}
\end{equation}
The two simple objects of $\mathsf{LMod}(\Vect_{\bbZ_2})$ are the $e$- and $m$-condensing boundaries of the 3d $\bbZ_2$ Dijkgraaf--Witten theory, and \eqref{eq:S3-IR-boundaries} agrees with \eqref{eq:boundary-cat} for $[\bar B\otimes A]_{H\cap K}=\Vect_{\bbZ_2}$. It is the counterpart of the category $\Vec$ of boundary conditions in example~\ref{ex:S3-lag-ir-Z2Z3}.
\end{enumerate}
More generally, the vacua are counted by the $G$-orbits on $G/H\times G/K$, i.e.\ by $H\backslash G/K$. The stabilizers $H\cap{}^{g}K$, together with the overlap of the twisted sectors of $\cL(A)$ and $\cL(B)$, are where the graded category $[\bar{B}\boxtimes A]_{H\cap K}$ of \eqref{eq:boundary-cat} enters on the Lagrangian side.
\end{examplebox}

\subsubsection{The RG 2-Functor \texorpdfstring{$\cF : \CUV \to \CIR$}{F : CUV -> CIR}}\label{subsec:lag-rg}
With this convention, the generalized forgetful 2-functor is the free module functor
\begin{equation}\label{eq:lag-free}
\begin{split}
    \mathsf{Free} \colon {}_{\cL(A)} {\fZ} &\longrightarrow {}_{\cL(A)} {\fZ}_{\cL(B)},\\
    X &\longmapsto X\otimes \cL(B)\, .
\end{split}
\end{equation}
The right $\cL(B)$-action is inherited from the regular right action of $\cL(B)$, while the left $\cL(A)$-action is inherited from $X$. This is the counterpart of \eqref{eq:warm-up-free}.

After identifying $\CUV$ with ${}_{\cL(A)}\fZ$ and the target ${}_{\cL(A)} {\fZ}_{\cL(B)} \simeq \Sigma M$ under the additional single vacuum assumption, this gives the generalized fiber $2$-functor
\begin{equation}
    \cF:\CUV \longrightarrow \Sigma M\, .
\end{equation}
Physically, a surface defect on the symmetry boundary is brought through the bulk interval and then terminated on the physical boundary by freely adjoining the $\cL(B)$-module structure.
This is the algebraic form of sending left modules to free bimodules. It is monoidal for the tensor product \eqref{eq:lag-CIR-fusion}, since $(X\otimes_{\cL(A)}Y)\otimes\cL(B)\simeq(X\otimes\cL(B))\otimes_{\cL(A),\cL(B)}(Y\otimes\cL(B))$, and it sends the unit $\cL(A)$ to the unit $\cL(A)\otimes\cL(B)$. In the bimodule presentation \eqref{eq:lag-CIR-bimod}, the same functor sends $Y\in{}_A(2\Vect_G^\pi)_A$ to the endofunctor $M\mapsto Y\boxtimes_AM$ of the boundary conditions ${}_A(2\Vect_G^\pi)_B$, the counterpart of \eqref{eq:RG-functor}.

\newl Together with remark~\ref{rmk:general-lagrangian}, this recovers the result of section~\ref{subsec:single-vacuum} in the Lagrangian algebra model: for any bosonic (AB-type) fusion 2-category $\cC\simeq{}_{\cL}\fZ$, with $\cL=\cL(H,\widehat{M}_H)$, taking the physical boundary to be $\cL(B)$ with $K=G$ and $B=\Vect_\Gamma^\psi$ the symmetry-extension algebra of section~\ref{subsec:single-vacuum}, the free-module 2-functor $X\mapsto X\otimes\cL(B)$ is a generalized fiber 2-functor on $\cC$. This fact stands in contrast to the corresponding two-dimensional problem presented in the introduction, where we observed that not every fusion category admits a fiber functor into $\Vect$.

\begin{examplebox}[The RG functor, the image of $X$ and the twist $\omega$][label=ex:S3-2lag-rg]
The RG functor is the free-module 2-functor
\begin{equation}
\begin{split}
  \cF=\mathsf{Free}:\ \CUV={}_{\cL(A)}\fZ&\longrightarrow\CIR={}_{\cL(A)}\fZ_{\cL(B)}\simeq\Sigma\cZ(\Vect_{\bbZ_2})\,,\\
  Y&\longmapsto Y\otimes\cL(B)\,,
\end{split}
\end{equation}
which sends the unit $\cL(A)$ to the unit $\cL(A)\otimes\cL(B)$. Since $\CIR$ is connected, both components of $\CUV$ land in its single component, as in \eqref{eq:S3-2RG}. We compute $\cF$ on lines and on surfaces, and then discuss the role of $A$, of the twisted sectors and of $\omega$, following examples~\ref{ex:S3-lag-rg} and~\ref{ex:S3-lag-rg-Z2Z3}.

\newl {\bf Lines.} On lines, $\cF$ sends a 1-morphism $y:\bone\to\cL(A)$ to $y\otimes\iota_B:\bone\to\cL(A)\otimes\cL(B)$, where $\iota_B$ is the unit of $\cL(B)$. In terms of equivariant objects, this is
\begin{equation}
  \big(\cL(A)_{\id}\big)^{S_3}\longrightarrow
  \big(\cL(A)_{\id}\boxtimes\cL(B)_{\id}\big)^{S_3}\,,\qquad
  Y\longmapsto Y\boxtimes 1_{\cL(B)}\,.
\end{equation}
Evaluate both sides on the representative label: the coset $H$ on the left, and the pair $(H,K)$ on the right. The source then consists of $\bbZ_2$-equivariant objects of $\cZ(\Vect_{\bbZ_2})$, while the target consists of objects equivariant under $\bbZ_2\cap\bbZ_3=\{\id\}$. The functor simply restricts the equivariance:
\begin{equation}
\begin{gathered}
  \Omega\cF:\ \cZ(\Vect_{\bbZ_2})^{\bbZ_2}\longrightarrow\cZ(\Vect_{\bbZ_2})\,,\\
  1^\pm\mapsto1\,,\quad em^\pm\mapsto em\,,\quad\eta\mapsto e\oplus m\,.
\end{gathered}
\end{equation}
This is exactly the forgetful functor \eqref{eq:S3-OmegaF} found with module 2-categories in example~\ref{ex:S3-2mod-rg}. Equivalently, $\Omega\cF$ is the de-equivariantization along the M\"uger center $\Rep(\bbZ_2)=\langle1^-\rangle$, so the Wilson line $1^-$ trivializes in the IR. Every simple line of $\cZ(\Vect_{\bbZ_2})$ appears in the image, so $\Omega\cF$ is surjective and $\Omega\CIR\simeq\cZ(A_1)$ is the modular closure \eqref{eq:modular-closure}. The phase is therefore minimal in the sense of section~\ref{subsec:minimality}. This is expected, since $S_3=HK$, $K$ is non-anomalous, $B_1=\Vect$ and $H\cap K=\{\id\}$.

\newl {\bf Surfaces: the image of $X$.} In the bimodule presentation \eqref{eq:S3-IR-boundaries}, a surface $Y\in\CUV\simeq{}_{\Ising}(2\Vect_{S_3})_{\Ising}$ acts on the boundary conditions $M$ of the IR theory by $M\mapsto Y\boxtimes_{\Ising}M$. Since $M\simeq\Ising\boxtimes_{A_1}M_\id\boxtimes B$, it suffices to read off the identity component, as in example~\ref{ex:S3-lag-rg-Z2Z3}. For $Y=X=\Ising\boxtimes_{A_1}(A_1)_a\boxtimes_{A_1}\Ising$,
\begin{equation*}
X\boxtimes_{\Ising}M\simeq\Ising\boxtimes_{A_1}(A_1)_a\boxtimes_{A_1}\Ising\boxtimes_{A_1}M_\id\boxtimes B\,,
\end{equation*}
whose identity component collects the triples $(h,h',k)\in\bbZ_2\times\bbZ_2\times\bbZ_3$ with $h\,a\,h'\,k=\id$. There are exactly two, $(\id,\id,a^2)$ and $(b,b,a)$, the same as for $E$ in \eqref{eq:S3-FE-bimod}, so
\begin{equation*}
\begin{split}
(X\boxtimes_{\Ising}M)_\id&\simeq\big(A_1\boxtimes_{A_1}A_1\boxtimes_{A_1}M_\id\big)\oplus\big(A_{-1}\boxtimes_{A_1}A_{-1}\boxtimes_{A_1}M_\id\big)\\
&\simeq M_\id\oplus M_\id\,,
\end{split}
\end{equation*}
using $A_{-1}\boxtimes_{A_1}A_{-1}\simeq A_1$, i.e.\ $\epsilon\otimes\epsilon=1\oplus g$. Hence
\begin{equation}\label{eq:S3-FX}
\cF(X)\simeq1\oplus1\,,\qquad \cF(\chi)\simeq1\oplus1\,,\qquad \cF(1)\simeq1\,,
\end{equation}
where $\cF(\chi)$ follows in the same way from the two terms $h\in\{\id,b\}$ of $(\Ising\boxtimes_{A_1}\Ising\boxtimes_{\Ising}M)_\id$. This is consistent with $X\otimes X\simeq\chi\oplus X$, since $(1\oplus1)\circ(1\oplus1)\simeq4\cdot1$. Physically, $X$ is the gauge-invariant combination of the surfaces $a$ and $a^2$ of $S_3$, the counterpart of $E$, which combines the lines $a$ and $a^2$ (example~\ref{CH on S3}). Since the physical boundary preserves $\bbZ_3\ni a,a^2$, both become trivial in the IR, and $\cF(X)$ is the trivial surface counted with the quantum dimension of $X$, just as $\cF(E)=2$ in example~\ref{ex:S3-lag-rg-Z2Z3}. In particular, $\cF(X)$ is not a condensation surface of the 3d $\bbZ_2$ gauge theory.

\newl {\bf Dependence on $A$.} Condensation surfaces do appear as images of the other objects of the same component. Repeating the computation for the objects $X_U$ of \eqref{eq:S3-XU},
\begin{equation}\label{eq:S3-FXU}
\cF(X_U)\simeq U\oplus\big(A_{-1}\boxtimes_{A_1}U\boxtimes_{A_1}A_{-1}\big)\,.
\end{equation}
Conjugation by $A_{-1}$ implements the $e$--$m$ duality of $\mathrm{TV}[\Vect_{\bbZ_2}]$. For $U=S_e$, the surface of $\mathrm{TV}[\Vect_{\bbZ_2}]$ on which $e$ condenses, this gives $\cF(X_{S_e})\simeq S_e\oplus S_m$. This is where the choice of $\bbZ_2$-graded fusion category $A$ enters. The computation of $\cF(X)$ only used $A_{-1}\boxtimes_{A_1}A_{-1}\simeq A_1$, which holds for any faithfully $\bbZ_2$-graded $A$, so $\cF(X)\simeq1\oplus1$ for every choice. The extension $A$ enters through conjugation by $A_{-1}$: in \eqref{eq:S3-FXU}, and in the $\bbZ_2$-action on $\cZ(A_1)$ that determines $\Omega\CUV=\cZ(A_1)^{\bbZ_2}$ and $\Omega\cF$. For instance, take $A=\Vect_{\bbZ_2\times\bbZ_2}$, graded by the second factor, so that $A_1=\Vect_{\bbZ_2}$ and $A_{-1}\simeq A_1$ as a bimodule and the action on $\mathrm{TV}[\Vect_{\bbZ_2}]$ is trivial. Then $\cF(X_{S_e})\simeq S_e\oplus S_e$, and $\Omega\CUV\simeq\cZ(\Vect_{\bbZ_2})\boxtimes\Rep(\bbZ_2)$ has eight simple lines instead of five. For $A=\Vect_{\bbZ_2}$, without $\mathrm{TV}[\Vect_{\bbZ_2}]$, $\Omega\CIR\simeq\Vect$ and $\cF$ is a genuine fiber 2-functor into $2\Vect$, the counterpart of the fiber functor of $\Rep(S_3)$ in example~\ref{ex:S3-lag-rg-Z2Z3}.

\newl {\bf Twisted sectors.} The objects $X$ and $1$ lie in different components of $\CUV$, so $\Hom_{\CUV}(X,1)=0$: in the UV, $X$ cannot end. In the IR, by \eqref{eq:S3-FX},
\begin{equation}
\Hom_{\CIR}\big(\cF(X),\cF(1)\big)\simeq\Omega\CIR\oplus\Omega\CIR\neq0\,,
\end{equation}
with one copy of the IR lines for each summand of $\cF(X)$. So $X$ acquires endpoints, the $X$-twisted sector of the IR theory. This is the 2-categorical counterpart of the $E$-twisted operators of example~\ref{ex:S3-lag-rg-Z2Z3}, counted by \eqref{eq:warm-up-trace}. In the SymTFT, these endpoints come from the bulk surfaces of flux $[a]$ that end on the physical boundary, i.e.\ the two copies of $S_{[a],\bbZ_3}$ in \eqref{eq:S3-LB-surfaces}, stretched across the slab and ending on $X$, which lies in the component $\bbZ_2a\bbZ_2$ reached by the flux-$[a]$ surfaces (example~\ref{ex:S3-2lag-uv}). Both the two copies of $S_{[a],\bbZ_3}$ and the two summands of $\cF(X)$ are labeled by $a,a^2\in K$.

\newl {\bf The twist $\omega$.} The twist enters exactly where the endpoints of $X$ do. It does not change $\cF$ on objects or on 1-morphisms, nor $\Omega\cF$: all of these were read off from identity components, on which $\omega$ is trivial. It enters through the labels $k\in K$ of the summands of $\cF(X)$. Composing summands multiplies their labels through $B_k\boxtimes B_{k'}\simeq B_{kk'}$, whose associativity constraint is $\omega$, so $\omega$ appears in the coherence data of $\cF$, namely its tensorator and the modification relating the two ways of composing three tensorators. In other words, $\omega$ plays the role of the SPT phase of the unbroken $\bbZ_3$, the analogue of the 3-cochain $\phi$ of the 3-group homomorphisms in section~\ref{sec:sym-ext}. In the SymTFT it is the associator of $\cL(B)$ on its flux-$[a]$ part (example~\ref{ex:S3-2lag-LB}), so it sits on the same bulk surfaces that provide the endpoints of $X$. The three classes $[\omega]\in \operatorname{H}^3(\bbZ_3;U(1))\simeq\bbZ_3$ thus give three distinct single-vacuum phases, i.e.\ three inequivalent $\CUV$-module 2-categories, with the same $\cF$ on objects and the same $\Omega\cF$. This parallels the classification of genuine fiber 2-functors by pairs $(K,\omega)$ \cite{Decoppet:2023bay}. It has no 1-categorical counterpart, since $\operatorname{H}^2(\bbZ_3;U(1))=0$ (example~\ref{ex:S3-lag-rg-Z2Z3}).
\end{examplebox}

\subsection{Example: \texorpdfstring{$p : Q_8 \to \bbZ_2^2$}{p : Q8 -> Z2^2} Revisited}\label{subsec:ex-Q8-revisited}
In section~\ref{subsec:ex-Q8} we produced a generalized fiber 2-functor for the anomalous symmetry $2\Vect^\pi_{\bbZ_2^2}$ by a symmetry-extension argument. We now recover the same 2-functor from the SymTFT. We follow the steps of the running $S_3$ example: the boundary data $(H,A)$ and $(K,B)$ of section~\ref{subsec:module-2-cats}, and then the Lagrangian algebras $\cL(A)$ and $\cL(B)$ of section~\ref{subsec:lagrangian-algebras}. Throughout, $G=\bbZ_2^2$ and $\pi$ is the anomaly \eqref{eq:q8-anomaly}.

\newl For the IR topological theory, we expect to obtain the 3d $\bbZ_2$ Dijkgraaf--Witten theory, whose topological lines form the braided fusion category $\cZ(\Vect_{\bbZ_2})$. All topological surfaces are condensation descendants, so the IR symmetry is the fusion 2-category
\begin{equation}
    \CIR \simeq \Sigma\cZ(\Vect_{\bbZ_2}) \simeq \End(2\Rep(\bbZ_2))\, .
\end{equation}
The RG 2-functor is therefore a monoidal 2-functor from the UV symmetry $\CUV \simeq 2\Vect_{\bbZ_2^2}^\pi$ to the IR symmetry,
\begin{equation}
    \cF \colon 2\Vect_{\bbZ_2^2}^\pi \to \End(2\Rep(\bbZ_2))\, .
\end{equation}

\begin{figure}[H]
    \centering
    \begin{tikzpicture}[scale = 1]
        \draw[gray, thick] (0,2) -- (1.5,2) node[anchor = south, black]{$\strut \fZ=\cZ_1\big(2\Vect^\pi_{\bbZ_2^2}\big)$} -- (3, 2);
        \draw[gray, thick] (-1,1) -- (2, 1);
        \draw[gray, thick] (0,0) -- (3, 0);
        \draw[gray, thick] (-1,-1) -- (2, -1);
        \draw[black, thick] (-1, -1) -- (-1, 1) -- (0, 2) -- (0, 0) -- (-1, -1);
        \node[anchor = north, align = center, font = \small] at (-0.5,-1.1) {physical\\ $\fB_{(\bbZ_2^2,\,\Vect^\psi_{Q_8})}$};
        \begin{scope}[shift={(3,0)}]
        \draw[black, thick] (-1, -1) -- (-1, 1) -- (0, 2) -- (0, 0) -- (-1, -1);
        \node[anchor = north, align = center, font = \small] at (-0.5,-1.1) {symmetry\\ $\fB_{\rm Dir}$};
        \node[anchor = west, font = \small] at (0.05,0.5) {$\CUV=2\Vect^\pi_{\bbZ_2^2}$};
        \end{scope}
        \draw[black] (6.0, 0.5) node{\huge$\strut\rightsquigarrow$};
        \begin{scope}[shift = {(8, 0)}]
            \draw[black, thick] (-1, -1) -- (-1, 1) -- (0, 2) -- (0, 0) -- (-1, -1);
            \node[anchor = north, align = center, font = \small] at (-0.5,-1.1) {$\TIR=\mathrm{DW}(\bbZ_2)$};
            \node[anchor = west, font = \small] at (0.05,0.5) {$\CIR\simeq\Sigma\cZ(\Vect_{\bbZ_2})$};
        \end{scope}
    \end{tikzpicture}
    \caption{The SymTFT sandwich for the symmetry-extension phase of section~\ref{subsec:ex-Q8}. The symmetry boundary is the Dirichlet boundary, supporting $\CUV=2\Vect^\pi_{\bbZ_2^2}$. The physical boundary $\fB_{(\bbZ_2^2,\,\Vect^\psi_{Q_8})}$ encodes the extension $\bbZ_2\to Q_8\xrightarrow{p}\bbZ_2^2$ together with the trivialization $\delta\psi=p^*\pi$. Compactifying the interval gives the 3d $\bbZ_2$ Dijkgraaf--Witten theory with a single vacuum.}
    \label{fig:Q8-sandwich}
\end{figure}
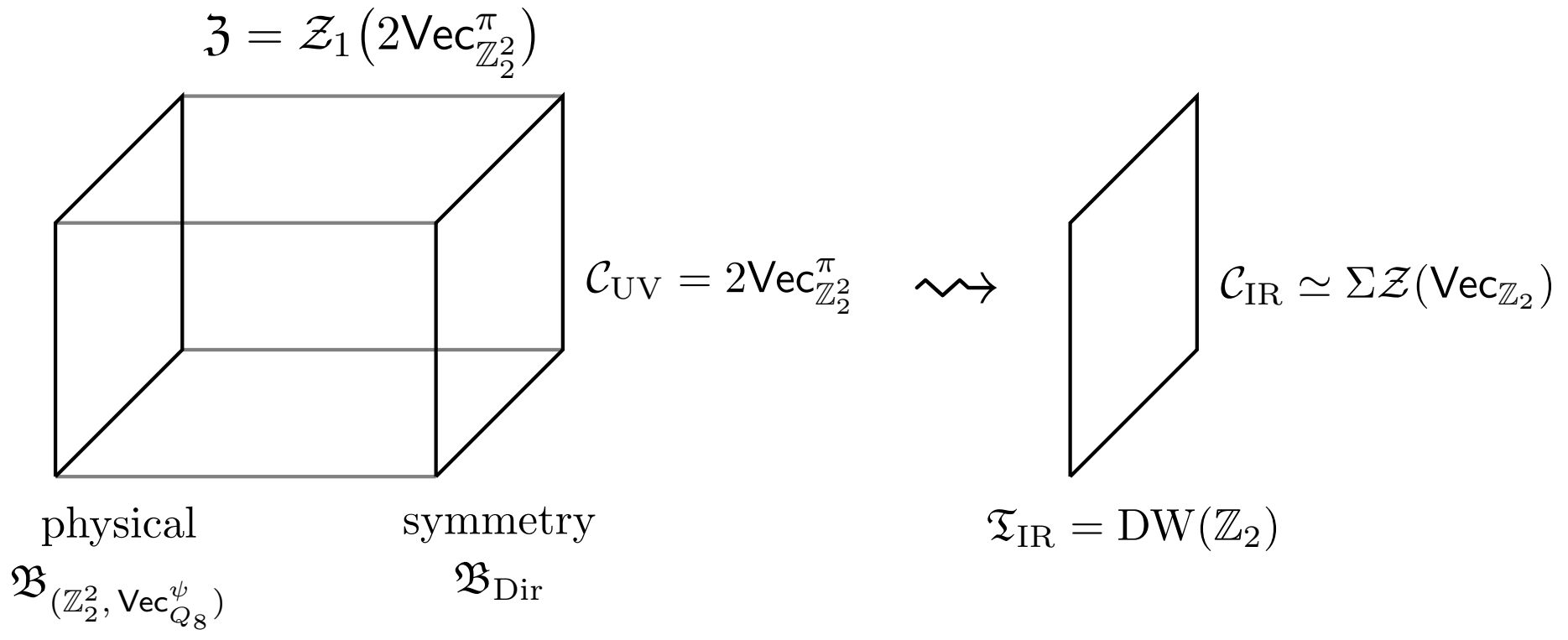

\newl {\bf The sandwich.} The SymTFT is the 4d $\bbZ_2^2$ Dijkgraaf--Witten theory twisted by $\pi$. Its topological surface and line defects form the braided fusion 2-category
\begin{equation}\label{eq:Q8-bulk}
    \fZ = \cZ_1\big(2\Vect_{\bbZ_2^2}^{\pi}\big) \, .
\end{equation}
Its topological boundaries are labeled by pairs $(H,A)$ as in section~\ref{subsec:symtft-defects-and-boundaries}, i.e.\ by Morita classes of algebras in $2\Vect_{\bbZ_2^2}^{\pi}$. Since $\CUV$ is the full anomalous $\bbZ_2^2$ symmetry, the symmetry boundary is the Dirichlet boundary $\fB_{\rm Dir}=\fB_{(\{\id\},\Vect)}$. The physical boundary encodes the symmetry extension. The fusion category $B=\Vect^{\psi}_{Q_8}$ is graded over $K=\bbZ_2^2$ through $p$, its neutral component is $\Vect_{\ker p}=\Vect_{\bbZ_2}$, and its associator is the 3-cochain $\psi$ of section~\ref{subsec:ex-Q8}. The condition $\delta\psi=p^*\pi$ is exactly what makes $B$ a $\pi$-twisted $\bbZ_2^2$-graded fusion category, i.e.\ an algebra in $2\Vect^\pi_{\bbZ_2^2}$. In summary,
\begin{align}
    \text{symmetry:}&\qquad H = \{\id\}\, ,\quad\;\;\, A = \Vect\, ,\quad\;\;\;\, A_1 = \Vect\, , \label{eq:Q8-A-boundary}\\
    \text{physical:}&\qquad K = \bbZ_2^2\, ,\quad\; B = \Vect_{Q_8}^{\,\psi}\, ,\quad B_1 = \Vect_{\bbZ_2}\, . \label{eq:Q8-B-boundary}
\end{align}
The resulting sandwich is shown in figure~\ref{fig:Q8-sandwich}. Compared with the running $S_3$ example (example~\ref{ex:S3-2lag-algebras}), the roles of the two boundaries are reversed. There, the symmetry boundary carries the non-trivial fusion category $A=\Ising$ and the physical boundary has $B_1=\Vect$. Here $A_1=\Vect$, and the topological order sits on the physical boundary, $B_1=\Vect_{\bbZ_2}$. As there, $G=HK$ and $H\cap K=\{\id\}$, so the IR has a single vacuum by \eqref{eq:single-vacuum}.

\newl {\bf Module 2-categories.} With the boundaries fixed, section~\ref{subsec:uv-symmetry} computes the UV symmetry 2-category and the interface 2-category $\cM$:
\begin{equation}
    \CUV \simeq \mathsf{Bimod}_{2\Vect_{\bbZ_2^2}^{\pi}}(\Vect) \simeq 2\Vect_{\bbZ_2^2}^{\pi}\,, \qquad
    \cM \simeq \mathsf{LMod}_{2\Vect_{\bbZ_2^2}^{\pi}}(\Vect) \simeq 2\Vect_{\bbZ_2^2}^{\pi}\,.
\end{equation}
So $\cM$ is the regular module, and the Dirichlet boundary has no non-trivial lines, $\Omega\CUV\simeq\Vect$. Section~\ref{subsec:ir-symmetry} computes the IR symmetry by compactifying the SymTFT with these two topological boundaries. Since $B$ is faithfully graded over $K=G$, a $B$-module in $2\Vect^\pi_{\bbZ_2^2}$ is determined by its neutral component, which is a module category over $B_1=\Vect_{\bbZ_2}$. Hence
\begin{equation}
\begin{split}
    \cN &\simeq \mathsf{RMod}_{2\Vect_{\bbZ_2^2}^{\pi}}(\Vect^\psi_{Q_8}) \simeq \mathsf{RMod}(\Vect_{\bbZ_2})\simeq 2\Rep(\bbZ_2)\,, \\
    \CIR &\simeq \End(\cN \boxtimes_{2\Vect_{\bbZ_2^2}^{\pi}} \cM) \simeq \End(\cN)\simeq \Sigma\cZ(\Vect_{\bbZ_2})\,,
\end{split}
\end{equation}
in agreement with \eqref{eq:boundary-cat} for $[\bar{B}\boxtimes A]_{H\cap K}=\bar{B}_1\simeq\Vect_{\bbZ_2}$. The two simple objects of $\cN$ are the $e$- and $m$-condensing boundaries of the 3d $\bbZ_2$ Dijkgraaf--Witten theory, as in \eqref{eq:S3-IR-boundaries}.

\newl {\bf The RG 2-functor.} By \eqref{eq:RG-functor}, a simple surface $V_a\in2\Vect^\pi_{\bbZ_2^2}$, $a\in\bbZ_2^2$, acts on $\cN$ by shifting the grading of a $B$-module by $a$. On neutral components, this is the relative tensor product with the graded component $B_a$, an invertible $B_1$-bimodule category. Since $\ker p$ is central in $Q_8$, and $\psi$ restricts trivially whenever two of its arguments lie in $\ker p$, $B_a\simeq B_1$ as a $B_1$-bimodule. Hence
\begin{equation}\label{eq:Q8-F-objects}
    \cF(V_a)\simeq 1\qquad\text{for all } a\in\bbZ_2^2\,,
\end{equation}
which reproduces the trivial homomorphism $\rho$ of section~\ref{subsec:ex-Q8}: the anomalous symmetry acts trivially on the IR surfaces. The non-trivial information sits in the coherence data of $\cF$, as for the twist $\omega$ in example~\ref{ex:S3-2lag-rg}. The tensorator $\cF(V_a)\otimes\cF(V_b)\Rightarrow\cF(V_{ab})$ is induced by the multiplication $B_a\boxtimes_{B_1}B_b\to B_{ab}$, i.e.\ by $\tilde a\,\tilde b=\varphi(a,b)\,\widetilde{ab}$ in $Q_8$. Its left $B_1$-module structure is twisted by $\psi(-1,\tilde a,\tilde b)=(-1)^{a_1b_1}=\hat\varphi(a,b)$. So the tensorator is the invertible line of $\cZ(\Vect_{\bbZ_2})\simeq\Vect_{K\times\hat K}$ labeled by $(\varphi(a,b),\hat\varphi(a,b))$, the pair of 2-cocycles of section~\ref{subsec:ex-Q8}. On the lifts themselves, $\psi(\tilde a,\tilde b,\tilde c)=\phi(a,b,c)$, and these values give the modification relating the two ways of composing three tensorators. This recovers the 3-group homomorphism $\cF^\times$ of section~\ref{subsec:ex-Q8}.

\newl {\bf Lagrangian algebras.} We now repeat the computation with Lagrangian algebras, following examples~\ref{ex:S3-2lag-algebras}, \ref{ex:S3-2lag-LB}, \ref{ex:S3-2lag-uv} and~\ref{ex:S3-2lag-ir}. For the symmetry boundary, $H=\{\id\}$ and $\cZ_H(A)=\Vect$, so \eqref{eq:LA-general} gives
\begin{equation}\label{eq:Q8-A-Lagrangian}
    \cL(A) \simeq \Fun\big(\bbZ_2^2,\Vect\big)\simeq \Vect^{\oplus \bbZ_2^2} \, .
\end{equation}
It is concentrated in degree $\id$, and $\bbZ_2^2$ permutes the four copies by translation. This is the condensation of all Wilson lines $\Fun(G)$, the 2-categorical counterpart of the Lagrangian algebra $\cL_{\{\mathrm{id}\}}=\Fun(G)$ of the Dirichlet boundary in section~\ref{subsec:warm-up-lagrangian}. By \eqref{eq:lag-CUV}, $\CUV\simeq{}_{\cL(A)}\fZ\simeq2\Vect^\pi_{\bbZ_2^2}$, and $\Omega\CUV\simeq(\cL(A)_\id)^{\bbZ_2^2}\simeq\Vect$.

\newl For the physical boundary, $K=G$, so the induction in \eqref{eq:LA-general} is trivial and the full center is the $\bbZ_2^2$-crossed center of $B$,
\begin{equation}\label{eq:Q8-B-Lagrangian}
    \cL(B) \simeq \cZ_{\bbZ_2^2}\big(\Vect^{\psi}_{Q_8}\big)\simeq\bigoplus_{a\in\bbZ_2^2}\cZ_{\Vect_{\bbZ_2}}(B_a) \simeq \bigoplus_{a\in\bbZ_2^2}\cZ(\Vect_{\bbZ_2})_a \, .
\end{equation}
The last step uses $B_a\simeq B_1$ as $B_1$-bimodules, as above. $\cL(B)$ is faithfully graded by $\bbZ_2^2$: every flux $a$ ends on the physical boundary, because $\bbZ_2^2$ is gauged there. On objects, the $\bbZ_2^2$-action is conjugation by the lifts, $\tilde a\mapsto\tilde g\,\tilde a\,\tilde g^{-1}$. It fixes the neutral component $\cZ(\Vect_{\bbZ_2})$ objectwise, since $\ker p$ is central, but it is non-trivial on the other components, since distinct non-trivial lifts anticommute in $Q_8$. The product is induced by the multiplication of $Q_8$, with associator $\psi$. So the data $(\varphi,\hat\varphi,\phi)$ found above is stored in the $\bbZ_2^2$-crossed structure of $\cL(B)$, just as $\omega$ is stored in $\cL(B)$ for the running example (example~\ref{ex:S3-2lag-LB}).

\newl As in example~\ref{ex:S3-2lag-ir}, $\cL(A)$ is supported in degree $\id$, so only $x=\id$ contributes to $(\cL(A)\otimes\cL(B))_\id$:
\begin{equation}
    \Omega\CIR\simeq\big(\cL(A)_\id\boxtimes\cL(B)_\id\big)^{\bbZ_2^2}\simeq\Big(\bigoplus_{g\in\bbZ_2^2}\cZ(\Vect_{\bbZ_2})^{(g)}\Big)^{\bbZ_2^2}\simeq\cZ(\Vect_{\bbZ_2})\,,
\end{equation}
since $\bbZ_2^2$ permutes the four copies freely and transitively. By \eqref{eq:lag-CIR-bimod}, $\CIR\simeq{}_{\cL(A)}\fZ_{\cL(B)}\simeq\End(\cN)\simeq\Sigma\cZ(\Vect_{\bbZ_2})$, and the RG 2-functor is the free-module 2-functor \eqref{eq:lag-free}, $Y\mapsto Y\otimes\cL(B)$. In the bimodule presentation it is the action on $\cN$ computed above.

\newl {\bf Minimal emergent lines.} Since $\Omega\CUV\simeq\Vect$, the functor $\Omega\cF:\Vect\to\cZ(\Vect_{\bbZ_2})$ is not surjective. The three non-trivial IR lines are all emergent, and the phase is not minimal in the sense of section~\ref{subsec:minimality}. This cannot be avoided. Since $H=\{\id\}$, a single vacuum forces $K=G$, so $B$ is a $\pi$-twisted $G$-graded extension of $B_1$, and $\Omega\CIR\simeq\cZ(\bar B_1)$. For $[\pi]\neq0$ we cannot have $B_1\simeq\Vect$, since then $B\simeq\Vect^{\psi}_{G}$ with $\delta\psi=\pi$. So $\Omega\CIR$ has global dimension $\dim(B_1)^2\geq4$, with equality exactly when $B_1\simeq\Vect^\alpha_{\bbZ_2}$. The extension by $Q_8$ attains this bound. It also has at least four simple lines: a non-trivial modular category with fewer than four simple objects has chiral central charge $c\not\equiv0$ mod $8$ \cite{Rowell:2007dge}, and so is not a Drinfeld center. This is the second strategy at the end of section~\ref{subsec:minimality}: $H\cap K=\{\id\}$ is kept and $B_1$ is made as small as possible. Among single-vacuum phases of the form considered in this section, the 3d $\bbZ_2$ Dijkgraaf--Witten theory therefore has the fewest emergent lines.

\section{Discussion and Outlook}\label{sec:discussion}

In this work we asked which fusion 2-categorical symmetries in $(2\!+\!1)$ dimensions admit a gapped phase with a unique vacuum, and how the symmetry acts in such a phase. We formalized the latter by a generalized fiber 2-functor $\cF:\CUV\to\Sigma\cB$, with $\cB$ the braided fusion category of lines of the IR topological order, which describes the RG flow of all topological defects. Our main result is that every bosonic (AB-type) fusion 2-categorical symmetry admits such a phase. This is in sharp contrast with $(1\!+\!1)$ dimensions, where a single vacuum forces the IR theory to be trivial, so that a single-vacuum phase exists only when the symmetry admits a fiber functor, and with genuine fiber 2-functors in $(2\!+\!1)$ dimensions, which exist only when the symmetry admits a non-anomalous complement. Unlike in $(1\!+\!1)$ dimensions, where a twist can compensate a non-trivial intersection $H\cap K$, this complement condition cannot be relaxed (section~\ref{subsec:symtft-rg-flow}).

\newl We gave two constructions of these phases.
For invertible symmetries $2\Vect_G^\pi$, a generalized fiber 2-functor arises from any symmetry extension that trivializes the anomaly, through a homomorphism of 3-groups (section~\ref{sec:sym-ext}).
This section was largely a review of \cite{Wang:2017loc}, phrased slightly differently.
For general bosonic symmetries, we constructed the phases from the SymTFT by pairing the symmetry boundary with a suitable physical boundary (section~\ref{sec:gen-fib-fun-symtft}). We computed the UV symmetry, the IR TQFT and the RG 2-functor explicitly, first using module 2-categories and then using Lagrangian algebras, in which the RG 2-functor becomes a free module functor. Section~\ref{sec:warm-up} serves as a review and develops the same constructions one dimension lower, where every step can be made completely explicit and the comparison isolates what is genuinely 2-categorical. Finally, we introduced minimal generalized fiber 2-functors, those producing no emergent 1-form symmetry, and determined when they exist: when the symmetry boundary admits a non-anomalous complementary boundary with no additional topological order, in which case the IR lines are canonically the modular closure of the UV 1-form symmetry.

\newl Conceptually, the mechanism is simple. The SymTFT always admits a topological boundary that, together with the symmetry boundary, leaves a single vacuum. The obstruction to a trivially gapped phase is not removed, but transferred to the IR topological order, which carries the anomaly in the form of emergent lines. Several directions remain open.

\begin{itemize}
\item \emph{Minimal phases beyond the minimality criterion.} When the conditions for a minimal generalized fiber 2-functor cannot be met, the two natural strategies, minimizing the intersection $H\cap K$ or the stacked topological order, compete, and we do not know whether a canonical smallest IR phase exists in general. A natural measure is the total dimension of $\Omega\CIR$. Since $\Omega\CUV\simeq\cZ(A_1)^H$ has Tannakian M\"uger center $\Rep(H)$, the question is also related to the existence of minimal non-degenerate extensions, whose obstruction is an anomaly in $\operatorname{H}^4(H;U(1))$; making this relation precise would sharpen our criterion.
\item \emph{Dynamics.} Our results classify which phases are allowed, not which are realized. Lattice models or field theories flowing to the single-vacuum phases constructed here, especially for anomalous and non-invertible symmetries, would test the RG 2-functors directly and connect to the symmetry-extension construction of symmetric gapped boundaries.
\item \emph{Higher dimensions and gapless phases.} The SymTFT arguments used here have natural generalizations to fusion $n$-categories and to gapless phases via generalized sandwich constructions. Whether every bosonic symmetry in $(d\!+\!1)$ dimensions, $d\geq3$, admits a unique-vacuum phase is an interesting question.
\end{itemize}


\acknowledgments
We thank Ingo Runkel and Andrea Grigoletto for useful discussions.
This work is funded by Villum Fonden Grant no.\ VIL60714.
In preparing this work we used the AI models Claude Opus and Claude Fable (Anthropic) to discuss mathematical concepts and to assist with writing and editing. 
All results and arguments were checked by the authors, who take full responsibility for the content of the paper.


\appendix
\section{Loop space and Suspensions of Fusion $n$-Categories}
\label{App:loops and suspensions}
\newl In this appendix we describe two basic operations on fusion $n$-categories that appear often throughout the main text.
These correspond to taking the loop space $\Omega$ and the suspension $\Sigma$ of a fusion $n$-category.
Together these operations allow one to move between the defect structures of theories in adjacent dimensions, and play a central role in the classification of topological orders and their boundary conditions.

\newl Given a fusion $n$-category $\mathcal{C}$ describing the topological defects of a $d$-dimensional theory, the loop space $\Omega\mathcal{C} := \End_{\mathcal{C}}(1)$ is the fusion $(n\!-\!1)$-category of defects of one lower dimension, obtained as the endomorphisms of the trivial defect in $\mathcal{C}$.
Concretely, $\Omega\mathcal{C}$ extracts the defects one codimension higher: if $\mathcal{C}$ is the fusion 2-category of surface defects in a $(2\!+\!1)$d theory, then $\Omega\mathcal{C}$ is the braided fusion category of line defects. 
The loop space operation can be iterated: $\Omega^k\mathcal{C}$ extracts defects of codimension $k$ higher than those in $\mathcal{C}$.

\newl The suspension $\Sigma\mathcal{C}$ is the left adjoint to $\Omega$, satisfying $\Omega\Sigma\mathcal{C} \simeq \mathcal{C}$. Concretely, $\Sigma\mathcal{C}$ is constructed in two steps \cite{Gaiotto:2019xmp}. One first forms the 
one-object delooping $\B\mathcal{C}$, the $(n\!+\!1)$-category with a single object $*$ and $\End(*) = \mathcal{C}$, and then performs the Karoubi completion (also known as condensation completion) by formally splitting all separable monads.
This completion adds all topological defects that can be obtained by condensing defects already present in $\mathcal{C}$.
The result $\Sigma\mathcal{C}$ is equivalently described as the Morita category of separable algebra objects in $\mathcal{C}$ and their bimodules. 
Physically, suspension adds a dimension: a fusion 2-category $\Sigma\mathcal{B}$ built from a braided fusion category $\mathcal{B}$ describes the defects of a theory one dimension higher where $\mathcal{B}$ controls the lines. 
A key fact is that every connected fusion 2-category (one with a single component) is of the form $\Sigma\mathcal{B}$ for a braided fusion category $\mathcal{B}$.

\section{3-Groups and their Homomorphisms}
\label{App:3-groups and homomorphisms}
\newl In the context of the present work, 3-groups embody the algebraic structure formed by the invertible topological defects of a $(2\!+\!1)$-dimensional theory.
These topological defects are the invertible surfaces, invertible lines, and invertible local operators, together with the rules for how they fuse and how they fail to fuse associatively. 
Just as a finite group $G$ is the same thing as a connected space $\B G$ with only $\pi_1$ non-vanishing, a 3-group $\mathbb{G}$ is the same thing as a connected space $\B\mathbb{G}$ whose only non-vanishing homotopy groups are $\pi_1$, $\pi_2$, and $\pi_3$. Unpacking this, a 3-group consists of the following data.

\begin{enumerate}
    \item {\bf Homotopy groups:} 
    \begin{equation}
    \Pi_0 := \pi_1(\B\mathbb{G}) \,, \qquad
    \Pi_1 := \pi_2(\B\mathbb{G}) \,, \qquad
    \Pi_2 := \pi_3(\B\mathbb{G}) \,.
\end{equation}
Here $\Pi_0$ is a possibly non-abelian finite group, while $\Pi_1$ and $\Pi_2$ are abelian. 
Physically, $\Pi_0$ labels the invertible surfaces (the 0-form symmetry), $\Pi_1$ labels the invertible lines (the 1-form symmetry), and $\Pi_2$ labels the invertible local operators, which in a unitary theory are just phases, $\Pi_2 = U(1)$ (or $\mathbb{C}^\times$).
\item {\bf Actions.} The group $\Pi_0$ acts on both $\Pi_1$ and $\Pi_2$ by automorphisms $\alpha_{j}$.
\begin{equation}
    \alpha_{j}:\Pi_{0}\longrightarrow \Aut(\Pi_j)\,.
\end{equation}
Unless mentioned, we assume $\alpha_{j}$ to be trivial.
Physically, this is the action of the 0-form symmetry on the lines and local operators. 
The group $\Pi_1$ acts trivially on $\Pi_2$.
\item {\bf Postnikov classes.} The three layers are glued together by two cohomology classes, which one may think of as the associator and the pentagonator of the corresponding fusion 2-category:
\begin{align}
    k_3 \in \operatorname{H}^3\big(\B\Pi_0;\, \Pi_1\big) \,, \qquad
    k_4 \in \operatorname{H}^4\big(\B\mathbb{G}_{\leq 1};\, \Pi_2\big) \,,
\end{align}
where $\mathbb{G}_{\leq 1}$ denotes the 2-group $\Pi_1[1]\cdot\Pi_0[0]$ obtained from the first two layers and $k_3$ alone. 
The class $k_3$ measures the failure of three $\Pi_0$-surfaces to fuse associatively, the mismatch being a $\Pi_1$-line. 
The class $k_4$ is the 't Hooft anomaly of the full 2-group symmetry $\mathbb{G}_{\leq 1}$. Its restriction to the 1-form part, $k_4|_{\B^2\Pi_1}\in \operatorname{H}^4(\B^2\Pi_1;\Pi_2)\cong \mathrm{Quad}(\Pi_1,\Pi_2)$, is a quadratic form on $\Pi_1$.
Physically it records the topological spins of the invertible lines, and hence the anomaly of the 1-form symmetry alone.
\end{enumerate}

\newl{\bf 3-group homomorphisms.}
A homomorphism of 3-groups $\cF^\times:\mathbb{G}\to\mathbb{G}'$ is a map of spaces $\B\mathbb{G}\to \B\mathbb{G}'$, and physically describes how the invertible defects of one theory are realized inside another. Unpacking layer by layer, it consists of:
\begin{enumerate}
    \item A group homomorphism $\rho_{0}:\Pi_0\to\Pi_0'$, together with $\rho_{0}$-equivariant homomorphisms $\rho_1:\Pi_1\to\Pi_1'$ and $\rho_2:\Pi_2\to\Pi_2'$.
    Here $\rho_0$ equivariance implies that the following diagram commutes
    \begin{equation}
\begin{tikzcd}
    \Pi_1 \arrow[r, "\rho_1"] \arrow[d, "\alpha_1(g)"'] & \Pi_1' \arrow[d, "\alpha'_1(\rho_0(g))"] \\
    \Pi_1 \arrow[r, "\rho_1"'] & \Pi_1'
\end{tikzcd}
\quad
\,,
\qquad 
\begin{tikzcd}
    \Pi_2 \arrow[r, "\rho_2"] \arrow[d, "\alpha_2(g)"'] & \Pi_2' \arrow[d, "\alpha'_2(\rho_0(g))"] \\
    \Pi_2 \arrow[r, "\rho_2"'] & \Pi_2'
\end{tikzcd}\,,
\end{equation}
for all $g\in \Pi_0$.

    \item A 2-cochain $\phi\in \operatorname{C}^2(\B\Pi_0;\Pi_1')$ trivializing the difference of the first Postnikov classes,
    \begin{equation}
        \delta\phi = (\rho_0)^* k_3' - (\rho_1)_{*}k_3 \,.
    \end{equation}
    In the physical example where $\mathbb G=\TwoVec_{G}^{\pi}$ and $\bbG'=\rm Pic(\cB)$ (the 3-group of invertible defects in the braided fusion category $\cB$) this 2-cocycle prescribes the symmetry fractionalization class of the $(G,\pi)$ symmetry realization on a 3d TFT whose underlying topological order is given by $\cB$.
    Physically, it records that the fusion of two $G$-surfaces is dressed by an invertible line in  $\cB$.
    \item A 3-cochain $\varphi\in \operatorname{C}^3(\B\mathbb G_{\leq 1};\Pi_2')$ trivializing the difference of the second Postnikov classes,
    \begin{equation}
        \delta\varphi = (\rho_0,\phi)^* k_4' - (\rho_2)_{*}k_4 \,.
    \end{equation}
    Here $(\rho_0,\rho_1,\phi)$ is understood as a map between 2-groups $B\bbG_{\leq 1}$ and $ B\bbG_{\leq 1}'$.
    Therefore pulling-back with respect to this map produces a class $(\rho_0,\rho_1,\phi)^{*}k_4'\in \operatorname{H}^{4}(B\bbG_{\leq 1};\Pi'_2)$.
    Meanwhile pushing forward $k_{4}$ by $\rho_2$ also produces a class $\operatorname{H}^{4}(B\bbG_{\leq 1};\Pi'_2)$.
    The difference of these classes is trivialized by $\varphi$.
    Physically in the context of homomorphism between  $\mathbb G=\TwoVec_{G}^{\pi}$ and $\bbG'=\rm Pic(\cB)$, this is an SPT phase for $\Pi_0$, and the existence of a solution says that the anomaly $k_4$ is cancelled by the anomaly $k_4'$ pulled back along the lower data.
\end{enumerate}
Two homomorphisms are equivalent if they are related by a homotopy, which shifts $\phi$ by a 1-cochain and $\varphi$ by a 2-cochain in the usual way.

\section{The Lyndon--Hochschild--Serre Spectral Sequence}\label{app:spec-seq}

Given a central extension
\begin{equation}
    K \to \Gamma \xrightarrow{p} G\, ,
\end{equation}
characterized by Postnikov data $\varphi \in \operatorname{Z}^2(G; K)$, the Lyndon--Hochschild--Serre (LHS) spectral sequence gives us a way to compute the group cohomology of $\Gamma$ from the cohomologies of $G$ and $K$. More precisely, there is a sequence of bigraded differential complexes
\begin{equation}
    E_r = \bigoplus_{p, q \geq 0} E^{p,q}_r\, ,\qquad d_r\, : \, E_r^{p,q} \to E_r^{p+r, q-r+1}\, ,\qquad d_r^2 = 0\, ,
\end{equation}
called ``pages", for $r \geq 2$ where
\begin{equation}
    E_{r+1} = H(E_r, d_r)\, ,\qquad E_2^{p,q} = \operatorname{H}^p(G; \operatorname{H}^q(K; U(1)))\, ,
\end{equation}
and for some finite $R$, the sequence stabilizes, $E_r = E_\infty$ for all $r \geq R$. The limit computes $\operatorname{H}^{n}(\Gamma; U(1))$ up to extension: there is a filtration $F^\bullet$ of $\operatorname{H}^{n}(\Gamma; U(1))$ whose associated graded is
\begin{equation}
   \operatorname{gr}\,\operatorname{H}^{n}(\Gamma; U(1)) = \bigoplus_{p+q = n} F^p/F^{p+1} \simeq \bigoplus_{p+q = n} E^{p,q}_\infty\, .
\end{equation}
In general $\operatorname{H}^{n}(\Gamma; U(1))$ is not the direct sum of these pieces; for instance $\operatorname{H}^{1}(\bbZ_4; U(1))\simeq\bbZ_4$ for $\bbZ_2\to\bbZ_4\to\bbZ_2$, while $E_\infty^{1,0}\simeq E_\infty^{0,1}\simeq\bbZ_2$. Notice, on each page the map $d_r$ vanishes automatically on $E_r^{p,r-2}$, since the bigrading is required to be non-negative. For example, when $r = 2$ this captures the component $E_2^{p, 0}$ which automatically lifts via
\begin{equation}
    p^* \, : \, \operatorname{H}^p(G; U(1)) \to \operatorname{H}^p(\Gamma; U(1))\, .
\end{equation}

\newl This spectral sequence formalism can be applied physically in the context of gauging subgroups of invertible gapped phases. Take a $\Gamma$-symmetric symmetry protected topological (SPT) phase classified by an $n$-cocycle $\psi \in \operatorname{Z}^n(\Gamma; U(1))$. A natural question to ask then, is if we gauge the subgroup $K$ to produce a $G$ symmetry, what $G$-symmetric gapped phase will we end up in?

\newl The answer turns out to be given naturally by the LHS spectral sequence. With respect to the filtration above, the class of $\psi$ has leading components $\psi^{p,q} \in E_\infty^{p,q}$ with $p+q = n$, each defined once the components of lower $p$ vanish. Gauging $K$ on the SPT phase $\psi$ produces a Dijkgraaf--Witten theory with gauge group $K$ and action $\psi|_K$. The $G$-action on that theory is determined by a homomorphism of higher groups which is trivial on objects, and whose higher data is captured in full by the components $\psi^{p,q}$ and the Postnikov class $\varphi$.

\newl As an example, consider $n = 3$ so we are working in three dimensions. Suppose moreover that $\psi|_K$ is trivial, as in the example of section~\ref{subsec:ex-Q8}. The Dijkgraaf--Witten theory then has a $K\times\hat{K}$ 1-form symmetry; for non-trivial $\psi|_K$ the group of invertible lines is in general different, e.g.\ $\bbZ_9$ for $K=\bbZ_3$ with a generator of $\operatorname{H}^3(\bbZ_3;U(1))$. The higher data of the $G$-symmetry then contains a $K\times\hat{K}$-valued symmetry fractionalization given by $(\varphi, \psi^{2,1})$, and an SPT phase given by $\psi^{3,0}$.

\section{Fusion 2-Categories, Algebras and Morita Theory}\label{app:2cat}

\newl In this appendix we collect the 2-categorical background used in the main text, principally in section~\ref{sec:symtft} and section~\ref{sec:gen-fib-fun-symtft}. Nothing here is new: we fix notation and record the statements from Douglas--Reutter \cite{Douglas:2018qfz} and follow-up works. We suggest the reader refer to the thesis \cite{Xu:2026thesis} of one of the authors for a more detailed exposition of this material.

\subsection{Fusion 2-Categories}\label{app:fusion2cat}

Let us first recall the notion of a fusion 2-category from Douglas and Reutter \cite{Douglas:2018qfz}.

\begin{definition}\label{def:fss-2cat}
    A finite semisimple 2-category $\cC$ is a linear 2-category with the following properties:
    \begin{itemize}
        \item $\cC$ is locally finite semisimple, i.e. for any two objects $X,Y \in \cC$, the hom category $\cC(X,Y)$ is a finite semisimple category;

        \item $\cC$ has direct sums and a zero object;

        \item $\cC$ has finitely many equivalence classes of simple objects;

        \item Every object in $\cC$ is a finite direct sum of simple objects;

        \item Every 1-morphism in $\cC$ has a left and right adjoint;

        \item Any separable monad in $\cC$ splits.
    \end{itemize}
\end{definition}

\begin{definition}\label{def:rigid-2cat}
    A monoidal 2-category is rigid if every object has a left and right dual.
\end{definition}

\begin{definition}\label{def:fusion-2cat}
    A multifusion 2-category is a finite semisimple 2-category $\cC$ equipped with a rigid monoidal structure such that the tensor product is bilinear.

    \newl A fusion 2-category is a multifusion 2-category $\cC$ such that the unit object $\mathbf{1} \in \cC$ is simple.
\end{definition}

\subsection{Drinfeld Center}\label{app:drinfeld-center}

\begin{definition}\label{def:half-braiding}
    A half-braiding in a monoidal 2-category $\cC$ consists of an object $X \in \cC$ together with:
    \begin{itemize}
        \item A family of invertible 1-morphisms $\beta^X_Y : X \otimes Y \to Y \otimes X$ for each object $Y \in \cC$;

        \item A family of invertible 2-morphisms
        \begin{equation}
            \beta^X_f := \quad \begin{tikzcd}
                {X \otimes Y}
                    \arrow[r, "\beta^X_{Y}"]
                    \arrow[d, "1_X \otimes f"']
                & {Y \otimes X}
                    \arrow[d, "f \otimes 1_X"]
                \\
                {X \otimes Z}
                    \arrow[r, "\beta^X_Z"']
                    \arrow[ur, Rightarrow, shorten >=1em, shorten <=1em]
                & {Z \otimes X}
            \end{tikzcd}
        \end{equation}
        for each 1-morphism $f : Y \to Z$ in $\cC$;

        \item A family of invertible 2-morphisms
        \begin{equation}
            R^X_{Y,Z} := \quad \begin{tikzcd}
                {X \otimes Y \otimes Z}
                    \arrow[dr, "\beta^X_Y \otimes 1_Z"']
                    \arrow[rr, "\beta^X_{Y \otimes Z}"]
                & {}
                    \arrow[d, Rightarrow, shorten >=0.5em, shorten <=0.5em]
                & {Y \otimes Z \otimes X}
                \\
                {}
                & {Y \otimes X \otimes Z}
                    \arrow[ur, "1_Y \otimes \beta^X_Z"']
                & {}
            \end{tikzcd}
        \end{equation}
        for each pair of objects $Y,Z \in \cC$;
    \end{itemize}
    subject to the coherence condition (3.9) of \cite{Xu:2026thesis}.
\end{definition}

\begin{definition}\label{def:half-braiding-morphisms}
    Suppose $(X,\beta^X,R^X)$ and $(Y,\beta^Y,R^Y)$ are half-braidings in a monoidal 2-category $\cC$.

    \newl A 1-morphism of half-braidings from $(X,\beta^X,R^X)$ to $(Y,\beta^Y,R^Y)$ consists of a 1-morphism $f : X \to Y$ in $\cC$ together with a family of invertible 2-morphisms
    \begin{equation}
        \beta^f_{Z} := \quad \begin{tikzcd}
            {X \otimes Z}
                \arrow[r, "\beta^X_Z"]
                \arrow[d, "f \otimes 1_Z"']
            & {Z \otimes X}
                \arrow[d, "1_Z \otimes f"]
            \\
            {Y \otimes Z}
                \arrow[r, "\beta^Y_Z"']
                \arrow[ur, Rightarrow, shorten >=1em, shorten <=1em]
            & {Z \otimes Y}
        \end{tikzcd}
    \end{equation}
    for each object $Z \in \cC$, subject to the coherence conditions (3.10) of \cite{Xu:2026thesis}.

    \newl Let $(f,\beta^f)$ and $(g,\beta^g)$ be two 1-morphisms from $(X,\beta^X,R^X)$ to $(Y,\beta^Y,R^Y)$.

    \newl A 2-morphism of half-braidings from $(f,\beta^f)$ to $(g,\beta^g)$ consists of a 2-morphism $\alpha : f \Rightarrow g$ in $\cC$ compatible with the $\beta$'s in the sense of the coherence conditions (3.11) of \cite{Xu:2026thesis}.
\end{definition}

\begin{definition}\label{def:drinfeld-center}
    The Drinfeld center of a monoidal 2-category $\cC$ consists of the following data:
    \begin{itemize}
        \item Objects: half-braidings $(X,\beta^X,R^X)$ in $\cC$;

        \item 1-morphisms: 1-morphisms of half-braidings $(f,\beta^f)$ in $\cC$;

        \item 2-morphisms: 2-morphisms of half-braidings $\alpha$ in $\cC$.
    \end{itemize}
    We denote the Drinfeld center of $\cC$ by $\cZ_1(\cC)$.
\end{definition}

\newl The Drinfeld center of a fusion 2-category $\cC$ is a braided fusion 2-category \cite{Decoppet:2022kcf, Kong:2019brm}. The braiding is induced simply by applying the half-braiding of one object to another. Moreover, there is a canonical forgetful 2-functor $\cZ_1(\cC) \to \cC$ which forgets the half-braiding data.

\subsection{Algebras and Modules}\label{app:algebras-modules}

The definitions below are the ones used throughout section~\ref{sec:gen-fib-fun-symtft}; in particular section~\ref{subsec:lagrangian-algebras} refers back to them rather than repeating them.

\begin{definition}\label{def:algebra-2cat}
    An algebra in a monoidal 2-category $\cC$ is an object $A \in \cC$ together with:
    \begin{itemize}
        \item A multiplication $m : A \otimes A \to A$;

        \item A unit $i : \mathbf{1} \to A$;

        \item An associator $\alpha : m \circ (m \otimes 1_A) \Rightarrow m \circ (1_A \otimes m)$;

        \item A left unitor $\lambda : m \circ (i \otimes 1_A) \Rightarrow 1_A$;

        \item A right unitor $\rho : m \circ (1_A \otimes i) \Rightarrow 1_A$;
    \end{itemize}
    satisfying the pentagon and triangle conditions of \cite[4.1.1]{Xu:2026thesis}.
\end{definition}

\begin{definition}\label{def:braided-algebra-2cat}
    A braided algebra in a braided monoidal 2-category $\cB$ is an algebra $(A,m,i,\alpha,\lambda,\rho)$ in $\cB$ together with a braiding $c : m \circ \beta_{A,A} \Rightarrow m$ satisfying the coherence conditions of \cite[4.1.5]{Xu:2026thesis}.
\end{definition}

\begin{definition}\label{def:module-2cat}
    A left module over an algebra $A$ in a monoidal 2-category $\cC$ is an object $M \in \cC$ together with:
    \begin{itemize}
        \item An action $\triangleright : A \otimes M \to M$;

        \item An associator $\alpha : \triangleright \circ (m \otimes 1_M) \Rightarrow \triangleright \circ (1_A \otimes \triangleright)$;

        \item A left unitor $\lambda : \triangleright \circ (i \otimes 1_M) \Rightarrow 1_M$;
    \end{itemize}
    satisfying the pentagon and triangle conditions of \cite[4.2.1]{Xu:2026thesis}.

    \newl Similarly, one can define a right module over an algebra $A$ in a monoidal 2-category $\cC$.
\end{definition}

\begin{definition}\label{def:bimodule-2cat}
    Let $A$ and $B$ be algebras in a monoidal 2-category $\cC$. An $(A,B)$-bimodule is an object $M \in \cC$ together with:
    \begin{itemize}
        \item A left $A$-module structure $(\triangleright, \alpha^L, \lambda^L)$;

        \item A right $B$-module structure $(\triangleleft, \alpha^R, \rho^R)$;

        \item An additional associator $\alpha \colon \triangleleft \circ (\triangleright \otimes 1_B) \Rightarrow \triangleright \circ (1_A \otimes \triangleleft)$ between the two composites $A \otimes M \otimes B \to M$
    \end{itemize}
    satisfying the pentagon and triangle conditions of \cite[4.2.3]{Xu:2026thesis}.
\end{definition}

\begin{definition}\label{def:local-module-2cat}
    A local module over a braided algebra $A$ in a braided monoidal 2-category $\cB$ is a right $A$-module $(M,\triangleleft,\alpha,\rho)$ together with a holonomy $h : \triangleleft \circ \beta_{A,M} \circ \beta_{M,A} \to \triangleleft$ satisfying the coherence conditions of \cite[4.7.2]{Xu:2026thesis}.
\end{definition}

\newl Higher morphisms between algebras and modules can be defined accordingly \cite[§4.1, §4.2, §4.7]{Xu:2026thesis}. Together they form various 2-categories, which we denote as follows:
\begin{itemize}
    \item $\mathsf{Alg}(\cC)$ is the 2-category of algebras in $\cC$;

    \item $\mathsf{Alg}_{br}(\cB)$ is the 2-category of braided algebras in $\cB$;

    \item $\mathsf{LMod}_\cC(A)$ or ${}_A \cC$ is the 2-category of left $A$-modules in $\cC$;

    \item $\mathsf{RMod}_\cC(A)$ or $\cC_A$ is the 2-category of right $A$-modules in $\cC$;

    \item $\mathsf{Bimod}_\cC(A,B)$ or ${}_A \cC_B$ is the 2-category of $(A,B)$-bimodules in $\cC$;

    \item $\mathsf{Mod}^{loc}_\cB(A)$ or $\cB_A^{loc}$ is the 2-category of local $A$-modules in $\cB$.
\end{itemize}

\newl By definition~\ref{def:bimodule-2cat}, an $(A,B)$-bimodule is an object $X\in\cC$ with a left $A$-action and a right $B$-action together with additional associativity coherence data, and that
\begin{equation}
{}_{A} (\cC_B)
\simeq
{}_{A} \cC_B
\simeq
({}_{A} \cC)_B\, .
\end{equation}
This equivalence is saying ``bimodules are right modules inside the 2-category of left modules'': after the left $A$-action has been fixed, adding a compatible right $B$-action is the same as forming a right $B$-module object in ${}_{A} \cC$.

\newl Unless stated otherwise, all algebras appearing in this paper are assumed to be separable and all modules are assumed to be finite semisimple, so that the 2-categories of modules, bimodules and local modules above, over fixed algebras, are again finite semisimple \cite{Decoppet}.

\subsection{Condensation, Full Centers and Lagrangian Algebras}\label{app:condensation}

Following the general idea of anyon condensation \cite{Kong:2013aya}, one can generalize it  to the 2-categorical setting using the 2-categorical notions we have just introduced.

\begin{definition}\label{def:lagrangian-algebra}
    A Lagrangian algebra in a non-degenerate braided fusion 2-category $\cB$ is a braided algebra $A \in \cB$ such that the 2-category of local modules $\mathsf{Mod}^{loc}_\cB(A)$ is equivalent to $\TwoVec$.
\end{definition}

\newl Physically, a braided algebra $A$ is Lagrangian if condensing it leaves no residual topological order, i.e., if its 2-category of local modules is trivial. 
The defects in the bulk are then recovered from the boundary as a Drinfeld center, $\cB\simeq\cZ_1(\cB_A)$ \cite{Decoppet:2023hkr}. 
This is the 2-categorical counterpart of the corresponding statement for Lagrangian algebras in non-degenerate braided fusion categories \cite{Davydov:2010wpa}, and the precise sense in which a Lagrangian algebra defines a topological boundary condition of the SymTFT.

\newl Suppose $\cC$ is a fusion 2-category. Its Drinfeld center $\cZ_1(\cC)$ is a braided fusion 2-category by appendix~\ref{app:drinfeld-center}. 
Then there is a canonical mechanism to construct Lagrangian algebras in $\cZ_1(\cC)$, generalizing the full center of \cite{Kong:2009inh,davydov2010centre}; the $2$-categorical centers used here are developed in \cite[Section 6.2]{Xu:2026center}.

\begin{definition}\label{def:central-commuting}
    Let $A \in \cC$ be an algebra in a fusion 2-category $\cC$. A commuting half-braiding of $A$ consists of a half-braiding $(X,\beta^X,R^X) \in \cZ_1(\cC)$ together with:
    \begin{itemize}
        \item A 1-morphism $f : X \to A$ in $\cC$;
        
        \item An invertible 2-morphism $\psi^f : m \circ (f \otimes 1_A) \Rightarrow m \circ (1_A \otimes f) \circ \beta^X_A$ in $\cC$,
    \end{itemize}  satisfying the coherence conditions (59) and (60) of \cite{Xu:2026center}.

    \newl For two commuting half-braidings $(X,\beta^X,R^X,f,\psi^f)$ and $(Y,\beta^Y,R^Y,g,\psi^g)$ over $A$, a 1-morphism between them consists of a 1-morphism of half-braidings \[(h,\beta^h) : (X,\beta^X,R^X) \to (Y,\beta^Y,R^Y)\] together with an invertible 2-morphism $\psi^h : f \Rightarrow g \circ h$ in $\cC$ satisfying the coherence condition (61) of \cite{Xu:2026center}.

    \newl For two 1-morphisms $(h,\beta^h,\psi^h)$ and $(k,\beta^k,\psi^k)$ between the same commuting half-braidings of $A$, a 2-morphism between them consists of a 2-morphism of half-braidings $\xi : (h,\beta^h) \Rightarrow (k,\beta^k)$ satisfying the coherence condition (62) of \cite{Xu:2026center}.
\end{definition}

\begin{definition}\label{def:full-center}
    For any algebra $A \in \cC$, its full center $Z(A) \in \cZ_1(\cC)$ is defined as the terminal object in the 2-category of commuting half-braidings of $A$.
\end{definition}

\newl Combining \cite[Theorem 6.12]{Xu:2026center} with \cite[Proposition 4.9.29]{Xu:2026thesis} (see also \cite[Proposition 6.3]{Xu:2024etale} for bosonic $\cC$) and the Morita invariance of the Drinfeld center (corollary~\ref{cor:morita-implies-center}), we have the following theorem.

\begin{theorem}\label{thm:full-center-lagrangian}
    For any connected separable algebra $A \in \cC$, its full center $Z(A) \in \cZ_1(\cC)$ is a Lagrangian algebra in $\cZ_1(\cC)$.
\end{theorem}

\newl In the main text we write $\cL(A):=Z(A)$ for the full center of the algebra $A\in2\Vect_G^\pi$ defining the boundary $\fB_{(H,A)}$. It is described explicitly in \eqref{eq:LA-general} of section~\ref{subsec:lagrangian-algebras}, and computed for the running example in examples~\ref{ex:S3-2lag-algebras} and~\ref{ex:S3-2lag-LB} and for the symmetry-extension example in section~\ref{subsec:ex-Q8-revisited}. The 1-categorical counterpart is the full center $\cL_{(H,\psi)}$ of $\bbC^\psi[H]\in\Vec_G$ in \eqref{eq:Lagrangian-induced} \cite{Kong:2009inh,davydov2010centre}.

\subsection{Morita Equivalences}\label{app:morita}

\begin{definition}\label{def:module-2category}
    A left module 2-category over a monoidal 2-category $\cC$ consists of a 2-category $\cM$ together with:
    \begin{itemize}
        \item A 2-functor $\triangleright : \cC \boxtimes \cM \to \cM$;

        \item An associator $\alpha_{X,Y,M} : (X \otimes Y) \triangleright M \to X \triangleright (Y \triangleright M)$ natural in $X,Y \in \cC$ and $M \in \cM$;

        \item A left unitor $\lambda_M : \mathbf{1} \triangleright M \to M$ natural in $M \in \cM$;

        \item An invertible 2-associator $\Pi_{X,Y,Z,M} : (1_X \triangleright \alpha_{Y,Z,M}) \circ \alpha_{X,Y\otimes Z,M}\circ (a_{X,Y,Z} \triangleright 1_M) \Rightarrow \alpha_{X,Y,Z\triangleright M} \circ \alpha_{X \otimes Y,Z,M}$ natural in $X,Y,Z \in \cC$ and $M \in \cM$, where $a$ is the associator of $\cC$;

        \item Invertible 2-unitors $\Lambda_{X,M} : \lambda_{X \triangleright M} \circ \alpha_{\mathbf{1},X,M} \Rightarrow l_X \triangleright 1_M$ and $E_{X,M} : (1_X \triangleright \lambda_M) \circ \alpha_{X,\mathbf{1},M} \Rightarrow r_X \triangleright 1_M$, natural in $X \in \cC$ and $M \in \cM$, where $l_X : \mathbf{1}\otimes X \to X$ and $r_X : X \otimes \mathbf{1} \to X$ are the unitors of $\cC$;
    \end{itemize}
    satisfying the higher associativity and unitality conditions of \cite[Definition 2.1.3]{Decoppet:2021ssm}.
\end{definition}

\newl Similarly, one can define a right module 2-category over a monoidal 2-category $\cC$. Bimodule 2-categories need additional higher data balancing the left and right actions, which we omit here.

\newl We will also need the relative version of module objects, a slight generalization of definition~\ref{def:module-2cat}.
If $A$ is an algebra in $\fZ$ and $\cC$ is a left $\fZ$-module 2-category (definition~\ref{def:module-2category} with $\fZ$ in place of $\cC$) with action
\begin{equation}
    \triangleright:\fZ\boxtimes\cC\to\cC,
\end{equation}
then a left $A$-module inside $\cC$ is an object $X \in \cC$ together with an action
\begin{equation}\label{eqn:relative_left_module_action}
    l^X \colon A\triangleright X \longrightarrow X
\end{equation}
satisfying the same associativity and unit coherence data as in definition~\ref{def:module-2cat}.

\newl Similarly, if $\cD$ is a right $\fZ$-module 2-category with action
\begin{equation}
    \triangleleft:\cD\boxtimes\fZ\to\cD,
\end{equation}
then a right $A$-module inside $\cD$ is an object $Y \in\cD$ with an action
\begin{equation}\label{eqn:relative_right_module_action}
    r^Y \colon Y \triangleleft A \longrightarrow Y
\end{equation}
satisfying the same associativity and unit coherence data as in definition~\ref{def:module-2cat}.

\newl 2-functors, 2-natural transformations, and modifications between module 2-categories can be defined accordingly \cite[Definitions 2.1.7, 2.1.9, 2.1.10]{Decoppet:2021ssm}. 
Together they form various 3-categories, which we denote as follows:
\begin{itemize}
    \item $\mathsf{LMod}(\cC)$ is the 3-category of left module 2-categories over $\cC$;

    \item $\mathsf{RMod}(\cC)$ is the 3-category of right module 2-categories over $\cC$;

    \item $\mathsf{Bimod}(\cC,\cD)$ is the 3-category of bimodule 2-categories between $\cC$ and $\cD$.
\end{itemize}

\begin{definition}\label{def:morita-equivalence}
    Two monoidal 2-categories $\cC$ and $\cD$ are Morita equivalent if their 3-categories of left module 2-categories $\mathsf{LMod}(\cC)$ and $\mathsf{LMod}(\cD)$ are equivalent.
\end{definition}

\newl Imposing the appropriate linearity, finiteness and semisimplicity conditions, we can define finite semisimple left, right, and bimodule 2-categories over fusion 2-categories. 
D\'ecoppet \cite{Decoppet:2021ssm, Decoppet, Decoppet:2022dnz-Morita, Decoppet:2022kcf} has generalized the following important results of Etingof--Nikshych--Ostrik \cite{Etingof:2002vpd, Ostrik2003} to the 2-categorical setting.

\begin{theorem}[{\cite[Theorem 5.3.4]{Decoppet:2021ssm}}]\label{thm:module-2cat-is-rmod}
    Any finite semisimple left module 2-category over a fusion 2-category $\cC$ is equivalent to the 2-category of right modules over a rigid algebra $A \in \cC$.
\end{theorem}

\begin{theorem}[{\cite[Theorem 5.4.3]{Decoppet:2022dnz-Morita}}]\label{thm:morita-criterion}
    Two fusion 2-categories $\cC$ and $\cD$ are Morita equivalent if and only if there exists a faithful separable algebra $A \in \cC$ such that $\cD \simeq \mathsf{Bimod}_\cC(A)$.
\end{theorem}

\begin{corollary}[{\cite[Theorem 2.3.2]{Decoppet:2022kcf}}]\label{cor:morita-implies-center}
    If two fusion 2-categories $\cC$ and $\cD$ are Morita equivalent, then their Drinfeld centers $\cZ_1(\cC)$ and $\cZ_1(\cD)$ are equivalent as braided fusion 2-categories.
\end{corollary}

\begin{warning}\label{warn:center-does-not-determine-morita}
    Etingof--Nikshych--Ostrik \cite[Theorem 3.1]{Etingof:2011weakly} have shown that the converse of corollary~\ref{cor:morita-implies-center} is true for fusion 1-categories. 
    However, it does not hold for fusion 2-categories.

    \newl The simplest counterexample is given by the fusion 2-categories $\mathsf{RMod}(\cM)$ for a modular tensor category $\cM$. 
    D\'ecoppet \cite[Theorem 3.1.4]{Decoppet:2022kcf} has shown that $\mathsf{RMod}(\cM)$ is Morita equivalent to $\TwoVec$ if and only if $\cM$ is Witt trivial, i.e.\ braided equivalent to the Drinfeld center of some fusion 1-category \cite{Davydov:2010wpa}. 
    There are many modular tensor categories that do not satisfy this condition, for example, from positive energy representations of affine Lie algebras.

    \newl On the other hand, D\'ecoppet \cite[Lemma 4.1.3]{Decoppet:2022kcf} has shown that $\cZ_1(\mathsf{RMod}(\cM)) \simeq \TwoVec$ whenever $\cM$ is a modular tensor category.
\end{warning}

\subsection{Bulk-Boundary Correspondence}\label{app:bulk-boundary}

One of the most important theorems in the theory of fusion 2-categories, combining the results of Deligne \cite{Deligne2002}, Johnson-Freyd--Yu \cite{Johnson-Freyd:2020ivj} and D\'ecoppet \cite{Decoppet:2022kcf}, is the following.

\begin{definition}\label{def:bosonic}
    A fusion 2-category $\cC$ is called bosonic if its transparent lines are Tannakian, i.e., if the symmetric fusion 1-category $\cZ_2(\Omega\cC)$ admits a (symmetric) fiber functor to $\Vect$ \cite[Definition 4.1.4]{Decoppet:2022kcf}.
\end{definition}

\begin{remark}\label{rmk:AB-EF-dictionary}
    The main body of the paper follows the physics convention and calls such a $\cC$ of \emph{all-boson (AB) type}; the two terms are used interchangeably. Likewise, the \emph{emergent-fermion (EF) type} fusion 2-categories of section~\ref{subsec:bosonic-fermionic}, whose transparent lines are super-Tannakian, are what we call \emph{fermionic} here.
\end{remark}

\begin{theorem}[{\cite[Theorem 4.1.6]{Decoppet:2022kcf}}]\label{thm:AB-center}
    For any bosonic fusion 2-category $\cC$, there exists a finite group $G$, a group 4-cocycle $\pi \in \operatorname{Z}^4(G;U(1))$ and a braided equivalence $\cZ_1(\cC) \simeq \cZ_1(\TwoVec^\pi_G)$.
\end{theorem}

\newl This structural theorem allows us to reconstruct bosonic fusion 2-categories using group-theoretical data; combined with the classification of connected and Lagrangian \'etale algebras in $\cZ_1(\TwoVec^\pi_G)$, it yields a classification of bosonic fusion 2-categories \cite{Xu:2024etale}. 
It is the input to the argument of section~\ref{subsec:single-vacuum}.

\begin{theorem}[{\cite[Proposition 4.9.29]{Xu:2026thesis}}]\label{thm:canonical-lagrangian}
    For any fusion 2-category $\cC$, there exists a canonical Lagrangian algebra $A \in \cZ_1(\cC)$, given by the full center of the monoidal unit, such that there exists a reconstruction $\cC \simeq \mathsf{RMod}_{\cZ_1(\cC)}(A)$.
\end{theorem}


\bibliographystyle{JHEP}
\bibliography{GenSym}


\end{document}